\documentclass[cernpreprint, atlasdraft=false, texlive=2023, UKenglish, texmf, orcidlogo]{atlasdoc}
 
\usepackage{atlaspackage}
\usepackage{atlasbiblatex}
\usepackage{atlasphysics}
 
\graphicspath{{logos/}{figures/}}
 
\usepackage{ANA-TRIG-2019-03-PAPER-defs}

\AtlasTitle{The ATLAS Inner Detector Trigger performance in pp collisions at 13\,TeV during LHC Run 2}

\AtlasAbstract{
The design and performance of the inner detector trigger for the high level trigger
of the ATLAS experiment at the Large Hadron Collider during the 2016--18 data taking period is discussed.
In 2016, 2017, and 2018 the ATLAS detector recorded $35.6~\mathrm{fb}^{-1}$,  $46.9~\mathrm{fb}^{-1}$,
and $60.6~\mathrm{fb}^{-1}$ respectively of proton--proton collision data at a centre-of-mass energy
of 13\,TeV. In order to deal with the very  high interaction multiplicities
per bunch crossing  expected
with the 13\,TeV collisions the inner detector trigger was redesigned during the
long shutdown of the Large Hadron Collider from 2013 until 2015.
An overview of these developments is provided and the
performance of the tracking in the trigger for the muon, electron, tau and $b$-jet signatures is discussed.
The high performance of the inner detector trigger with these
extreme interaction multiplicities  demonstrates how the inner detector tracking continues to lie at the
heart of the trigger performance and is essential in enabling
the ATLAS physics programme.
}

\author{The ATLAS Collaboration}
 
\AtlasRefCode{TRIG-2019-03}
 
\PreprintIdNumber{CERN-EP-2021-076}

\AtlasJournalRef{ Eur. Phys. J.  C 82 (2022) 206 }
\AtlasDOI{10.1140/epjc/s10052-021-09920-0}

\hypersetup{pdftitle={ATLAS document},pdfauthor={The ATLAS Collaboration}}
 
\newif\ifnotevar
\notevarfalse

\newif\ifstwocolumn
\stwocolumntrue
 
\newcommand{\figbreak}{\\}
 
\begin{document}
 
\maketitle
 
\tableofcontents

 
\section{Introduction}

The trigger system is an essential component of any collider experiment -- especially so for a hadron collider --
as it is responsible for deciding whether or not to keep an event from a given bunch-crossing interaction
for later study. During \runi (2009 to early 2013) of the Large Hadron Collider (LHC)~\cite{Evans:2008zzb},
the ATLAS detector~\cite{PERF-2007-01}
trigger system~\cite{PERF-2011-02,TRIG-2016-01,TRIG-2018-05,TRIG-2012-03,ATLAS-CONF-2017-061,PERF-2013-06,TRIG-2012-01,TRIG-2018-08}  
operated efficiently at instantaneous
luminosities of up to $8\times\lumi{e33}$ and primarily at centre-of-mass energies, $\sqrt{s}$, of
\SI{7}{\TeV} and \SI{8}{\TeV}.
In \runii (2015--18) the increased centre-of-mass energy of \SI{13}{\TeV},
higher luminosity and increased number of proton--proton interactions per
bunch-crossing (\pileup) required a significant upgrade to the trigger system. This was
necessary to avoid the required processing time exceeding that available
when running with the trigger thresholds that are needed to satisfy the
ATLAS physics programme. For this reason, the first long shutdown between LHC
\runi and \runii was used to improve the trigger system
with almost no component left untouched.

Final states involving
electrons,
muons,
taus,
and $b$-jets~\cite{PERF-2017-01,PERF-2014-05,PERF-2013-06,PERF-2014-06,FTAG-2018-01,PERF-2012-04},
provide important signatures for many precision measurements
of the Standard Model and searches for new physics. The ability of the ATLAS trigger to process
information from the inner detector (ID)~\cite{ATL-DAQ-PUB-2013-002} to reconstruct particle
trajectories is an essential requirement for the efficient triggering of these objects at manageable
rates. Typical rate rejection factors possible using tracking information from the ID trigger
range very approximately from 50 to 100 for the electron and \bjet triggers, and from 3 to 10 for the muon and
tau triggers. Without this rate rejection from the ID trigger
it would not be possible to achieve the goals of the ATLAS physics programme.
The ID trigger must therefore be able to reconstruct tracks with high efficiency across the
entire range of possible physics signatures, under all expected circumstances.
This challenge is exacerbated by the extremely high track and hit occupancies in the ID that
arise for the large \pileup multiplicity
present during LHC running at the highest operation intensities.
 
The performance of the ATLAS trigger system for the lower luminosity running in 2015
has been reported previously~\cite{TRIG-2016-01}, and included a brief description of the
ID trigger design.
In this paper the full details of the ID trigger design and implementation are presented, followed by a discussion
of the execution time of the
different components of the ID trigger and the full performance
in terms of efficiency and resolution from the 2016--2018 running period.

\section{The ATLAS detector, trigger and data acquisition systems}
 
\label{section:two}

\newcommand{\AtlasCoordFootnote}{
ATLAS uses a right-handed
coordinate system with the origin at the nominal interaction point
(IP) at the centre of the detector and the \(z\)-axis along the beam
direction; the \(x\)-axis pointing from the IP to the centre of the LHC
ring, and the \(y\)-axis pointing upwards. Cylindrical coordinates
\((r,\phi)\) are used in the transverse plane, \(\phi\) being the
azimuthal angle around the \(z\)-axis. The pseudorapidity is defined
in terms of the polar angle, \(\theta\), as \(\eta = -\ln
\tan(\theta/2)\). Angular distance is measured using
$\Delta R \equiv \sqrt{(\Delta\eta)^{2} + (\Delta\phi)^{2}}$.}
 
The ATLAS detector~\cite{PERF-2007-01} at the LHC is a multi-purpose
particle detector with a forward--backward symmetric cylindrical
geometry and a near 4$\pi$ coverage in solid angle
around the collision point.\footnote{\AtlasCoordFootnote}
It consists of an inner tracking detector
surrounded by a thin superconducting solenoid, providing
a \SI{2}{\tesla} axial magnetic field,
electromagnetic and hadronic calorimeters, and a muon spectrometer
incorporating large superconducting toroidal magnets. 
Owing to the cylindrical geometry, subdetector components generally
consist of a barrel region with central pseudorapidities close to zero,
and two endcap components, one each in the forward regions with
large absolute pseudorapidity.
 
\begin{figure}[thp]
\centering{\includegraphics[width=0.999\textwidth]{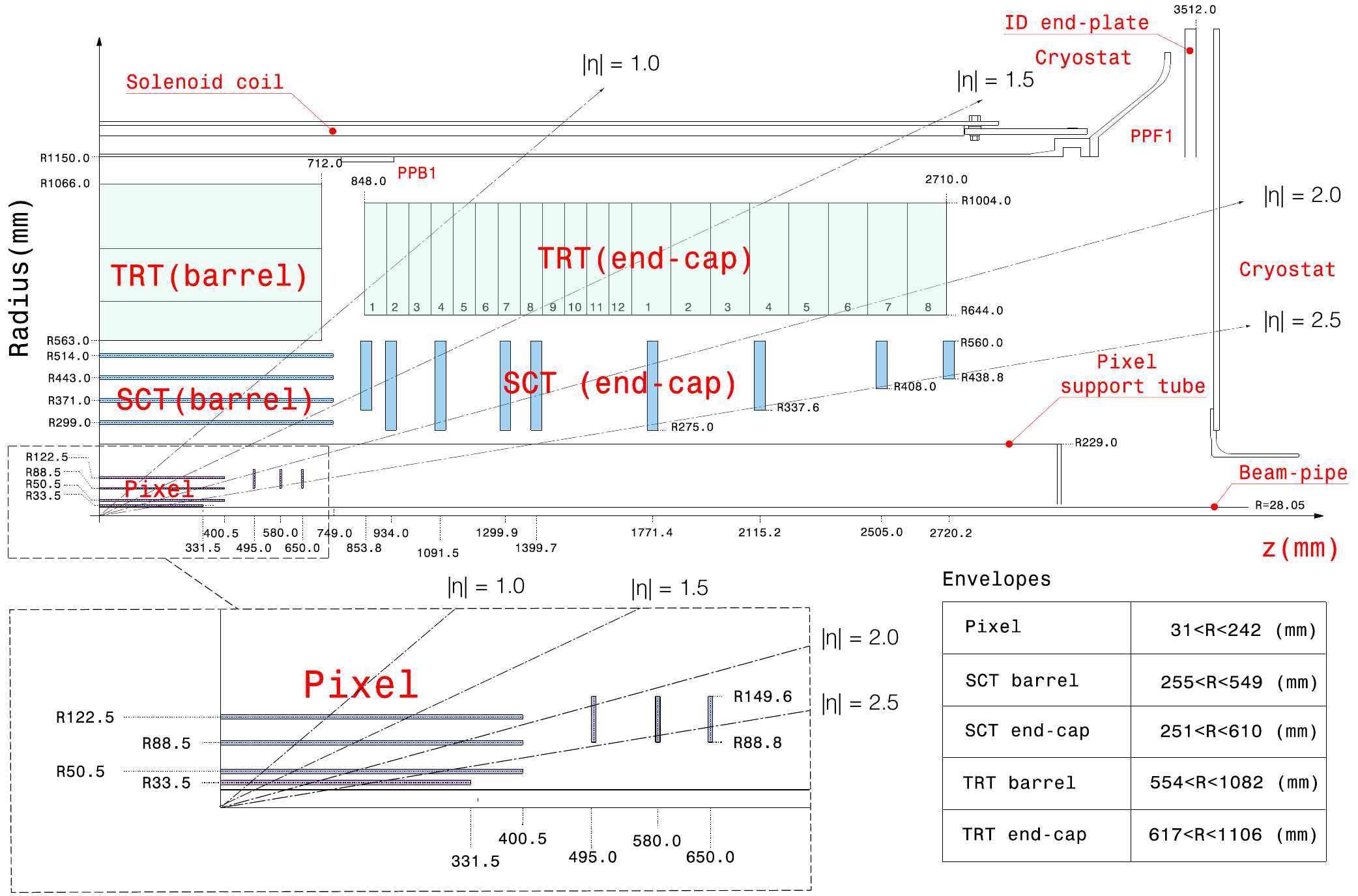}}
\caption{A cross section of one quadrant of the ATLAS inner detector, showing the pixel, SCT and TRT barrel and endcap sections. The pixel detector
includes the insertable b-layer.}
\label{fig:id}
\end{figure}
 
The inner detector, shown schematically in Figure~\ref{fig:id}, allows the precise reconstruction of
charged-particle trajectories in the pseudorapidity range
\(|\eta| < 2.5\). It consists of three subsystems: the high granularity
silicon pixel detector, the silicon microstrip tracker, and the
transition radiation tracker.
The pixel detector
consists of  three barrel layers, and three layers in each endcap, and an insertable
B-layer (IBL)~\cite{ATLAS-TDR-2010-19,PIX-2018-001}
closest to the interaction point, installed before \runii.
The IBL spans the entire
pseudorapidity range of the inner detector. The pixel
detector typically provides four measurements per track.
The pixel detector is surrounded
by the silicon microstrip tracker (SCT)~\cite{Aad_2008,AHMAD200798}, which consists of
double layer silicon strip detectors, four in the barrel,
and nine in each endcap.
These provide axial hits, or azimuthal hits in the endcaps, and
small-angle stereo hits used to provide improved spatial resolution in the
longitudinal direction along the modules. The SCT typically provides eight measurements
per track. The silicon detectors are complemented by
the transition radiation tracker (TRT)~\cite{Abat:2008zza,ATL-INDET-PUB-2014-001,IDET-2015-01}, consisting of multiple layers
of straw cylindrical drift tubes filled with either a Xe--CO$_2$--O$_2$, or
Ar--CO$_2$--O$_2$ mixture, and
which enables radially extended track reconstruction up to
\(|\eta| = 2.0\).

The calorimeter system covers
the pseudorapidity range \(|\eta| < 4.9\). Within the region
\(|\eta|<3.2\), electromagnetic calorimetry is provided by barrel and endcap
high granularity lead/liquid-argon (LAr) calorimeters, with an
additional thin LAr presampler covering \(|\eta| < 1.8\) to correct
for energy loss in material inside the calorimeter radius. Hadronic
calorimetry is provided by the steel/scintillator-tile calorimeter,
and two copper/LAr hadronic endcap calorimeters. The more
forward regions have copper/LAr and tungsten/LAr calorimeter
modules optimised for electromagnetic and hadronic measurements
respectively.
 
The muon spectrometer (MS)~\cite{ATLAS-TDR-10} is composed of
separate trigger
and high-precision tracking chambers measuring the deflection of muons
in a magnetic field generated by superconducting air-core toroids.
The field integral of the toroids ranges between \SI{2.0}{T\hspace{0.1mm}m} and
\SI{6.0}{T\hspace{0.1mm}m} across most of the detector.
Precision chambers cover the region \(|\eta| < 2.7\) and are
complemented by cathode-strip chambers in the
forward region.
The muon trigger
system~\cite{PERF-2011-02,TRIG-2016-01,TRIG-2018-01} covers the range
\(|\eta| < 2.4\) with resistive-plate chambers
in the barrel, and thin-gap chambers in the endcap
regions.
 
Interesting events are selected to be recorded by the
first level trigger system (Level~1, L1)~\cite{ATLAS-TDR-12,PERF-2011-02}
implemented in custom hardware,
followed by selections made by algorithms implemented in software in the
high level trigger (HLT)~\cite{PERF-2011-02,TRIG-2016-01}. The L1 trigger
processes  events at the \SI{40}{\MHz} bunch crossing rate. Events are accepted at a rate
below \SI{100}{\kHz},
which the HLT further reduces in order to record events
to permanent storage at approximately \SI{1.2}{\kHz}~\cite{Aad:2020wji}.


\newcommand{\floatbarra}{}


The ATLAS trigger and data acquisition (TDAQ) system used during \runi~\cite{PERF-2011-02,DAQRun1Paper}
consisted of a multi-level trigger system with a hardware, pipelined Level 1
trigger and two large farms of commodity CPUs, one for the
Level 2 (L2) software trigger, and the second for the third level, event filter (EF)
processing. In \runi the separate L2 and EF farms collectively constituted the
HLT system.
For the L1 trigger operation, a reduced granularity subsample of the data from the
calorimeters and the muon spectrometer are
read out and processed in a custom \readout path and sent to the off-detector hardware
L1 processors. In parallel, the data from the entire detector are stored in
custom on-detector pipelined \readout buffers (ROBs)~\cite{ATLAS-TDR-16} ready to be read out
in the eventuality of a L1 trigger accept decision. Upon such a decision,
the pipelines are stopped and the  data for the corresponding bunch crossing
is read out from the ROBs to the off-detector \readout subsystem, ready
for distribution to the farm of the HLT processing nodes. Because of the extremely
high data volume in the SCT and pixel detectors, these can be read out
only following an L1 accept and therefore the first stage at which tracks could be reconstructed
in the silicon layers during \runi was at Level 2. The coarse granularity
muon spectrometer or calorimeter data
read out for the L1 trigger are used to reconstruct objects of interest,
such as potential track segments in the muon spectrometer from
muon candidates, or high energy clusters reconstructed in the calorimeter. These objects
are used to define regions of interest (\roi) in the detector which are worth
reconstructing with the data read out at the full granularity. Using \roi typically
reduces the amount of data to be transferred and processed in the HLT to between
2\%--6\% of the total data volume~\cite{TRIG-2012-03}.
 
For \runii, the LHC increased the centre-of-mass energy in $pp$ collisions
from \num{8} to \SI{13}{\TeV}, with a reduction of the nominal bunch spacing
from \num{50}\,\ns to \SI{25}\,\ns~\cite{ATLAS-CONF-2019-021} and an increase in
the beam intensity per bunch crossing. This increase lead to a significant increase in the
mean \pileup interaction multiplicity per bunch crossing, with correspondingly higher track and hit
multiplicities.
 
The increase in beam energies meant that the trigger rates were on average
a factor of 2 to 2.5 times larger for the same luminosity and with the same
trigger criteria~\cite{TRIG-2016-01}.
In order to prepare for the significantly higher expected rates in \runii,
several trigger upgrades were implemented during the 2013--15 long shutdown, which are
summarised here.

For \runii, the separate L2 and EF farms were combined into
a single, homogeneous HLT farm~\cite{HLTSteering:2008}.
During \runi, Level 2 requested only partial event data to be sent over the network, while the event filter operated on the full event information assembled by separate farm nodes
dedicated to event building~\cite{Ferrer:2007rp}.
For \runii, merging the separate farms
into a single homogeneous farm allowed for better resource sharing and an
overall simplification of both the hardware and software.
To achieve higher \readout and output rates, the \readout subsystem, data
collection network, and data storage system were upgraded~\cite{ATLAS-TDR-023}.
The on-detector
front-end electronics and detector-specific \readout systems were not
changed for \runii operation in any significant way.
 
During \runii, the typical mean processing time for the complete HLT processing per event
was between 0.4 and 0.8\,s for the higher \pileup at the start of a fill,
decreasing to around 200~ms for lower \pileup.


\floatbarra

 
\section{The inner detector trigger}
 
The HLT is the first level where information from the silicon detectors
in the ID is available in the trigger. The track reconstruction in the HLT is then
performed either
within an \roi identified at L1 or for the full detector.
Throughout \runi and \runii, the HLT ran  progressively more complex
algorithms to reduce the rate of events to be processed successively throughout the HLT.
Starting from the initial input rate determined by the L1 trigger accept rate, the fastest
algorithms were executed first to reduce that rate at which these more complicated algorithms
needed to run.
To achieve this, the reconstruction in the HLT generally consists of a two-stage approach
with a fast track reconstruction stage to reject the less interesting events, followed  by
a slower, precision reconstruction stage -- the precision tracking -- for those events
remaining after the fast reconstruction.
For both \runi and \runii,
each of the initial \rois used in the HLT is wedge shaped,
extending in $z$ along the full 450~mm interaction region at the \beamline and opening
out in $\eta$ and $\phi$. This large extent in $z$ is required since the $z$ position of
the interaction is not known before the tracking itself has been executed.

During the 2013--15 LHC long shutdown, many changes were implemented in the ID trigger
to better handle the changing run conditions expected in \runii. To
motivate these changes, Section~\ref{sec:runi} briefly summarises the tracking strategy
used during \runi and is followed in Section~\ref{sec:evo} by an explanation of the
developments implemented for \runii and why they were needed. The subsequent sections
describe in detail the structure and algorithms used in the ID trigger during \runii.

\subsection{ The inner detector trigger for \runi}
\label{sec:runi}

For \runi the ID tracking in the trigger~\cite{KONSTANTINIDIS2006166,ATLAS-CONF-2010-014-local,PERF-2011-02}
ran the fast reconstruction stage on the dedicated L2 CPU farm and the precision tracking
stage on the dedicated Event Filter CPU farm.
This division meant that the information passed between the different
trigger levels was essentially limited to the \roi within which the reconstruction should run,
necessitating that all stages in the track reconstruction, including any data preparation,
be performed independently in each of the two trigger levels.
 
At L2 the tracking consisted of several alternative custom pattern recognition algorithms
and tracking algorithms, which were all preceded by a fast data preparation stage
to read out any data from the pixel detector and SCT, and prepare the \spacepoints.
To optimise the tracking efficiency, the best-performing
pattern recognition algorithm was chosen independently for each signature.
Following the pattern recognition, track candidates were then identified using
a fast Kalman filter track fit~\cite{l2kalman} and the tracks could be extended into
the TRT.
 
In the event filter, the precision tracking stage (referred to as EFID during \runi) ran
offline tracking algorithms adapted for use in
the trigger. Starting with the offline data preparation, the tracking continued with the
offline pattern recognition, followed by the ambiguity solver~\cite{Cornelissen:2008zzc} which
resolves any ambiguity with respect to duplicated hits or hits falsely attributed to tracks,
ranking the tracks and rejecting those where the track quality is deemed insufficient.
A more detailed exposition of the ID trigger used during \runi can be found elsewhere~\cite{ATLAS-CONF-2010-014-local}.

\newcommand{\figtauroi}{
\begin{figure*}[t!]
\centering
\includegraphics[width=0.99\textwidth]{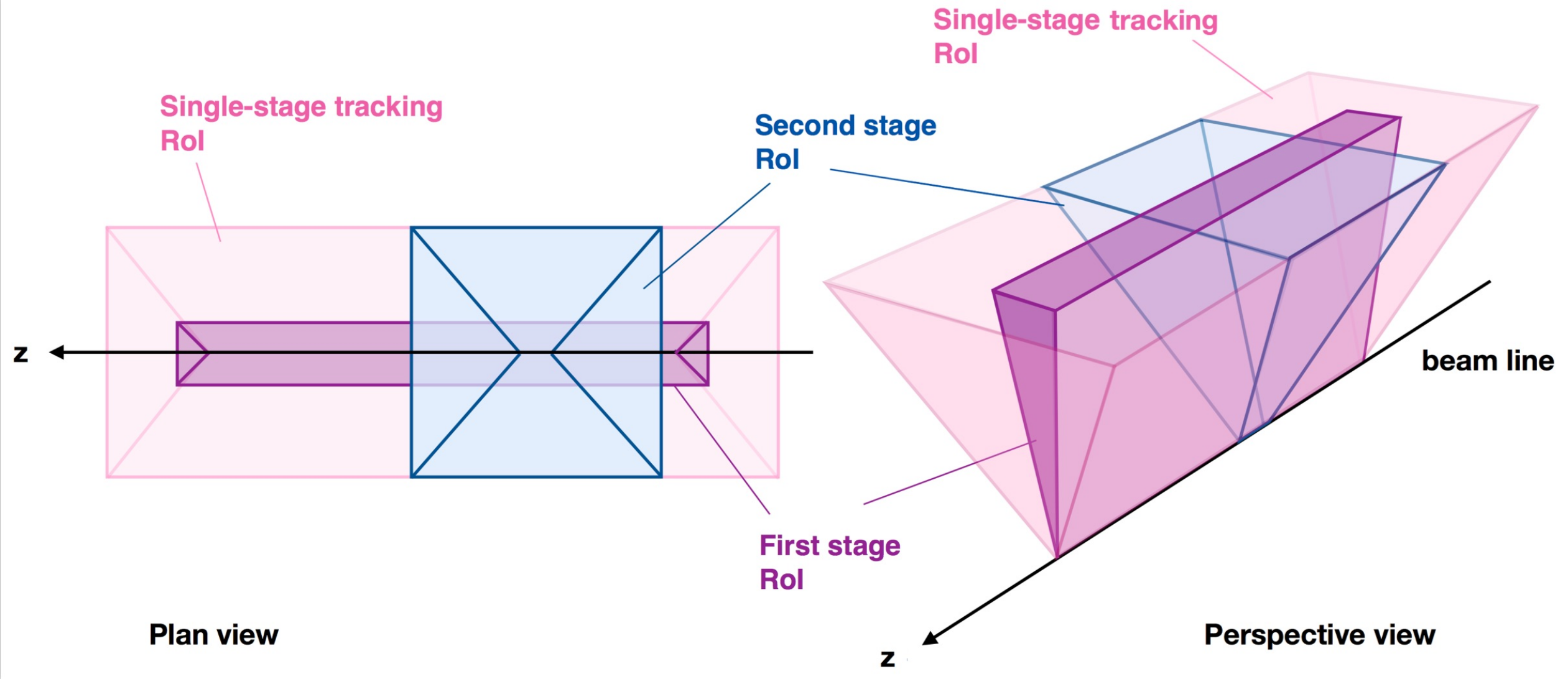}
\caption{ Schematic illustrating the \rois from the single-stage and
two-stage tau lepton trigger tracking, shown in plan view ($x$--$z$ plane) along
the transverse direction and in perspective view. The $z$-axis is along the
\beamline. The combined tracking volume of the $1^{\mathrm{st}}$ and
$2^{\mathrm{nd}}$ stage \roi in the two-stage tracking approach is
significantly smaller than the \roi in the single-stage tracking scheme.  }
\label{fig:tauroi}
\end{figure*}
}
 
\newcommand{\figsuperroi}{
\begin{figure*}[t!]
\centering
\includegraphics[width=0.99\textwidth]{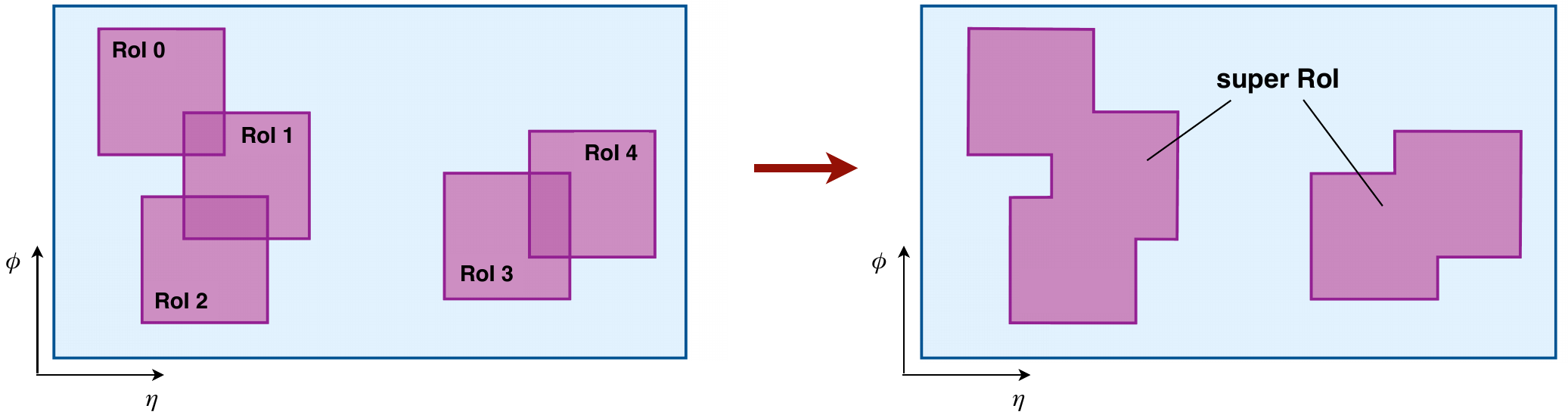}
\caption{ Schematic illustrating how the \superroi for the vertex tracking is created
from all the trigger jets reconstructed with  $\et > 30$\gev. The \roi around
each initial jet direction is fully extended in $z$ along the \beamline.}
\label{fig:superroi}
\end{figure*}
}
 
\newcommand{\figftfdesign}{
\begin{figure*}[thp]
\centering
\subfloat[]{\includegraphics[height=0.3\textwidth]{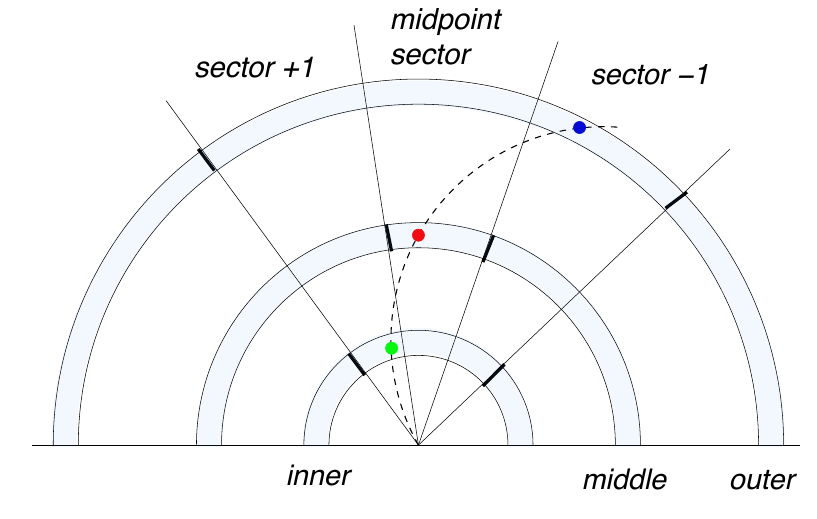}\hspace{3mm}}
\subfloat[]{\includegraphics[height=0.3\textwidth]{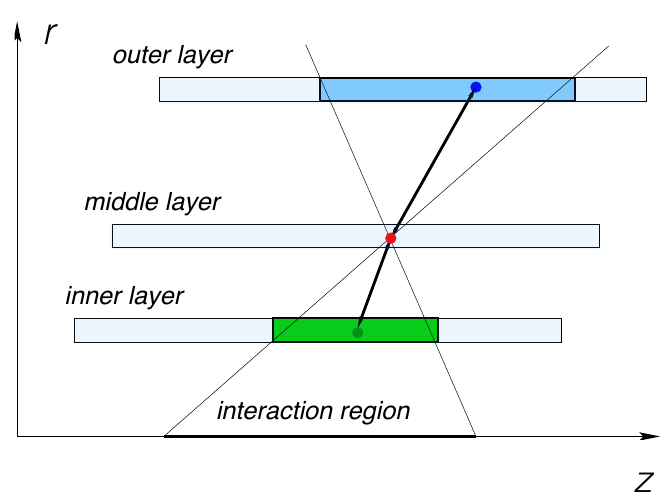}}
\caption{ Schematic illustrating (a) the track seed formation in radial bins and
azimuthal sectors, where the {\em midpoint sector} for a triplet is the sector containing the midpoint,
and where the inner and outermost
hits are allowed to originate in the same sector, or the the adjacent sectors,
and (b) track seed formation in the $r$--$z$ plane, where the extrapolated $z$ position of the triplet
at the beamline can be restricted to within a specific $z$ region.}
\label{fig:ftf_design}
\end{figure*}
}
 
\newcommand{\figsinglestage}{
\begin{figure}[t]
\centering{\includegraphics[width=0.4\textwidth]{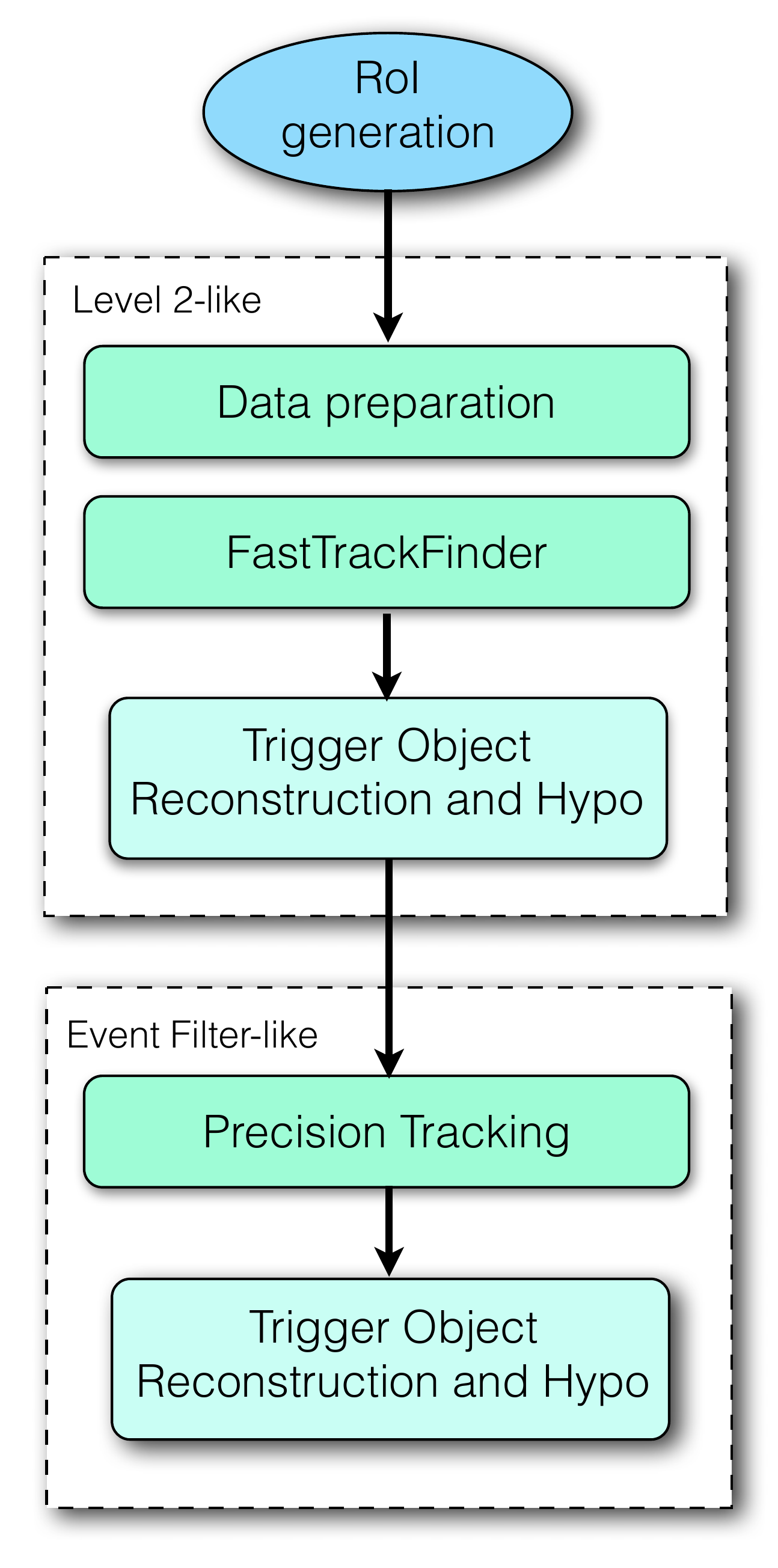}}
\caption{ Schematic illustrating the single-stage tracking for a single \roi. Here {\em single-stage} refers to the
single processing of a specific \roi, rather than the separate steps in the track processing itself. Selection can
be applied in hypothesis algorithms (hypos) which reduce the rate for processing the later steps. }
\label{fig:singlestage}
\end{figure}
}

\newcommand{\figmultistage}{
\begin{figure}[tp]
\centering{\includegraphics[width=0.4\textwidth]{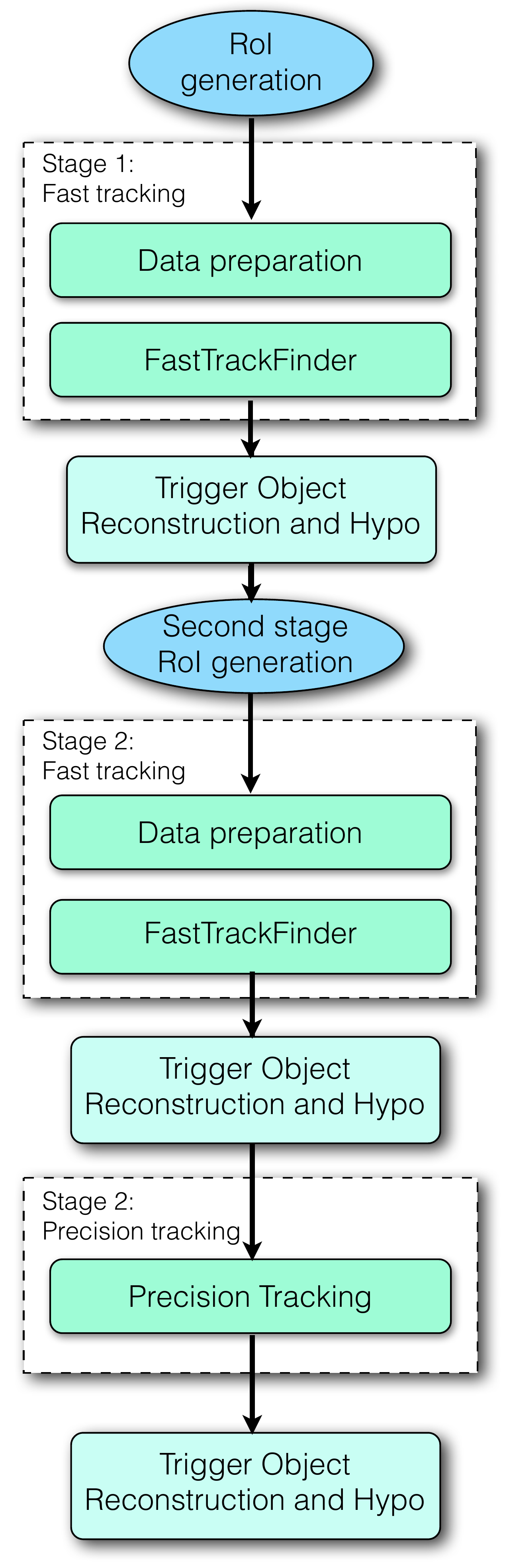}}
\caption{  Schematic illustrating the \multistage tracking. Here \multistage refers to the
multiple passes of the track processing over specific parts of the detector, perhaps with an updated, or
otherwise modified \roi. As in the single-stage case, selection can
be applied in hypothesis algorithms (hypos) between any of the tracking steps in the two stages. }
\label{fig:multistage}
\end{figure}
}

\newcommand{\figone}{
\begin{figure}[tp]
\centering
\includegraphics[width=0.5\textwidth]{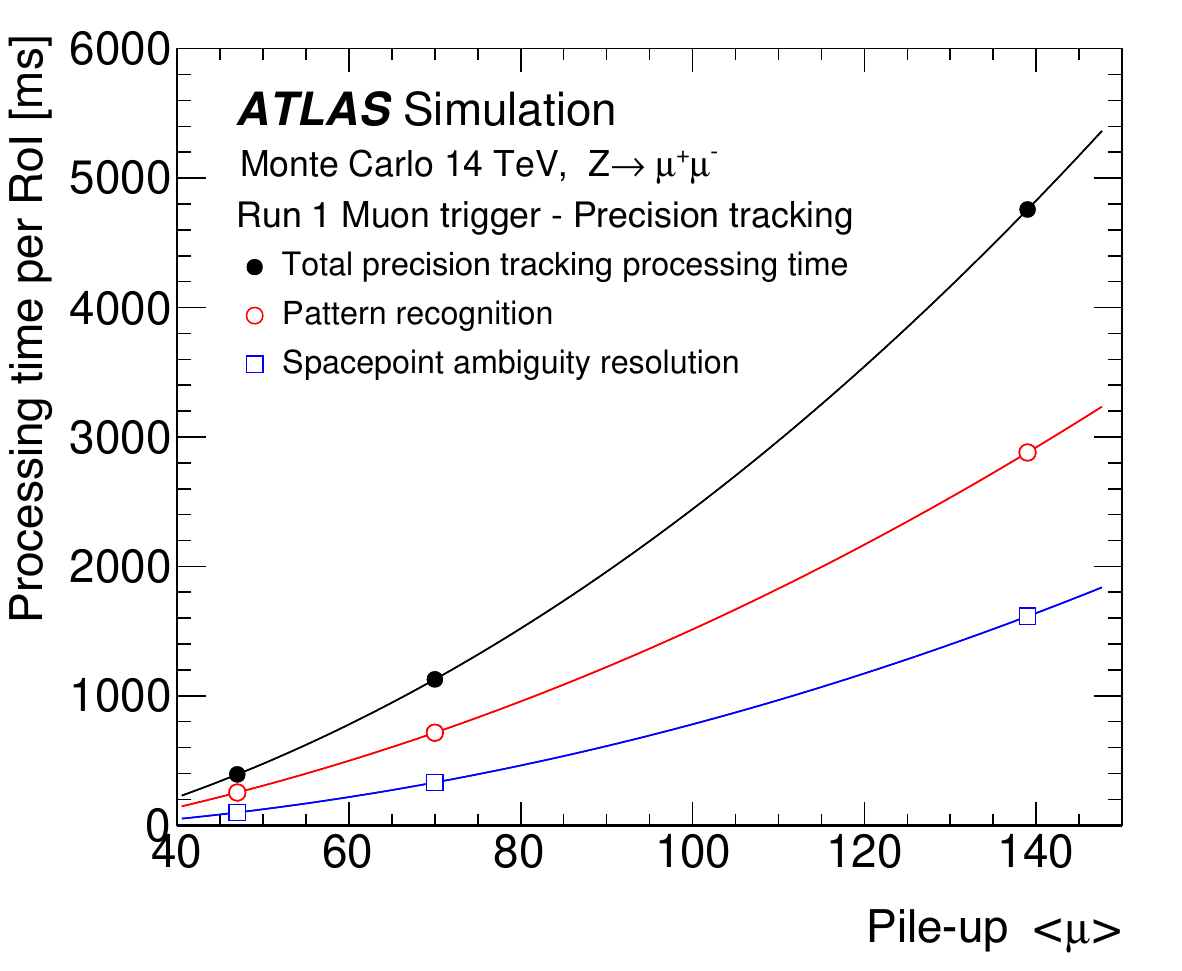}
\caption{
The total processing time for the \runi\ precision tracking timing
as a function of \pileup interaction multiplicity.}
\label{fig:runoneEF}
\end{figure}
}

\newcommand{\figtwo}{
\begin{figure}[tp]
\centering
\subfloat[]{\includegraphics[width=0.49\textwidth]{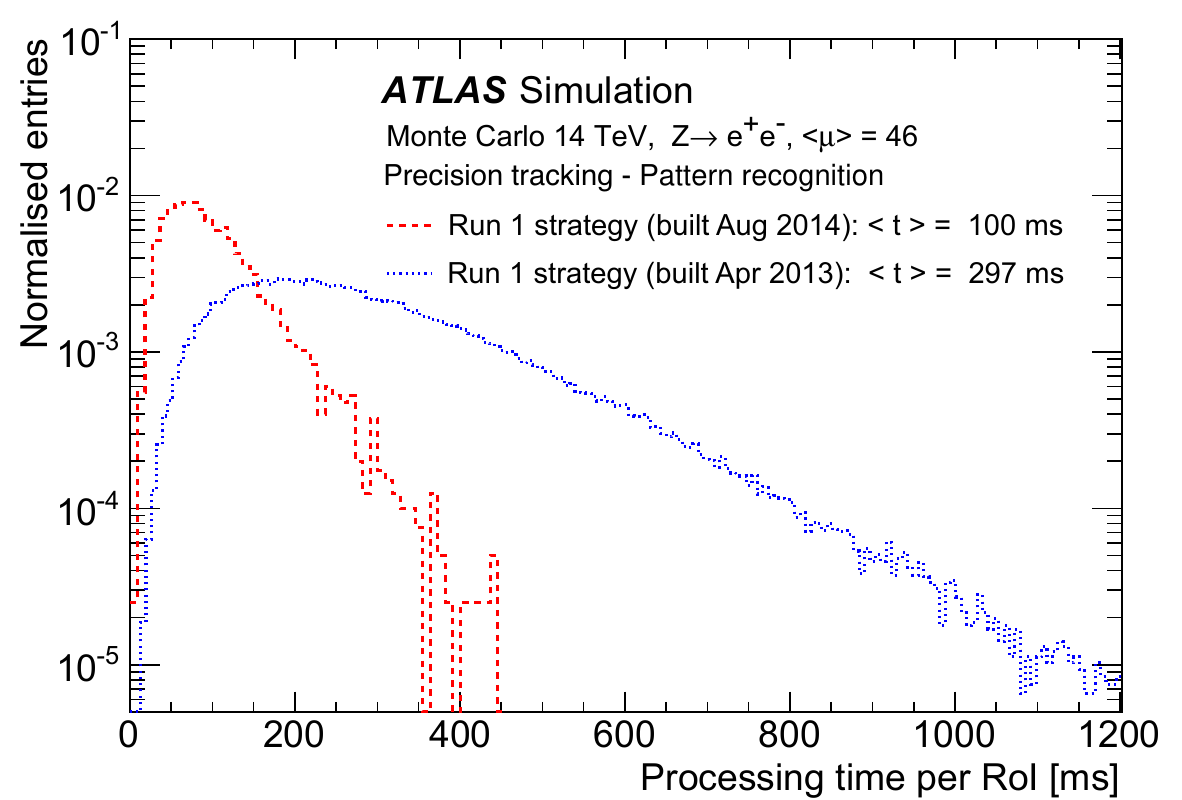}}
\subfloat[]{\includegraphics[width=0.49\textwidth]{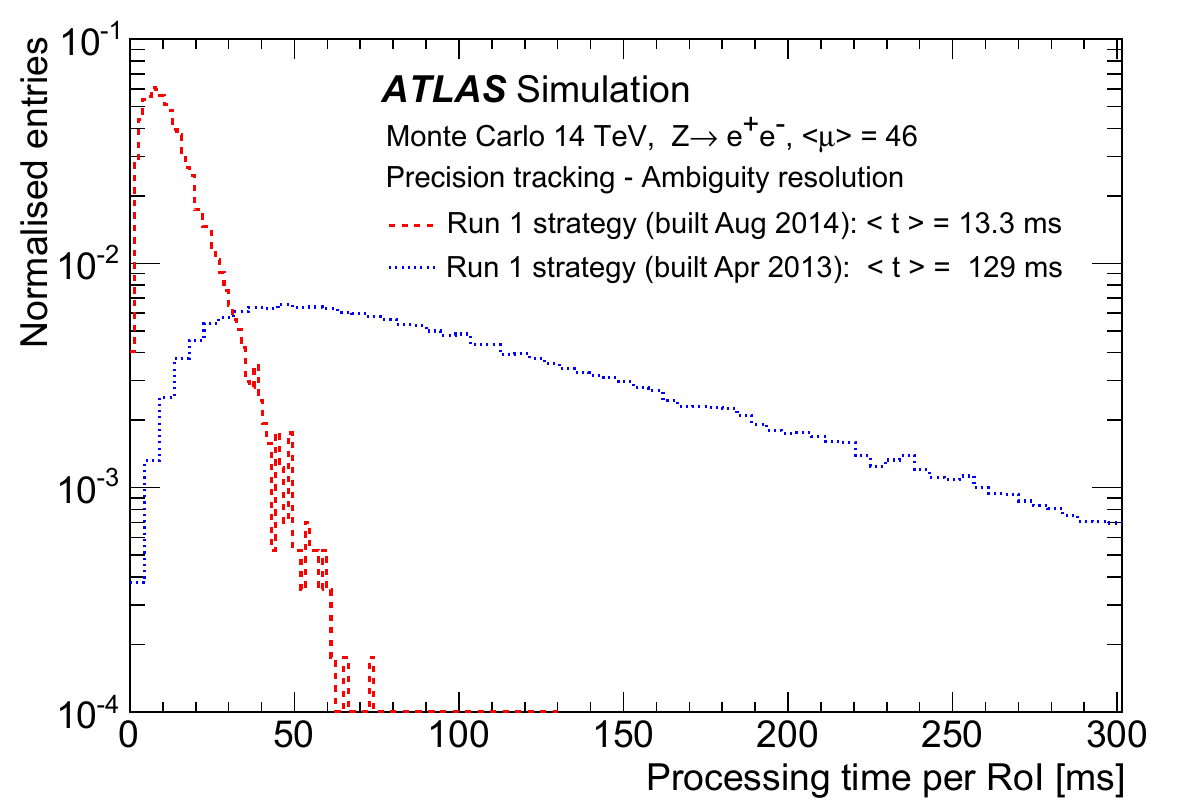}}
\caption{The processing time for the \runi\ precision tracking pattern recognition stage (a), and
the processing time of the ambiguity solver stage from the \runi\ precision tracking (b).
In both cases the processing time is shown  before and after the improvements
to the algorithms implemented during the 2013--15 long shutdown.}
\label{fig:runoneEFcomponents}
\end{figure}
}

\newcommand{\figthree}{
\begin{figure}[tp]
\centering
\subfloat[]{\includegraphics[width=0.49\textwidth]{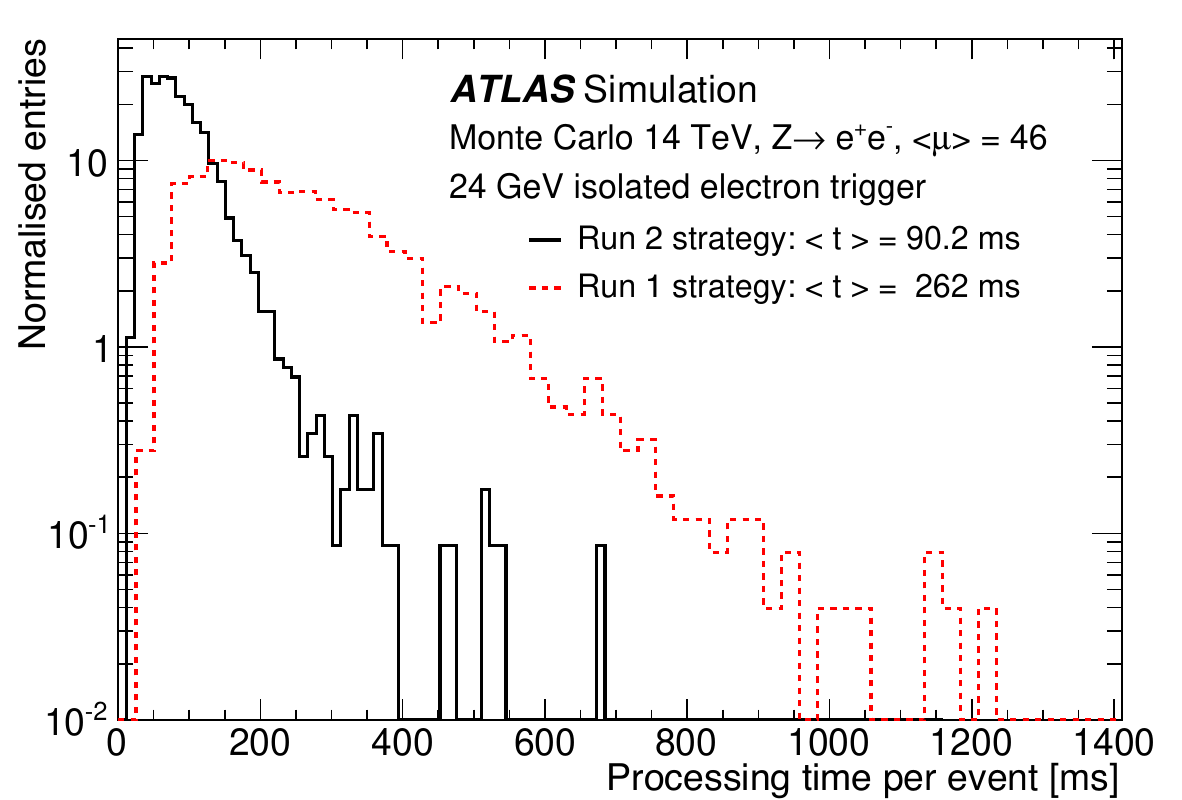}}
\subfloat[]{\includegraphics[width=0.49\textwidth]{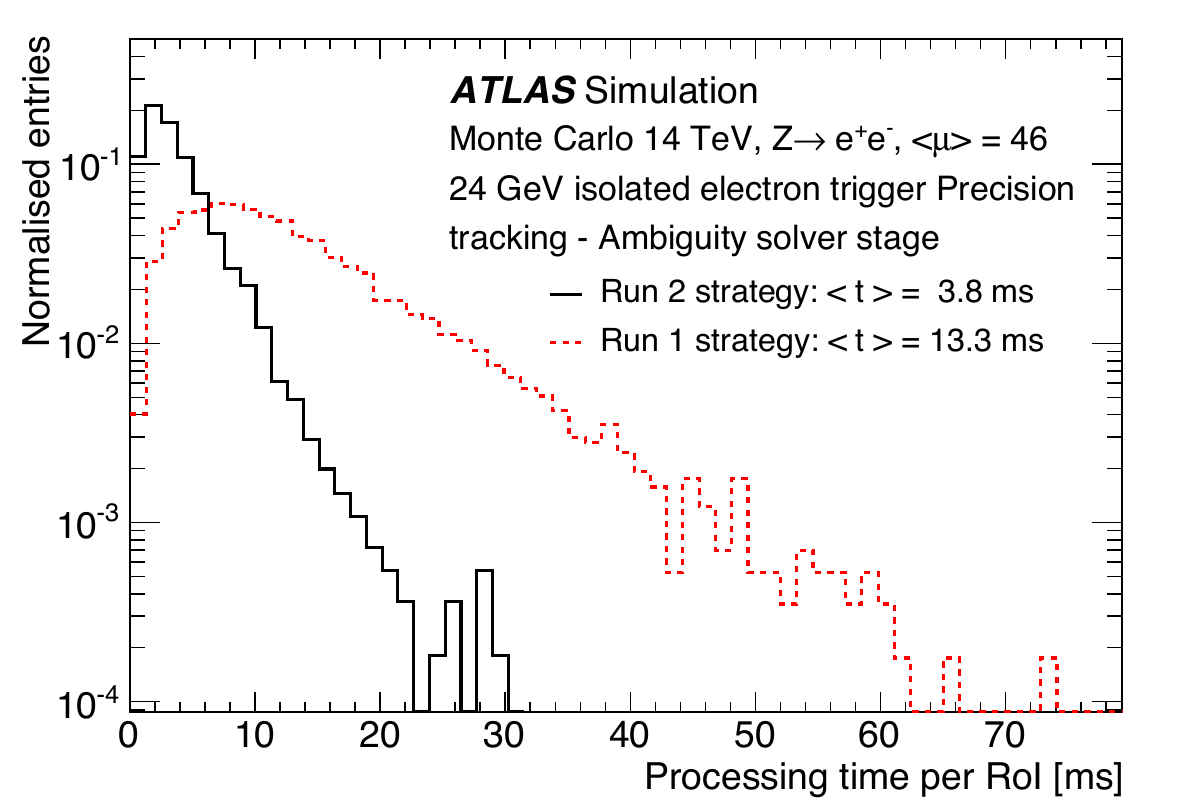}}
\caption{The reduction in the latency for the \runii\ ID trigger strategy with respect to the \runi\ strategy:
(a) the total time for the isolated 24\gev\ electron trigger with the \runi\ and \runii\ strategies;
(b) the time taken by the ambiguity solver in the \runi\ and \runii\ strategies. In all cases,
the improved offline code implemented late in 2014 has been used in the precision tracking stage. }
\label{fig:runtwoegamma}
\end{figure}
}
 
\subsection{Evolution of the inner detector trigger strategy from \runi to \runii}
\label{sec:evo}

The general purpose in running the event filter during \runi was to confirm the objects
reconstructed and selected by Level 2, and to improve on the precision of their reconstruction,
so allowing for the application of tighter selection to further reduce the rate.
For the \runi trigger, the events processed by the event filter were necessarily required to have
passed the Level 2 selection and consequently, running any algorithm in the event filter that was
more efficient than its Level 2 counterpart   was in some sense unnecessary,
as any event that had not passed the Level 2 stage would have already been discarded.
However, the longer processing times available for the offline algorithms executed in the \runi
event filter afforded by the lower processing rate meant that the EF tracking was indeed more
efficient than that in Level 2, but that this additional efficiency could not be exploited
as the events would have already been rejected by the Level 2 trigger.

As mentioned in Section~\ref{section:two}, for \runii, the previously separate
L2 and EF processing stages were merged to instead run as part of the same process on
individual
HLT nodes within the HLT farm.
This merging allows information from
algorithms that previously ran in the L2 reconstruction stage to more easily
be used later in the processing chain, thus removing any
need to run separate, duplicated data preparation and pattern recognition
stages for the tracking.

The performance of the offline tracking
algorithms as used in the precision tracking for the EF trigger in \runi  was studied in detail
under \runii conditions to determine which aspects of the offline tracking could
be retained for the ID trigger in \runii.
Figure~\ref{fig:runoneEF} shows the time taken by the different stages of the EF
tracking from the \runi\ ATLAS muon trigger as a function of the mean \pileup interaction
multiplicity, \mubar,
expected during \runii and beyond, using simulated $Z\to \mu^+\mu^-$ events with a 14\tev
centre-of-mass energy.\footnote{The sample was generated before the limitations on the LHC operating energy
were fully understood.}
Under these \runii conditions, the pattern recognition stage of the offline tracking can be seen
to constitute more than 60\% of the total time of the \runi\ precision tracking with a \pileup
of 70 interactions per bunch
crossing and does not scale linearly with the \pileup multiplicity.

In addition to the constraints on the trigger system in \runii, there are strong
constraints on the time taken to perform the offline reconstruction. For this
reason an extensive programme of software optimisation~\cite{Salzburger:2015sgq,Styles:2015mba}
was undertaken between
\runi and \runii. This provided tangible benefits for the ID trigger for those
algorithms shared with the offline reconstruction.
Improvements in the execution time resulted from improvements to the computing
infrastructure\emdash namely the switch to running on a 64-bit, rather than a 32-bit
kernel, and using a newer compiler: GCC 4.8 rather than 4.3\emdash and
the replacement of the CLHEP~\cite{CLHEP} linear algebra library, by the Eigen library~\cite{Eigen}.
 
\figone

These consequently led to a reduction in the latency of both the pattern recognition and
track fitting stages for offline tracking used in the \runi\ precision tracking. The effect on
the processing times following these improvements for their modified use in the trigger
can be seen in Figure~\ref{fig:runoneEFcomponents}. This shows a clear improvement in the \runi
pattern recognition strategy and ambiguity solver processing times by factors of approximately
three and ten respectively, when running on simulated $Z\to e^+e^-$ events.

\figtwo

Despite this significant improvement in processing time, the offline pattern recognition was still
too time consuming to be used in the trigger with the increase in the \pileup multiplicity
expected for \runii.
As a consequence, the chosen strategy for the ID trigger for \runii
was to reuse the output from the pattern recognition for the fast tracking by using the identified
tracks and hits to seed the ambiguity solver directly,
so completely removing the need to run the offline pattern recognition stage in the trigger.

The result of the improvement
can be seen in Figure~\ref{fig:runtwoegamma}
which shows, for the same simulated $Z\to e^+e^-$ events,  the total time required for
the full 24\gev isolated electron trigger including the calorimeter
reconstruction\emdash in this case comparing the \runii strategy described above,
with that from \runi, but with the upgraded offline tracking algorithms.

\figthree
 
Even with the complete reconstruction of the signature,
including the time to run the calorimeter and the \spacepoint reconstruction,
the \runii\ strategy is still approximately three times faster than the \runi\ strategy.
 
Also illustrated in
Figure~\ref{fig:runtwoegamma} is the execution time for the improved ambiguity solver code with both the
\runi and \runii strategies. The \runii\ strategy is seen to be approximately three times faster
than that used in \runi. This is
a consequence of the track selection and reduction having been already performed by the fast tracking
so that the ambiguity solver is not required to perform as much computation as it would if running
on the greater number of track candidates that would arise from the offline pattern recognition.
 
\floatbarra


\subsection{The inner detector trigger in \runii}

For \runii, the division of
the tracking into \emph{fast tracking} and \emph{precision tracking}
stages was retained: the fast tracking stage consisting of a newly implemented, trigger-specific
pattern recognition stage, based on a hybrid of those used at L2 during \runi, and
the precision tracking stage again relying heavily on offline tracking algorithms, but seeded with
information from the fast tracking stage~\cite{ATL-DAQ-PUB-2013-002,TRIG-2016-01}.
As in  \runi, the inner detector data preparation reconstructs clusters,
and \spacepoints from the silicon clusters, using the information from the pixels and the SCT strips~\cite{IDET-2013-01}. 
The clusters and spacepoints are used for the remainder of the track reconstruction. Since the fast and precision tracking steps
were both running as aspects of the same process on single CPU nodes, they can share
a single silicon data preparation step.
 
\figsinglestage

The generic \runii tracking strategy is shown schematically in Figure~\ref{fig:singlestage}.
This illustrates how the tracking is split into separate steps, which can be separated
by additional non-tracking-related algorithms to reconstruct additional features, and hypothesis algorithms
to select objects which satisfy certain selection criteria. These are used to reduce the rate of \rois
to be processed between the \FTF and \PTF steps.

\subsubsection{Data preparation}
 
The pixel and SCT data preparation consists of decoding hits in the silicon modules of the
inner detector from a byte stream format, grouping these hits into clusters~\cite{IDET-2013-01}\emdash
necessary because particles deposit charge in several adjacent silicon cells\emdash and forming
\spacepoints from the clusters. These represent points in three-dimensional
space which are used for the pattern recognition, the constituent clusters themselves
being used for the actual track fitting.
 
The data preparation employs the \roi mechanism, described later in this section, which  allows  the  trigger
to  request only the data from those silicon modules falling within an \roi, saving both
processing time, and data transfer bandwidth~\cite{boisvert2003new}.  Following the L1 accept the
data for each detector are read out to the ROB system, but to reduce the bandwidth only those ROBs
containing data corresponding to an \roi need be interrogated for read-out to the HLT.
 
The computational complexity of the decoding algorithms for both the pixel and SCT detector subsystems is
linear with respect to the number of data words, containing the encoded detector data from the subsystem.
The decoders iterate over ROB data fragments,
and over the data words within each fragment. While looping through an individual fragment, the algorithm maintains
information regarding which module it is decoding, assigning raw data objects (RDOs)\emdash in this case,
decoded hits\emdash to an
appropriate in-memory container.
Error information can also be encoded in the byte stream,
but this is a fringe case which is not discussed here.
Once the data words have been decoded
and RDOs created and mapped to their appropriate modules, the clustering algorithm takes over,
grouping adjacent
hits within a module into clusters, referred to as reconstruction input objects.
The clustering algorithms for the pixel and SCT detectors
are similar in principle, but differ significantly in complexity. Since it operates in only one dimension,
the SCT clustering algorithm complexity is linear with respect to the number of hits per module.
It consists of a single loop over the SCT strips, accumulating those which are adjacent and active
into clusters. The pixel clustering algorithm operates in two dimensions, and so with quadratic complexity,
consisting of a double loop over each module, generating clusters for each hit and merging clusters with shared
hits where necessary.
 
Finally, the pixel and SCT clusters are converted to \spacepoints
by means of a simple geometric transformation. The pixel clusters are rotated according to the module orientation
and then offset by the module position in the global ATLAS coordinate system. The cluster positions in the ATLAS coordinate
system are obtained using the module geometry determined by the offline alignment procedure~\cite{IDTR-2019-05-local}.
In the SCT case, one of the modules in each
back-to-back module pair is rotated at a 40~mrad stereo angle to provide longitudinal information along the module.
Clusters from
each side of the module are combined into pairs and two-dimensional local positions are derived from these
combinations, after which the rotation and translation procedure is identical to that for the pixels. During the decoding
the data are checked for errors to ensure increased robustness against detector failures, and the decoded data from each
module are cached to remove the need to decode again, should they
be requested multiple times.
 
The TRT data preparation is performed only for the precision tracking in the \rois for which the TRT track
extension is included. As in the pixel and SCT case, the data preparation consists of
the TRT byte stream decoding and creation of RDOs with the drift time information. The drift time measurements
are then converted to the drift distances which are stored in the output TRT reconstruction objects.
 
\subsubsection{Fast tracking}

For \runii,
the fast track finder~(FTF)~\cite{ATL-DAQ-PUB-2013-002} was developed
to provide track candidates for use early in the trigger.
These track candidates are then used to seed the precision tracking stage. Since the precision tracking
provides the track candidates used for the final objects selection in the trigger,
the FTF design prioritises track finding efficiency over purity. During the FTF pattern recognition, a
search for triplets of
\spacepoints ({\em track seeds}) is performed in bins of $r$ and approximately 50 sectors in $\phi$,
where triplets may also be formed using hits in adjacent
sectors at more positive, or more negative, $\phi$, as illustrated schematically in
Figure~\ref{fig:ftf_design}. The innermost hits from the initial triplets must originate in the pixel detector.
Triplet formation starts from a {\em middle} \spacepoint and
selects {\em outer} and {\em inner} \spacepoints at
larger and smaller radii, respectively. Also illustrated in Figure~\ref{fig:ftf_design}, the inner and outer
pairs of \spacepoints must be compatible with the nominal interaction region along the \beamline.
In the case of tracking performed within an \roi, this region can be replaced by a restricted $z$-region of the \roi
along the \beamline if some information about the $z$ position of the interaction is
available\emdash the \roi itself is used to pass external information about the expected
track $z$ interval to the track seeding algorithm.
The triplet track parameters $\phi_0$, transverse momentum \pt, and
transverse impact parameter at the point of closest approach
to the \beamline $d_0$, are estimated using a conformal transformation
\cite{Yepes1996582}, with the transformation centre placed at the middle \spacepoint and
applying cuts on $d_0$ and \pt. The track transverse momentum parameter from the
track reconstruction is signed by the track charge.
In practice, $z$ information for the refinement of the \roi  at the \beamline is only
available with high enough resolution after some version of the tracking itself has run.

\figftfdesign
 
Initial track candidates are then formed from the track seeds using a simple
track finding algorithm which extends the track candidates into further layers to find
additional hits, using the offline track-following algorithm~\cite{PERF-2015-08},
but with a modified
configuration for faster execution. The track-following algorithm
extends track candidates into adjacent layers to identify additional hits,
refitting the track candidates as new hits are added.
An algorithm to remove duplicate tracks which share track seeds
is applied, retaining those tracks of higher quality.
These preliminary tracks are then passed to a fast Kalman
filter track fitter~\cite{Fruhwirth:1987fm}.
To reduce the processing time, hits from the TRT are not used in the FTF.
 
During the initial track finding, track candidates that have too large an absolute value
of \dzero are rejected in order to keep the contribution from fake tracks
to a manageable level. For the muon signature, where the muon candidates are
used to seed the $B$-physics signature which reconstructs hadronic resonances
with a displaced secondary vertex, the maximum allowed $|\dzero|$ is 10~mm.
For all other signatures, a maximum of 4~mm is used.
 
For some signatures, discussed later, a two-stage tracking strategy is used, running
additional tracking stages in an updated \roi. In such a case, where the $z$ range of
the \roi along the \beamline is restricted,
the FTF triplet seed formation employs an additional selection where the individual
\spacepoint doublets used to construct the triplets are rejected if they do not
point back approximately to the restricted region in $z$.

\subsubsection{Precision tracking}
 
The precision tracking stage takes the FTF tracks as input, and applies a
version of the offline tracking algorithms~\cite{Cornelissen:2008zzc,ATLAS-CONF-2010-072}
configured to run online in the trigger~\cite{ATLAS-TDR-04,ATLAS-TDR-05}.
This includes a sophisticated algorithm for rejection of duplicate track candidates
and clusters previously assigned to tracks but which lie too far from the track
trajectory. In addition, the track candidates are extended into the TRT
in an attempt to select TRT hits at larger radii to improve the track
momentum resolution. The final ID track fit is then performed
using a more precise global $\chi^2$ fitter algorithm~\cite{Cornelissen:2008zza} from code used
for offline reconstruction.
These stages are combined as aspects of the offline ambiguity solver algorithm discussed in Section~\ref{sec:runi}.
 
Since the rate of event processing for the precision tracking is generally significantly
lower than for the fast tracking, more detailed handling of the detector conditions
is possible. This includes better compensation for detector effects such as inactive sensors
or calibration corrections. The resulting precision tracks are therefore much closer in
performance to the offline tracks than the fast tracks.
By necessity, the precision tracking must generally make use of the detector conditions
as they are known during the initial data taking for the run, and, unlike the offline tracking,
cannot benefit from the improved knowledge of those conditions obtained during the subsequent
offline processing.
One exception where information must be updated during the data taking concerns knowledge
of the \beamline position. As the subsequent offline beamspot information is not available,
a custom \beamline vertex trigger is used, which runs the fast tracking, and then uses these
tracks to run a fast vertex algorithm to establish the interaction region position in the transverse
plane as a function of the $z$ position itself. This is then used to update the \beamline
position used online whenever this position has moved by a predetermined amount~\cite{PERF-2011-02,Aad:2020wji}.
 
Since the precision tracking uses the tracks and clusters identified by
the fast tracking, the precision tracking efficiency, by construction, cannot
exceed that of the fast tracking.
Overall, the primary purpose of the precision tracking is to perform a higher quality fit to improve
the purity and quality of the trigger tracks relative to those tracks reconstructed offline.

\subsubsection{Vertex reconstruction}
\label{sec:vertexalg}
 
During \runii, two vertex algorithms were used online: a histogramming based algorithm, and
the offline iterative vertex algorithm~\cite{Ferrari:2007nr,PERF-2015-01}.
Typically, trigger signatures use only the offline vertex algorithm. For the case of the \bjet
trigger~\cite{ATL-COM-DAQ-2020-064} however, both vertex algorithms were employed to maximise
the vertex finding efficiency.
The simple histogramming algorithm
functions by first histogramming the $z_0$
position for the point of closest approach to the \beamline of each track. It then calculates
the vertex $z$ position using the mean of the bin centres weighted by the number
of tracks in each bin for the group of adjacent bins within the 1~mm sliding window which contain
the largest number of tracks. All tracks passing some basic quality selection are used and are
weighted equally.
The second algorithm, based on the offline vertex finder,
with some minor modifications to run online,
iteratively clusters tracks to determine the positions of vertex candidates, sorting the
resulting vertex candidates by the sum of the squared transverse momenta of the tracks assigned
to each vertex.
 
In both cases, the vertex algorithm runs only on tracks that have been reconstructed in the
relevant \roi of the track finding.
For the leptonic triggers the offline based algorithm is usually executed using the precision
tracks from the final stage of processing. For the \bjet trigger, both vertex algorithms are
executed using tracks from a specific additional vertex tracking stage before the dedicated tracking
in the \bjet \rois.

\subsection{Multistage tracking}
\label{sec:multistage}
 
Although the fast and precision tracking run in distinct stages or steps -- often separated
by additional non-tracking algorithms and event rejection -- running both algorithms sequentially in a
single \roi is in general considered to be processing in a {\em single tracking stage}.
This is because there is only a single pass through the full set of tracking algorithms
over any given \roi.
Where multiple passes of aspects of the tracking
are intentionally performed over similar, but not identical, regions of the detector, and where the
second pass is in a different \roi constructed to overlap with, extend, or update the \roi of the first pass,
this is referred to as
{\em \multistage} tracking. In this case, each set of steps within a specific \roi constitutes a
single {\em tracking stage}. Such a case might be where the first stage runs the fast tracking in a
narrow \roi, and
a subsequent stage again runs the fast tracking in a new \roi, along the same direction,
but wider. This is illustrated for a generic \multistage signature in Figure~\ref{fig:multistage}.
 
\figmultistage

For the hadronic tau trigger, it is useful to run the tracking in a larger \roi
than that used, for instance, in the electron trigger, to allow for the opening angle of the tracks from three-prong tau
decays~\cite{PERF-2013-06,PERF-2014-06}.
For \runi, this was done by using a single \roi, wider in \etaphi but fully extended along
the \beamline, which consequently was one of the most time-consuming aspects of the tracking in \runi.
To limit the tracking CPU usage in the  wider \roi, a {\em two-stage} processing approach was implemented for
the tau trigger in \runii.
In the first stage, the position of the tau event vertex along the \beamline is identified by executing
the fast tracking in a narrow \roi with a full width of 0.2 in both $\eta$ and $\phi$, directed towards the
tau candidate calorimeter cluster,
but fully extended along the \beamline in the range $|z|<\SI{225}{\mm}$ to identify the leading \pt tau decay
tracks. The second stage executes the fast tracking again, followed by the precision tracking for the tracks found
in this second fast tracking stage,
but this time in a wider \roi with a full width 0.8 in both $\eta$ and $\phi$, centred on the $z$
position of the leading track identified by the first stage and limited to $|\Delta z|<\SI{10}{\mm}$
relative to this leading track.
The \rois from these different single-stage and two-stage strategies are illustrated schematically in
Figure~\ref{fig:tauroi}.
 
\figtauroi

For comparison purposes, during the commissioning for the \runii data taking, the tau lepton triggers
were also executed in a single-stage mode, similar to that used for \runi~\cite{TRIG-2011-02}, running
the fast track finder only once in a wide \roi in $\eta$ and
$\phi$ and fully extended along the \beamline, followed by the precision tracking again in this same
\roi.
As a result of the time saved in the processing of
the second stage because of the narrow $z$ extent along the \beamline, the actual \etaphi
width of the second stage \roi can be made larger than would otherwise be possible running only
in a single stage with the full $z$ extent.

In \runi, the
$b$-jet trigger ran the full tracking individually in the full $z$-width \rois centred on jets from the trigger~\cite{TRIG-2012-01}
with a wide \etaphi  extent.
However, for \runii, the increased beam energy
and interaction multiplicity meant that this approach was inappropriate, resulting in too many \rois with too large an
overlap such that the degree of processing of overlapping parts of the detector became prohibitive.
 
Consequently, for the \runii $b$-jet trigger~\cite{ATL-COM-DAQ-2020-064}  a new
\multistage tracking strategy was adopted with a new
initial tracking stage, running tracking specifically to identify the
likely event vertex $z$ position for use in the second stage.
For the second stage, separate \rois about each jet axis are used as in \runi,
but with each \roi specified more tightly at the \beamline about the $z$ vertex position
identified in the first stage. For the tracking in the initial vertex stage, all jets identified
by the jet trigger with transverse energy $\et>\SI{30}\gev$ are considered, and tracks are
reconstructed with the fast
tracking algorithm in a narrow region with full widths of 0.2 in $\eta$ and $\phi$ around the jet axis
for each jet, but with $|z|<\SI{225}{\mm}$ along the \beamline.\footnote{In 2016 the constituent \roi
full width in $\eta$ and $\phi$ was 0.4.}
In order to prevent the multiple processing
of regions of the detector that overlap, before running the fast tracking, the \rois
about each jet axis are first aggregated into a single {\em \superroi} for the event.
This \superroi is then
used by the data preparation of the first stage vertex tracking
to determine which detector elements\emdash in this case individual silicon modules from the
pixel and SCT detectors\emdash should be read out for the subsequent processing.
This happens only once per event for the single \superroi, which can include non-overlapping regions.
This is illustrated in Figure~\ref{fig:superroi}, which shows the aggregation
of the individual narrow jet \rois into a single, more complicated
region used by the data preparation. With only the data from these detector
elements corresponding to the \rois about the jet directions, the tracking then runs as
if it were processing the entire detector.

Following this stage, the tracks identified in the \superroi are used
for the primary vertex reconstruction~\cite{Meloni_2016}.
This vertex is used to define wider \rois about
each jet axis, with $|\Delta\eta|<0.4$ and $|\Delta\phi|<0.4$ with respect to the jet axis,
but with $|\Delta z|<\SI{10}{\mm}$
relative to the primary vertex $z$ position. These \rois are then used for the second-stage
reconstruction which again runs the fast tracking but this time in the wider $\eta$
and $\phi$ regions about the jets. This is then followed by the precision
tracking, secondary vertexing and $b$-tagging algorithms.
For the vertex stage, different vertex algorithms are available.
As in the case of the tau \multistage tracking, the time saved by running the jet
tracking in very narrow \rois in $z$ about the vertex position allows the jet tracking
stage to execute in a wider \roi than would be possible with a single-stage processing.
 
\figsuperroi

In \runii this same approach was
also adopted for some of the standard jet triggers in order to be able to run the
global sequential calibration~\cite{JETM-2018-05}. This makes use of the information from the precision track
reconstruction in the jets to improve the jet energy resolution measured with respect to the
offline jets. This allows a tighter selection
on the jet transverse momentum~\cite{PERF-2014-02,Aad:2020flx},
closer to that used offline, thus reducing the rate from the jet trigger.


\section{Inner detector trigger timing}

\newcommand{\figmuontime}{
\begin{figure}[!th]
\centering
\subfloat[]{\includegraphics[width=0.49\textwidth]{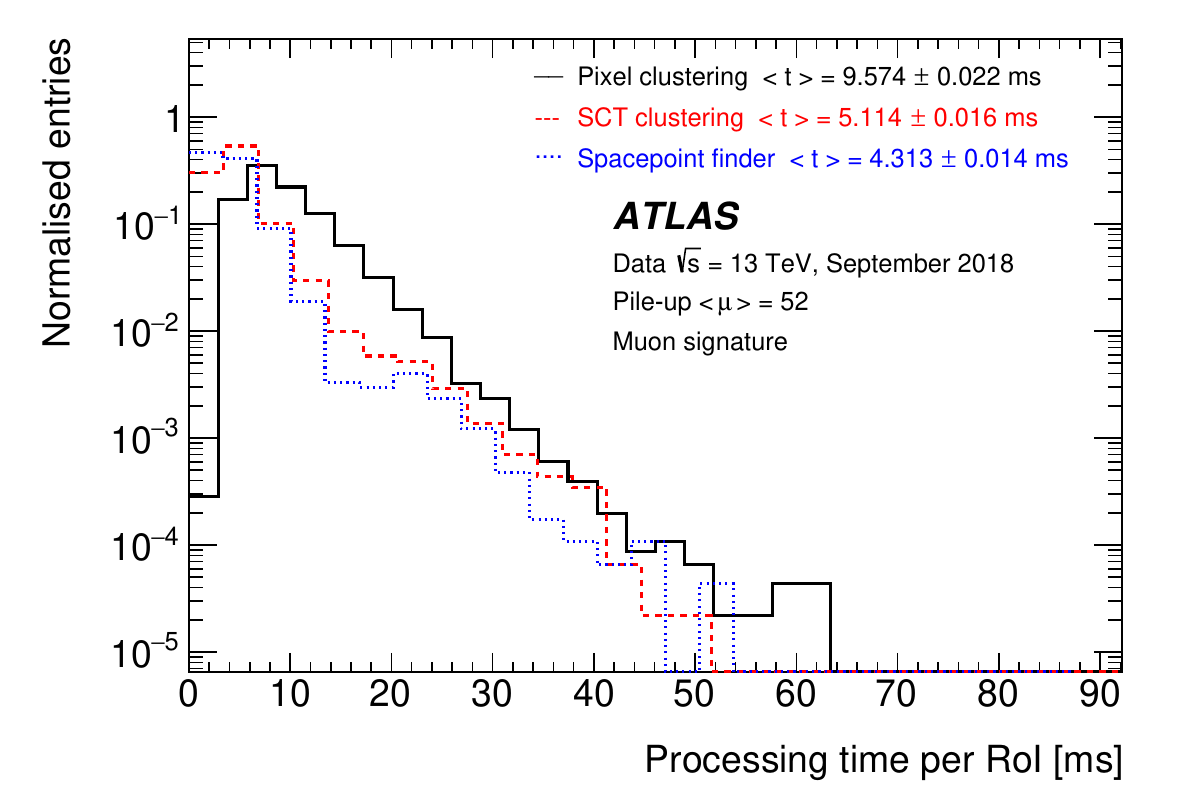}\label{fig:muontimea}}\figbreak
\subfloat[]{\includegraphics[width=0.49\textwidth]{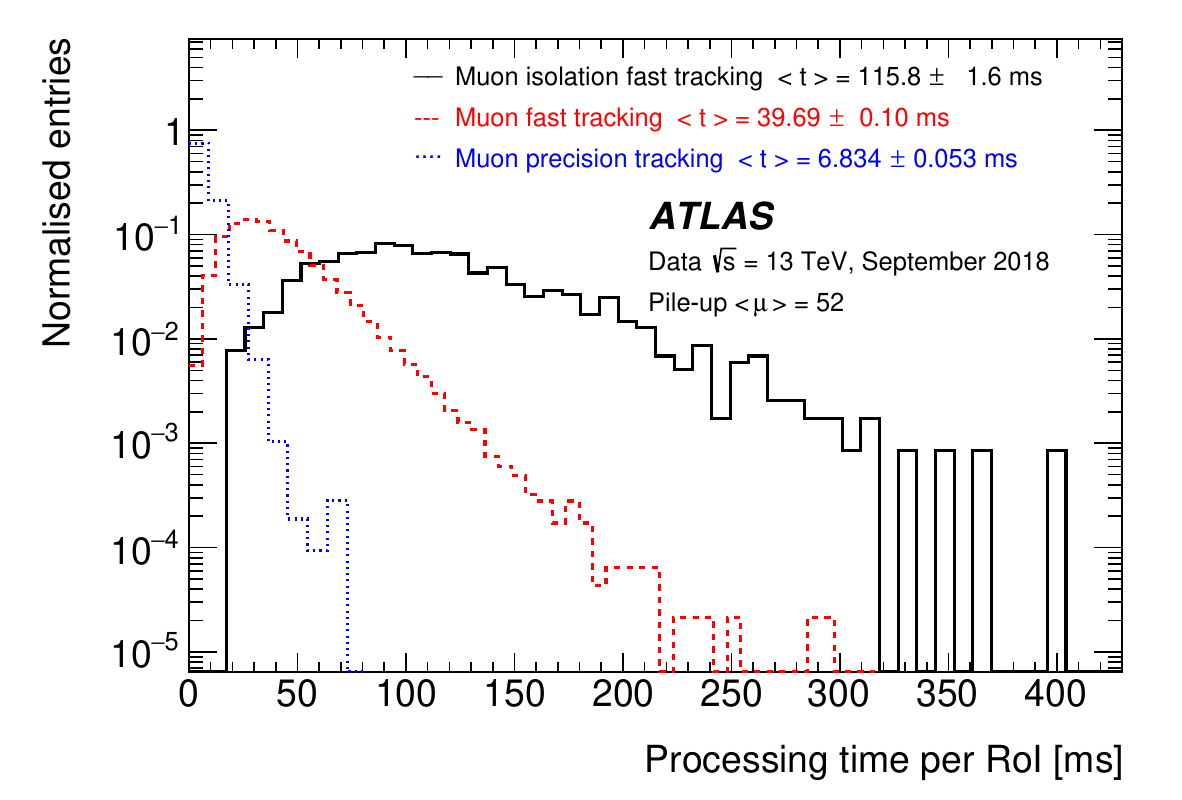}\label{fig:muontimeb}}\\
\subfloat[]{\includegraphics[width=0.49\textwidth]{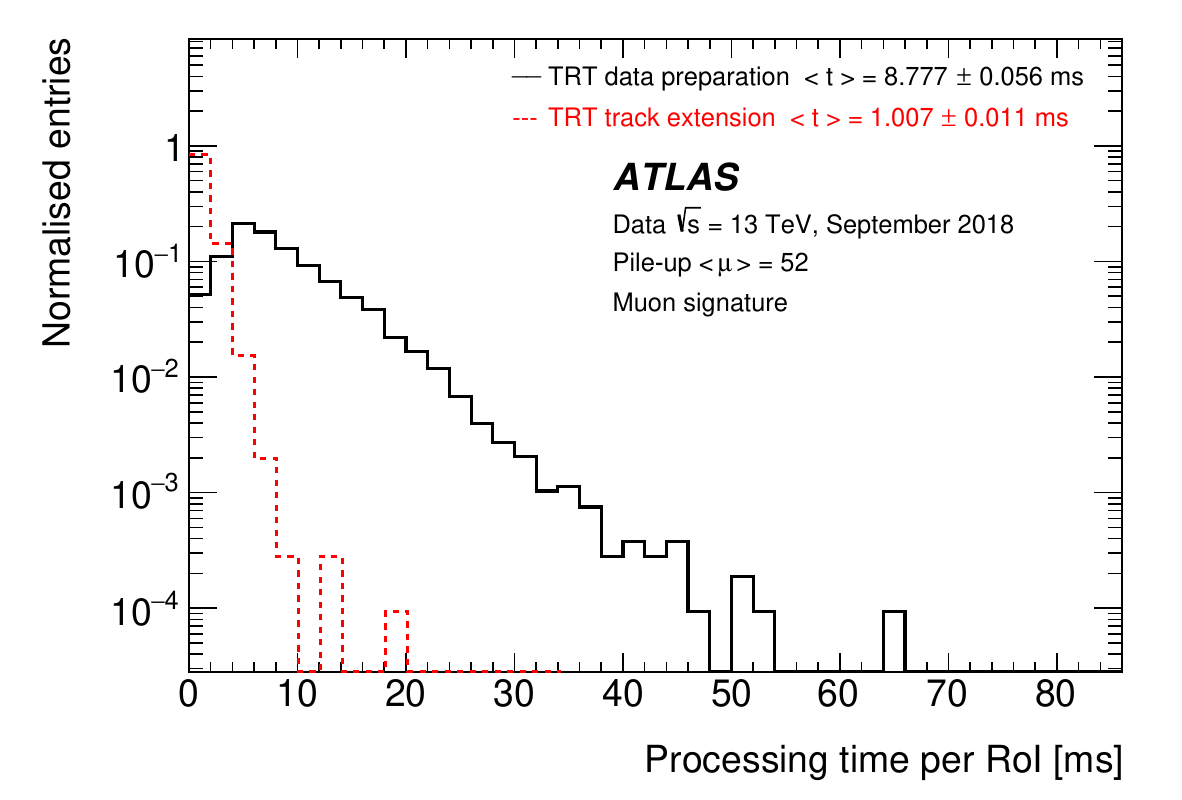}\label{fig:muontimec}}
\caption{The processing time for muon tracking related algorithms;
(a) the data preparation times,
(b) the tracking algorithm times, and (c) the TRT data processing.
}
\label{fig:muontime}
\end{figure}
}

\newcommand{\figjettime}{
\begin{figure}[!th]
\centering
\subfloat[]{\includegraphics[width=0.49\textwidth]{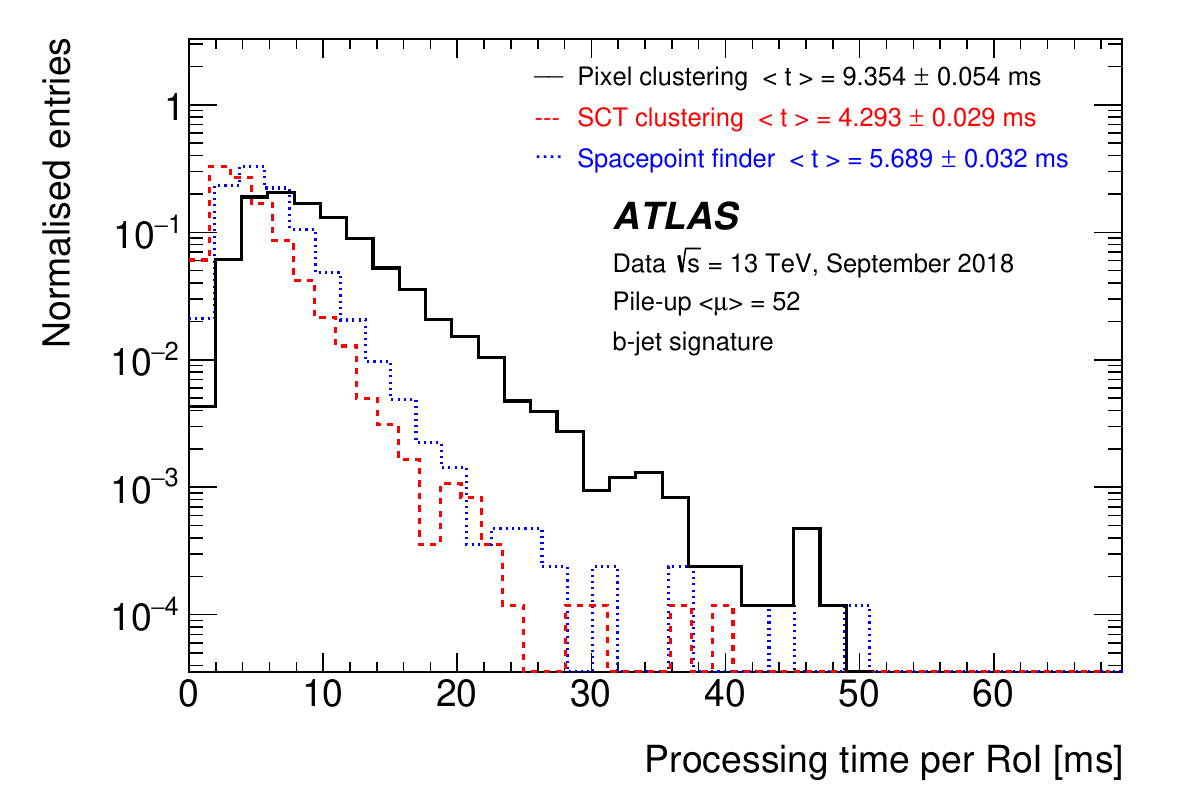}\label{fig:jettimea}}\figbreak
\subfloat[]{\includegraphics[width=0.49\textwidth]{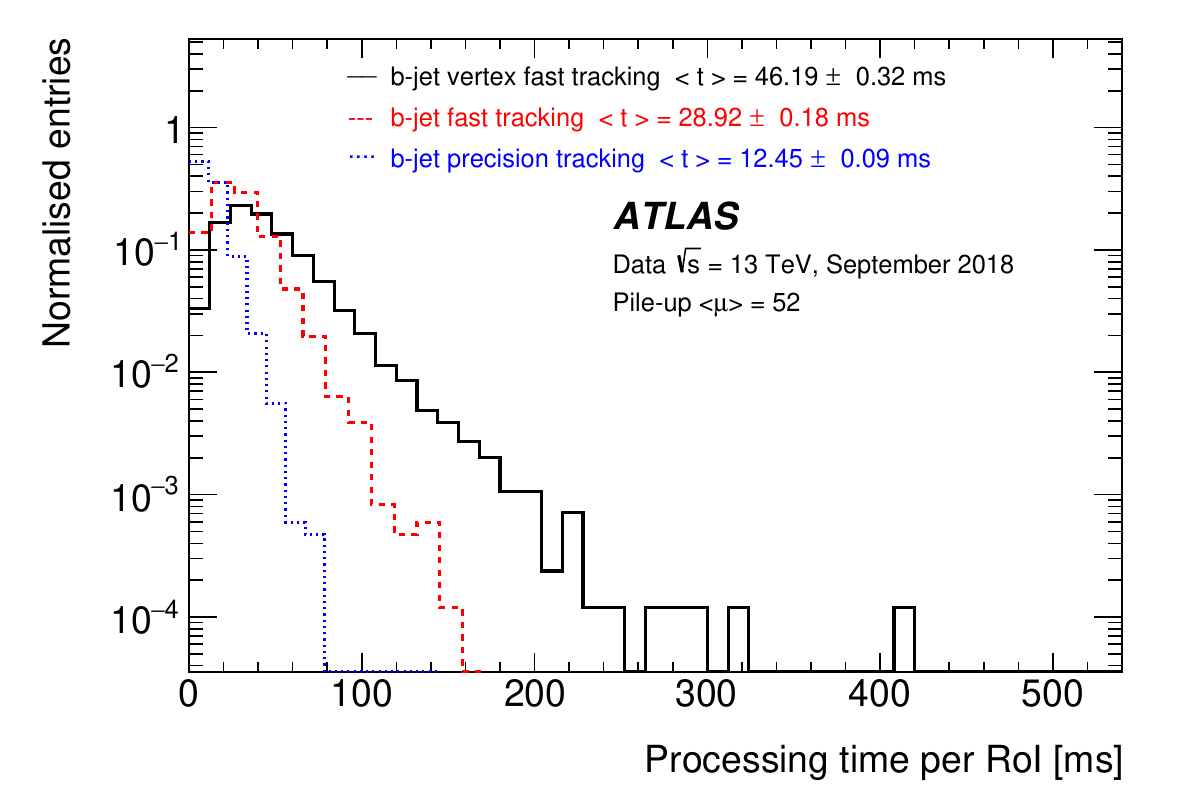}\label{fig:jettimeb}}\\
\subfloat[]{\includegraphics[width=0.49\textwidth]{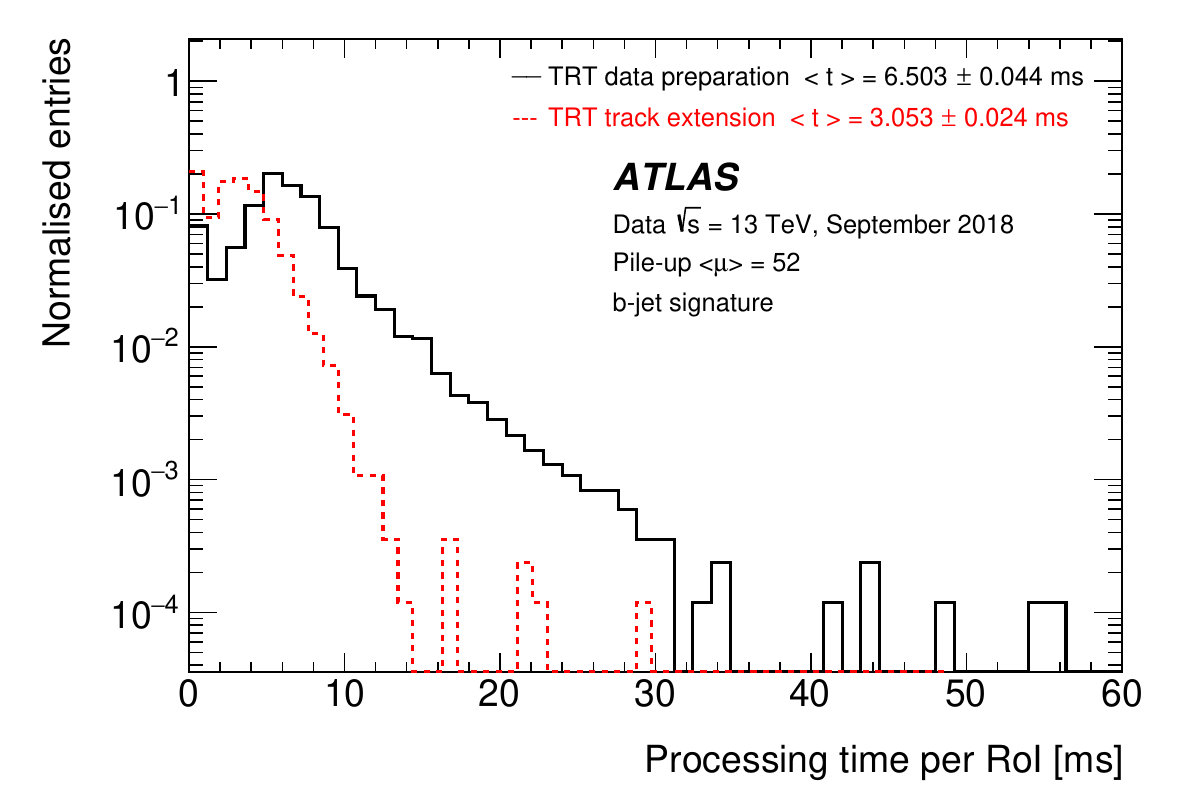}\label{fig:jettimec}}
\caption{The processing time for \bjet tracking related algorithms;
(a) the data preparation times,
(b) the tracking algorithm times, and (c) the TRT data processing.
}
\end{figure}
}

\newcommand{\figegammatime}{
\begin{figure}[!th]
\centering
\subfloat[]{\includegraphics[width=0.49\textwidth]{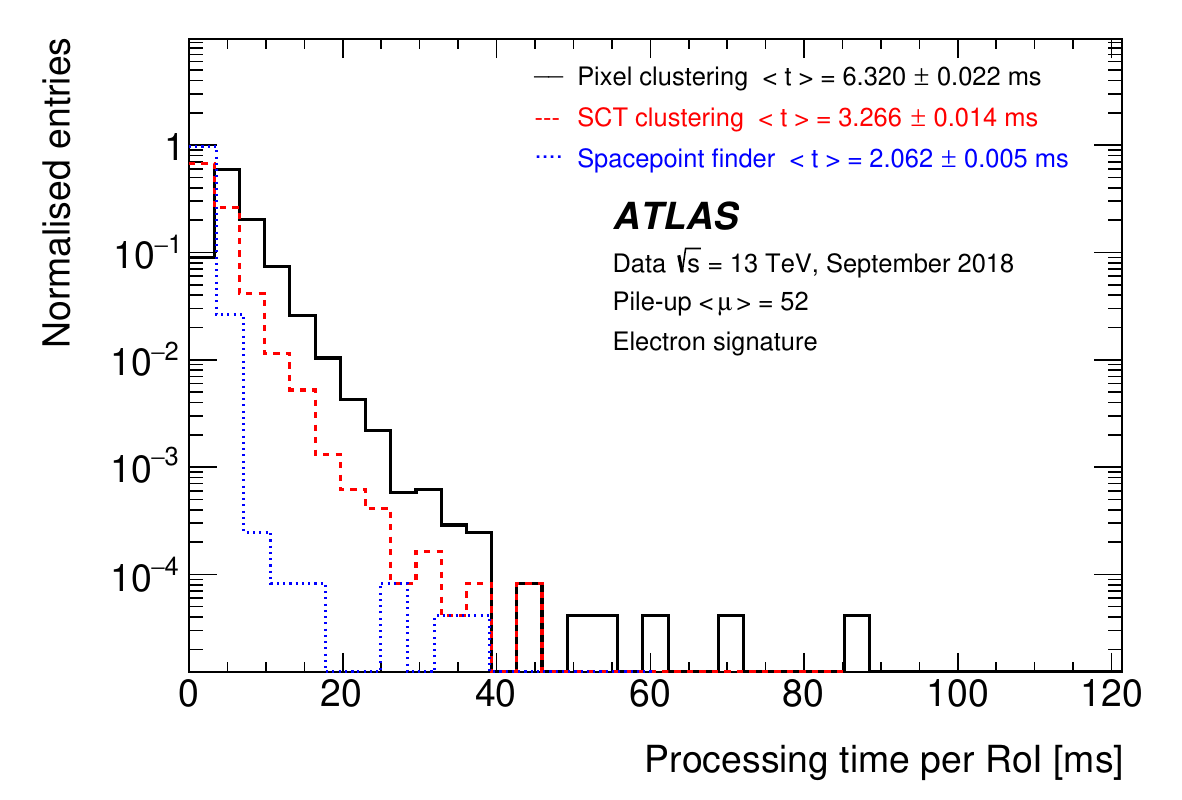}}\figbreak
\subfloat[]{\includegraphics[width=0.49\textwidth]{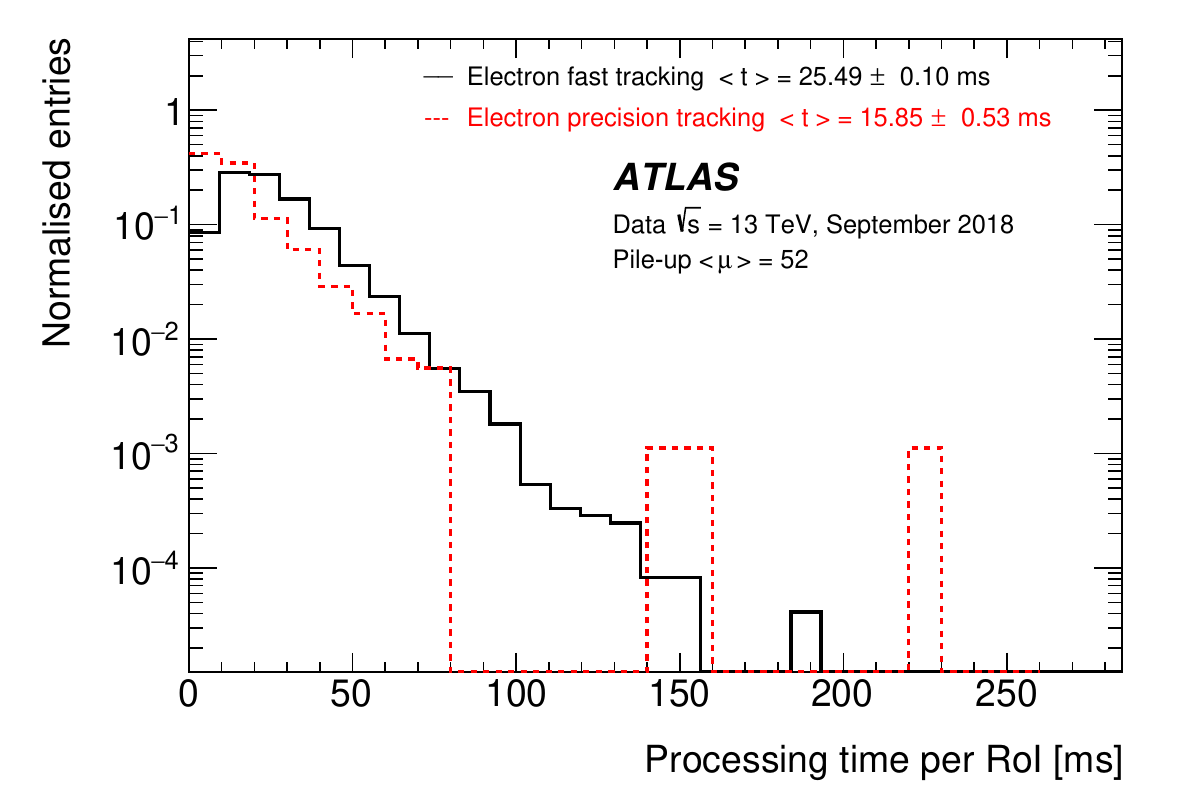}}\\
\subfloat[]{\includegraphics[width=0.49\textwidth]{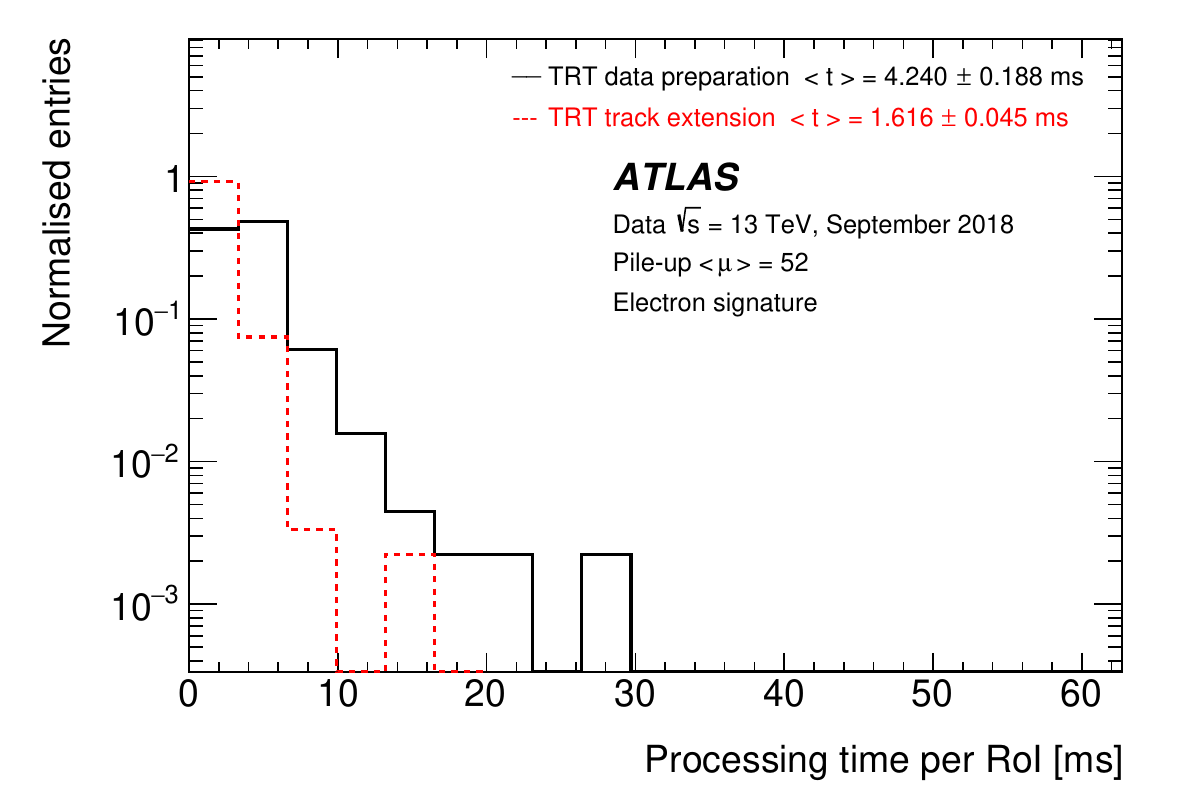}}
\caption{The processing time for electron tracking related algorithms;
(a) the data preparation times,
(b) the tracking algorithm times, and (c) the TRT data processing. }
\label{fig:egammatime}
\end{figure}
}

\newcommand{\figtaustagetime}{
\begin{figure}[!th]
\centering
\subfloat[]{\includegraphics[width=0.5\textwidth]{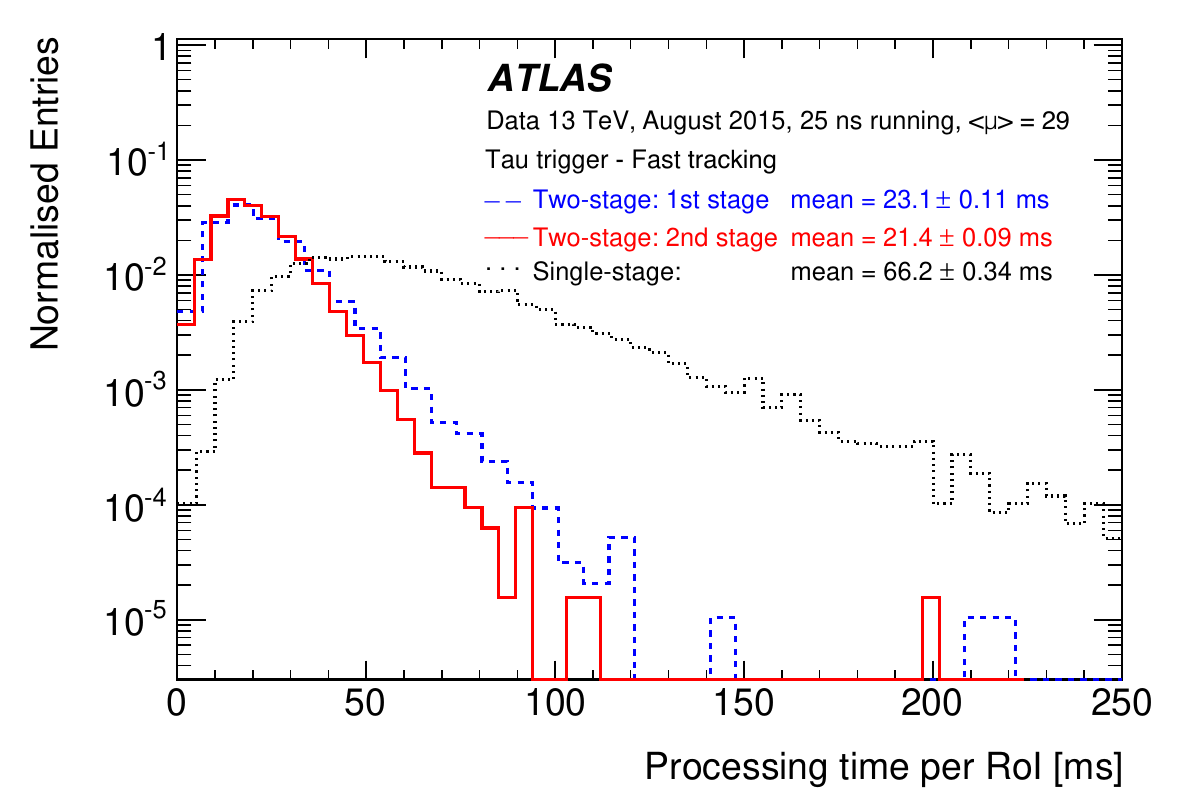}\label{fig:tautwostagea}}
\subfloat[]{\includegraphics[width=0.5\textwidth]{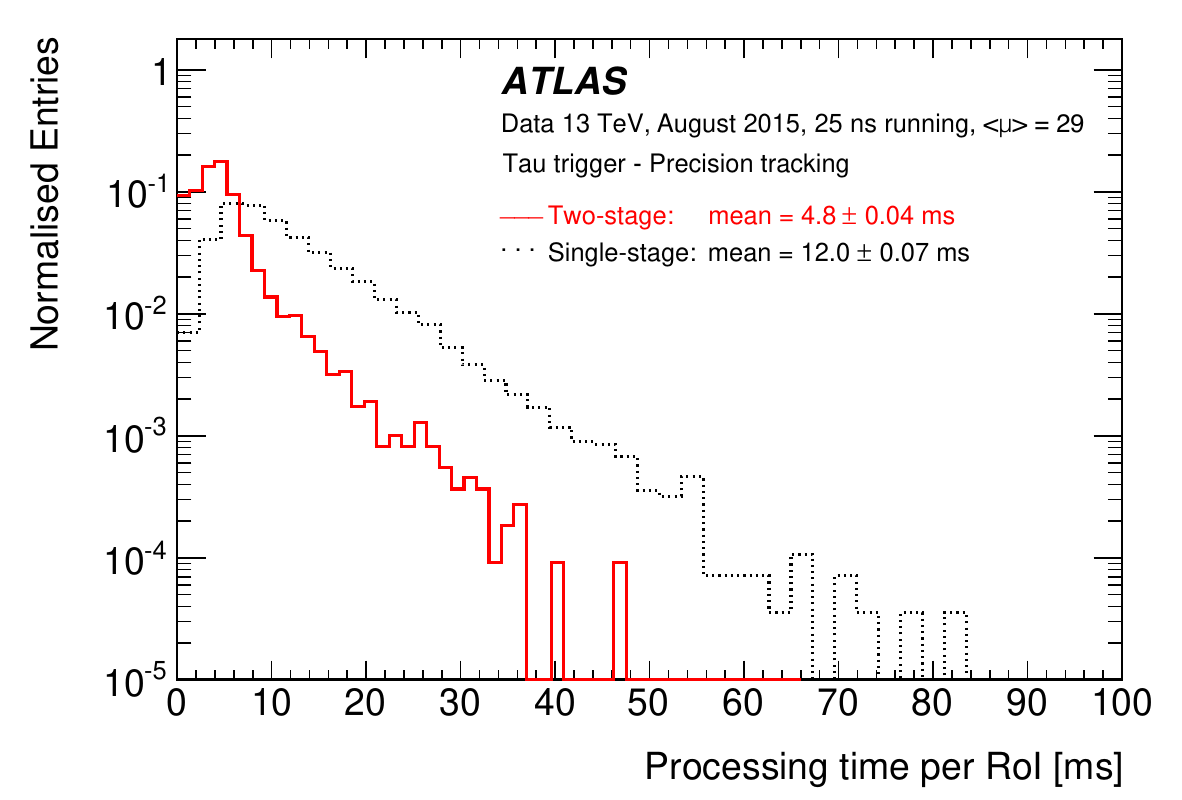}\label{fig:tautwostageb}}
\caption{The ID trigger tau tracking processing time for (a) the fast track
finder and (b) the precision tracking comparing the single-stage and two-stage
tracking approach.}
\label{fig:tautwostage}
\end{figure}
}

\newcommand{\figtautime}{
\begin{figure}[!th]
\centering
\subfloat[]{\includegraphics[width=0.49\textwidth]{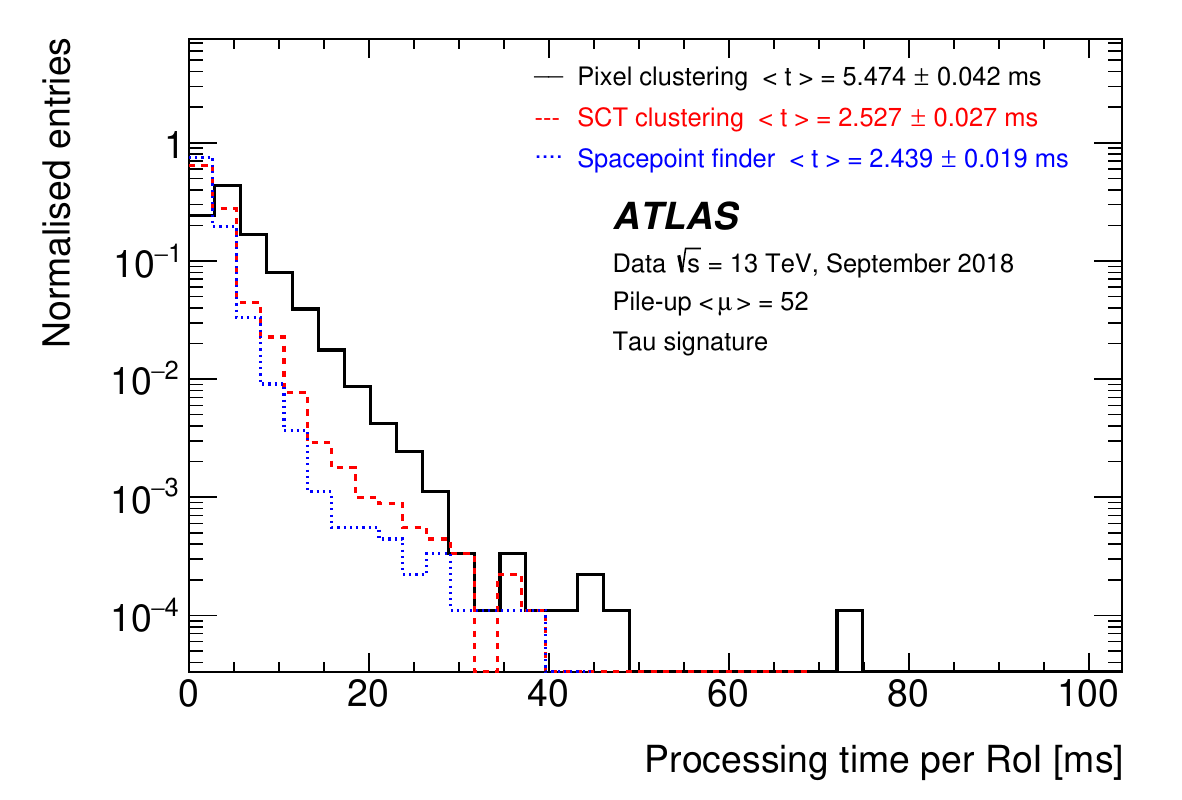}}\figbreak
\subfloat[]{\includegraphics[width=0.49\textwidth]{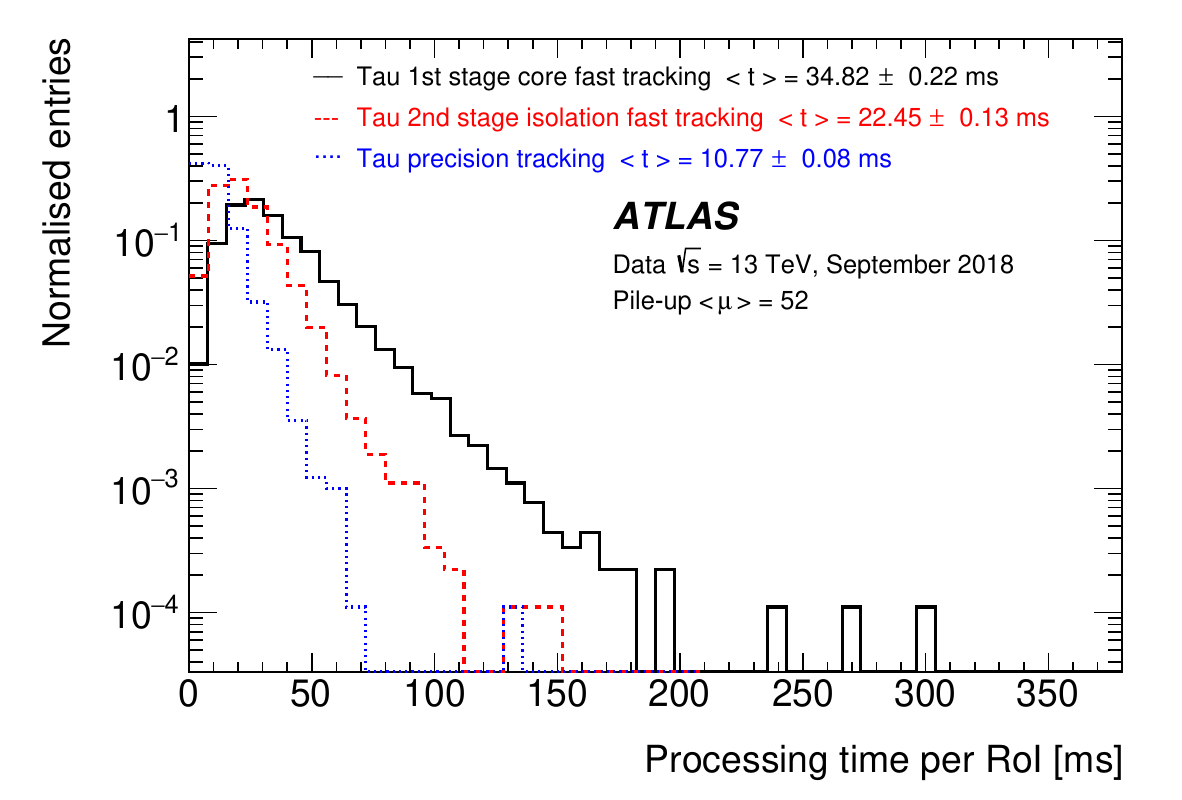}}\\
\subfloat[]{\includegraphics[width=0.49\textwidth]{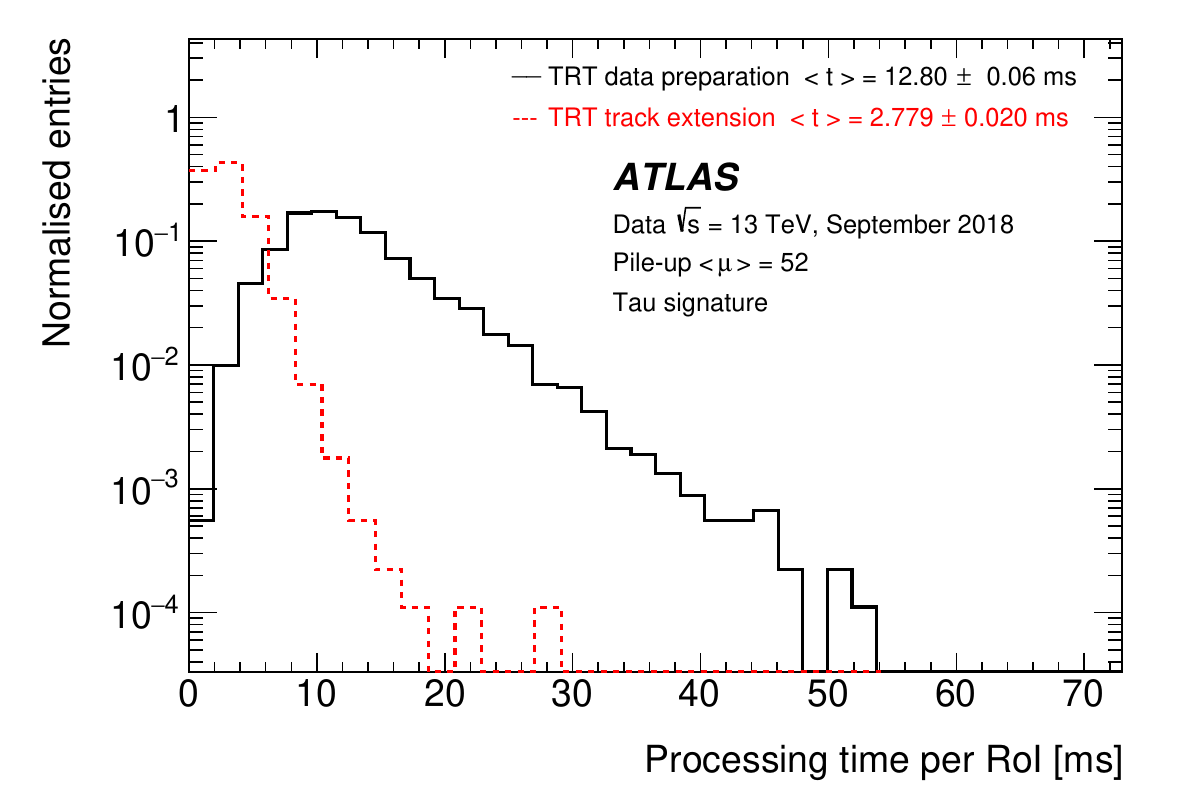}}
\caption{The processing time for tau tracking related algorithms;
(a) the data preparation times,
(b) the tracking algorithm times, and (c) the TRT data processing. }
\label{fig:tautime}
\end{figure}
}

\newcommand{\figvtxtime}{
\begin{figure*}[!tph]
\centering
\subfloat[]{\includegraphics[width=0.49\textwidth]{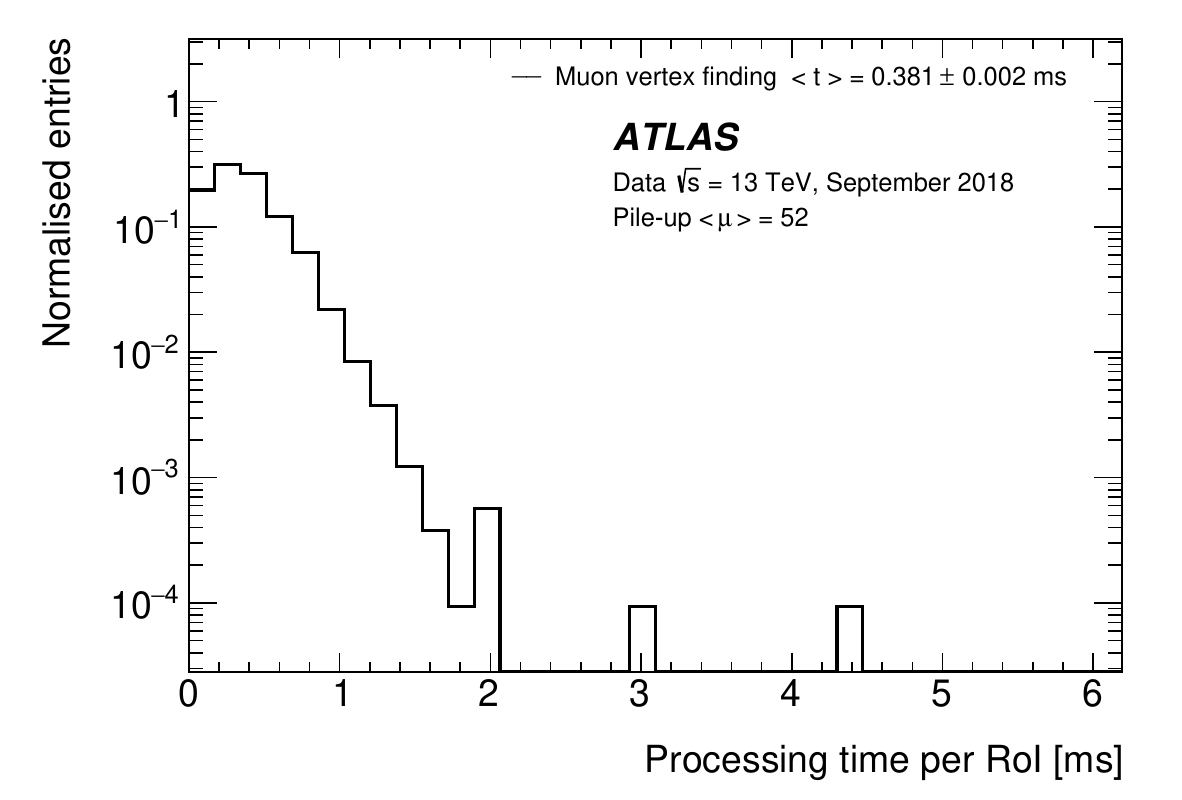}}
\subfloat[]{\includegraphics[width=0.49\textwidth]{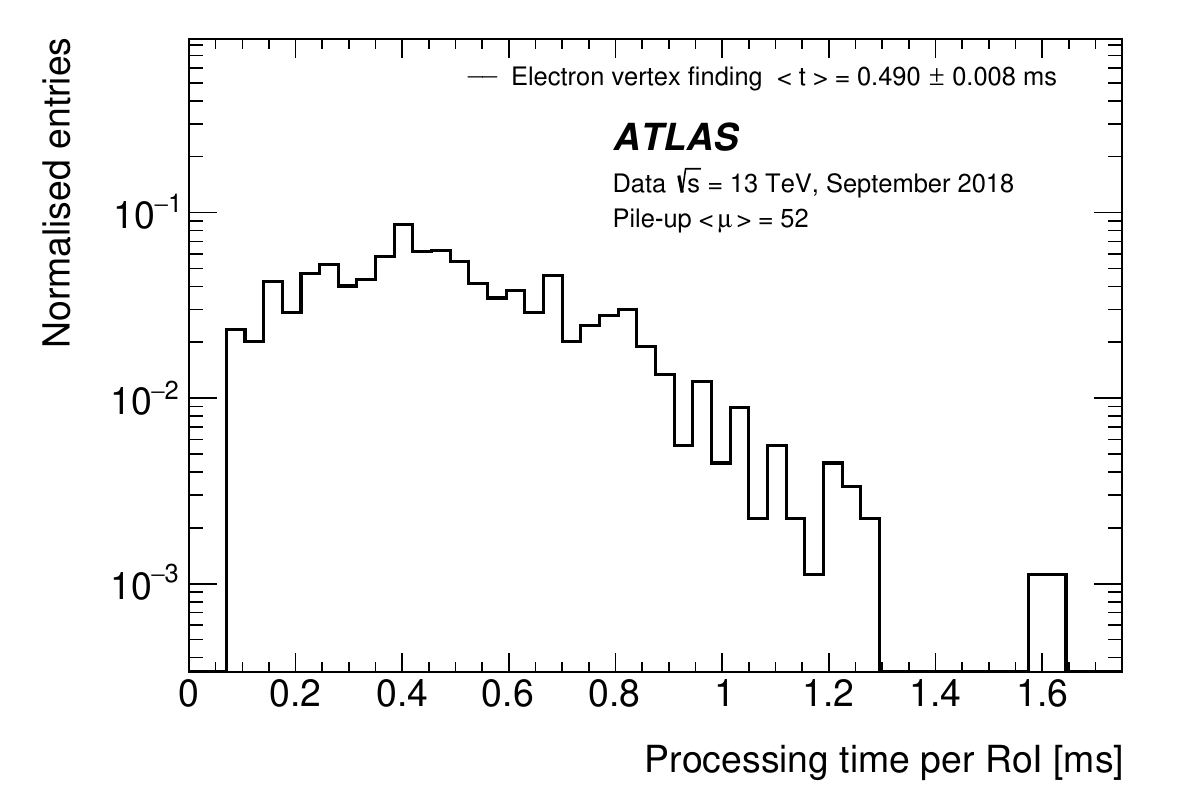}}\\
\subfloat[]{\includegraphics[width=0.49\textwidth]{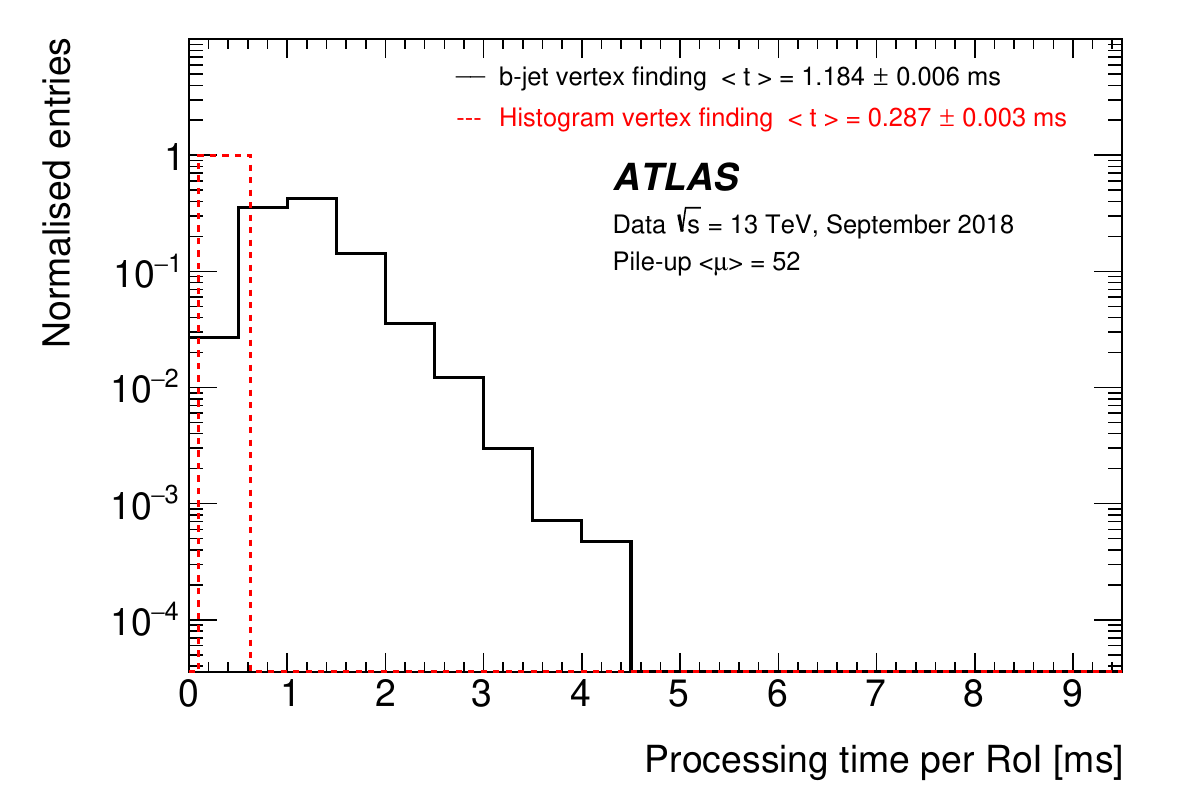}\label{fig:vtxtimec}}
\subfloat[]{\includegraphics[width=0.49\textwidth]{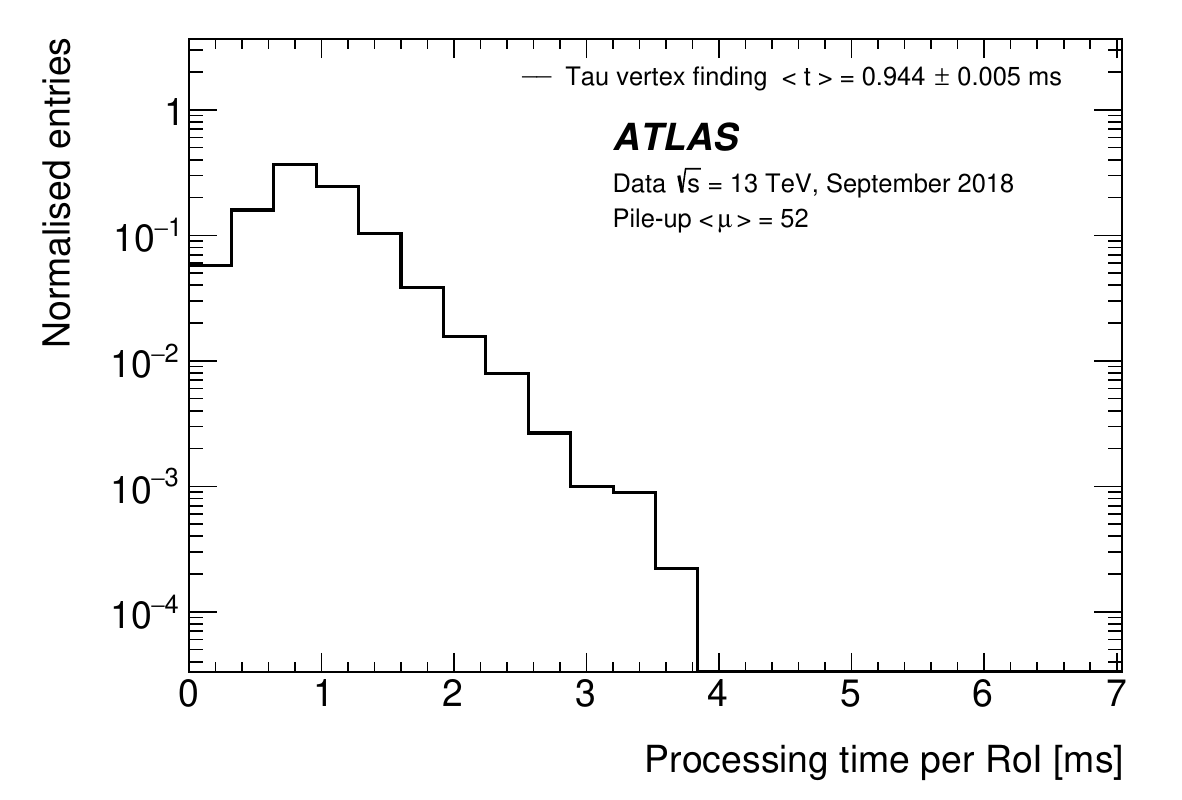}}
\caption{The processing times for trigger vertex algorithms;
(a) and (b) for the muon and electron signatures,
(c) for the \bjet vertex tracking, and
(d) for the tau signature. }
\label{fig:vtxtime}
\end{figure*}
}

\newcommand{\figpileuptime}{
\begin{figure*}[!t]
\centering
\subfloat[]{\includegraphics[width=0.49\textwidth]{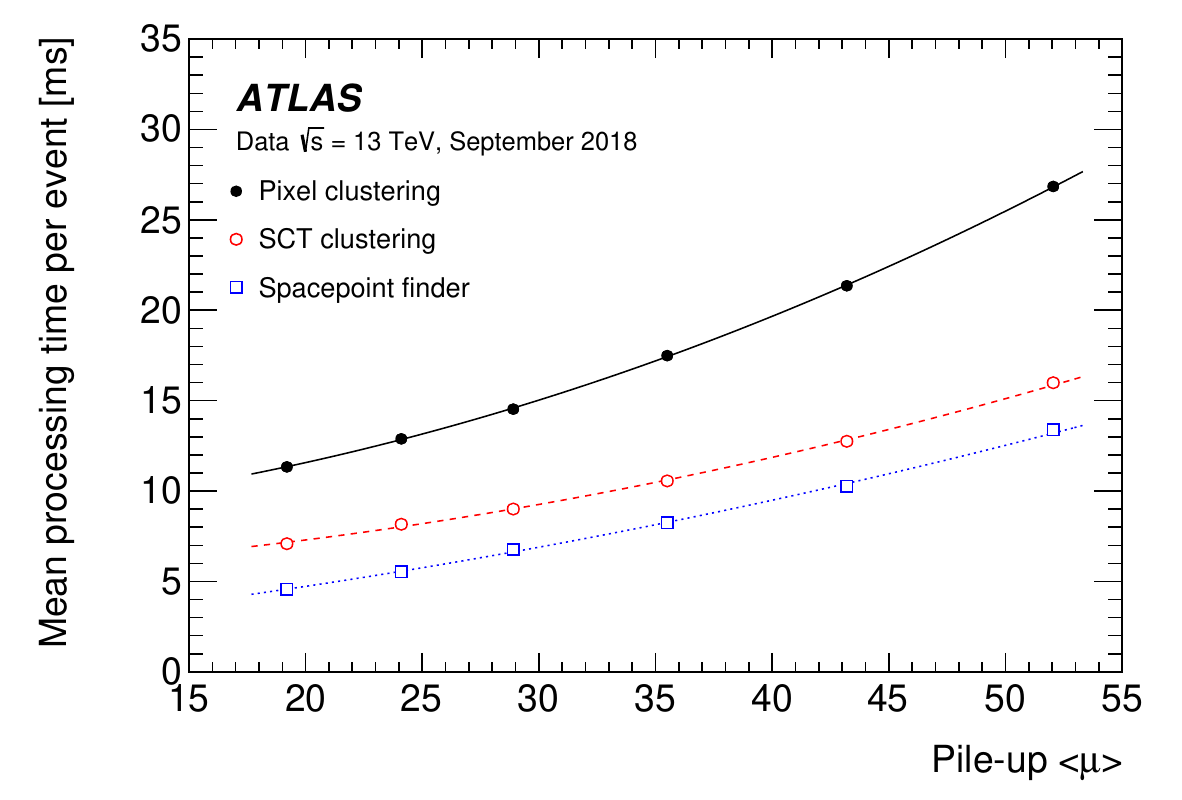}}
\subfloat[]{\includegraphics[width=0.49\textwidth]{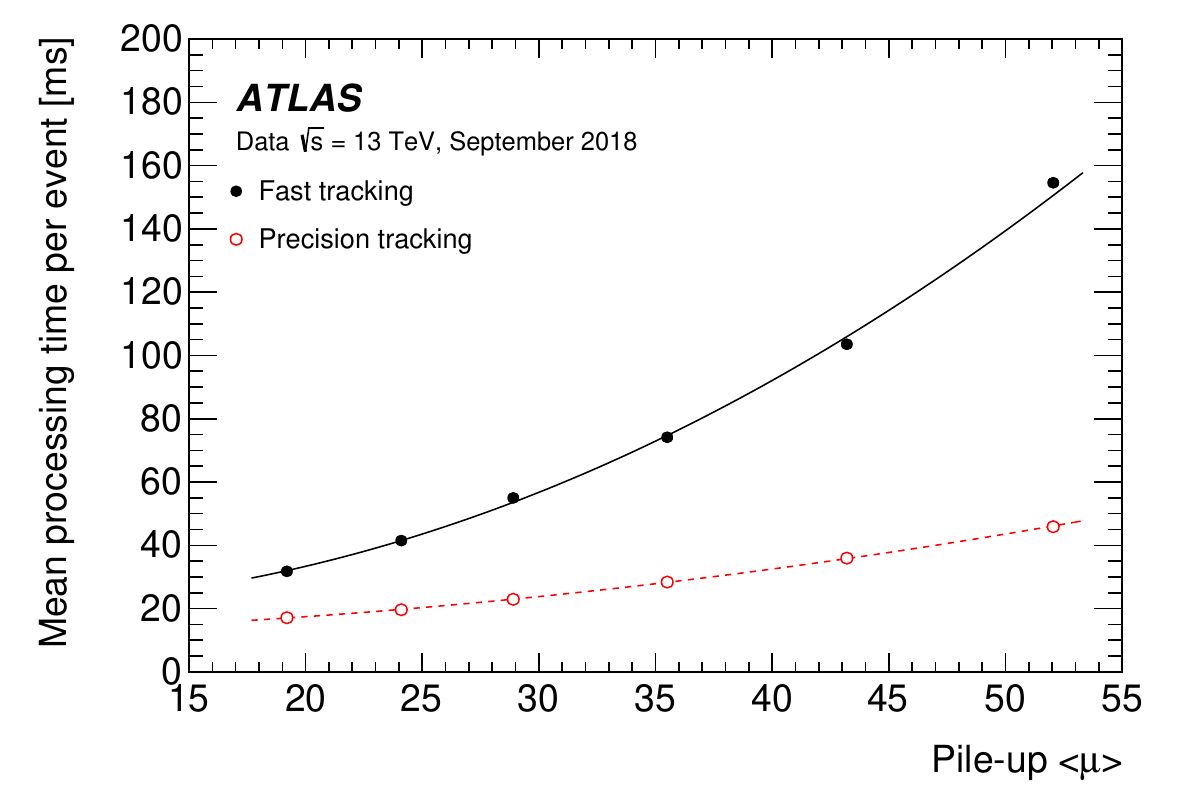}}\\
\subfloat[]{\includegraphics[width=0.49\textwidth]{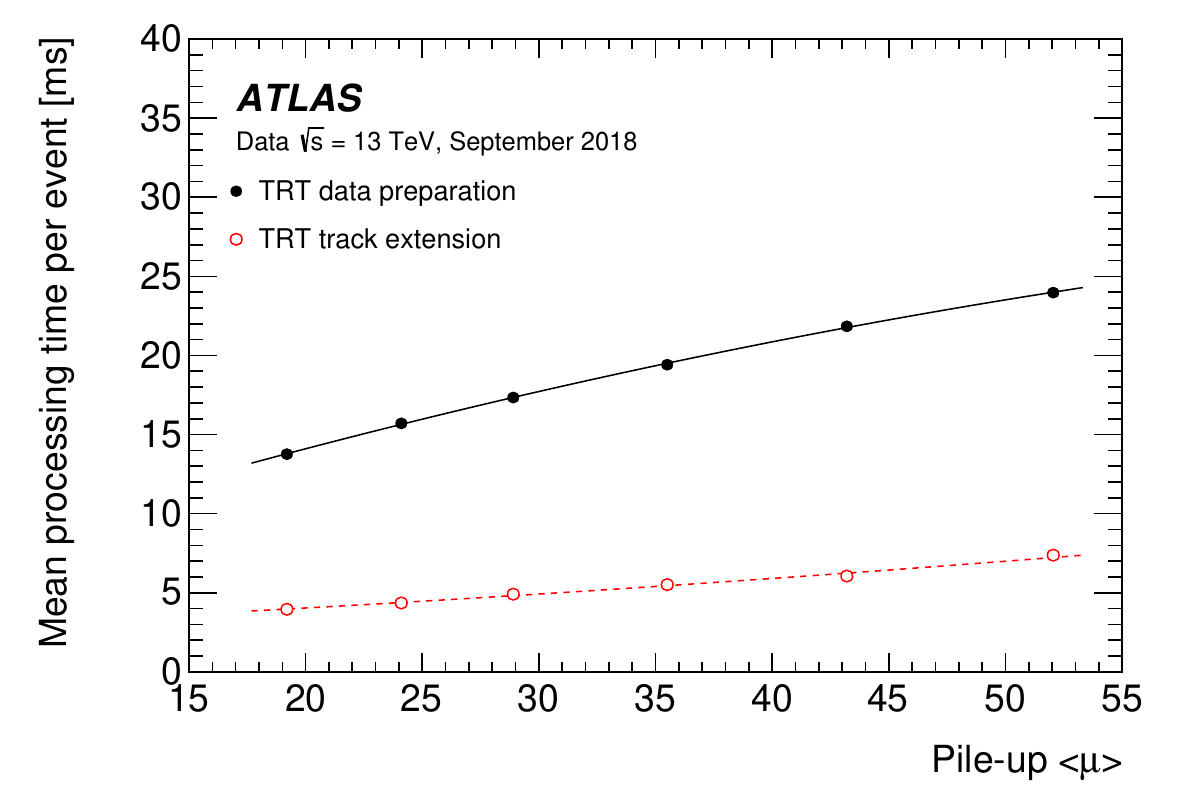}}
\subfloat[]{\includegraphics[width=0.49\textwidth]{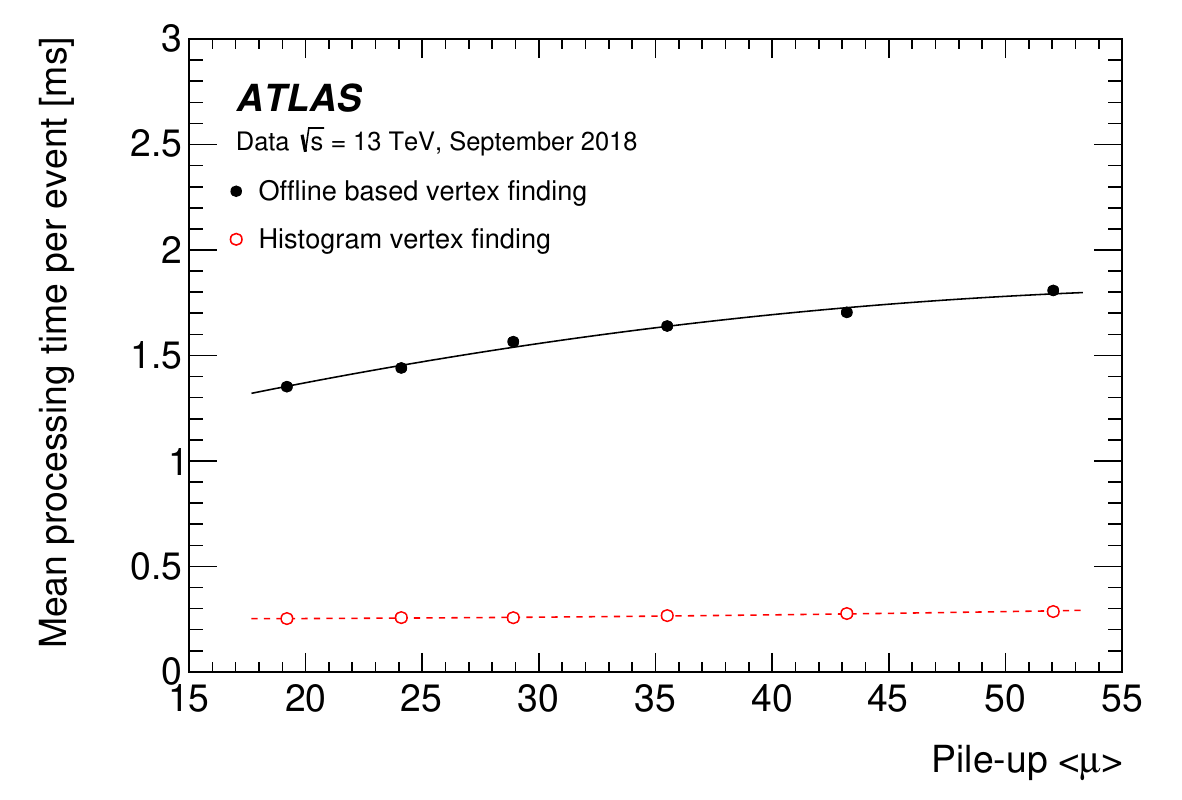}}
\caption{The mean of the total processing time per event for the tracking related algorithms;
(a) the data preparation for the silicon detectors,
(b) the fast, and precision tracking,
(c) the TRT data preparation and track extension, and
(d) the vertexing.}
\label{fig:pileuptime}
\end{figure*}
}

In this section, timing measurements for the tracking related algorithms used in the
inner detector trigger in \runii are presented. Timing measurements for the different tracking
related algorithms were taken online,
using the trigger cost monitoring framework~\cite{Aad:2020wji}.
Unless otherwise stated, the timing measurements were
recorded during a typical physics run,
on Saturday, September 29th, 2018 with a mean \pileup interaction multiplicity of 52 at the
start, and 19 at the end of the run. The full timing distributions are presented for
the start of the run only.

The retrieval and data preparation for the silicon detectors prior to running
the tracking algorithms themselves
is typically reasonably fast, but is still a significant contribution to the overall
processing time. The subsequent fast track reconstruction then precedes the
execution of hypothesis algorithms which combine the data from the calorimeters or muon
spectrometer with the ID tracks to further reduce the rate. Following this reduction, the
precision tracking stage is executed,
and the precision tracks can be extended into the TRT, which requires the TRT data preparation and
track extension algorithms. Subsequent algorithms may use the tracks in combination with clusters
reconstructed in the calorimeter,
or perform a combined fit using the information from the muon spectrometer and the inner detector. A
discussion of some of these other algorithms can be found elsewhere~\cite{ATL-LARG-PUB-2008-002,PERF-2013-01}.

A  vertex finding algorithm can also be executed using the tracks from any stage
in the processing, typically following the precision tracking, but also using the
tracks from the fast track finder, for example, following the $b$-jet vertex tracking.
In the case of the \multistage tracking, the data preparation
may be executed a second time in the wider regions of the detector. However,
the times for the data preparation are not recorded separately for the
different stages, and so for this analysis are aggregated into a
single distribution for each specific trigger.

\subsection{Algorithm execution time  in the  muon trigger}
 
\figmuontime
 
The principal processing times for the algorithms in the muon signature are shown in
Figure~\ref{fig:muontime} for the standard full \roi width of 0.2 in both $\eta$ and $\phi$.
Some triggers run a second isolation tracking stage in a wider \etaphi region,
after the first
round of precision tracking, to determine whether tracks are isolated with respect to
other tracks from the interaction.
In this case the data preparation and fast track finder are executed a
second time but in a wider \roi in $\eta$ and $\phi$,
but still extended by the full length in $z$ along the \beamline.
The precision tracking then runs a second time on the tracks in this larger \roi.
As in the case of the data preparation, information to distinguish between the
first and second execution of the precision tracking within any specific muon trigger
is not available so only the precision tracking time of the first pass is shown here.
 
The \spacepoint preparation is generally very fast, as can be seen in
Figure~\ref{fig:muontimea}, each algorithm having a mean processing time per
\roi of less than 10~ms.
 
Figure~\ref{fig:muontimeb} illustrates that the fast tracking
in the first stage processing
is reasonably fast with a mean
execution time of 40~ms and a maximum at approximately 300~ms.
The precision tracking has a mean execution time of less than 7~ms.
 
The second stage muon isolation fast track processing with the fully extended $z$ range,
and wider \etaphi size is significantly slower, with a mean processing time
of approximately 116~ms.
This is only executed on \rois where a trigger muon candidate has already been
confirmed such that the rate should be significantly lower than for the first stage \roi
processing.
The TRT extension is only performed for the precision tracks
and the execution time from Figure~\ref{fig:muontimec} is under 10~ms.

One additional factor in the processing of the muon trigger is that prior to the
execution of the first stage processing with the fast tracking, the \roi is determined
by the muon spectrometer trigger~\cite{TRIG-2012-03}, which calculates an updated
\roi position based on
the first processing of the spectrometer data in the HLT. Partly because of the
toroidal field in the muon spectrometer, the resolution of the tracks in $z$ extrapolated
back to the \beamline for
the spectrometer-only muon candidates is not good enough to refine the $z$ position
of the \roi, so the full extent in $z$ along the \beamline is required. Additionally,
in the transition regions between the barrel and endcap detectors, with pseudorapidities near $\pm 1$,
the $\eta$
resolution of the spectrometer-only candidates is poor, and so the \rois for the ID
tracking can be made commensurately wider to ensure that the \spacepoints for the
ID track are contained within the \roi.

\subsection{Algorithm execution time  in the   electron trigger}

The processing for the electron triggers is slightly less complex than for the muons,
since it does not run a second stage. As such both the fast and precision tracking
run in the same \roi, produced after the first calorimeter reconstruction algorithm.
The \roi dimensions are the same as for the first stage from the muon signature. Between
the fast tracking and precision tracking, additional reconstruction algorithms combining the
calorimeter and tracking information to produce electron candidates can be executed.
Hypothesis algorithms are then executed to select on these electron candidates to reduce the rate
before the precision tracking. The principal algorithm processing times
for the different steps in the sequence can be seen in Figure~\ref{fig:egammatime}.
 
\figegammatime
 
Here the data preparation stages are very quick\emdash typically less than 7~ms.
The mean processing time in data from late 2018 was approximately 26~ms for the fast tracking
and under 16~ms for the precision tracking. The full TRT extension following the precision
tracking took under 6~ms.

\subsection{Algorithm execution time  in the  tau trigger}

As an example of the benefits afforded by using the \multistage approach for the tau processing
discussed in Section~\ref{sec:multistage},
the time taken by the tau tracking in both the single-stage and two-stage
variants is shown in
Figure~\ref{fig:tautwostage}. This is from an ATLAS data-taking run with 25~ns bunch spacing
taken during the detector commissioning from 2015 where the mean \pileup interaction multiplicity
was 29.
For subsequent production luminosity
running in 2016 and onwards, only the two-stage version was running for all tau triggers.
Figure~\ref{fig:tautwostagea} shows the
processing times per \roi for the fast tracking stages: individually for the first and
second stages of the two-stage tracking,
and for the single-stage
tracking with the original wider \roi in $\eta$, $\phi$ and $z$. The fast tracking in the
single-stage has a mean execution time of approximately \SI{66}{\ms}, with a
very long tail. In contrast, the first stage tracking with an \roi that is wide
only in the $z$ direction has a mean execution time of \SI{23}{\ms}, driven predominantly
by the narrower \roi width in $\phi$. The second stage tracking, although wider in $\eta$
and $\phi$ even than the \roi used in the single-stage, takes only \SI{21}{\ms} on
average because of the significant reduction in the \roi $z$-width along the \beamline.
Figure~\ref{fig:tautwostageb} shows
a comparison of the processing time per \roi for the precision tracking. The precision tracking
in the  two-stage strategy executes faster, with a mean of \SI{4.8}{\ms} compared to \SI{12}{\ms} for
the single-stage tracking.
 
\figtaustagetime
 
\figtautime
 
The execution times for the tracking related algorithms for the tau signature from
production luminosity running late in 2018 are shown in Figure~\ref{fig:tautime}.
In this case the \pileup interaction multiplicity, with a peak mean value of
approximately 52 interactions per bunch crossing, is significantly larger than for the
2015 data-taking from the previous figure.
In addition, at the start of running
in 2017, the $z$-width of the second stage tau \roi was reduced further.
Here the first stage fast track finder execution time in the full $z$-width \roi is
35~ms and that of the second stage is 23~ms. The precision tracking takes 10~ms.
 
A significant time\emdash approximately 13~ms\emdash is taken by the calculation of the
TRT drift circles for the extension of the precision tracks into the TRT.

\subsection{Algorithm execution time in the  \bjet trigger and vertexing}
 
The \bjet trigger runs as a \multistage trigger with regard to the tracking.
Before the $b$-jet specific parts of the trigger are executed, the trigger runs
as a standard jet trigger, running the jet finding in the full detector~\cite{TRIG-2012-01}.
 
\figjettime
 
\figvtxtime
 
Following this, the first stage of the $b$-jet trigger takes the collection
of jets from the full detector, and creates a separate \roi about each
jet with $\et>30$\gev. These are aggregated into a \superroi as discussed
in Section~\ref{sec:multistage}. The data preparation and fast tracking
for the vertexing runs in this \superroi, followed by a vertex algorithm.
The vertex position is then used to refine the $z$ position of each jet \roi,
which are also widened in $\eta$ and $\phi$.
 
Subsequently, the data preparation, fast tracking and precision tracking
are executed in each jet \roi individually. The precision tracks are then
used for the subsequent $b$-tagging algorithms.  For the standard jet
triggers with the global sequential calibration, the tracks are used for
the additional correction of the jet \et before the $b$-tagging is performed.
 
The distribution of processing time for the ID data preparation
can be seen in Figure~\ref{fig:jettimea}, which includes the data preparation
for both the vertex tracking stage and the second, \roi based, \bjet
tracking stage.
The mean execution time for
each algorithm is under 10~ms.

Due to the larger volume of the detector to be processed, the
vertex tracking for the fast tracking is the longest of the tracking stages,
with a mean execution time of 46~ms, shown in Figure~\ref{fig:jettimeb}.
However, caching of the objects reconstructed by the trigger
means that the vertex tracking need be executed only once per event, with the
output being shared between all jet triggers. The subsequent tracking within the individual
jet \rois has a mean execution time of 29~ms and 12~ms per jet \roi for the fast and
precision tracking algorithms respectively. The TRT extension for the
precision tracks takes a total of approximately 10~ms per jet \roi, as
illustrated in Figure~\ref{fig:jettimec}.

Compared to the tracking itself, the execution time for the vertexing is
rather fast. Shown in Figure~\ref{fig:vtxtimec}, the execution time for the
offline based vertex algorithm is a little over 1~ms, whereas the
histogram based vertex algorithm takes less than 0.5~ms. Also shown in
Figure~\ref{fig:vtxtime} are the times for the execution of the vertex
algorithm in the muon, electron and tau signatures. In each case of the
single-\roi leptonic triggers the
offline based algorithm takes less than 1~ms.

\floatbarra
 
\subsection{Total tracking time per event}
 
The total time spent by the tracking related algorithms in the event processing from
this same typical run from September 2018 can be seen in
Figure~\ref{fig:pileuptime}.

This figure shows the average of the total execution time per event for each
algorithm of the
ID trigger, and how this varies as the mean \pileup
interaction multiplicity changes throughout the run. The total time for each algorithm
is dependent on the details of the trigger menu, which determines how often and for which
trigger thresholds the tracking will be executed.
 
\figpileuptime
 
The execution times for the data preparation in the silicon detectors are clearly seen
to decrease for falling \pileup multiplicities. As expected from the execution time
for each algorithm for the different signatures, the data preparation for the pixel detector
is the slowest, taking 27~ms overall in the trigger per event for the high interaction
multiplicities, with $\mubar=52$, at the start of the fill.
 
The fast tracking is clearly the longest with a mean of approximately 155~ms
per event in total at high interaction multiplicities, and has a strong non-linear behaviour
with increasing multiplicities. The precision tracking in contrast has a more linear dependence,
and takes only a third as long as the fast tracking in total. This is due to the
time per call being lower, and the precision tracking
being executed less often due to the early rejection in the
trigger, which terminates processing before running the precision tracking algorithm.

The TRT data preparation and track extension are both reasonably fast and exhibit a more
linear behaviour, since they are only executed on tracks from the precision tracking.
 
Finally, the vertex algorithms are seen to be very fast overall. The histogramming algorithm
is in principle only executed once per event and, as in the previous section, is seen to
have a mean time per call at high \pileup of around 0.3 ms, exhibiting only a slight
dependence on the \pileup interaction multiplicity. In contrast, the offline vertex algorithm also runs
once per event for the \bjet vertexing, but in addition is executed once per \roi for the other
signatures. The execution time shows a small dependence on the
interaction multiplicity, and consists mostly of the time taken by the single execution
in the \bjet vertex tracking.
 
Overall, the total time spent in all the tracking related algorithms discussed here is
approximately 290~ms per event at high \pileup multiplicities, decreasing to a little
more than 90~ms for the lower multiplicities.



\section{High-level trigger tracking performance}

\subsection{Data selection}
 
For the performance analyses of the muon and electron signatures, the full available integrated
luminosity for the 2016--18 running period is used. This is possible since the processing for these
signatures did not change significantly during  the data taking period.
 
For the case of the tau and \bjet analyses, which ran two-stage tracking with a modified
second stage \roi, significant changes were made to the reconstruction of either the first,
or the second stages over the three years. These included modifications to the second stage
seed finding, and so only the results for 2018 are presented in full detail.
For each of the three years, 2016, 2017, and 2018 individually, brief comparisons of the
efficiencies in each year are presented.
For the tau signature, the full available integrated luminosity from each year is used. For the \bjet
analyses only a subsample from each year, sufficient to provide a small statistical
uncertainty is used.
 
In all cases, only events where the data quality was determined to be good for physics
analyses are used~\cite{DAPR-2018-01}.

The tracking efficiency for the trigger with respect to the offline tracking
is determined using a number of support triggers.
These triggers are essentially identical to the physics triggers,
and operate by reconstructing the tracks in the trigger
as normal, but then selecting on the objects reconstructed in the
muon spectrometer or calorimeter only, with no selection on
the tracking information from the inner detector.
In this way, it is possible to estimate the efficiency of the tracking,
unbiased by the ID track reconstruction itself.
One caveat is that because the trigger does not select on the
inner detector track information, the prescale for the triggers
generally needs to be kept high to restrict the rate. Since the trigger
objects used for the selection do not include any tracking, the fraction
of background selected by the trigger will of necessity
be significantly higher than the fully selecting triggers, {\em e.g.} electron candidates
from these triggers will have been selected without any requirement on
track multiplicity, instead being based purely on the clusters reconstructed in the
calorimeter.
As a consequence there
may be a large contribution from narrow QCD jets which resemble an
electromagnetic cluster. For each of the analyses, any such background is
reduced by the selection of a reference sample using
the full offline reconstructed objects, against which the performance
is measured. Consequently, the sample of good offline objects, including
the offline tracks,  that are used to monitor
the trigger may be statistically limited in regions of phase space.
 
\subsection{Track selection}
 
For the measurement of the efficiency and resolution, the ensemble of
inner detector trigger tracks within the \roi are first matched to the selected offline
reference objects using a modified  solution to the
stable marriage problem~\cite{GS62}\emdash the closest matches  taken only from those
tracks which match within a loose preselection cone of size
$\Delta R = \sqrt{(\Delta\eta)^2+(\Delta\phi)^2} = 0.05$ around an offline
reconstructed track. For any given offline track, the closest matching trigger
track is chosen as a match if, and only if, it does not match more closely any other
offline track.
 
Efficiencies are then measured by taking the correlated ratio of the number of
offline reference objects that have such a matched trigger track to the total number of
offline objects  passing the selection.
 
Since the residual distributions themselves may have very long non-Gaussian tails, the
resolution is estimated by taking the root-mean-square of the central 95\% of the
residual distribution. This is then scaled by the inverse of the root-mean-square of the
central 95\% of a Gaussian distribution with unit standard deviation,
such that for any Gaussian residual distribution the true width would be obtained.

Reference objects for the analyses are  chosen to satisfy the full offline selection
criteria for such objects, similar to those used for  physics analyses, but with additional
selection. These are discussed in detail later in the text.

In all cases, offline candidate tracks are required to lie within the \roi used for the trigger
reconstruction, since tracks outside the \roi would not be reconstructable.  An additional
requirement ensures that offline tracks are used only if the transverse impact parameter
for the point of closest approach to the \beamline, and the $z$ position relative to any
relevant offline vertex, are both less than 10~mm.
 
For the remaining discussion no distinction is made between positively
and negatively charged particles for the transverse momentum and only the absolute
value is considered.

 
\subsection{Tracking in the muon signature}

For the muon signature the HLT uses triggers~\cite{TRIG-2018-01,TRIG-2016-01}
with muon transverse momentum reconstruction down to 4\gev.
The performance of the trigger tracking is therefore studied using a range of support
triggers covering the region down to this threshold.
 
\begin{figure*}[tp]
\centering
\subfloat[]{\includegraphics[width=0.5\textwidth]{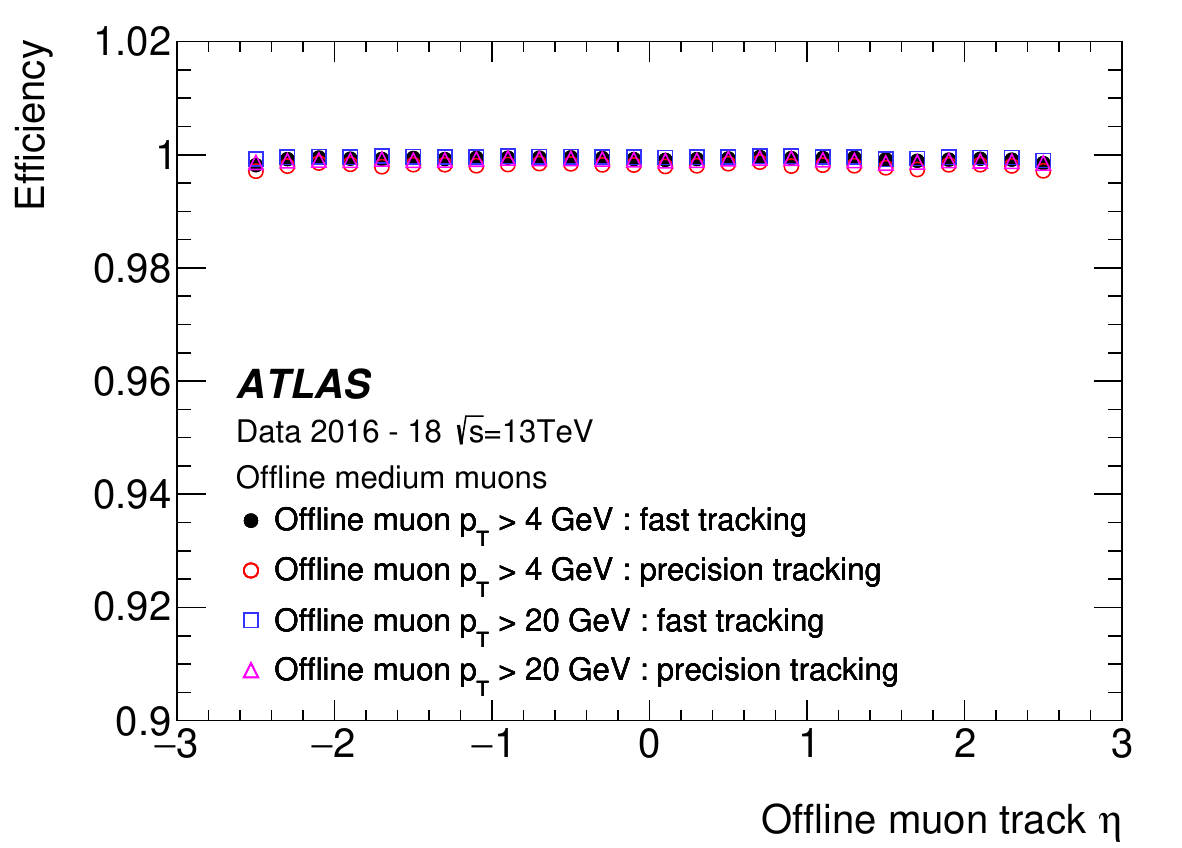}}
\subfloat[]{\includegraphics[width=0.5\textwidth]{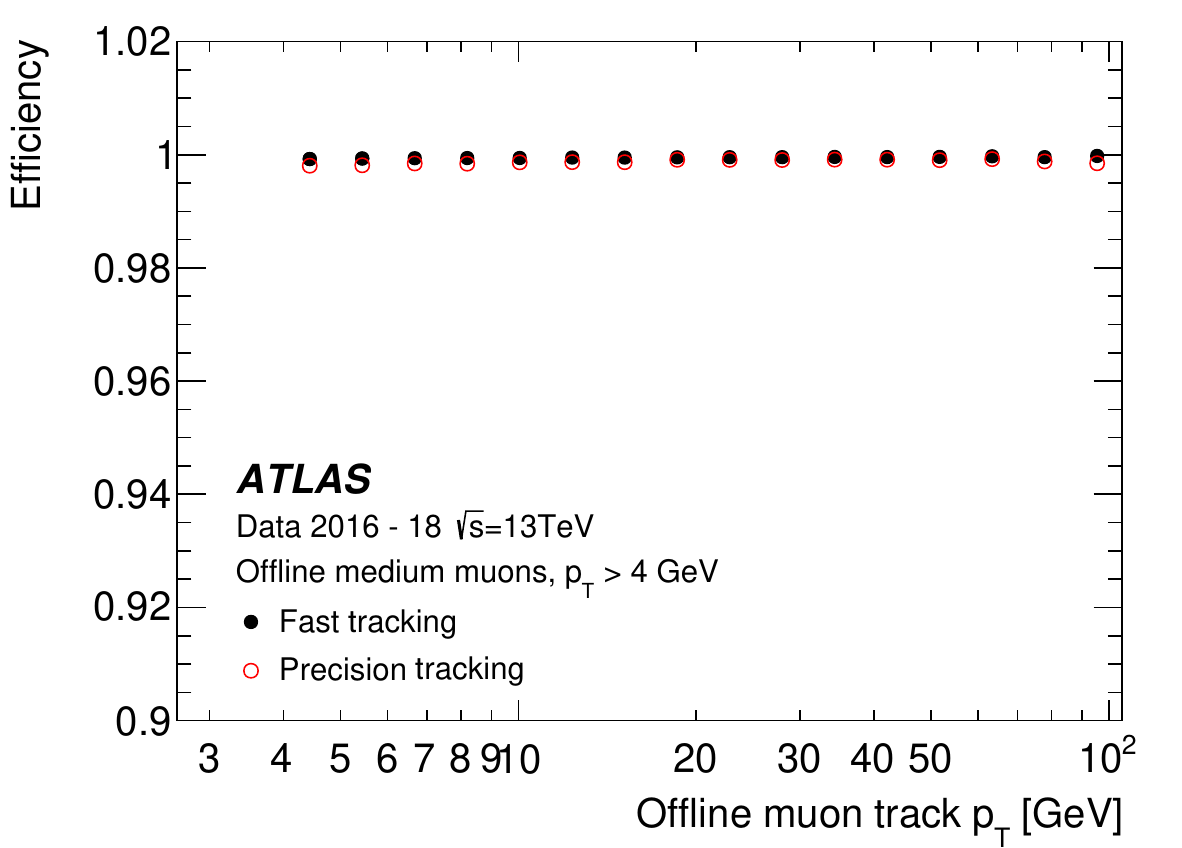}}\\
\subfloat[]{\includegraphics[width=0.5\textwidth]{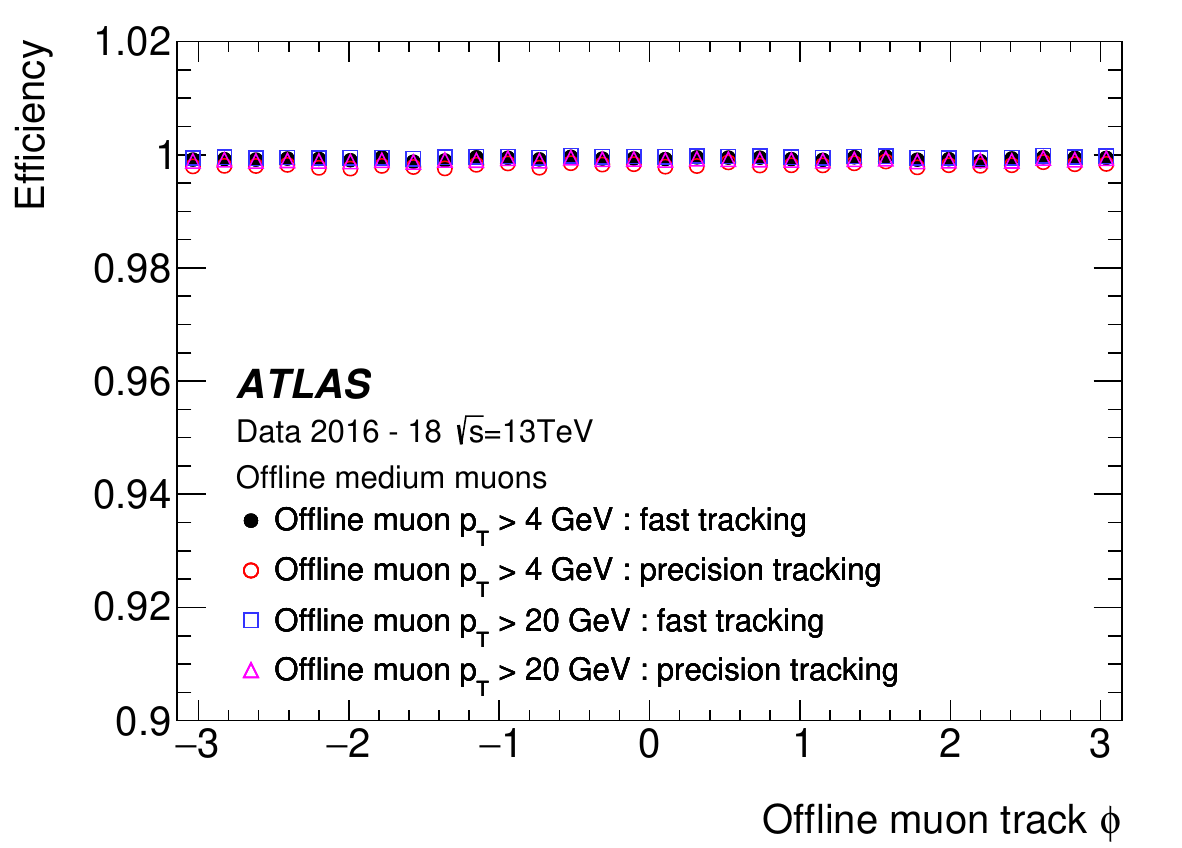}}
\subfloat[]{\includegraphics[width=0.5\textwidth]{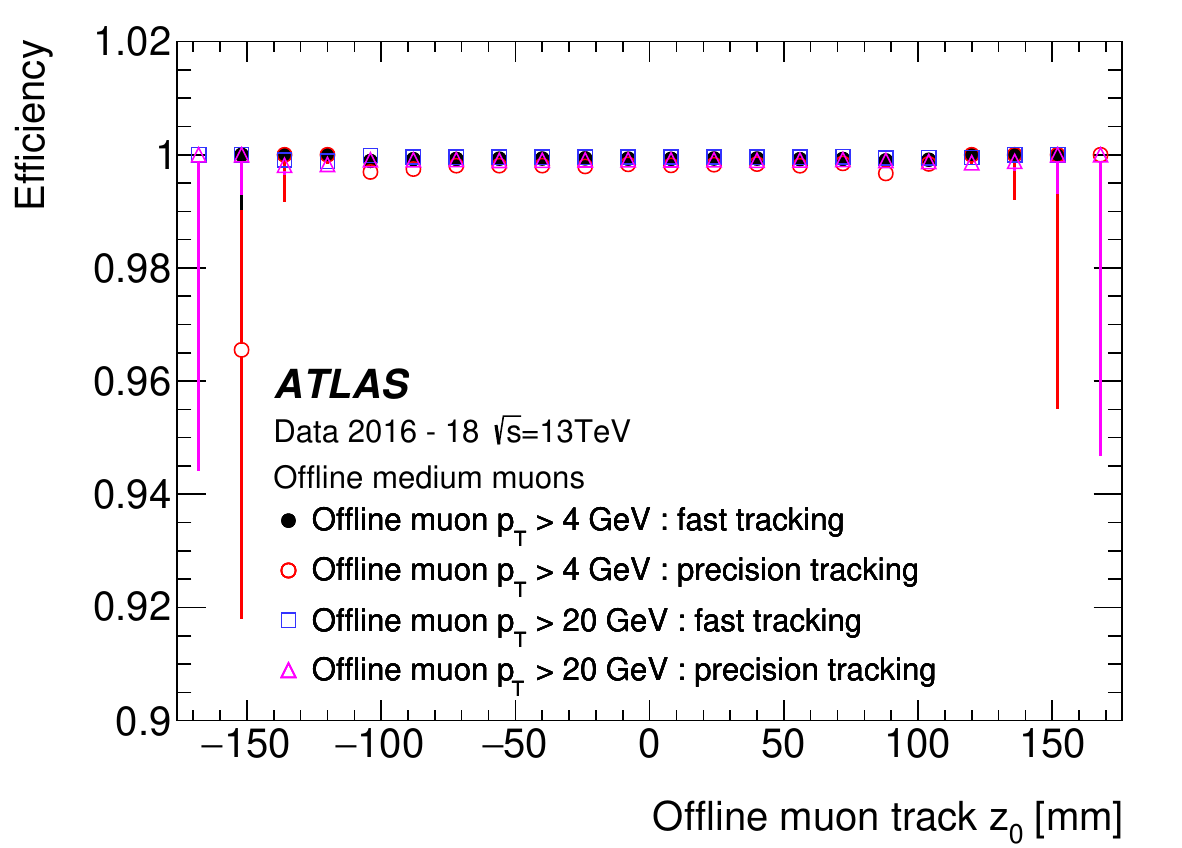}}\\
\subfloat[]{\includegraphics[width=0.5\textwidth]{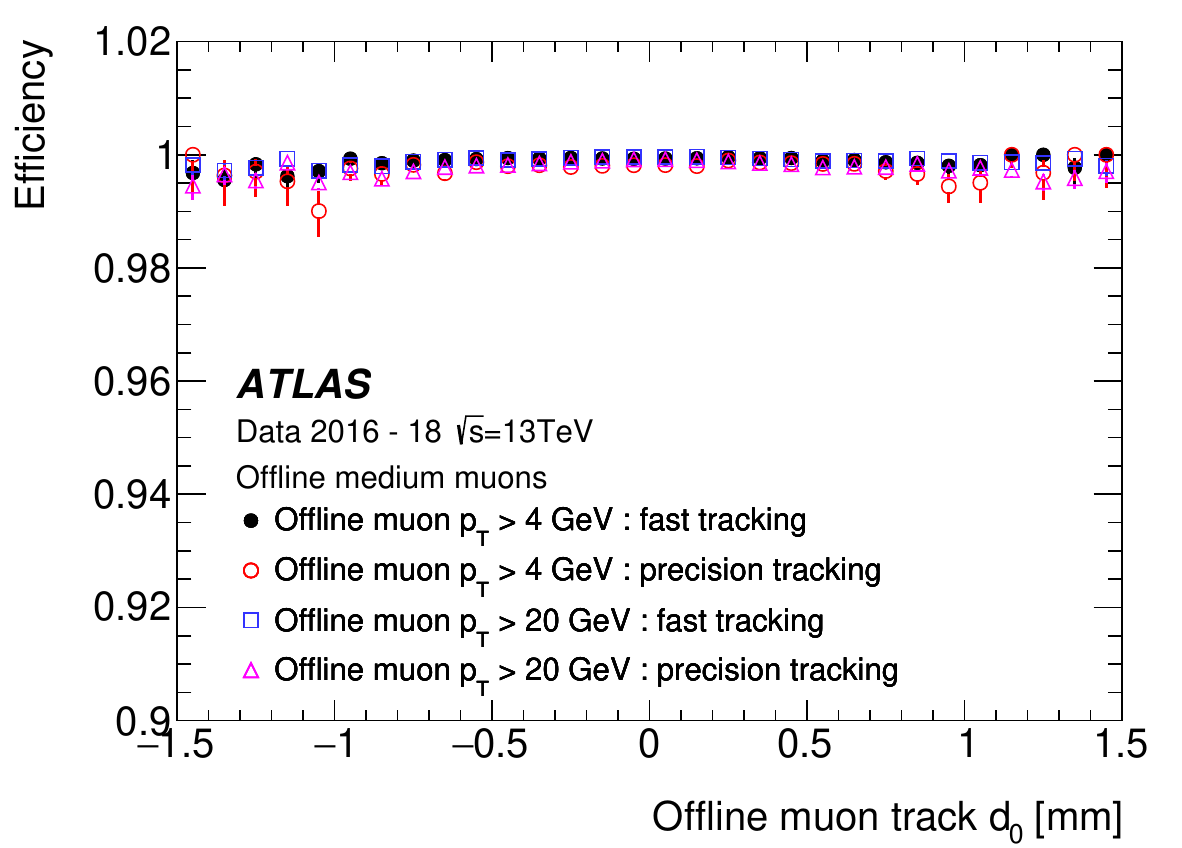}}
\subfloat[]{\includegraphics[width=0.5\textwidth]{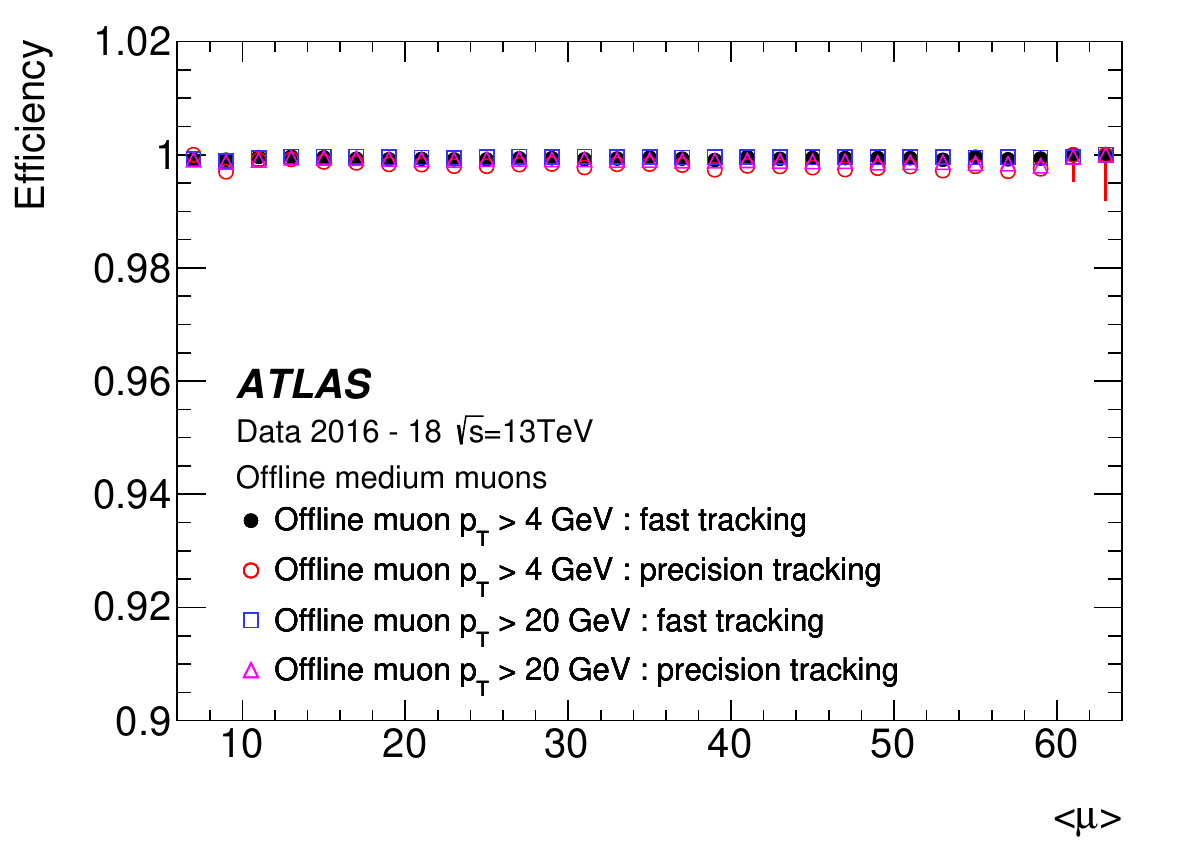}}
\caption{
The ID tracking efficiency for muons selected by the \SI{4}\gev and \SI{20}\gev  muon support triggers,
with respect to medium offline muon candidates with $\pt>\SI{4}\gev$ or $\pt>\SI{20}\gev$.
The efficiency is shown as a function of:
(a) the offline reconstructed muon $\eta$,
(b) the offline reconstructed muon $\pt$,
(c) the offline reconstructed muon $\phi$,
(d) the offline reconstructed muon $z_{0}$,
(e) the offline reconstructed muon $d_{0}$, and
(f) the mean number of \pileup interactions, \mubar.
Efficiencies are shown for both the fast tracking and precision tracking algorithms.
Statistical Bayesian uncertainties are shown, using a 68.3\% confidence interval
and uniform beta function prior.}
\label{fig:muoneff}
\end{figure*}

Figure~\ref{fig:muoneff} shows the tracking efficiency of the ID
trigger  for medium quality offline muon candidates~\cite{PERF-2015-10}.
The efficiency integrated over \pt is shown for
the fast tracking and the  precision tracking for two representative
thresholds for the offline muon selection: $\pt>\SI{4}\GeV$ corresponding to the
lowest trigger threshold, and  $\pt>\SI{20}\GeV$.  The efficiency is shown
as a function of the offline muon candidate track parameters
and the mean number of \pileup interactions per event.
The selection of the offline muon candidates is slightly biased by
the muon spectrometer trigger selection near the threshold, but this does
not affect the measurements of the inner detector trigger performance.
The efficiency is shown for both the fast and precision tracking and is
significantly better than 99\%. The efficiency is observed
to be constant as a function of the mean \pileup interaction multiplicity and other
variables, with perhaps a slight decrease in efficiency for tracks with
a large impact parameter.

\begin{figure*}[p]
\centering
\subfloat[]{\includegraphics[width=0.5\textwidth]{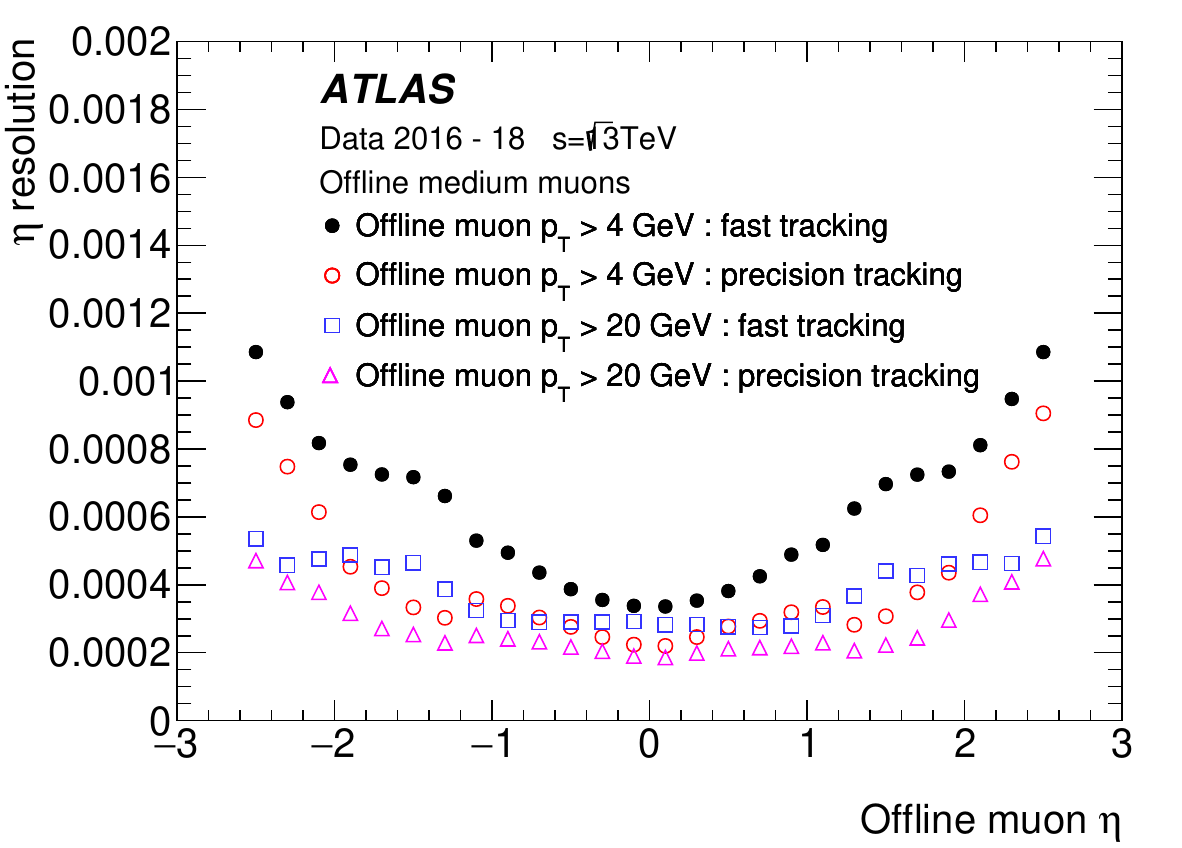}}
\subfloat[]{\includegraphics[width=0.5\textwidth]{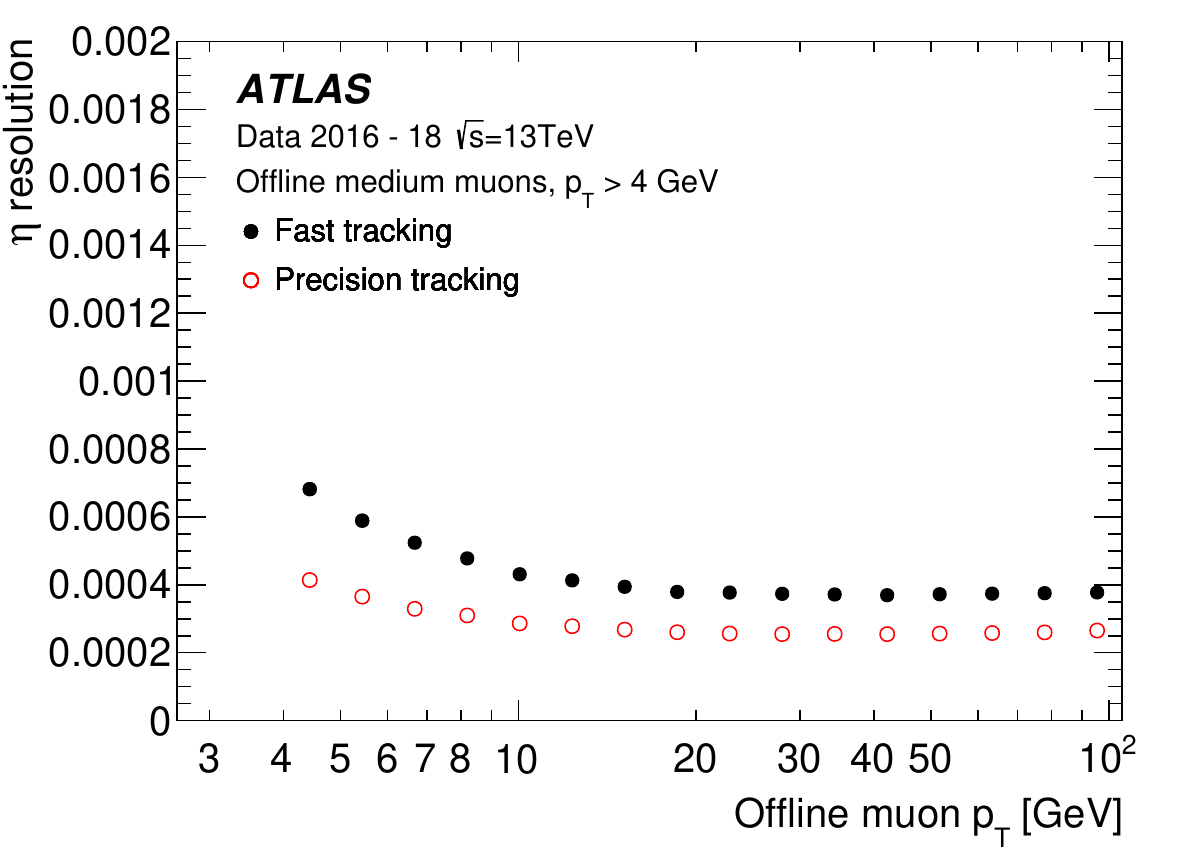}}\\
\subfloat[]{\includegraphics[width=0.5\textwidth]{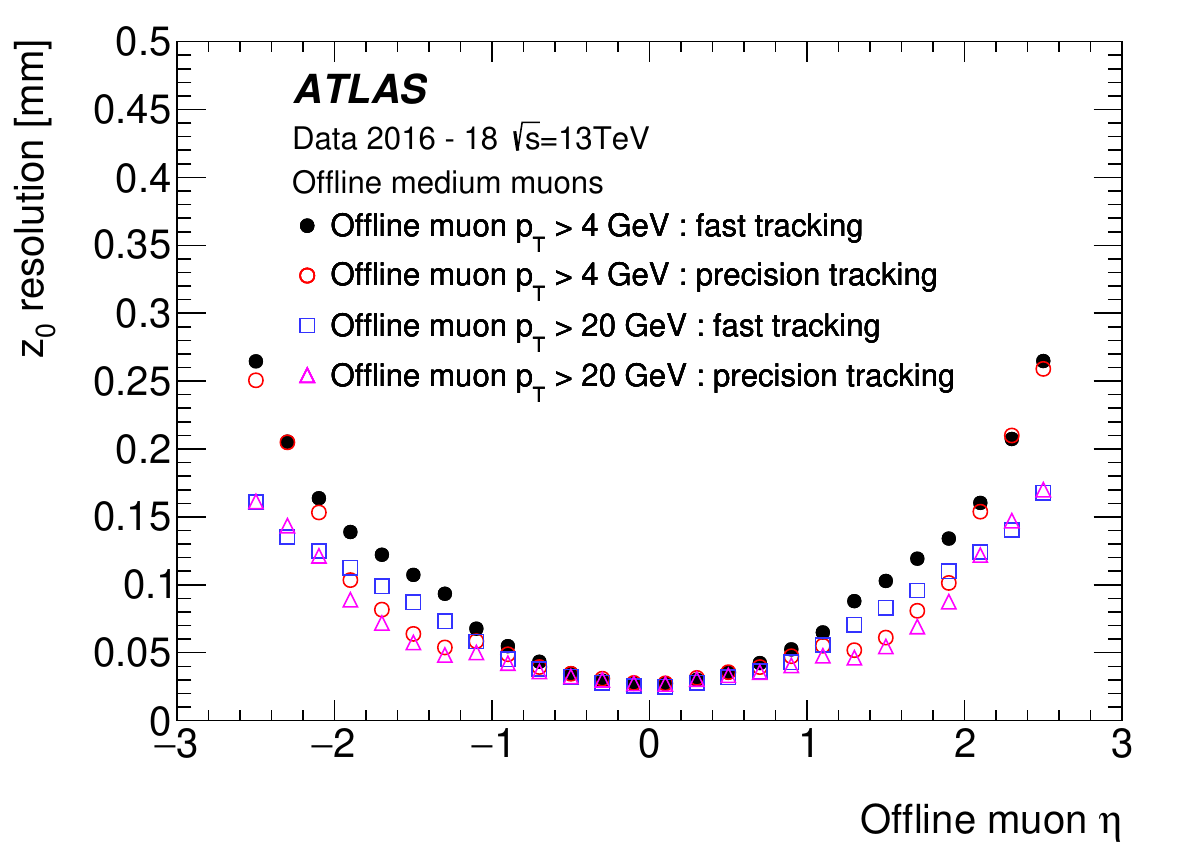}}
\subfloat[]{\includegraphics[width=0.5\textwidth]{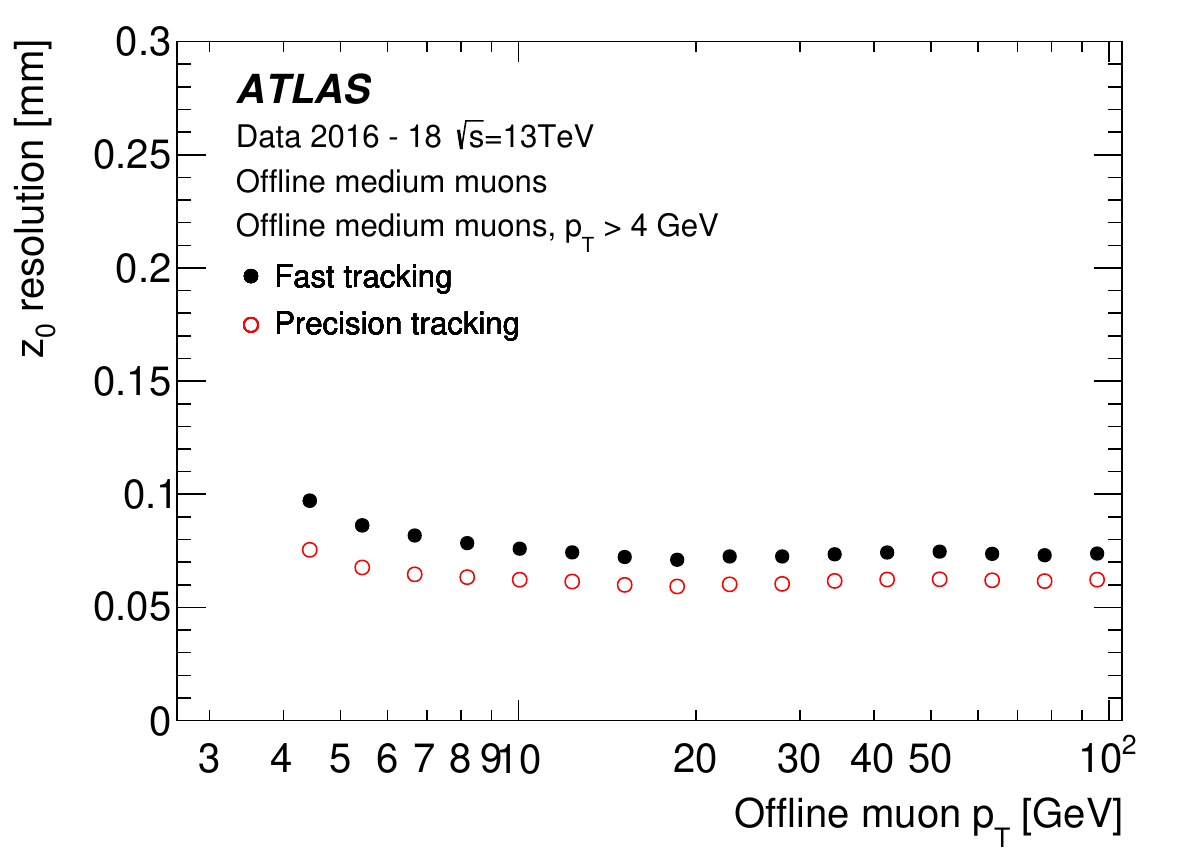}}\\
\subfloat[]{\includegraphics[width=0.5\textwidth]{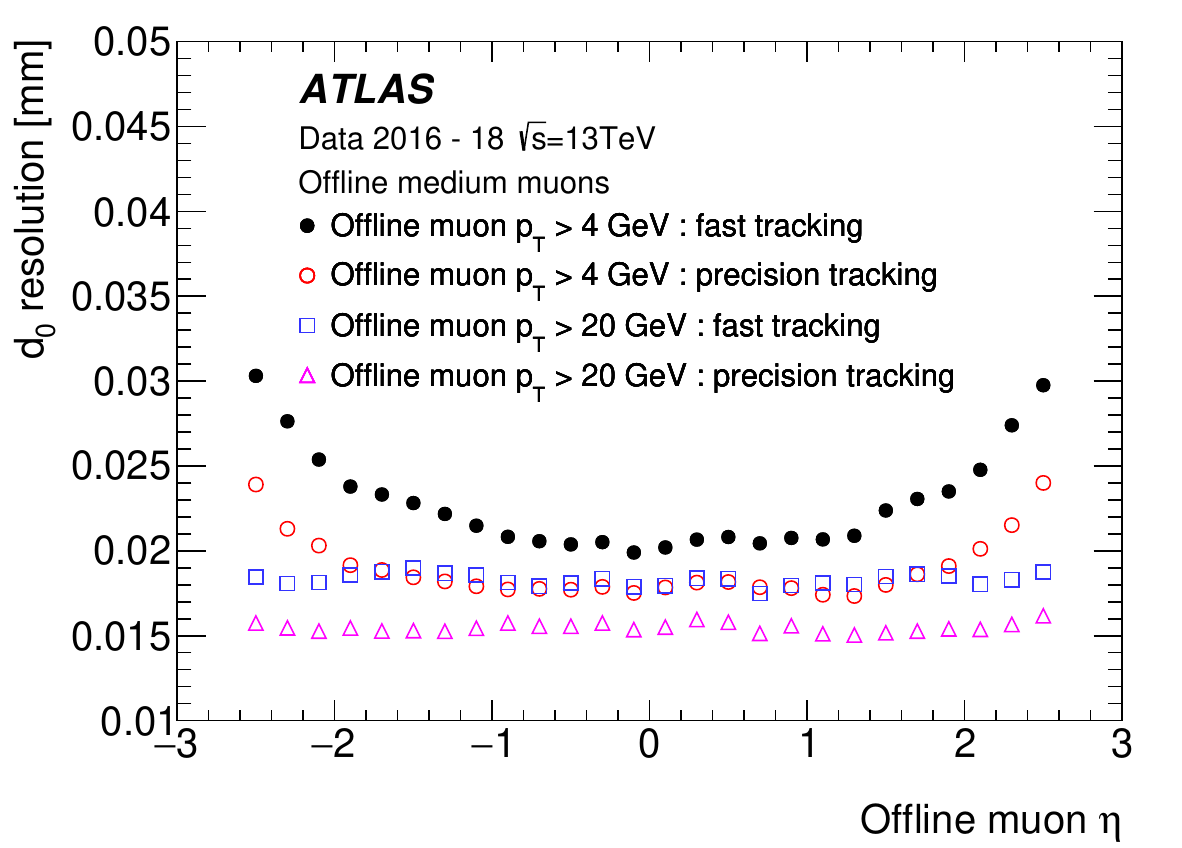}}
\subfloat[]{\includegraphics[width=0.5\textwidth]{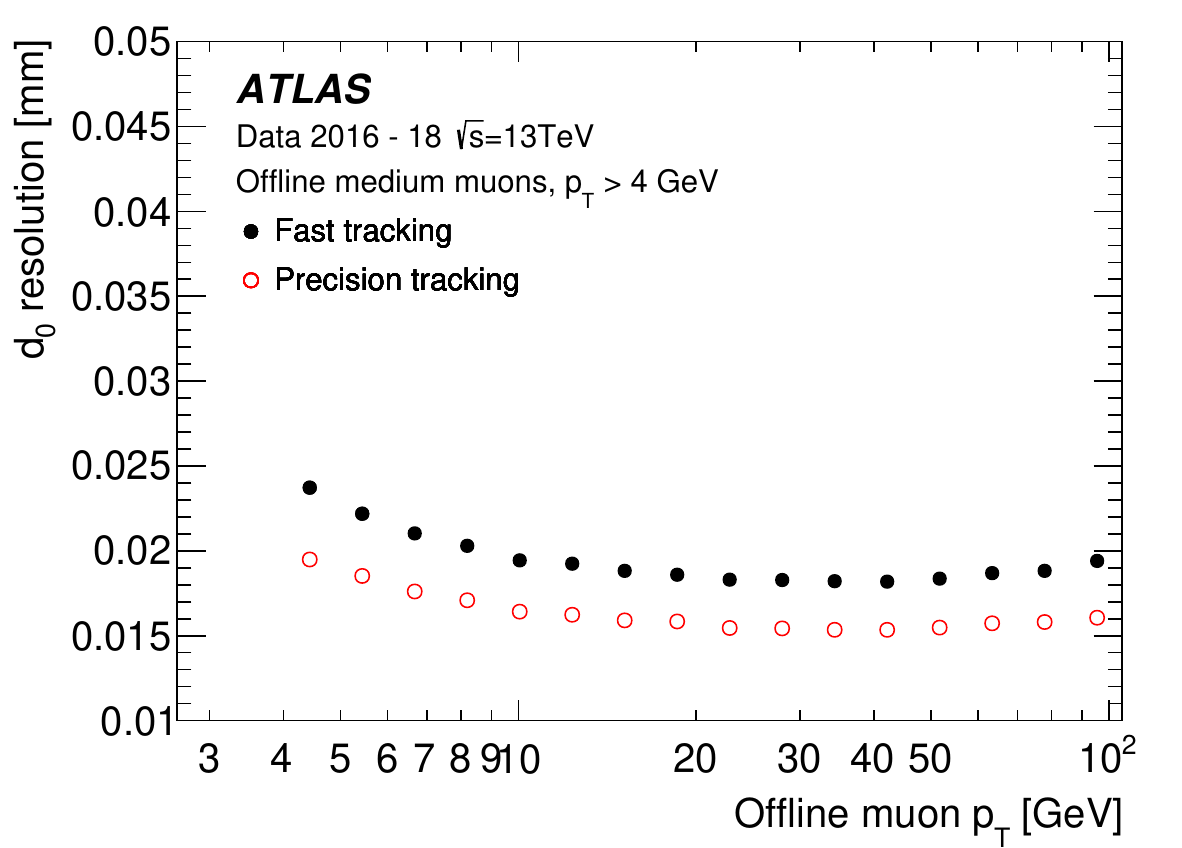}}
\caption{ The track  resolution for pseudorapidity ($\eta$), track
$z$-vertex position, and transverse impact parameter  with respect
to the \beamline, $d_{0}$, as a function of offline track
pseudorapidity and transverse momentum. The resolutions are shown
for muons selected by the 4\gev and 20\gev muon
support triggers for both the fast tracking and precision tracking
algorithms. }
\label{fig:muonres}
\end{figure*}

Shown in Figure~\ref{fig:muonres}
are the resolutions for the trigger tracks in  $\eta$, $z_{0}$, and
$d_{0}$  with respect to the offline muon candidate pseudorapidity on the left, and
with respect to the offline muon candidate  \pt on the right.
The resolution with respect to $\eta$ is shown for the precision tracking and the fast
tracking for the same two representative thresholds,  $\pt>\SI{4}\GeV$ and \SI{20}{\GeV},
used in the efficiency determination.
The resolution is seen to deteriorate somewhat for larger pseudorapidities, largely
due to the larger amount of material through which the muons must pass as they leave the beam pipe
and traverse the inner detector. The endcap silicon detectors are arranged perpendicular to the beam
to partially ameliorate this, the effect of which can be seen for absolute pseudorapidities larger than 1.2
which is approximately the boundary between the barrel and endcap silicon detectors.

\newcommand{\emuonfive}{
\begin{figure}[thp]
\centering
\subfloat[]{\includegraphics[width=0.5\textwidth]{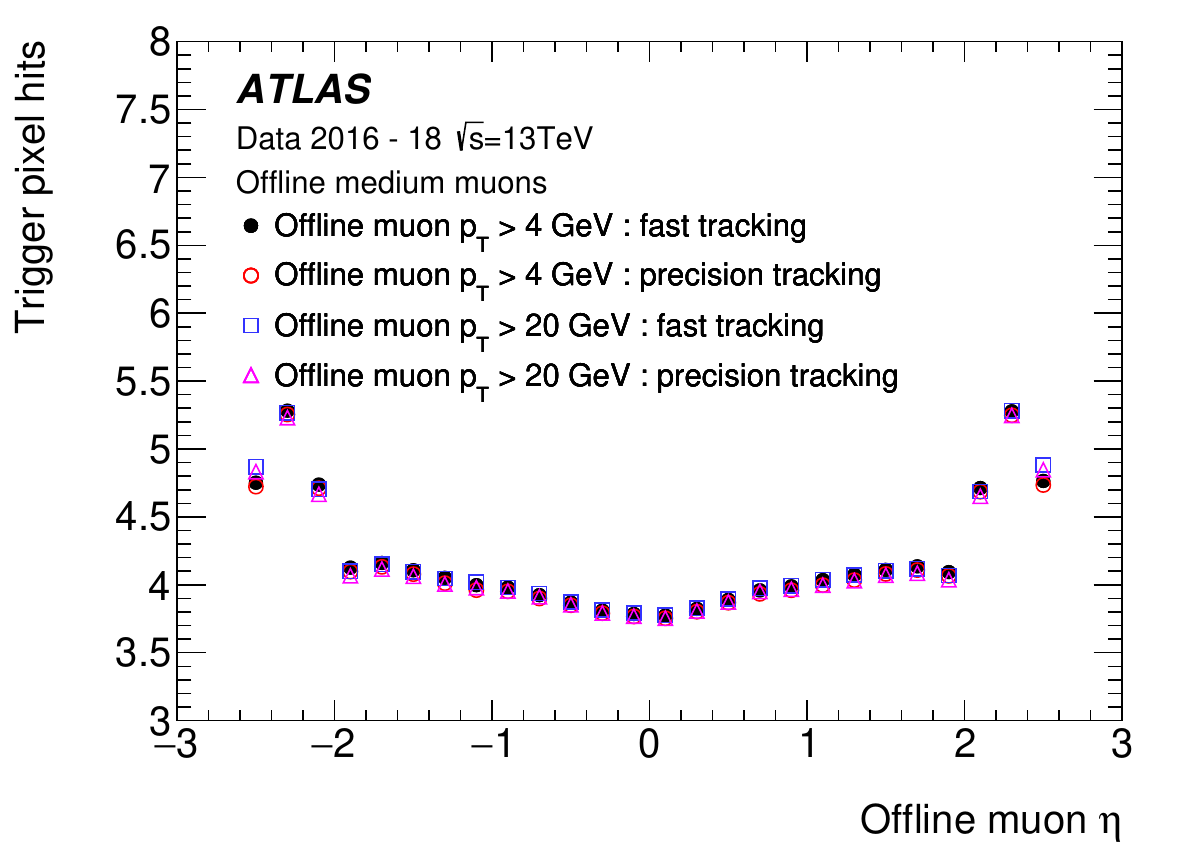}}\figbreak
\subfloat[]{\includegraphics[width=0.5\textwidth]{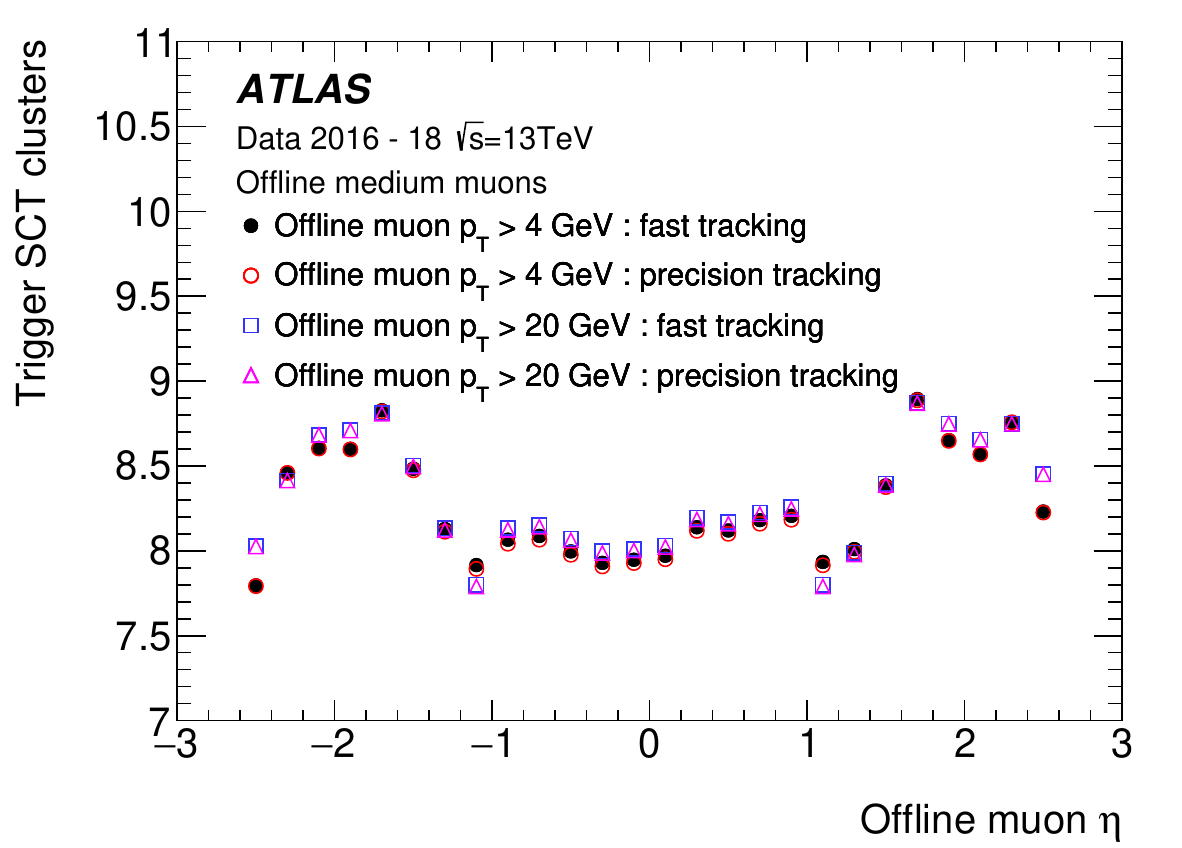}}\\
\subfloat[]{\includegraphics[width=0.5\textwidth]{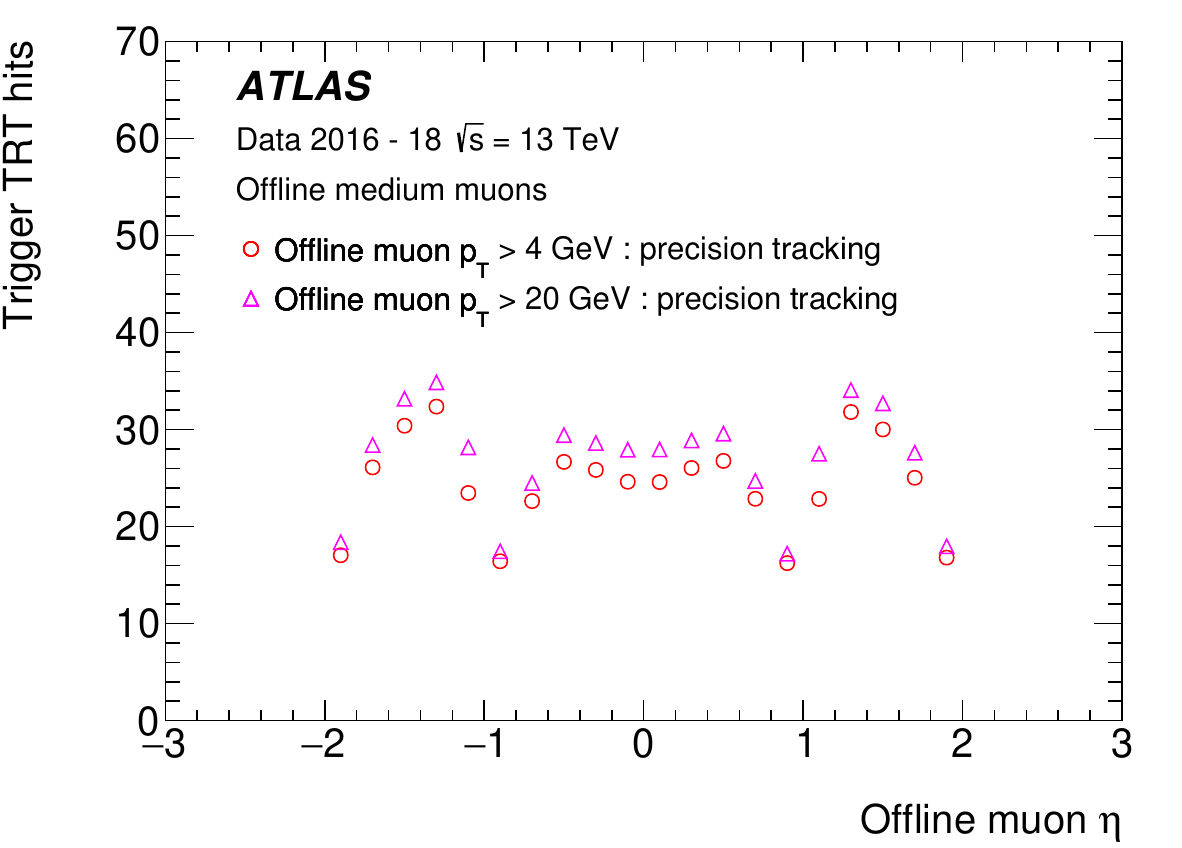}}
\caption{The mean multiplicities for (a) pixel, (b) SCT,  and (c) TRT  hits on trigger tracks matched to
offline tracks as a function of the matching offline
track pseudorapidity. Shown are the hit multiplicities for the precision and fast tracking for the pixel and SCT.
The TRT extension does not run for the FTF tracks and so the multiplicity for the FTF is not shown in (c).}
\label{fig:muonspfive}
\end{figure}}

\emuonfive
 
These effects on the resolutions can be further understood by examination of the mean
number of hits on the tracks.
The mean multiplicity of pixel, SCT and TRT hits on trigger muon candidate tracks
which are matched to offline muon candidates can be
seen in Figure~\ref{fig:muonspfive} as a function of the pseudorapidity of the offline
muon candidate. For the TRT multiplicity, only the precision tracks are shown since the
TRT extension is not performed for the FTF tracks.
 
For central pseudorapidities there are  approximately four pixel hits per track originating from
the four pixel barrel layers. For $|\eta|>1.9$, which delimits approximately the end of the
acceptance for the outer pixel barrel layer,
the multiplicity increases as the tracks transition into the
endcap detectors. It should be noted that the acceptance of the IBL and the innermost of the
three original pixel barrel layers extends beyond $|\eta|=2.5$. This is the reason
that the mean pixel cluster multiplicity for tracks at large absolute pseudorapidity is greater than four.
 
Because of the much larger radius of the outer SCT layer, the transition between barrel and
endcap for tracks in the SCT begins much earlier in pseudorapidity, around $|\eta|=1.2$.
The four double layers of the barrel are responsible for the mean of eight SCT clusters in the barrel
region, and the nine additional wheels in each endcap allow the tracks to maintain a mean of eight
of more clusters out to the maximum acceptance at $|\eta|=2.5$.
It is largely the transition from SCT barrel to
endcap which is responsible for the structure in the resolutions as the tracks transition from
pure barrel hits, to a combination of barrel and endcap hits, and finally, to purely endcap hits
for $|\eta|>1.6$. The small asymmetry between positive and negative pseudorapidities is due
to small asymmetries in the detector conditions regarding the detector response and non-operational
modules.
 
The structure of the TRT is somewhat different\emdash there are two barrel
sections with only axial
straws providing only $r$--$\phi$ hit information -- and two TRT endcaps which provide
only $z$--$\phi$ information. The $z$ position for the end of the TRT barrel is approximately
the same as for the SCT barrel, and the endcaps extend to approximately the same maximum
$z$ position as the SCT endcaps, such that
tracks with $|\eta|>2.1$ have few or no TRT hits. The drop in hit multiplicity for $|\eta|$ around 0.8
is predominantly due to the transition from the barrel to the endcap TRT
at $\pm$800\,mm in $z$.
A small number of tracks at all pseudorapidities
also have no TRT hits. This is more likely for tracks with low transverse momentum, such that the
contribution of tracks with no TRT hits reduces the apparent mean hit multiplicity over all tracks and is
responsible for the lower mean TRT multiplicity for the lower \pt muon trigger.
 
In general, the track hit multiplicities are broadly similar for all signatures
for full-length tracks. There is a notable exception, however, for electron candidates where the hit
multiplicities can be affected by bremsstrahlung. This is discussed in Section~\ref{sec:egamma}.

\label{sec:muontnp}
 
In addition to the standard performance triggers, which select
on tracks reconstructed using the muon spectrometer only,
there are additional dimuon performance triggers. These
select events with two muon candidates consistent with both muons
arising from the decay of a $Z$ boson, by selecting candidates with
a dimuon mass within a window around 91\gev with the range $70 < m_{ll} < 110$\gev.
For these triggers,
the two muons must each be in a different \roi. One of the \rois\emdash the
{\em tag} \roi\emdash selects on the full muon candidate including the full selection
on the combined spectrometer and inner detector muon candidate,
and the other\emdash the {\em probe} \roi\emdash selects solely on the muon candidate
reconstructed in the spectrometer, without using any inner detector information.
Selecting on the combined muon candidate in the tag \roi
increases the purity of the selected events.
The spectrometer-only selection in the probe \roi  allows the
efficiency to be measured for the probe muon candidate, unbiased by the
track reconstruction in the inner detector itself.
The presence of the first, reasonably pure
fully selected tag muon reduces the rate of such events and the
selection of the two muon candidates having to form a $Z$ boson candidate
greatly enhances the purity of the spectrometer-only probe muon
candidates, particularly at high transverse momenta.
 
For the fully selecting, tag leg, the muon trigger selects muon candidates
with $\pt>13$\gev. The  probe leg, selecting on the muon candidates
reconstructed in the spectrometer only, also selects candidates with
transverse momenta greater than 13\gev.
Unbiased measurements are therefore possible from these triggers for
muon transverse momenta greater than 13\gev. The data used for this
analysis correspond to the full 2018 integrated luminosity.

\newcommand{\muoninvmass}{
\begin{figure}[thp]
\centering
\includegraphics[width=0.5\textwidth]{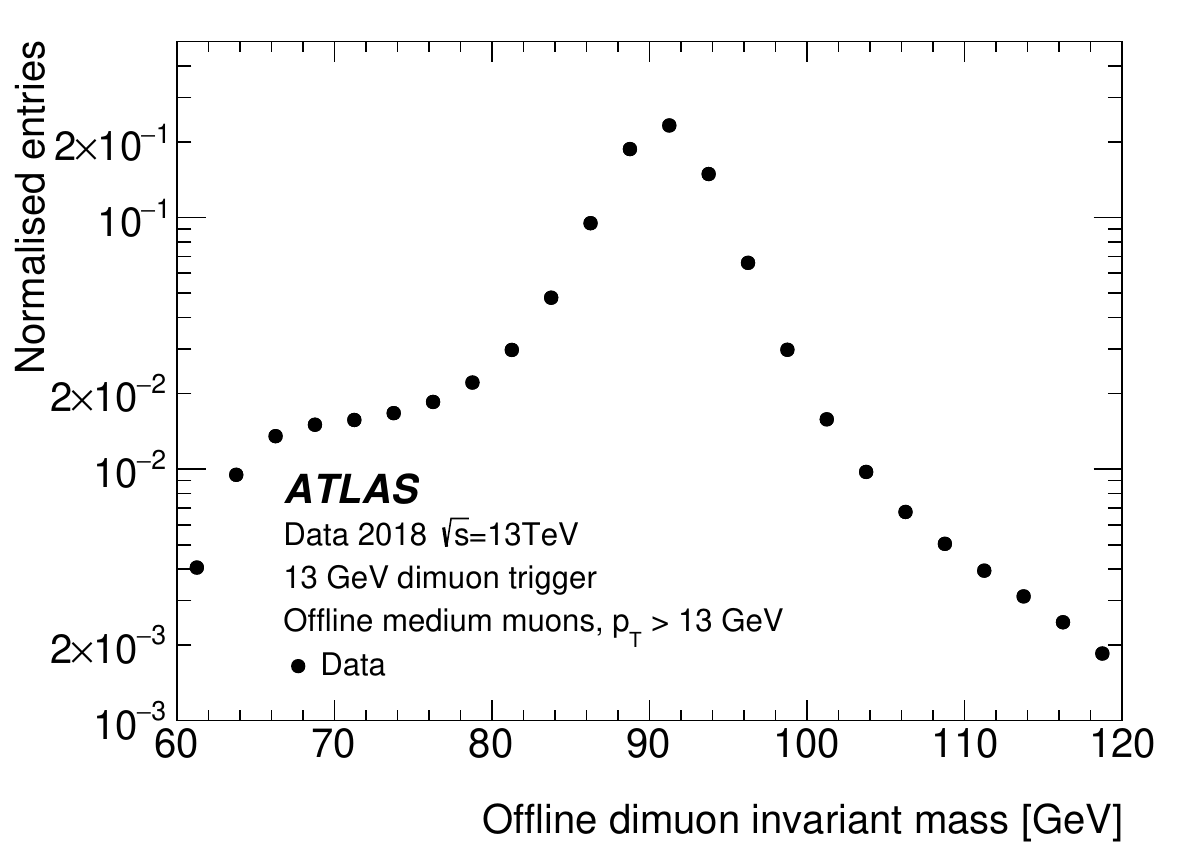}
\caption{The offline dimuon invariant mass from events passing the analysis selection from the
dimuon performance trigger.
For the performance triggers, the RoIs used by the fast and precision tracking are the
same and as such the offline dilepton candidates chosen for the analysis of the fast and
precision tracking are identical.}
\label{fig:muinvmass}
\end{figure}
}
 
\muoninvmass
 
The dilepton invariant mass distribution for offline candidates
for the events passing the
medium offline muon selection, with no additional background rejection, for the dimuon performance
trigger but with no corrections for trigger efficiency or detector response
can be seen in Figure~\ref{fig:muinvmass}.

\begin{figure*}[thp]
\centering
\subfloat[]{\includegraphics[width=0.5\textwidth]{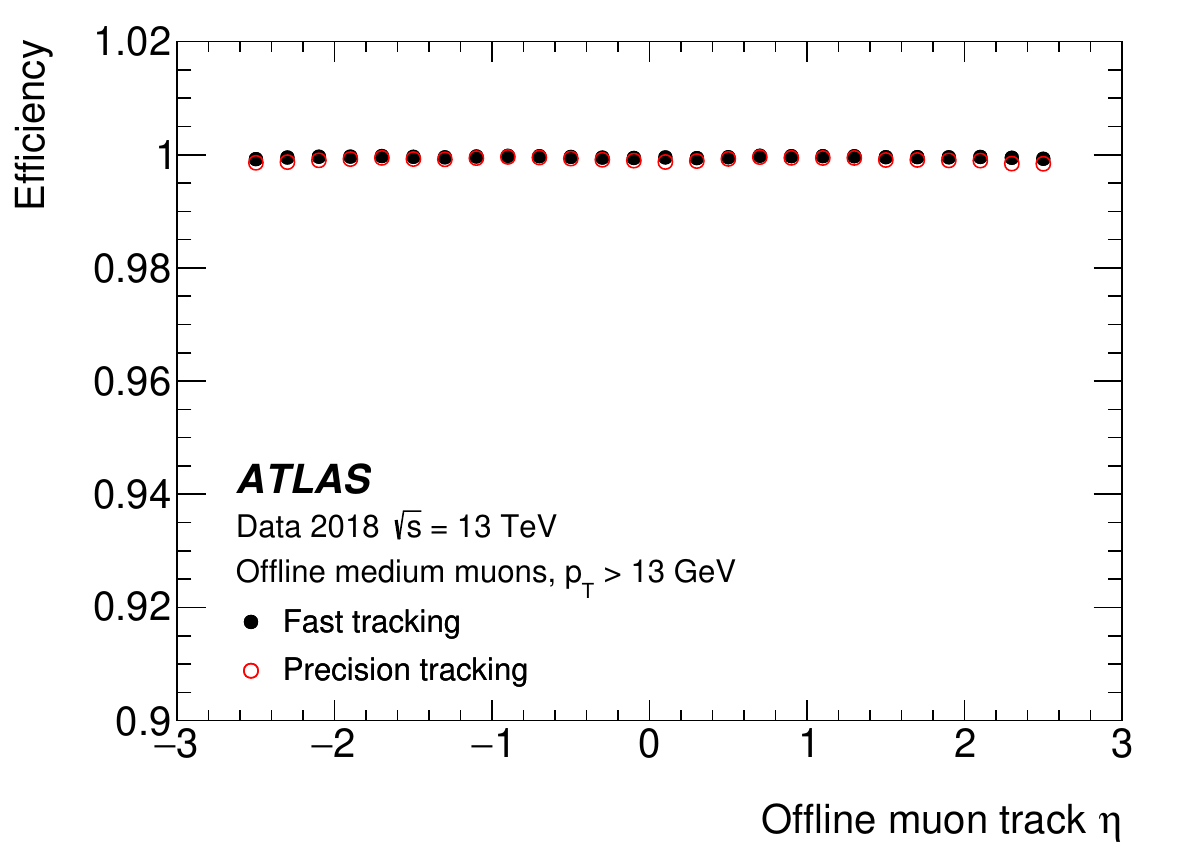}}
\subfloat[]{\includegraphics[width=0.5\textwidth]{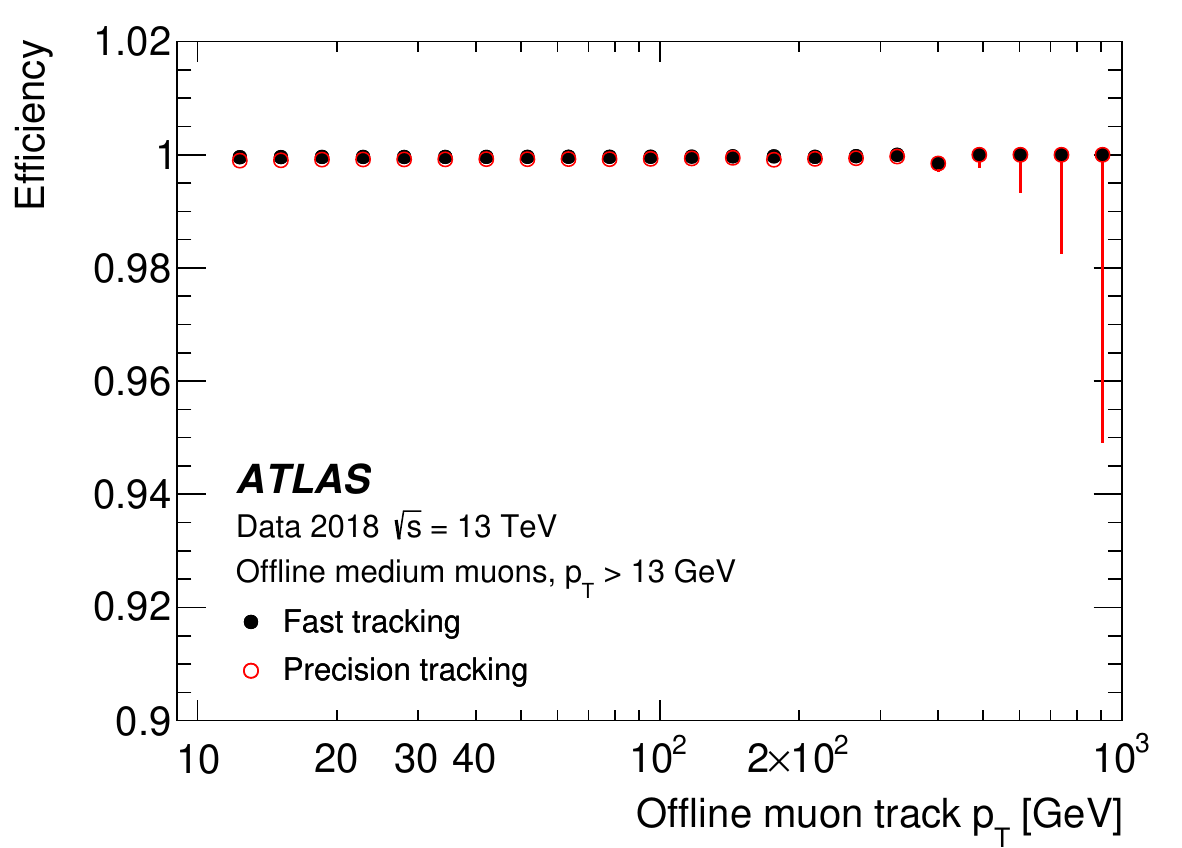}}\\
\subfloat[]{\includegraphics[width=0.5\textwidth]{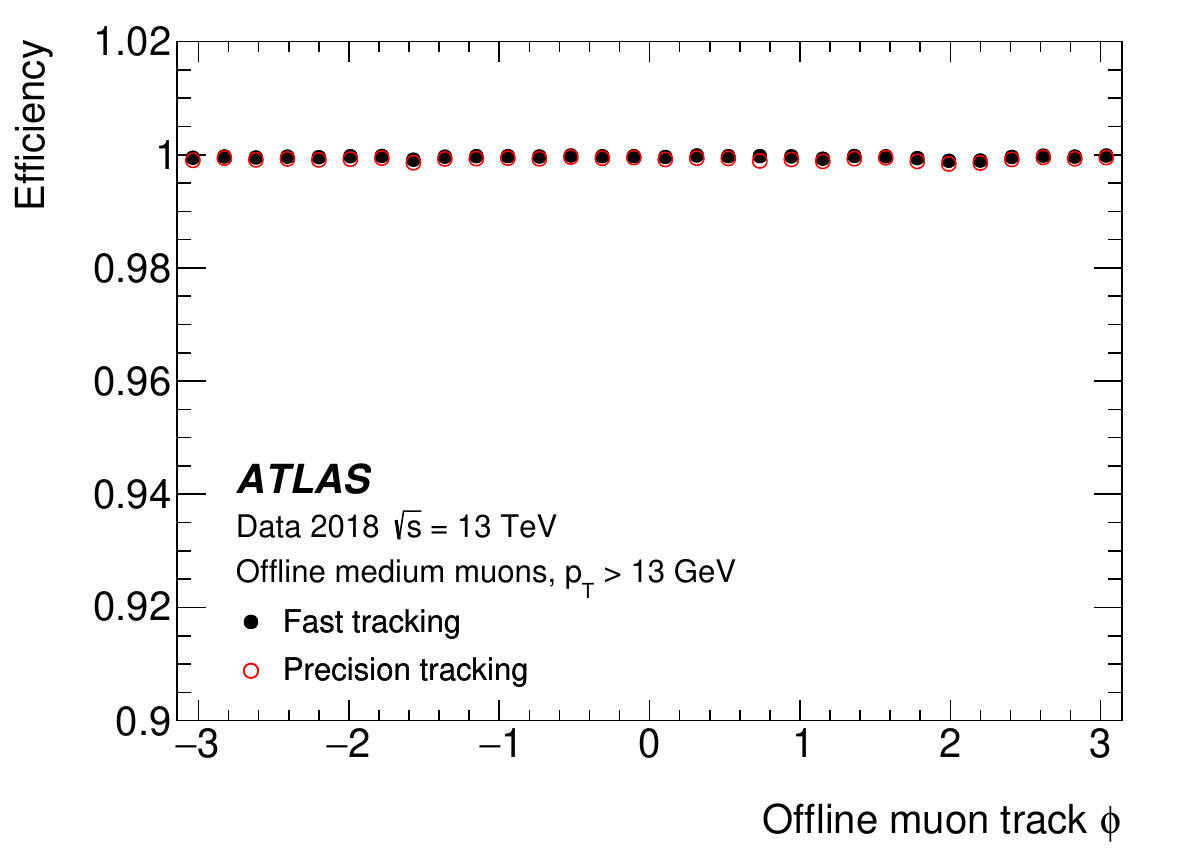}}
\subfloat[]{\includegraphics[width=0.5\textwidth]{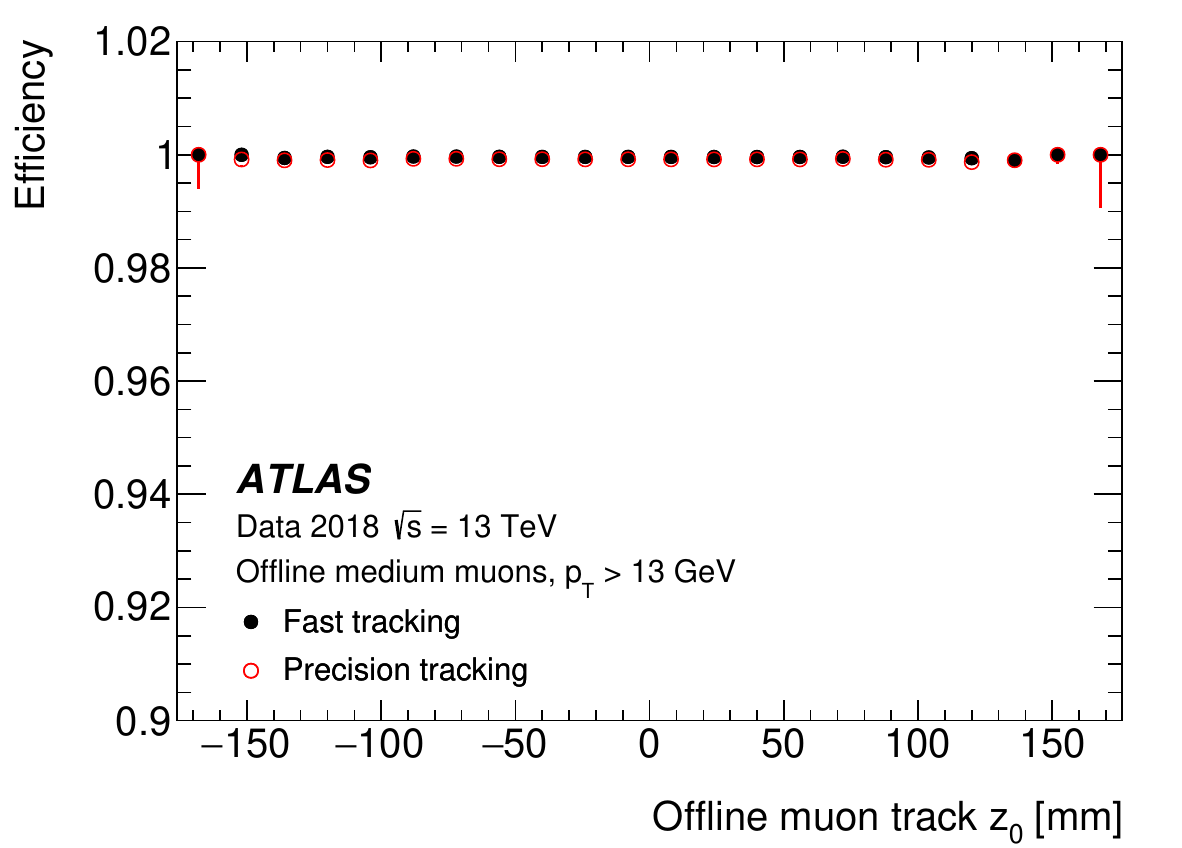}}\\
\subfloat[]{\includegraphics[width=0.5\textwidth]{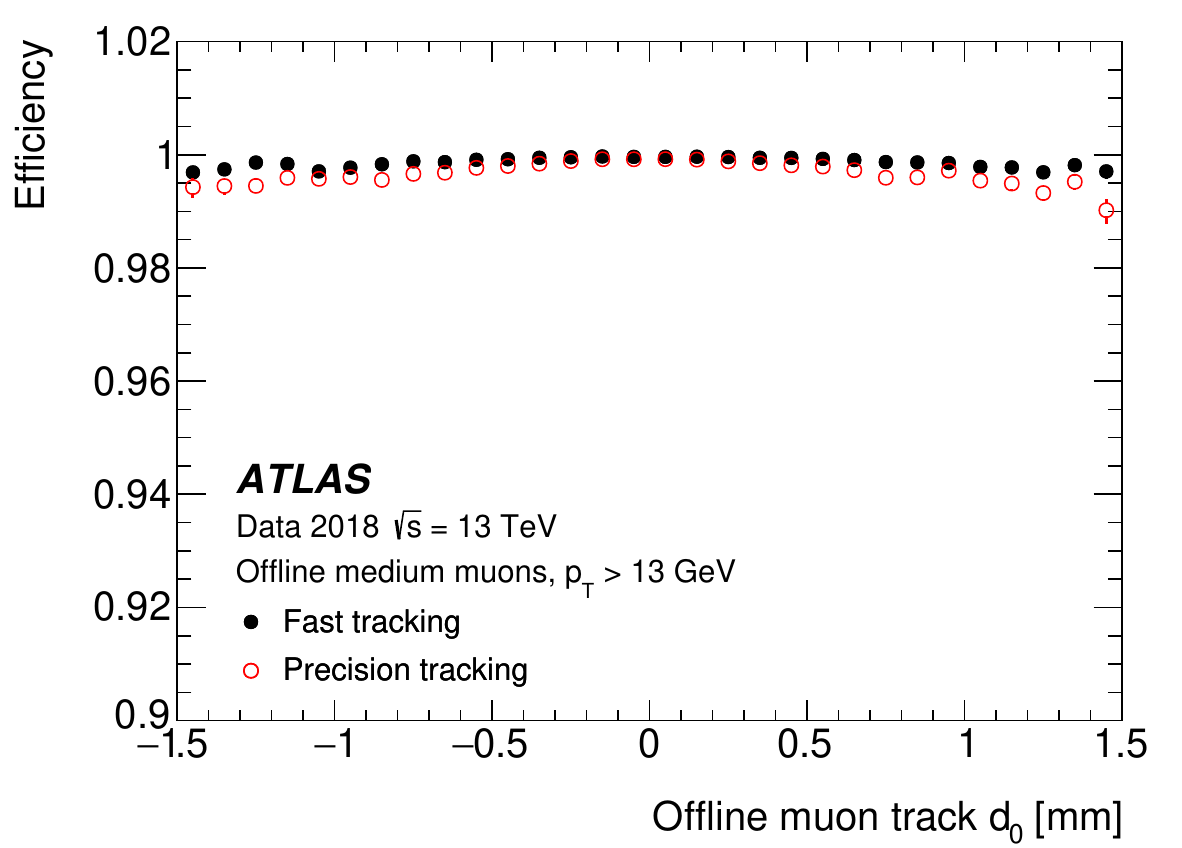}}
\subfloat[]{\includegraphics[width=0.5\textwidth]{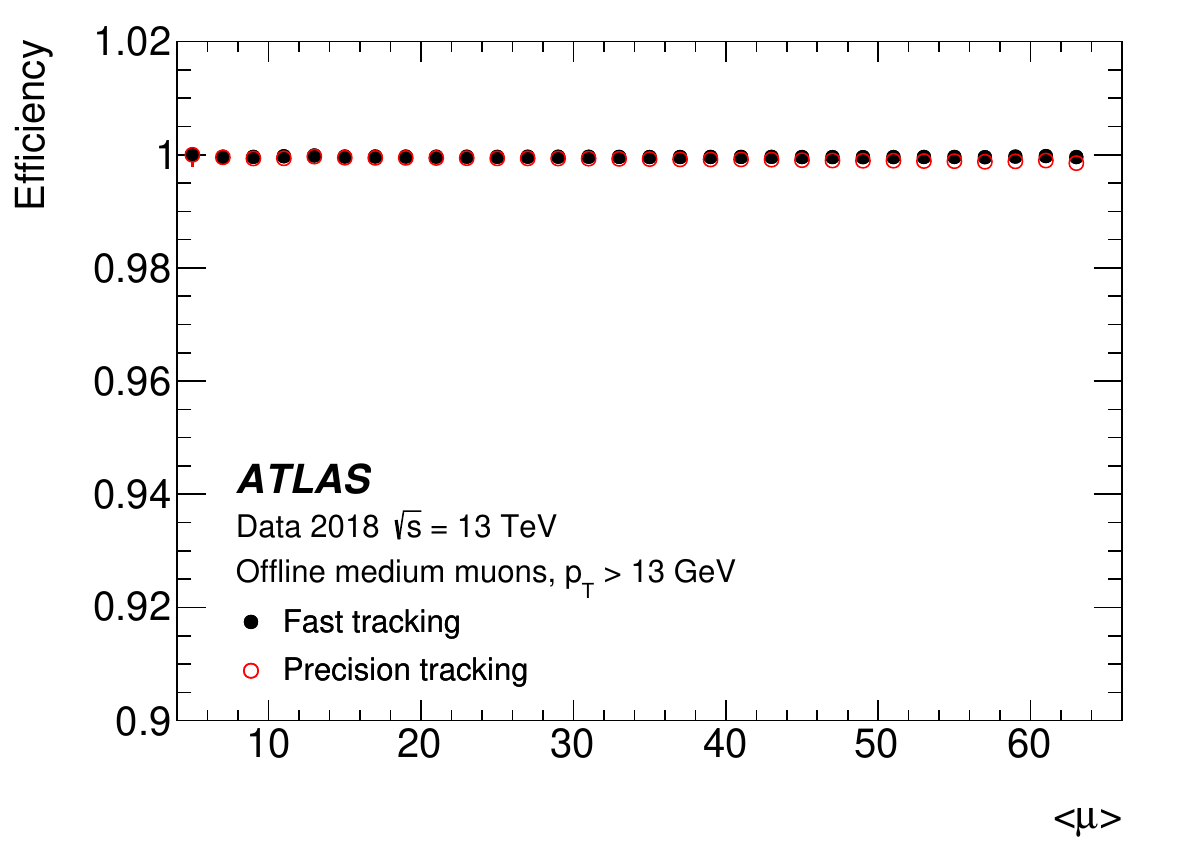}}
\caption{The muon track finding efficiency versus
(a) the offline muon track pseudorapidity,
(b) the offline muon track transverse momentum,
(c) the offline muon track azimuthal angle,
(d) the offline muon track $z_0$ position along the \beamline,
(e) the offline muon track transverse impact parameter, $d_0$, measured with respect to the \beamline, and
(f) the mean \pileup interaction multiplicity, all measured using the dimuon performance trigger.
Statistical Bayesian uncertainties are shown, using a 68.3\% confidence interval
and uniform beta function prior.}
\label{fig:muetnpeff}
\end{figure*}

The efficiency as a function of the mean \pileup interaction multiplicity, \mubar, and the
offline track variables can be seen in Figure~\ref{fig:muetnpeff}.
With this analysis,
the statistical uncertainties are very much reduced with respect to the analyses presented earlier in this section.
The efficiency for both the fast and precision tracking can be seen to approach 100\%
with small statistical uncertainties even  at large transverse momenta, approaching 1\tev.
The efficiency
versus the transverse impact parameter clearly shows a very small decrease of slightly less than
0.5\% for the fast tracking for $d_0$ values of 1.5~mm, and slightly larger than 0.5\% for the
precision tracking. This is consistent with the suggestion from Figure~\ref{fig:muoneff}.
The efficiency for both the precision and fast tracking
is again seen to be approximately constant with respect to changes in the $z_0$ position, but
with a very small loss of efficiency for the precision tracking
at high \pileup interaction multiplicity consistent with Figure~\ref{fig:muoneff}.

{\begin{figure*}[thp]
\centering
\subfloat[]{\includegraphics[width=0.5\textwidth]{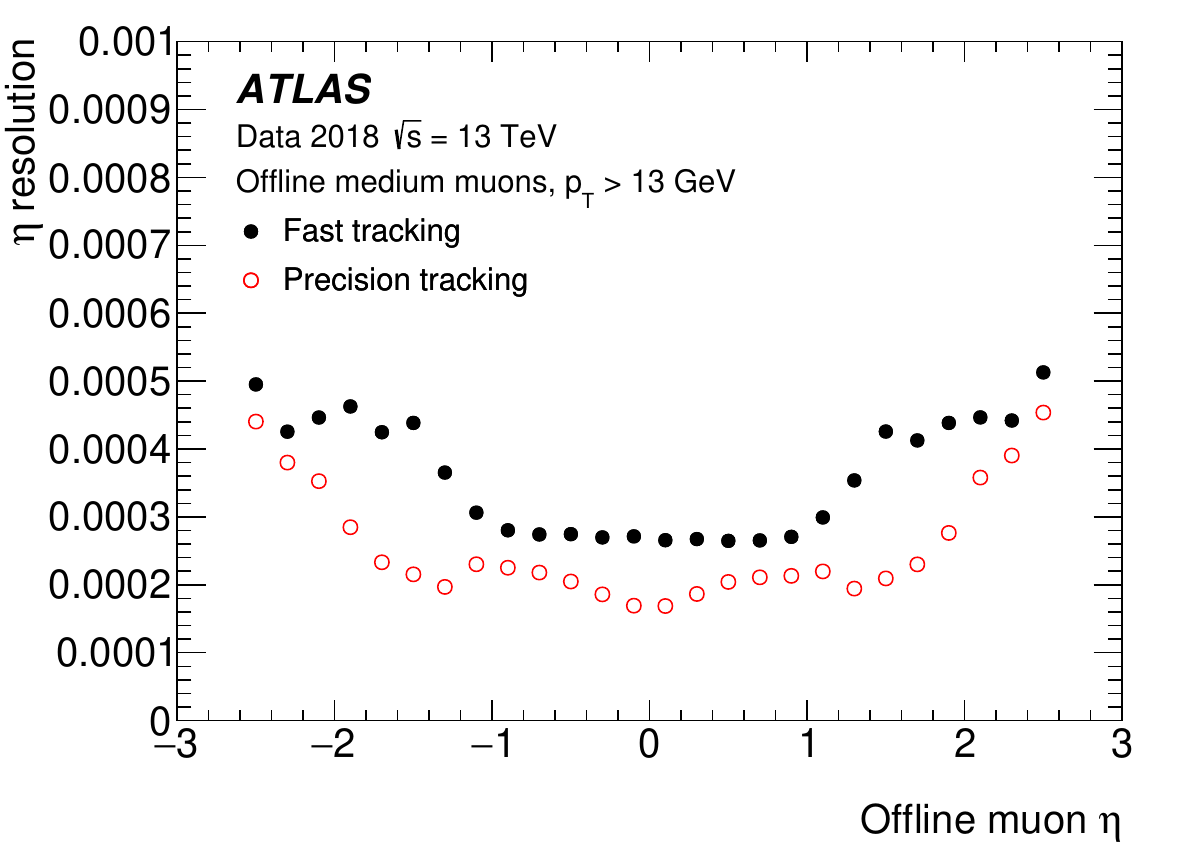}}
\subfloat[]{\includegraphics[width=0.5\textwidth]{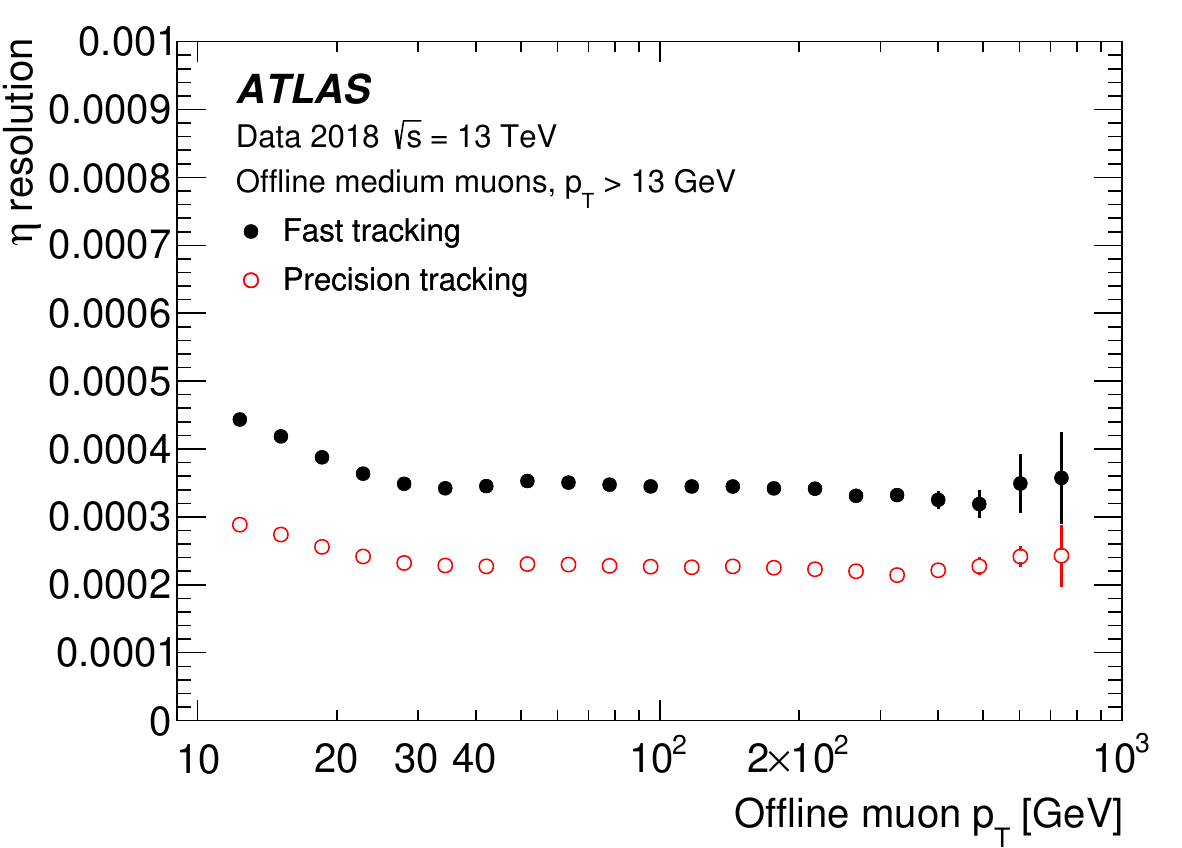}}\\
\subfloat[]{\includegraphics[width=0.5\textwidth]{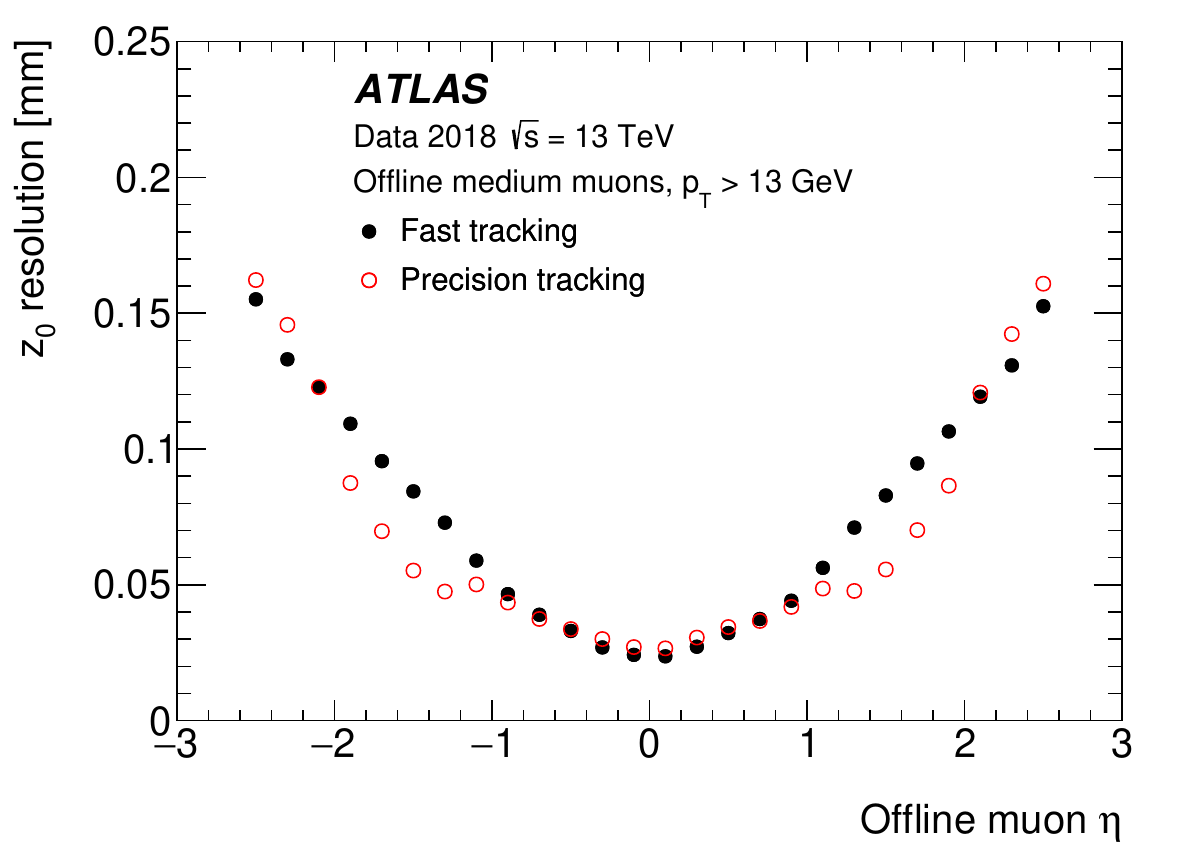}}
\subfloat[]{\includegraphics[width=0.5\textwidth]{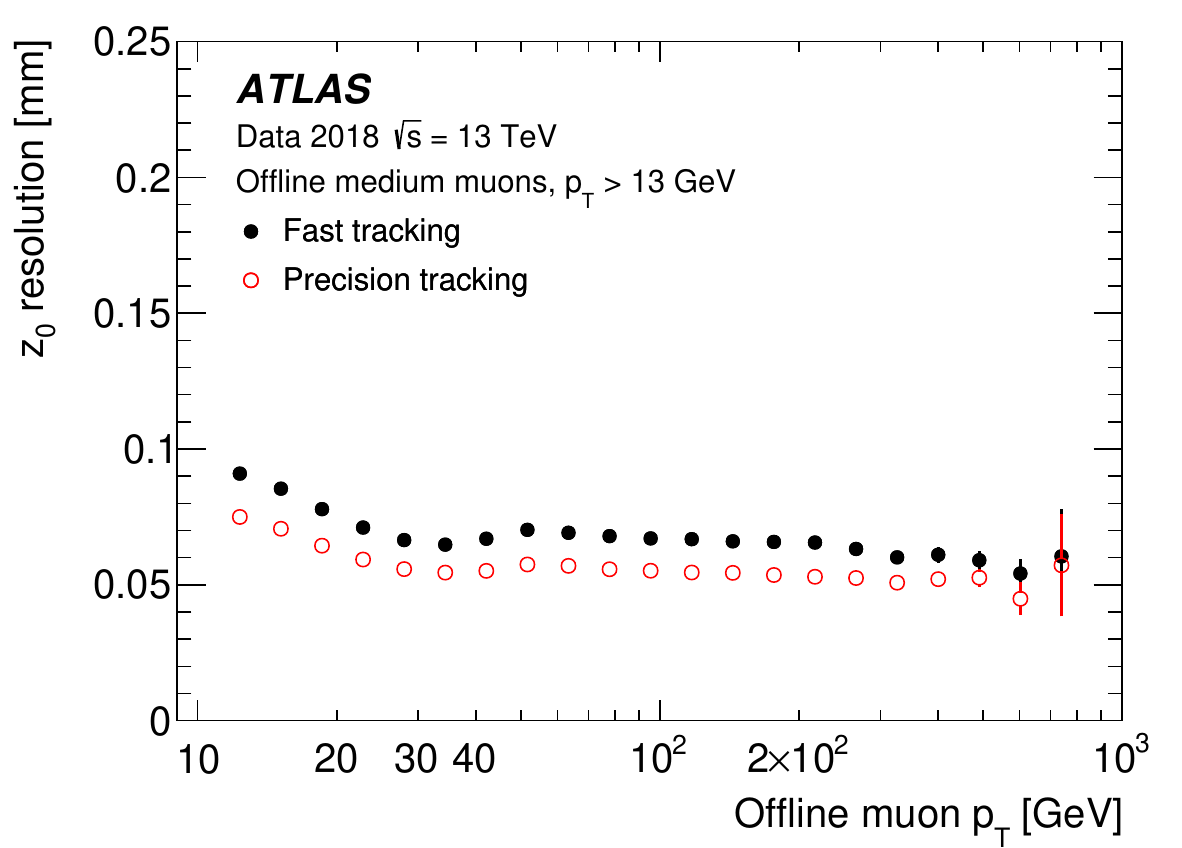}}\\
\subfloat[]{\includegraphics[width=0.5\textwidth]{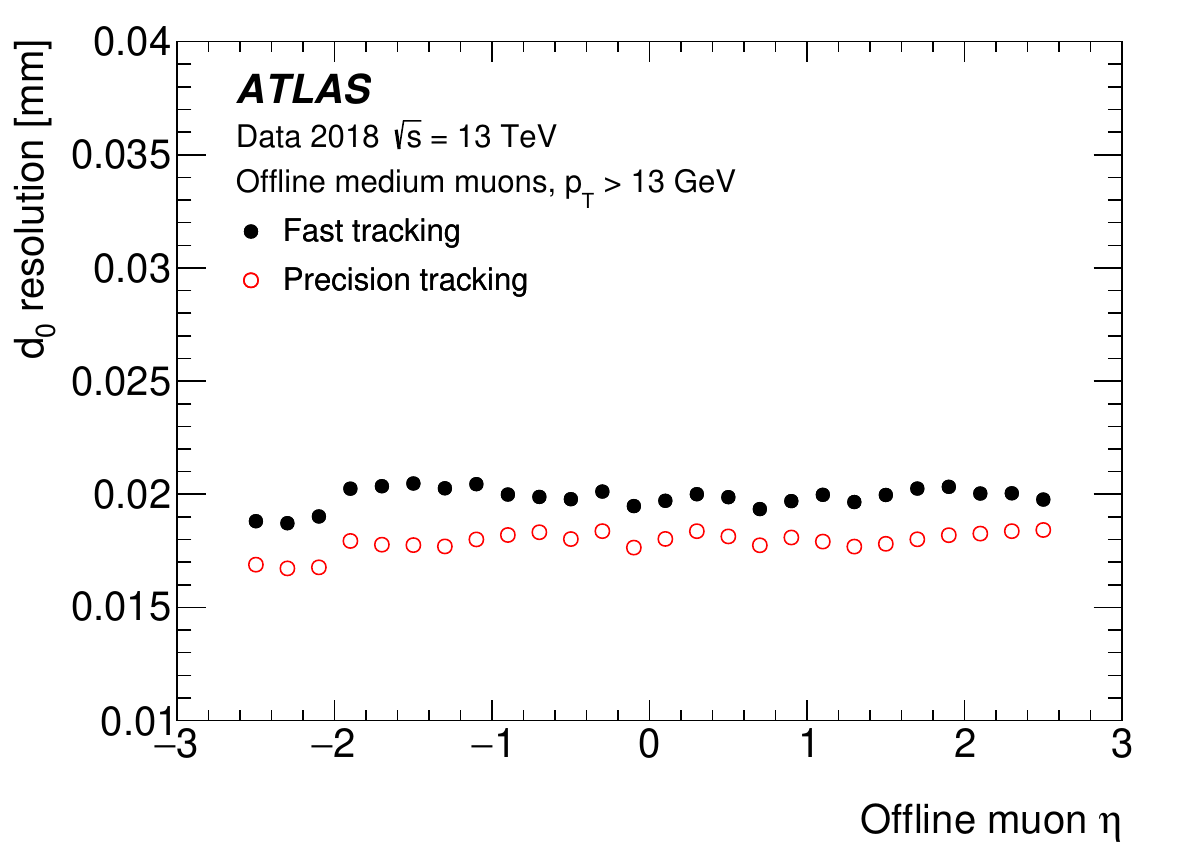}}
\subfloat[]{\includegraphics[width=0.5\textwidth]{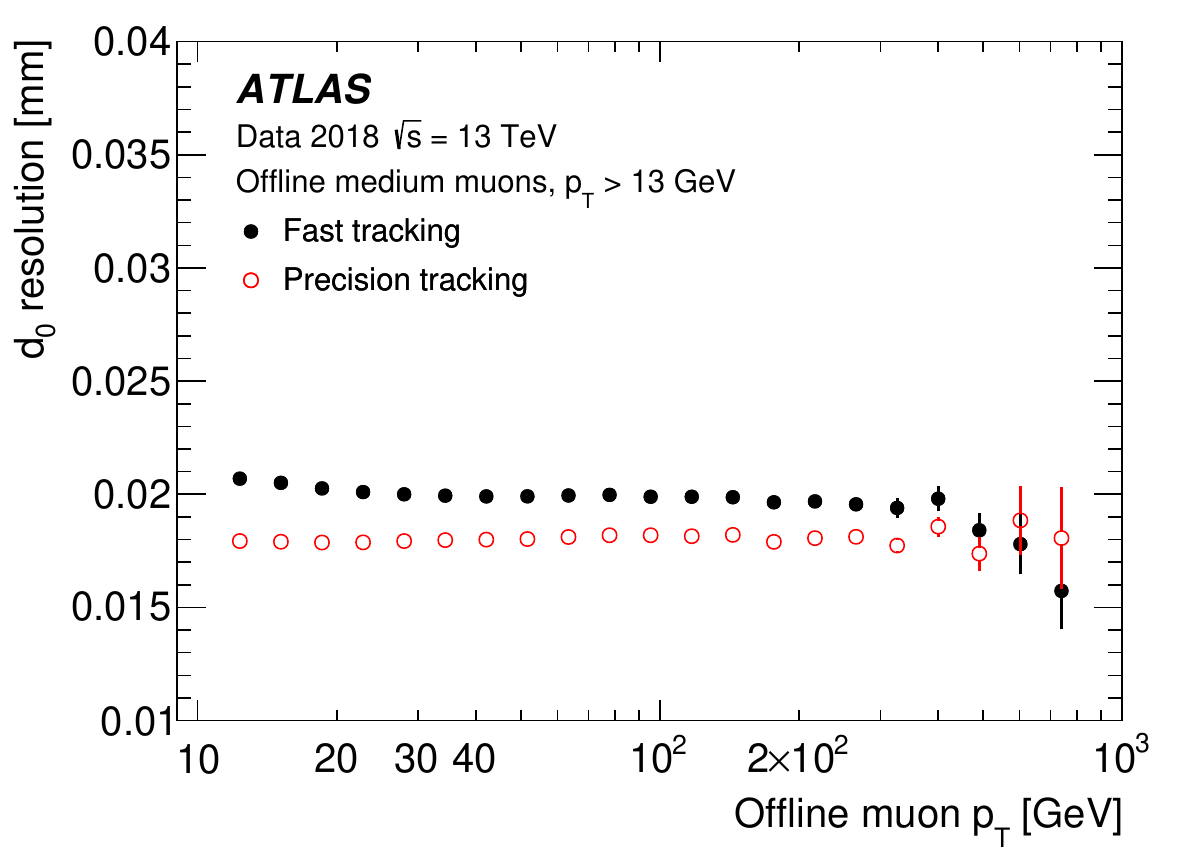}}
\caption{The resolution in the pseudorapidity and inverse transverse momentum
as a function of offline track pseudorapidity and transverse momentum. The resolutions are shown for
muons selected with the $\pt>\SI{15}{\GeV}$ muon requirement from the dimuon support triggers, for both the
fast tracking and precision tracking algorithms.
}
\label{fig:muegtnpres}
\end{figure*}}

The resolution in the track pseudorapidity, track $z$ position along the \beamline, and the transverse
impact parameter for these data are  shown in
Figure~\ref{fig:muegtnpres} as functions of the offline track  pseudorapidity and transverse momentum.
The performance of the resolution
at large transverse momentum, with the exception of $d_0$,  is seen to be generally consistent with that
seen in Figure~\ref{fig:muonres} but again with significantly smaller uncertainties, most notably at
large track transverse momenta. With regards to the resolution in $d_0$, the dilepton analysis is
performed using only the 2018 data where the mean \pileup interaction multiplicity is larger than for
the combined 2016-2018 data. Since the transverse impact parameter resolution has some dependence on the
\pileup interaction multiplicity, the resolution
seen in Figure~\ref{fig:muegtnpres} is somewhat worse than that shown in Figure~\ref{fig:muonres} for
the combined 2016-2018 data.

\floatbarra


 
\subsection{Tracking in the electron signature}
\label{sec:egamma}
 
\newcommand{\egammafigeff}{
{\begin{figure*}[th]
\centering
\subfloat[]{\includegraphics[width=0.5\textwidth]{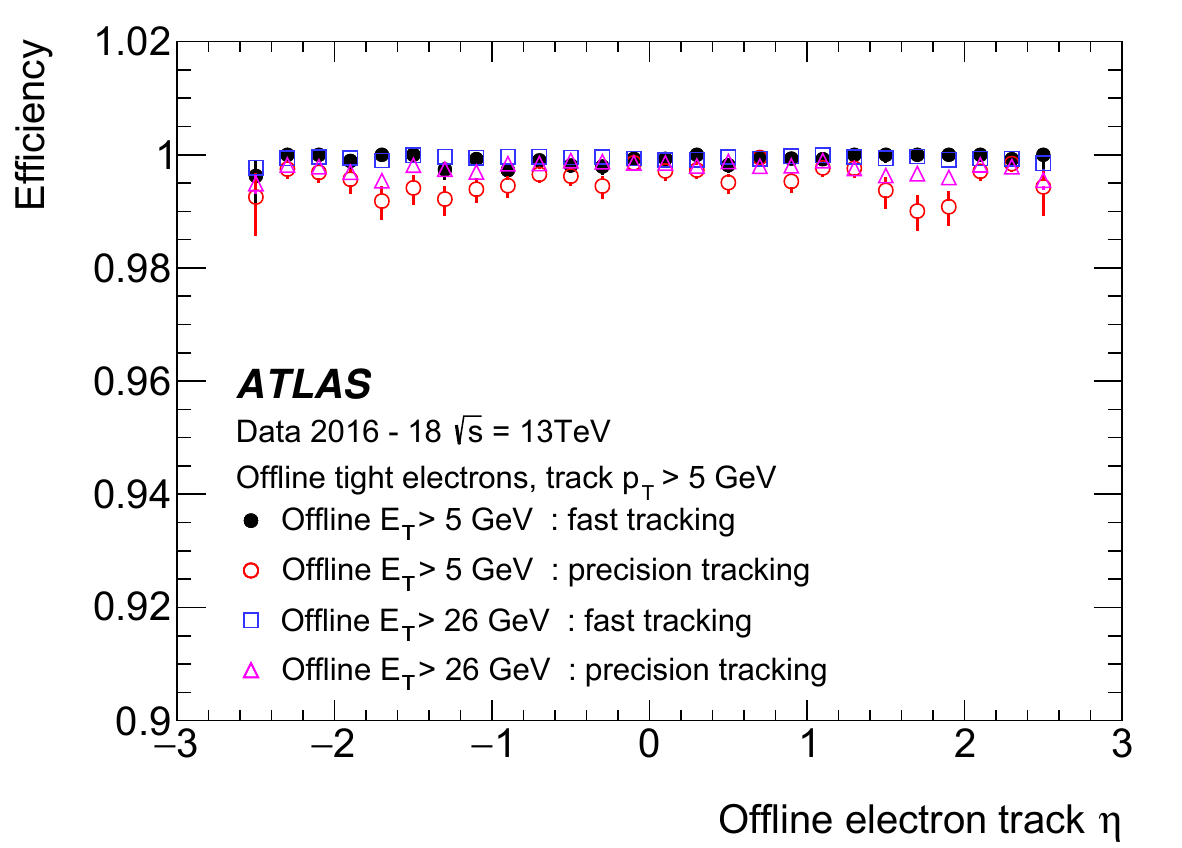}}
\subfloat[]{\includegraphics[width=0.5\textwidth]{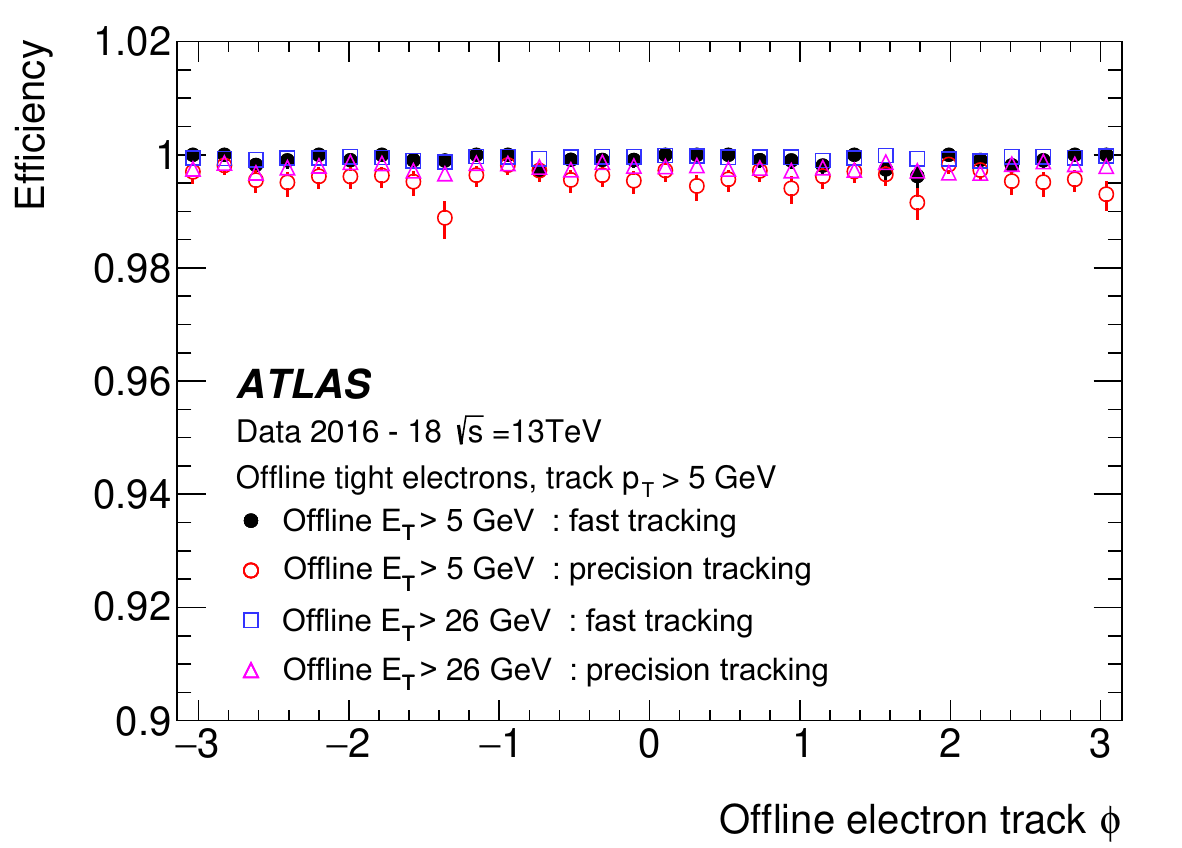}}\\
\subfloat[]{\includegraphics[width=0.5\textwidth]{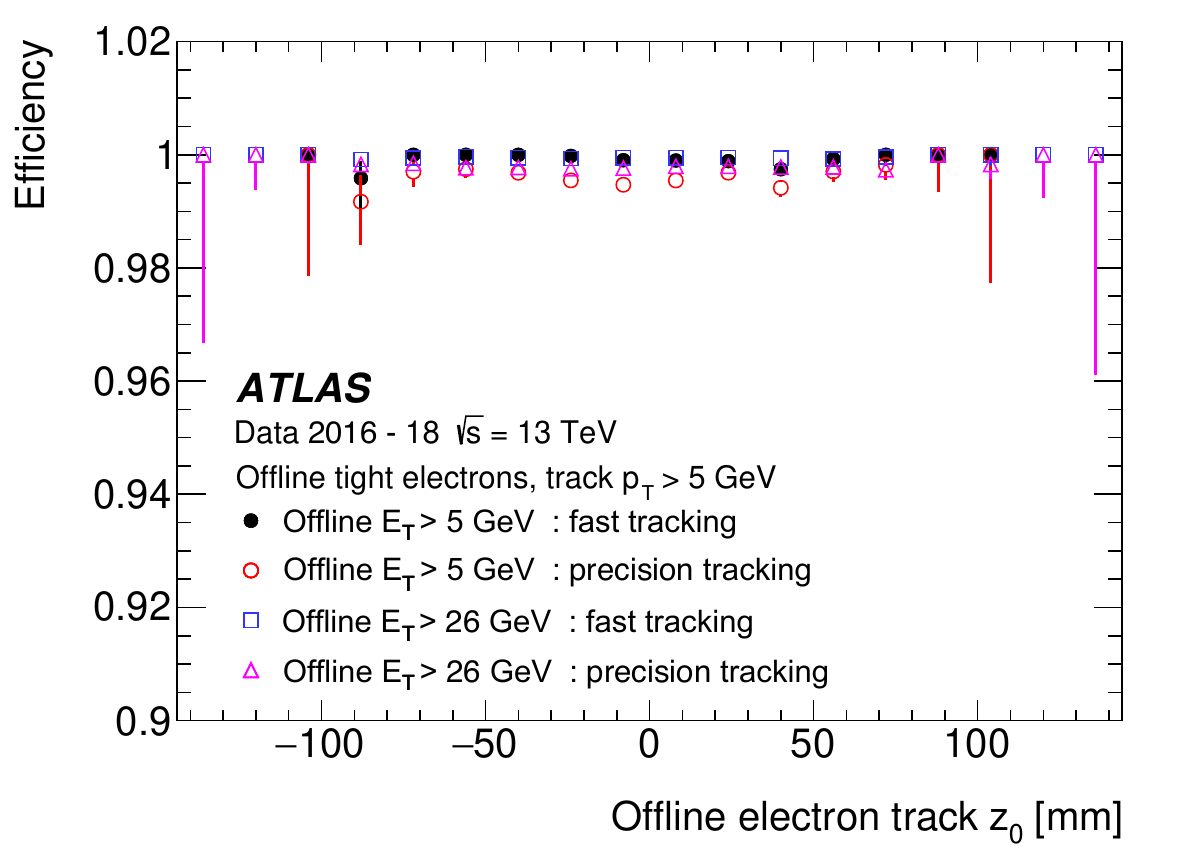}}
\subfloat[]{\includegraphics[width=0.5\textwidth]{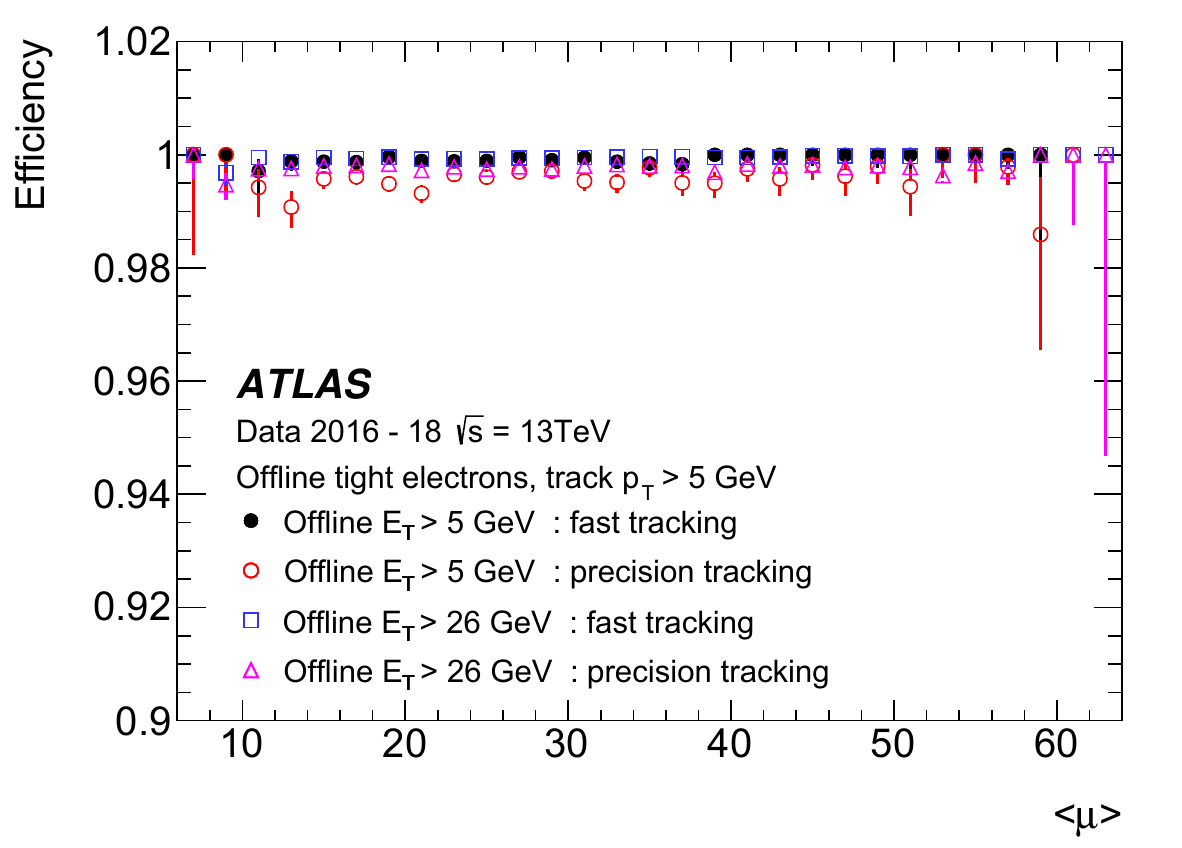}}
\caption{The ID trigger tracking efficiency for electrons selected by the \SI{5}{\GeV} and \SI{26}{\GeV}
electron triggers, with respect to offline electron candidates with $\et>\SI{5}{\GeV} $ for the 5 GeV
trigger and $\et>\SI{26}{\GeV}$ for the 26 GeV trigger. The efficiency is shown as a function of
(a) the offline reconstructed electron track $\eta$,
(b) the offline reconstructed electron track $\phi$,
(c) the offline reconstructed electron $z_0$,
(d) the mean number of \pileup vertices, \mubar.
Efficiencies are shown for both the fast tracking and precision tracking algorithms.
Statistical Bayesian uncertainties are shown, using a 68.3\% confidence interval
and uniform beta function prior.}
\label{fig:egeff}
\end{figure*}}
}
 
\egammafigeff

Because of the significant background from QCD jet events, a tight offline
likelihood identification~\cite{PERF-2017-01} is used for the reference electron selection for
the electron signature analysis.
The offline electron candidate track is required to have
at least two pixel hits, an IBL hit if passing through at least one active IBL module,
and at least four clusters in the SCT.
These requirements were
chosen to ensure better reconstruction in the pixel detector for offline tracks
by eliminating extremely poorly reconstructed bremsstrahlung
candidates where the hits in the pixel detector were missed.
Analyses for several triggers are presented, each employing a
kinematic selection of $|\eta|<2.5$ and $\pt > \SI{5}{\GeV}$ for the
offline electron candidate tracks.
In addition, there is a selection on the electron candidate transverse energy, \ET.
For the candidates from the 5\gev trigger a selection of $\ET>5$\gev is applied.
For the  10\gev trigger the selection is
$\ET>10$\gev, and $\ET>26$\gev for the 26\gev trigger.
For the analyses with the harder \ET selection, the same 5\gev track \pt selection
is used. Consequently, the fraction of the electron candidate energy that can be radiated as
bremsstrahlung for candidates near the trigger threshold for the higher
\ET threshold triggers is larger than for the lower \ET threshold triggers.
As an additional requirement, to remove offline candidates where the track \pt has been badly
overestimated, offline  candidates with $\ET/\pt<0.8$ are removed, where \ET is measured in
the calorimeter, and \pt is from the offline track.
 
\newcommand{\figegres}{
{\begin{figure*}[th]
\centering
\subfloat[]{\includegraphics[width=0.5\textwidth]{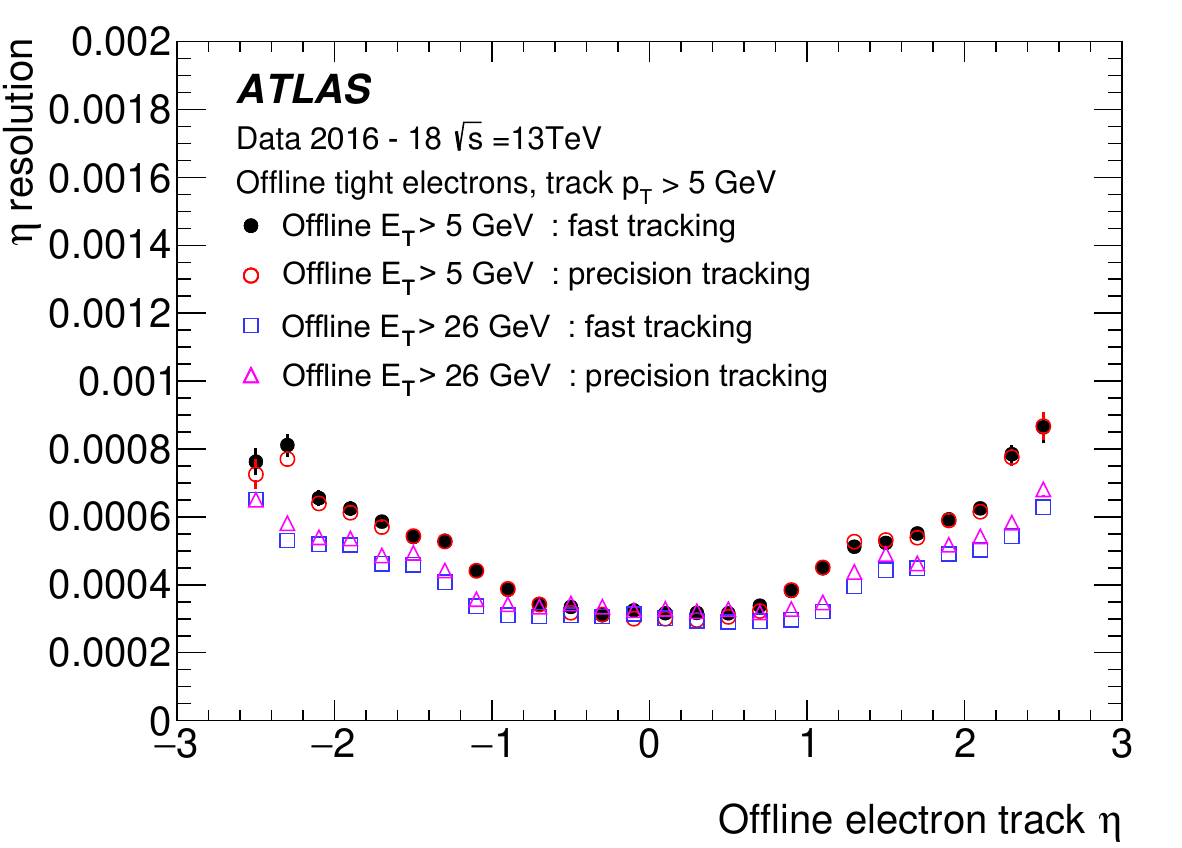}}
\subfloat[]{\includegraphics[width=0.5\textwidth]{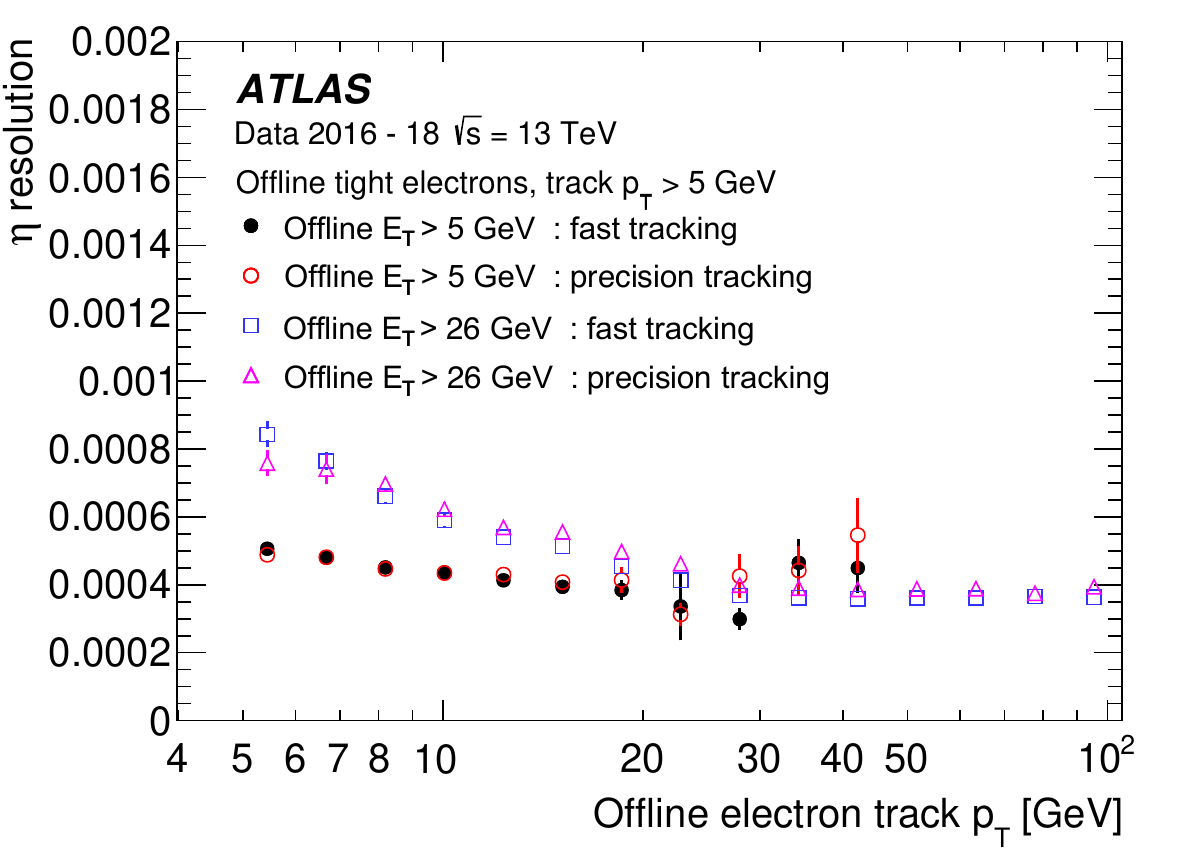}}\\
\subfloat[]{\includegraphics[width=0.5\textwidth]{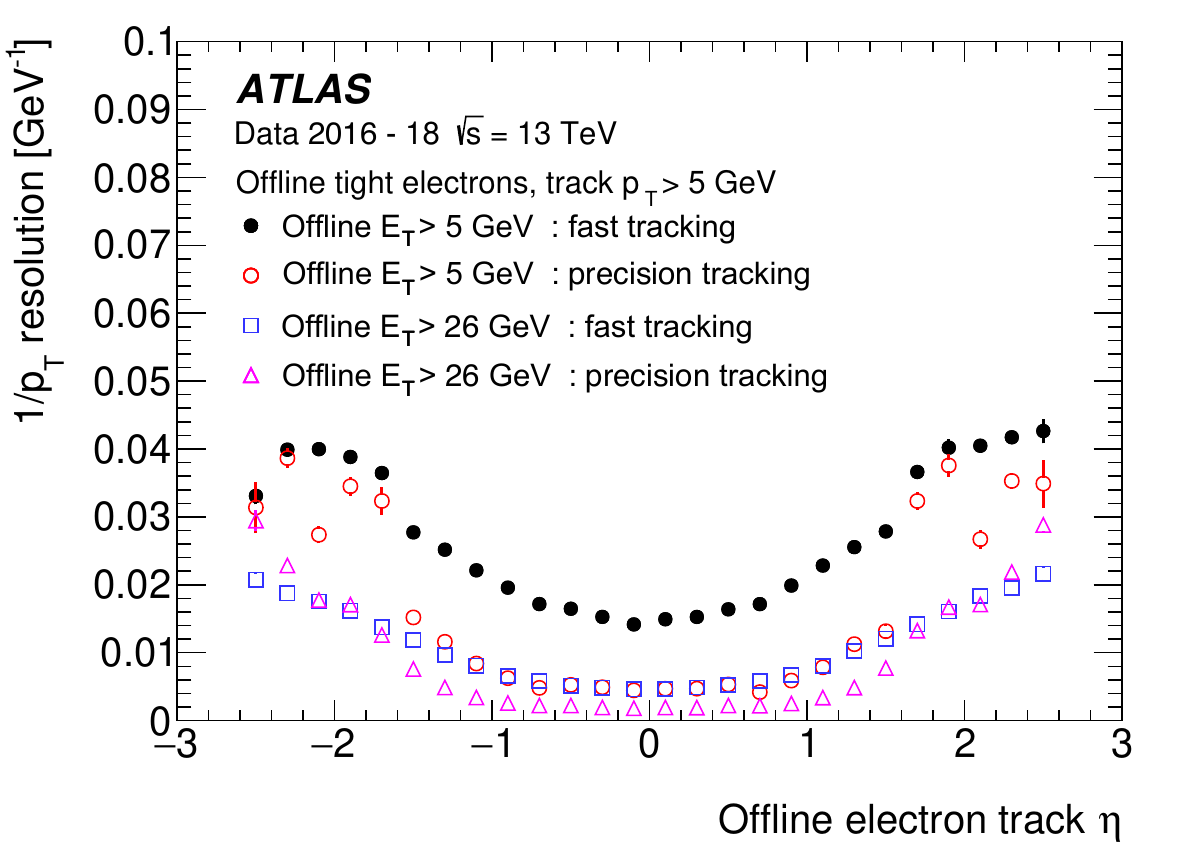}}
\subfloat[]{\includegraphics[width=0.5\textwidth]{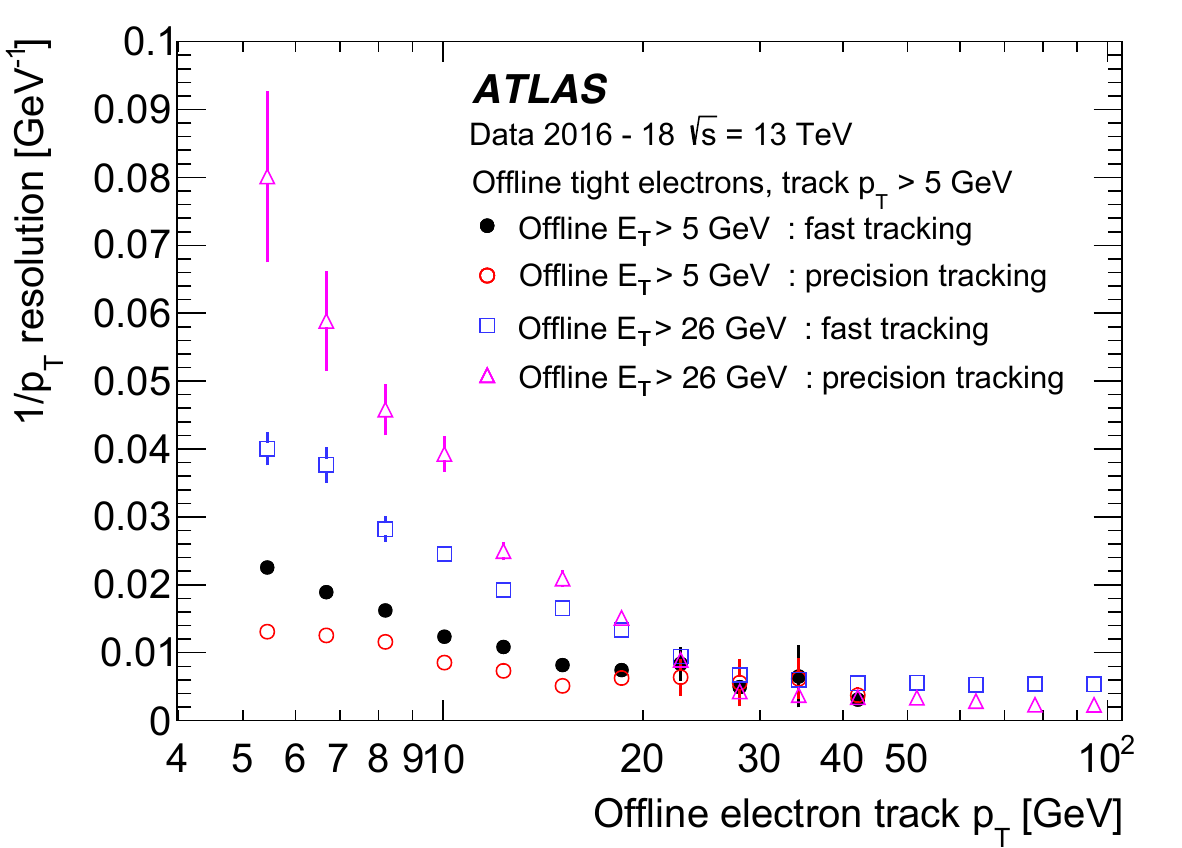}}
\caption{The resolution for pseudorapidity
and inverse transverse momentum
as a function of offline track pseudorapidity and transverse momentum.
The resolutions are shown for electrons selected by the \SI{5}{\GeV} and \SI{26}{\GeV} electron support triggers for both the fast tracking and precision tracking algorithms.
}
\label{fig:egres}
\end{figure*}}
}
 
\newcommand{\figetovpt}{
{\begin{figure*}[th]
\centering
\subfloat[]{\includegraphics[width=0.5\textwidth]{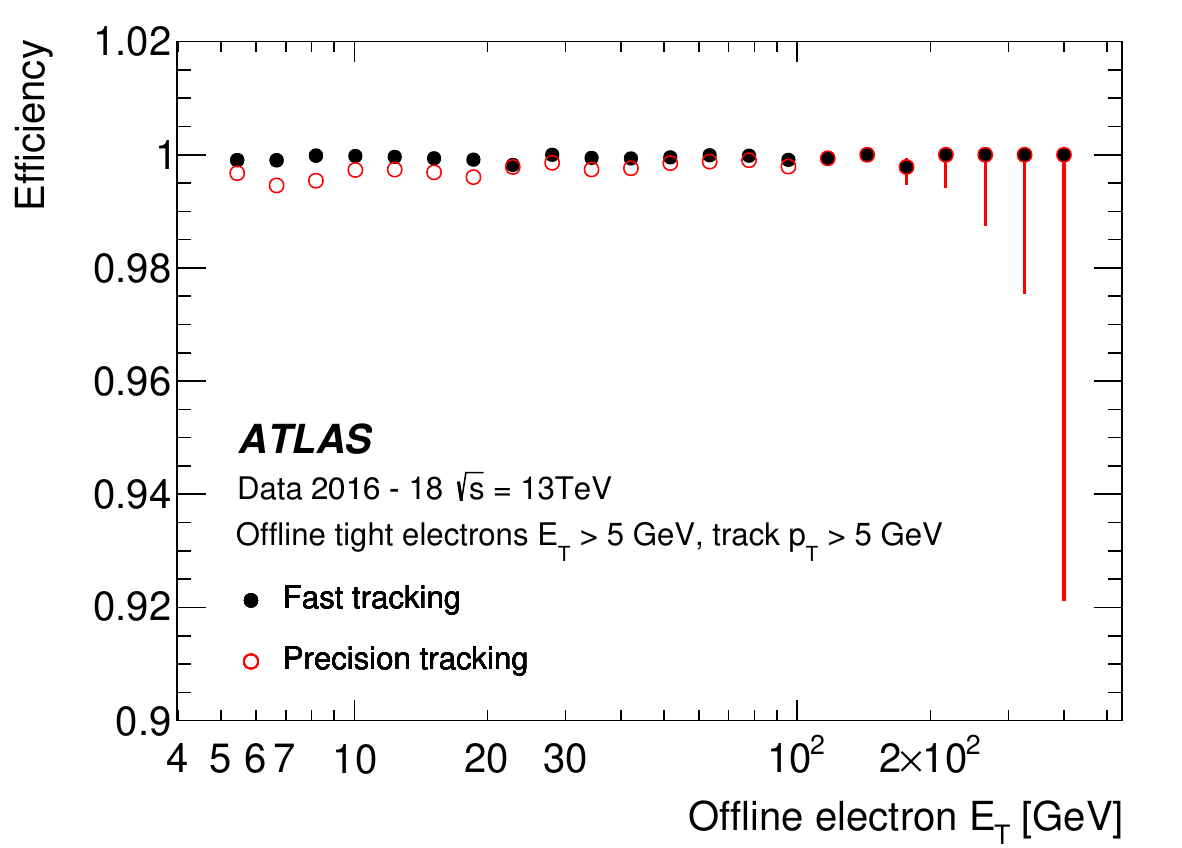}}
\subfloat[]{\includegraphics[width=0.5\textwidth]{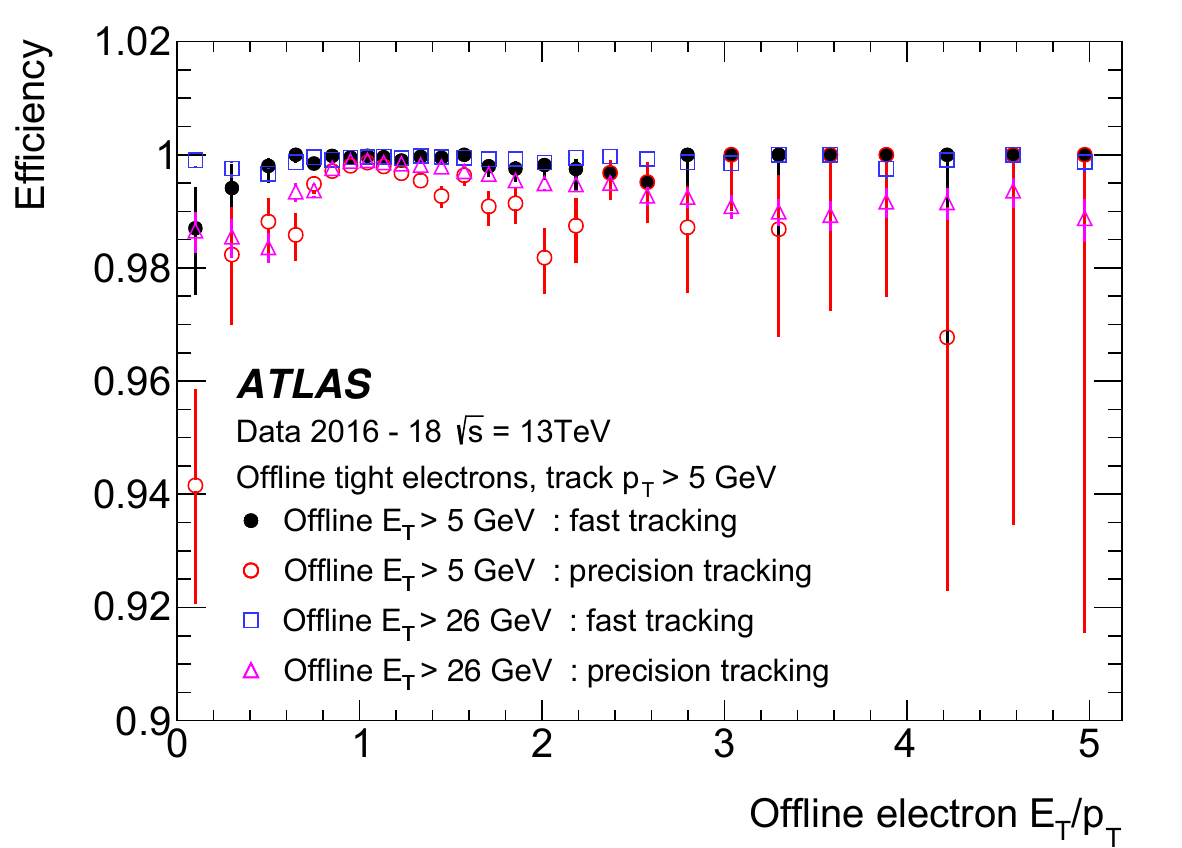}}\\
\subfloat[]{\includegraphics[width=0.5\textwidth]{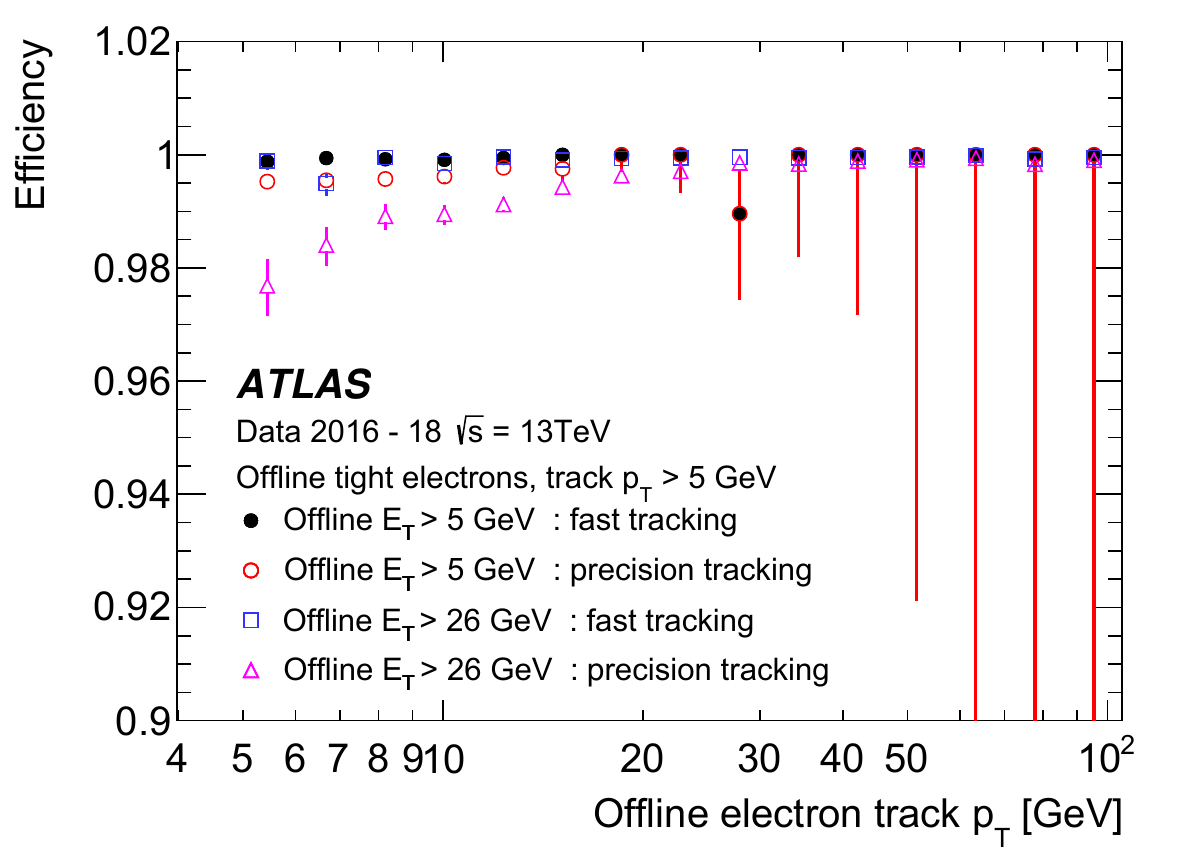}}
\subfloat[]{\includegraphics[width=0.5\textwidth]{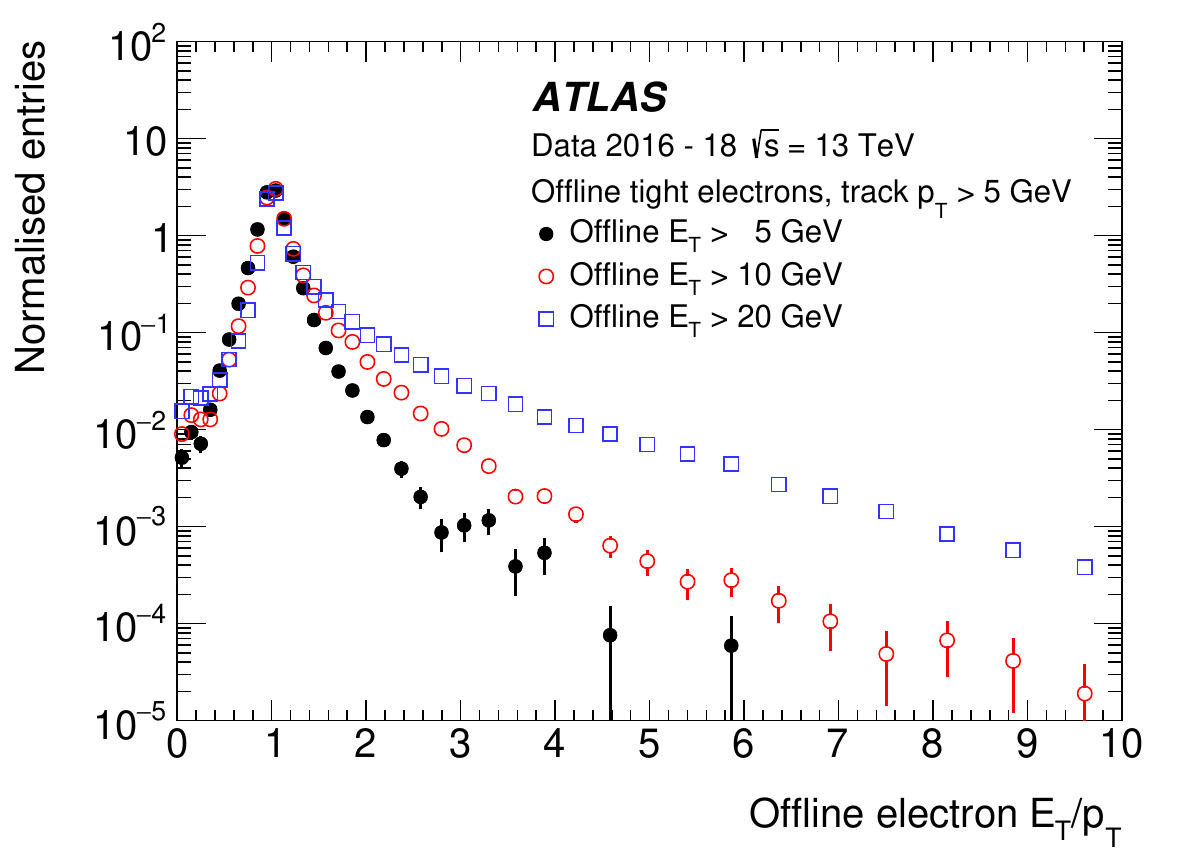}}
\caption{The efficiency with respect to
(a) the offline electron transverse energy, \ET,
(b) the offline electron $\ET/\pt$, and
(c) the offline electron track \pt. Shown in (d) is the distribution of $\ET/\pt$ for the offline electron candidates.
Results are shown for electrons selected by the \SI{5}{\GeV}, \SI{10}{\GeV} and \SI{26}{\GeV} electron support triggers for panel (a), and
by the \SI{5}{\GeV}, \SI{26}{\GeV} support triggers for panels (b) and (c). For the
efficiencies versus offline electron candidate \ET and \pt, only candidates where $\ET/\pt>0.8$ are used.
Statistical Bayesian uncertainties are shown, using a 68.3\% confidence interval
and uniform beta function prior.}
\label{fig:etovpt}
\end{figure*}}
}
 
\figegres

Figure~\ref{fig:egeff} shows the ID tracking efficiency for the
\SI{5}{\GeV} and \SI{26}{\GeV} electron triggers as a
function of the $\eta$, $\phi$, and $z_{0}$ values of the offline electron candidate track,
and as a function of the mean \pileup multiplicity.  The tracking
efficiency is measured with respect to offline electron tracks with $\pt>\SI{5}{\GeV}$
for trigger electron candidates with either $\et>5$\gev or $\et>26$\gev, from the
\SI{5}{\GeV} and \SI{26}{\GeV} electron support triggers. The offline electtro tracks
are those from tight offline electron candiates. The efficiencies of the fast track
finder and precision tracking exceed 99\% for all pseudorapidities.

\figetovpt
 
Figure~\ref{fig:egres} shows the resolutions of the track pseudorapidity
and $1/\pt$  as a function of either the $\eta$ or $\pt$ of the offline track from
the offline electron candidates.
For this analysis, the same selection on the electron candidate \et and \pt discussed
previously is applied.
Consequently, for events with
a 5\gev selection, there is no phase space for the electron candidates
near the threshold to radiate any bremsstrahlung photons. For the 26\gev
electrons however, there is significant phase space available for radiation.
As such, for the 26\gev threshold trigger, tracks which have \pt far below 26\gev
will correspond to electrons that have undergone significant radiation.
The resolutions for these
tracks will be significantly worse than for the tracks with the same \pt
from the 5\gev trigger, but that have not undergone such
emission. For the 5\gev threshold, the resolution of the inverse \pt
for the fast tracking is slightly worse than for the precision tracking.
As expected it is seen to
deteriorate at larger $\eta$ for both the higher and lower threshold triggers.
Despite the significantly
worse resolution at low \pt for the tracks from the 26\gev trigger, the
resolution, when integrated over \pt, is still significantly better than for the
lower \pt threshold since only a relatively small fraction of events from the 26\gev
trigger have low \pt tracks. Therefore, the usual improvement of the resolution with
increasing \pt is enough to provide the observed significantly better overall resolution.

The resolution for the precision tracking at low $\pt$ and large pseudorapidities
from the 26\gev trigger is worse than for the fast tracking as a
consequence of the different SCT hit multiplicities observed on each type of trigger
track. This is discussed in more detail later in this section.

This difference between the degree of bremsstrahlung for tracks between 5 and
26\gev from the different triggers is also responsible for the differences
in the efficiency as a function of the track \pt observed in
Figure~\ref{fig:etovpt}. For this figure the $\et/\pt$ selection is relaxed.
Here, it is clear that the track finding efficiency
for all electron candidates as a function of the \et of the offline
electron candidate is consistently high. For the track candidates from the
26\gev trigger but with tracks with  \pt below 26\gev there is significant
radiation, and the track reconstruction is less efficient, as there
may be significant kinks in the electron trajectory. Also shown is the distribution of
the ratio of the offline electron transverse energy measured in the
calorimeter to the offline track transverse momentum. The long tail to
values greater than unity represents bremsstrahlung candidates.
Also seen in the
figure is the clear tail with values of \et/\pt less than one,
which results from electron candidates where the track \pt is larger than the
cluster \et from the calorimeter. This is a consequence of the offline \pt measurement
migrating to larger values when the track \pt itself is less well reconstructed.
In this case, the track cannot realistically have significantly
more energy than the cluster, and so any larger \pt value
would therefore represent a less reliable track
\pt measurement.
The efficiency as a function of \et/\pt shows a clear decrease, with
a loss of around 1\% for tracks where the track \pt is 60\% higher than
might be physically possible. The efficiency for tracks which have not radiated is well
over 99\%, and the efficiency for tracks that have radiated well over half
of their original energy is still greater than 98\%.
 
The selection on $\ET/\pt>0.8$ is not strictly required for measurements
which integrate over the transverse momentum of the tracks, but is essential
for an unbiased measurement versus the offline track $\pt$ itself. Therefore,
for consistency this selection is applied everywhere, with the exception
of the study of the $\ET/\pt$ variable and the corresponding efficiency.

To complement the standard performance triggers, which
do not select on the tracking, as with the muon trigger, there are additional dilepton
performance triggers in the electron signature which select events with two electron candidates consistent with both electrons
originating from the decay of a $Z$ boson. This allows a tag and probe analysis largely similar
to that discussed in Section~\ref{sec:muontnp}. For these triggers, one of the electrons
is fully selected\emdash using the full selection including the tracking information
to increase the purity of the selected candidates\emdash and the selection of the
other electron is based solely on the calorimeter reconstruction in the trigger.
 
\newcommand{\figeginvmass}{
\begin{figure}[thp]
\centering
\includegraphics[width=0.5\textwidth]{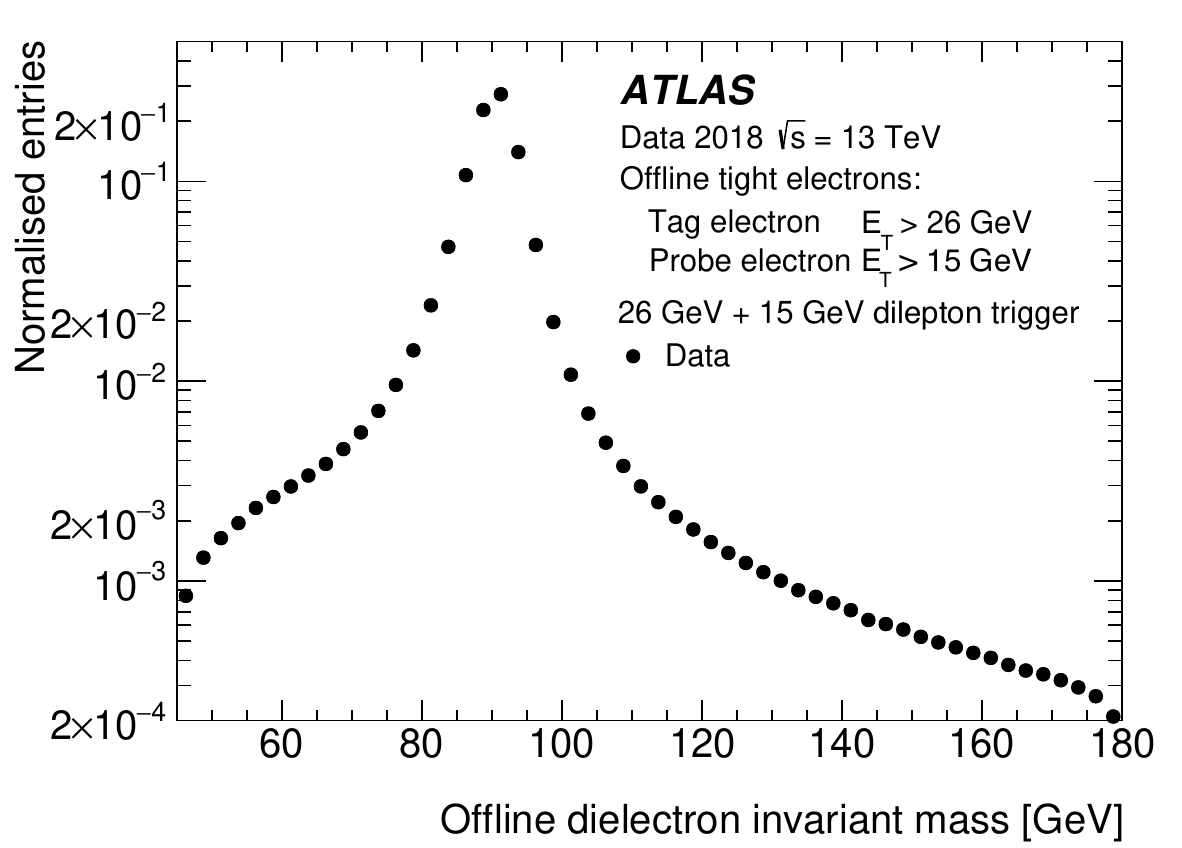}
\caption{The uncorrected offline dilepton invariant mass in events
from the
dilepton performance trigger that  pass the analysis selection. For the performance triggers, the RoIs used
by the fast and precision tracking are the same and so the offline dilepton candidates chosen for the analysis
of the fast and precision tracking are identical. }
\label{fig:elinvmass}
\end{figure}
}

This allows the efficiency of the tracking to again be measured for the
electron candidate selected using only the calorimeter reconstruction, unbiased
by the track selection itself.
The presence of the first, reasonably pure
fully selected {\em tag} electron reduces the rate of events with two
electron candidates and the
selection of the two electron candidates having to form a $Z$ boson candidate
greatly enhances the purity of the calorimeter-only {\em probe} electron
candidate. The purity of electron candidates in such events from these
triggers is therefore high and the triggers themselves have a reasonably
high rate.
However, because the triggers are limited to $Z$ candidates, and have a reasonably
high threshold for the electon selection in the calorimeter, they do not have
the same full range in transverse energy as the standard performance triggers.

For the following analysis such a double-electron performance trigger which selects on
a $Z$ boson candidate is used, with the fully selected electron
candidate passing the $\ET>26$\gev trigger. In this case the
calorimeter-only probe electron candidate is required to have $\ET>15$\gev.
As such, unbiased measurements are possible from this trigger for electron
candidates above this $15$\gev threshold.
The data used for this analysis correspond to the
full 2018 integrated luminosity.
 
\figeginvmass
 
The dilepton invariant mass distribution for offline candidates
for the events passing the
offline selection for the combined 26\gev and 15\gev dilepton performance trigger but
with no corrections for trigger efficiency or detector response
can be seen in Figure~\ref{fig:elinvmass}.

\begin{figure*}[thp]
\centering
\subfloat[]{\includegraphics[width=0.5\textwidth]{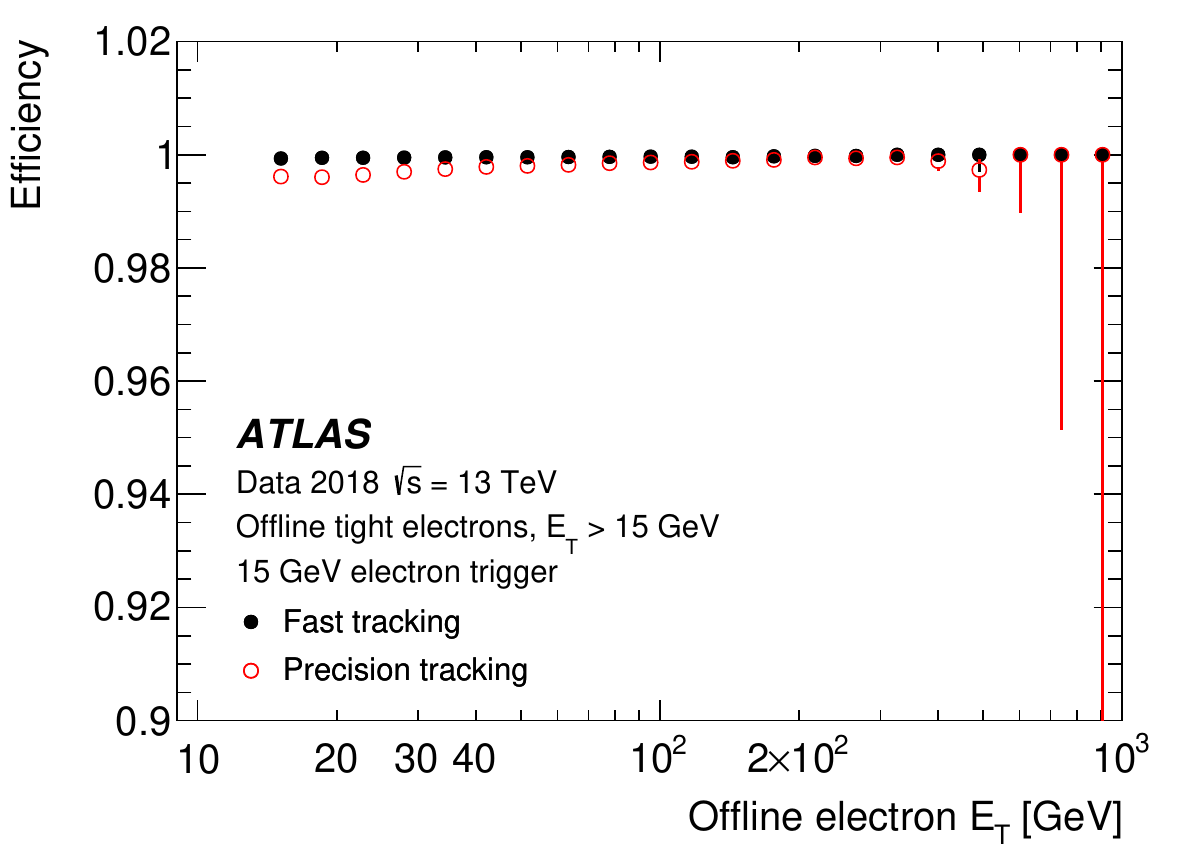}}
\subfloat[]{\includegraphics[width=0.5\textwidth]{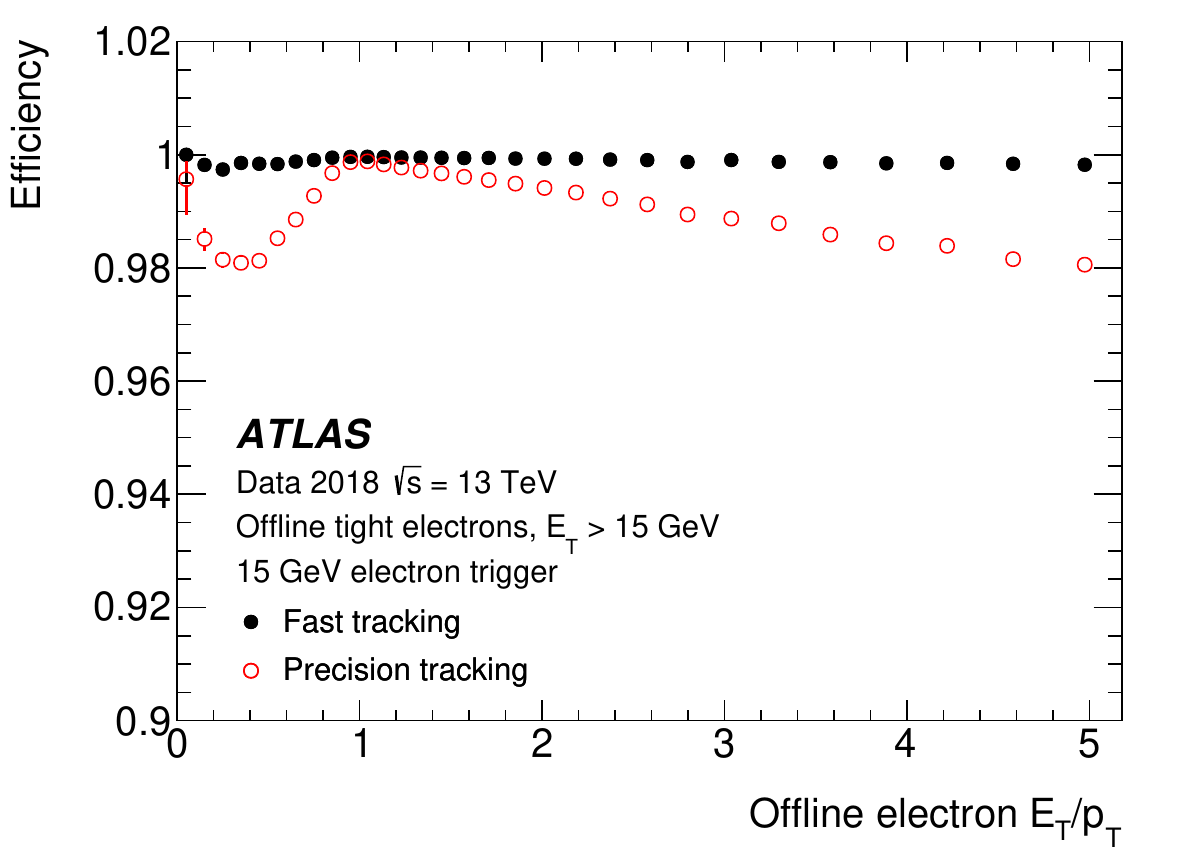}}\\
\subfloat[]{\includegraphics[width=0.5\textwidth]{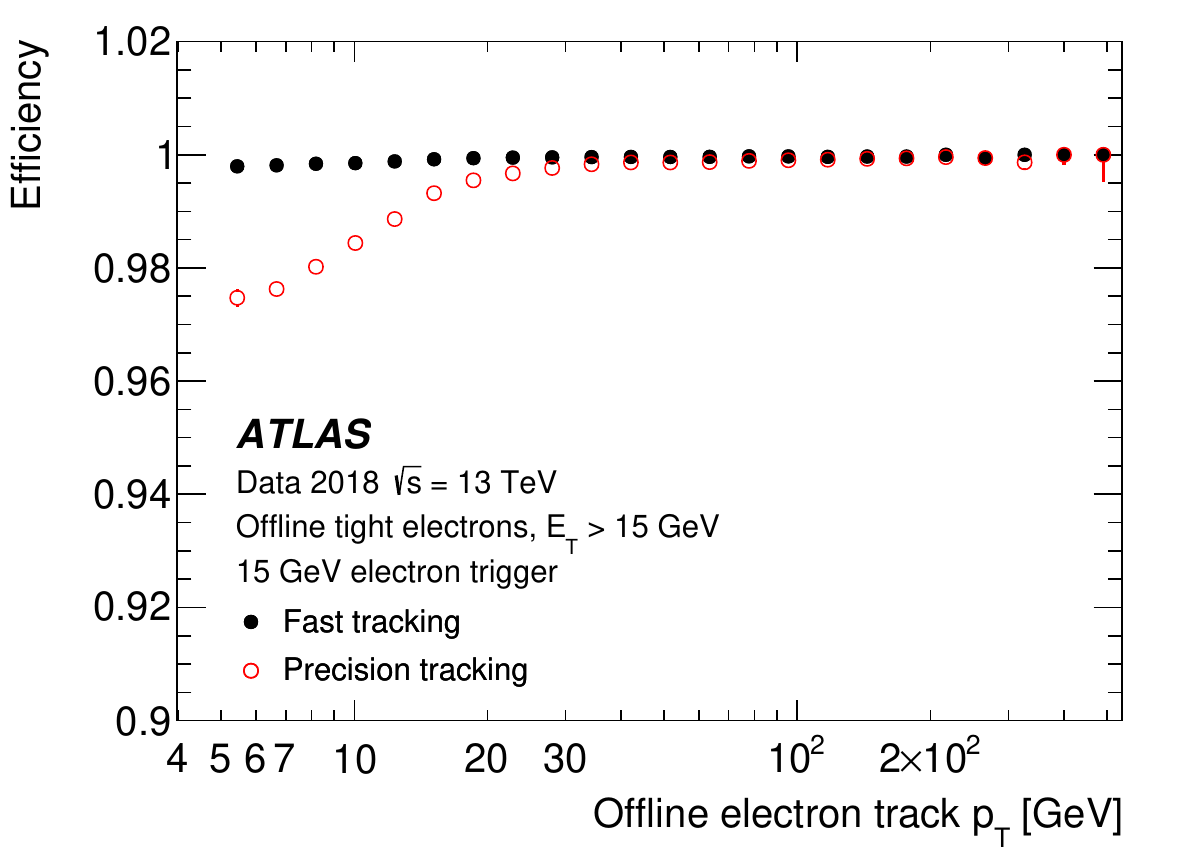}}
\subfloat[]{\includegraphics[width=0.5\textwidth]{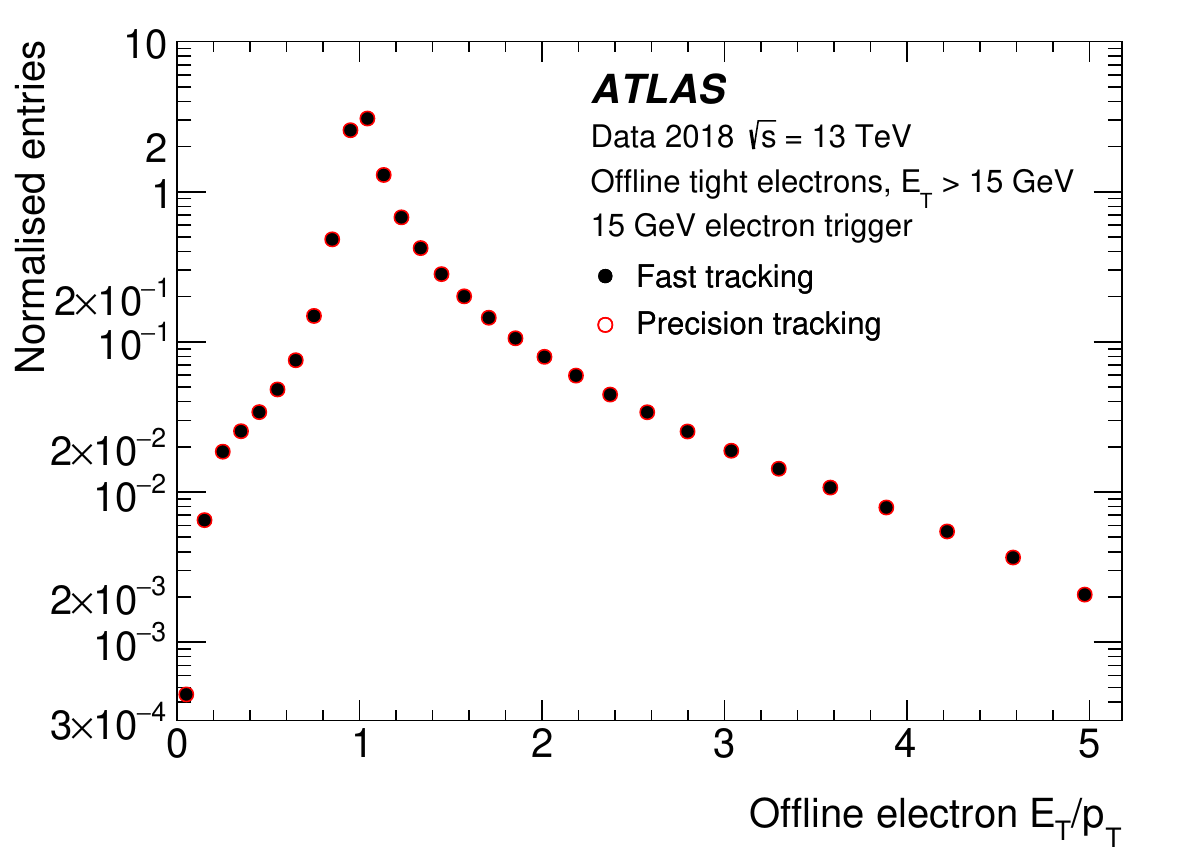}}
\caption{The efficiency versus transverse energy of the electron candidate from the tag and probe analysis of the \SI{15}\gev trigger
with respect to
(a) the offline electron transverse momentum,
(b) the  offline electron $\ET/\pt$, and
(c) the offline electron track \pt.
Shown in (d) is the distribution of $\ET/\pt$ for the offline electron candidates. For the
efficiencies versus offline electron candidate $\ET$ and $\pt$, only candidates
with $\ET/\pt>0.8$ are used.}
\label{fig:etnpet}
\end{figure*}

Figure~\ref{fig:etnpet} shows the efficiency for the offline electron candidates versus the
offline electron candidate \ET and
track \pt, and versus the ratio $\ET/\pt$, together with the $\ET/\pt$ distribution itself, determined using the
tag and probe trigger.
The reduction of the efficiency for candidates with significantly overestimated
track transverse momenta can again be seen but with higher statistical precision. For the remainder
of the analysis, those candidates with $\ET/\pt < 0.8$ are once again removed to avoid
biases from the track \pt measurement at high track \pt.
 
The efficiency as a function of $\ET/\pt$ is seen to behave in a manner consistent with that from
the inclusive analyses presented earlier in this section,
but now with greatly reduced statistical uncertainties.
It is reassuring to see that the efficiency for the fast tracking remains very high for all such candidates,
and is still seen for the precision tracking to be higher than 98\% even for candidates that have radiated up to 80\%
of their energy.
 
\newcommand{\figetnpeff}{
\begin{figure}[thp]
\centering
\subfloat[]{\includegraphics[width=0.5\textwidth]{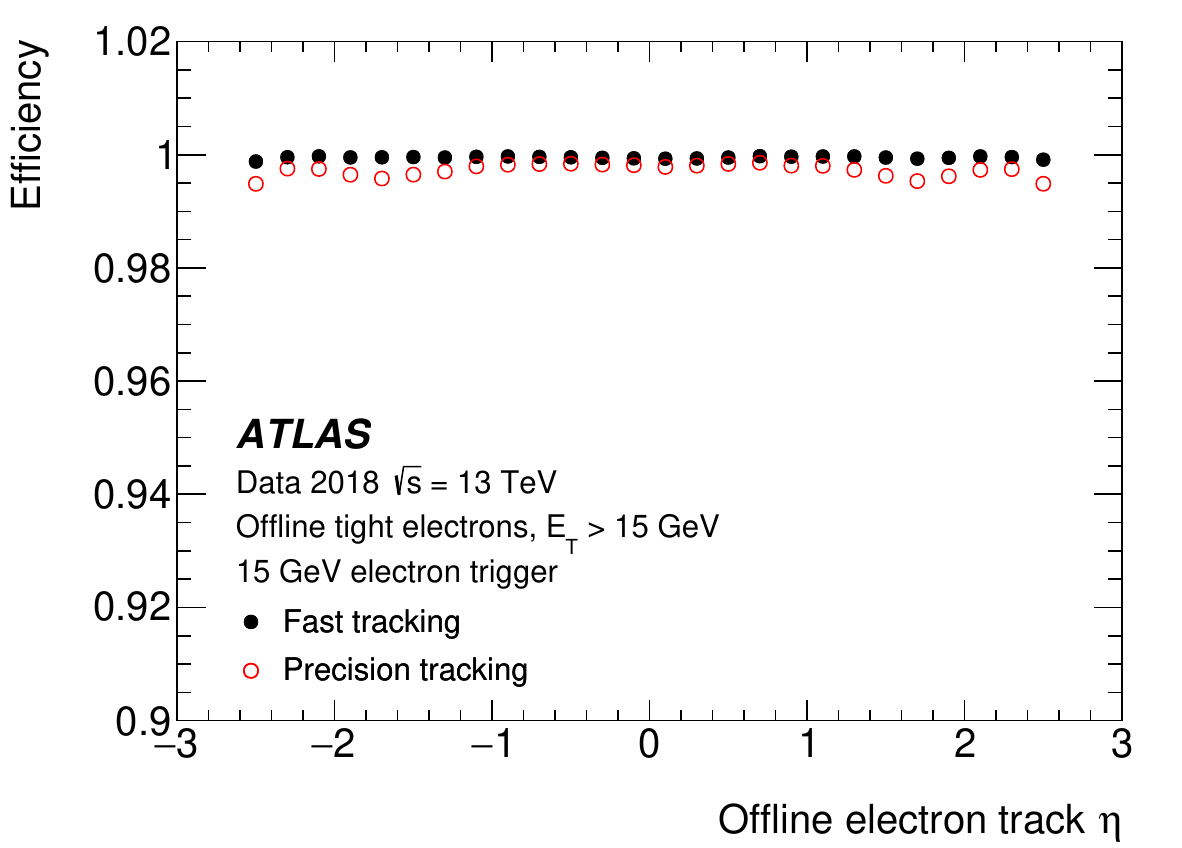}}\figbreak
\subfloat[]{\includegraphics[width=0.5\textwidth]{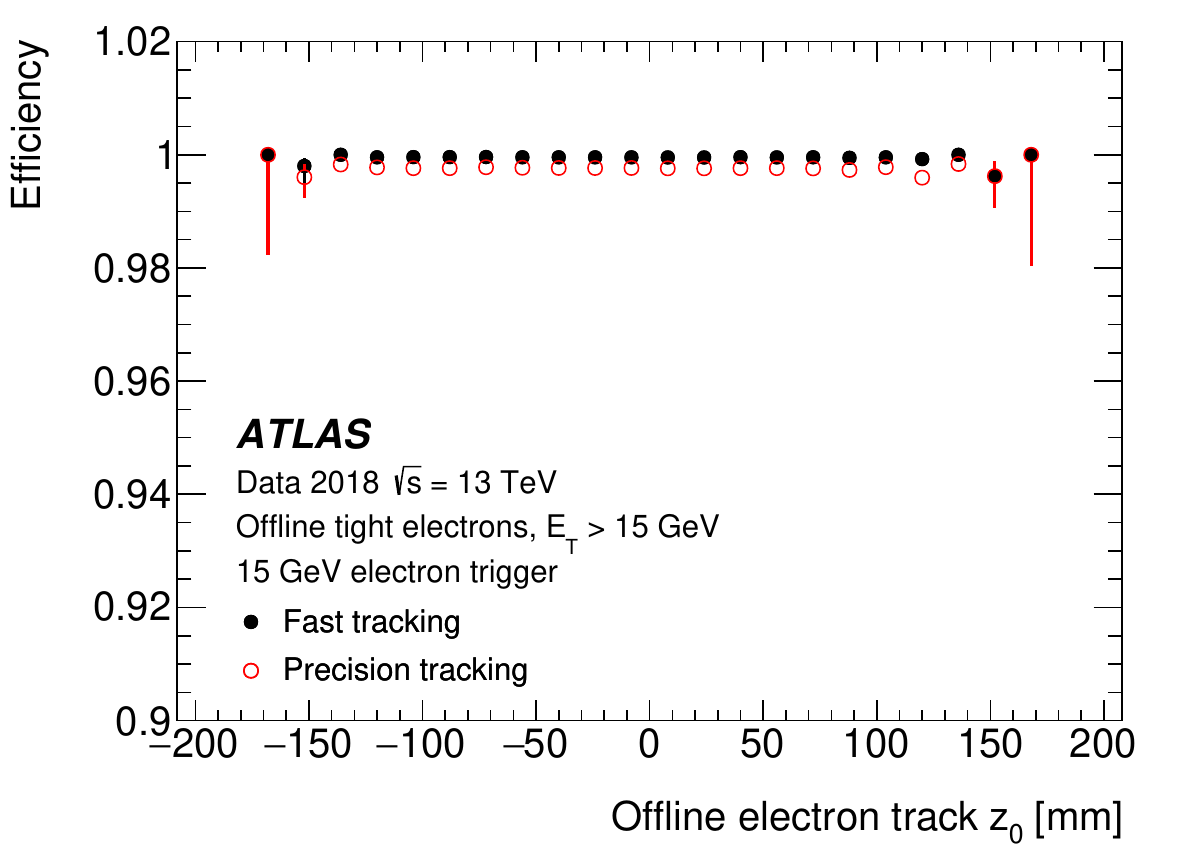}}\\
\subfloat[]{\includegraphics[width=0.5\textwidth]{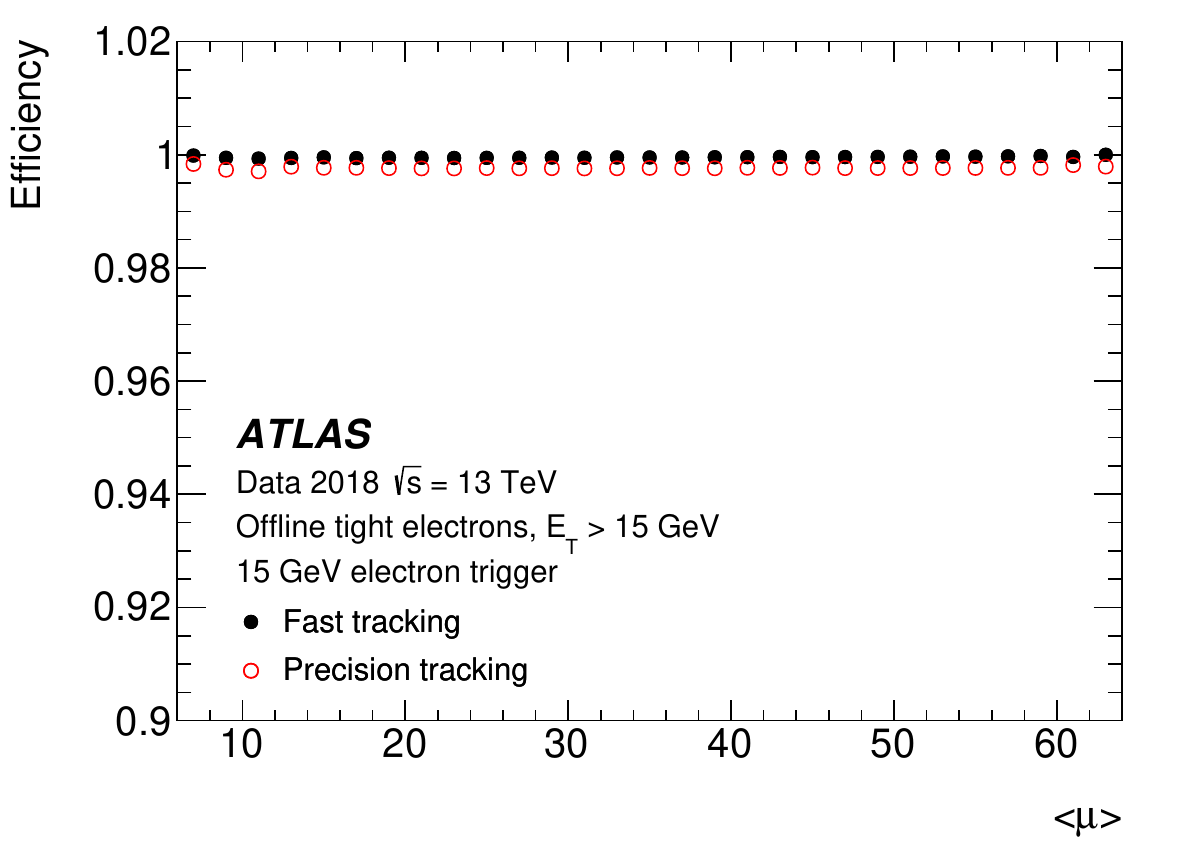}}
\caption{The electron track finding efficiency versus (a) the offline electron track pseudorapidity, (b) $z$ position at the
\beamline, and (c) the mean \pileup interaction multiplicity measured using the dilepton trigger.}
\label{fig:etnpeff}
\end{figure}
}
 
The efficiency as a function of \ET remains very high even for very high energies up to
1\tev seen here,  with a reduction of only approximately 0.5\% near the 15\gev trigger threshold.
The  efficiency versus the track $\pt$ is also constant at large $\pt$, as seen in the earlier analyses.
Candidates with \pt between 5\gev and 15\gev consist almost entirely of candidates that have undergone significant bremsstrahlung
and yet are still reconstructed with efficiencies greater than 97\% for the precision tracking, and better
than 99.5\% for the fast tracking. It should be stressed that in this region between 5\gev and 15\gev, it is not
that the overall electron tracking efficiency is reduced,  but rather that the selection on the electron candidate \ET
is acting to preferentially reject any candidate that has not radiated.
This efficiency for candidates with transverse momenta below 15\gev therefore represents that only for
those candidates that have undergone bremsstrahlung.

The efficiency with respect to the offline track pseudorapidity and $z$ position along the \beamline, together with the
efficiency versus the mean \pileup interaction multiplicity can be seen in Figure~\ref{fig:etnpeff}. The efficiency versus
pseudorapidity shows a clear 0.5\% inefficiency for the precision tracking with pseudorapidity in the range $1.2<|\eta|<2$
in the SCT endcaps. This is consistent with what can be observed from
Figure~\ref{fig:egeff} but with the higher precision illustrating this
more clearly.
The efficiency for both the precision and fast tracking is again seen to be approximately constant with respect to
changes in the $z_0$ position and \pileup interaction multiplicity.
 
\figetnpeff

{\begin{figure*}[thp]
\centering
\subfloat[]{\includegraphics[width=0.5\textwidth]{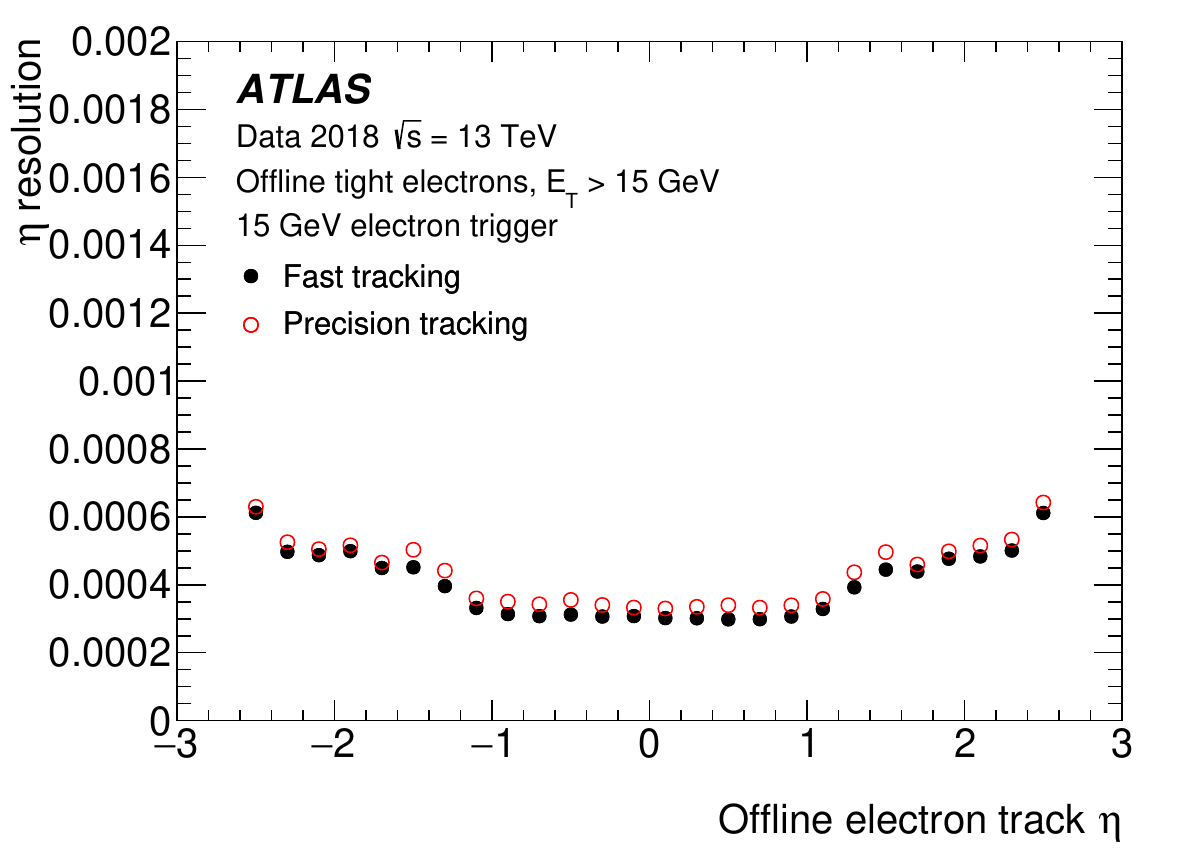}}
\subfloat[]{\includegraphics[width=0.5\textwidth]{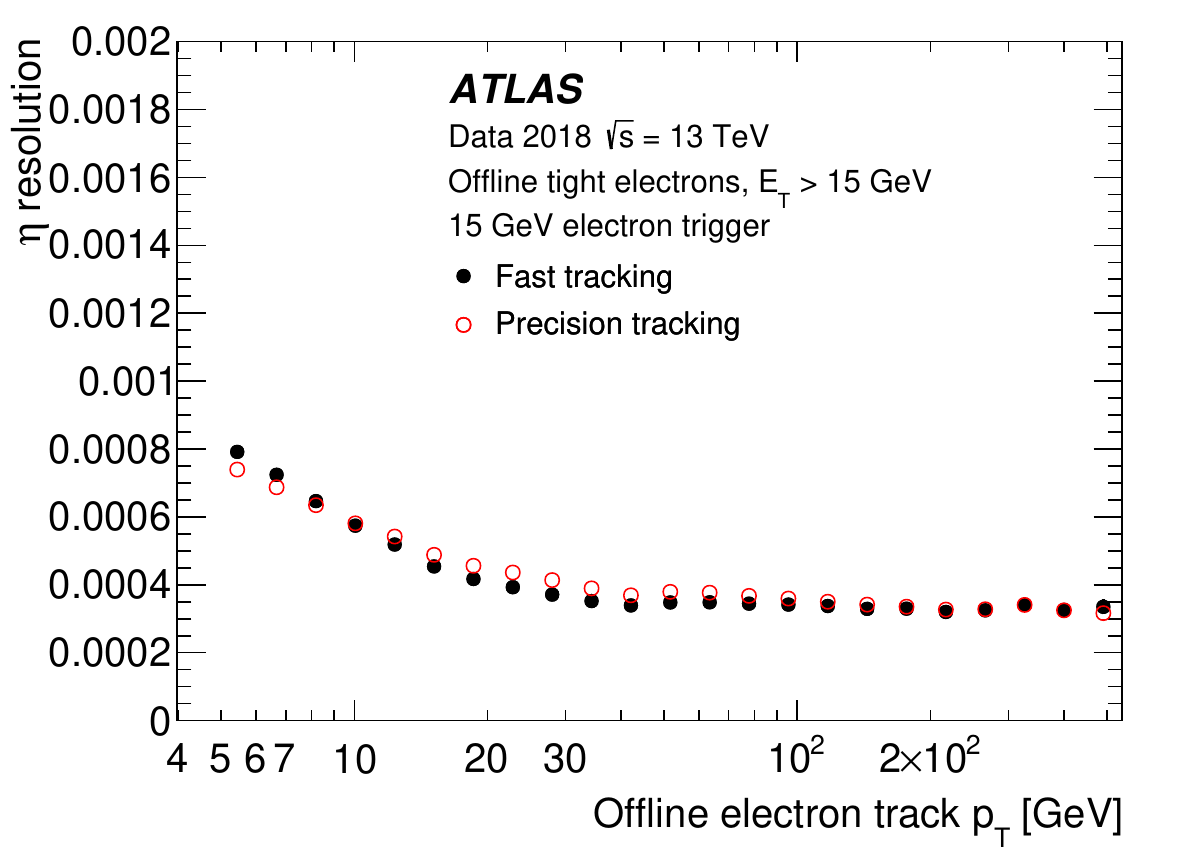}}\\
\subfloat[]{\includegraphics[width=0.5\textwidth]{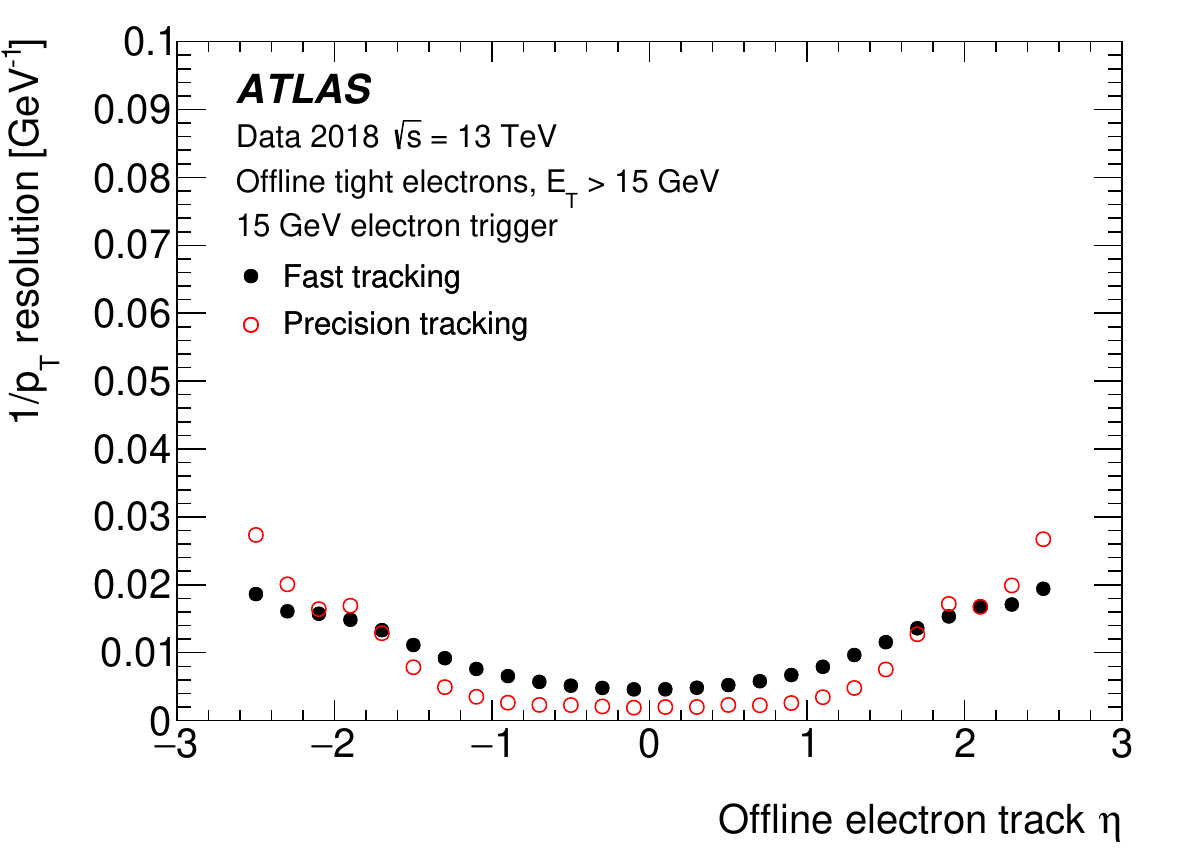}}
\subfloat[]{\includegraphics[width=0.5\textwidth]{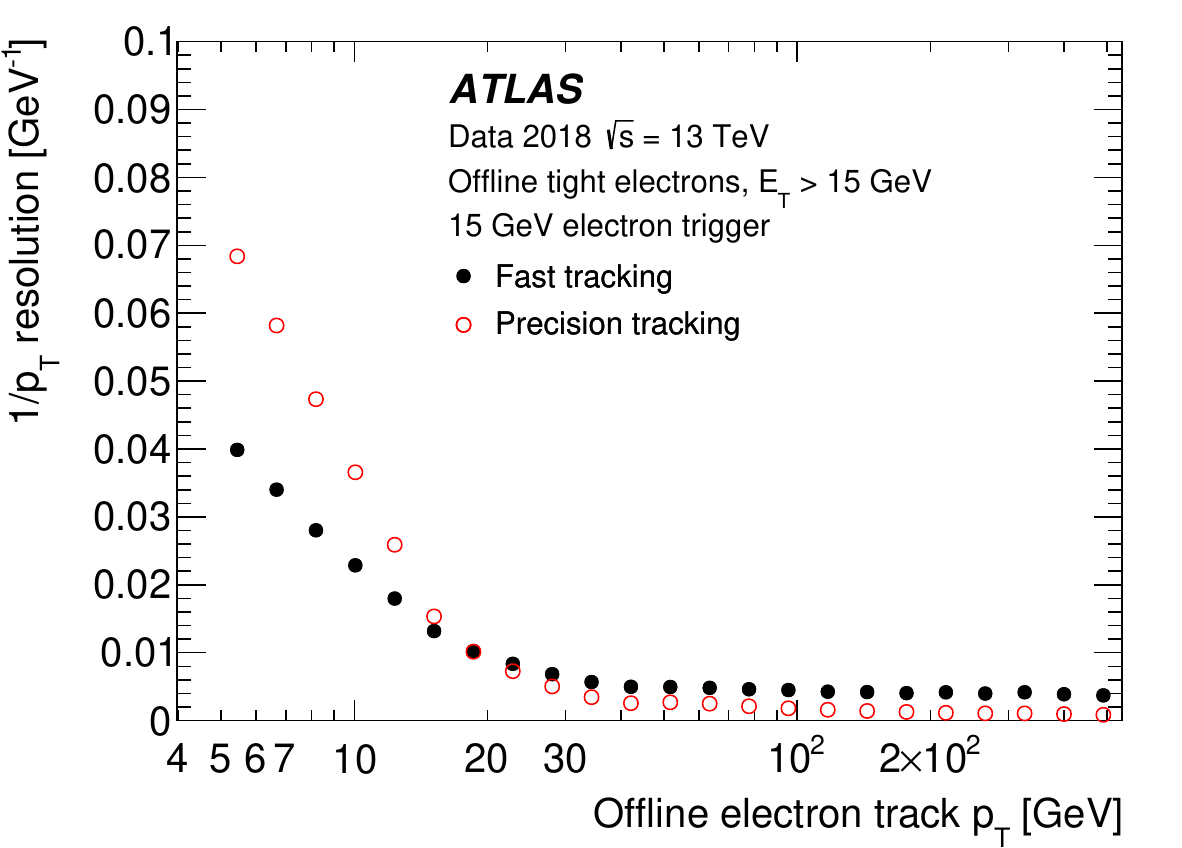}}
\caption{The resolution in the pseudorapidity (a) and (b), and the
inverse transverse momentum (c) and (d)
as a function of offline track pseudorapidity and transverse momentum.
The resolutions are shown for electrons selected by the $\et>\SI{15}{\GeV}$ electron
requirement from
the dilepton support triggers, for both the
fast tracking and precision tracking algorithms.
}
\label{fig:egtnpres}
\end{figure*}}

\newcommand{\egammasp}{
{\begin{figure*}[tp]
\centering
\subfloat[]{\includegraphics[width=0.5\textwidth]{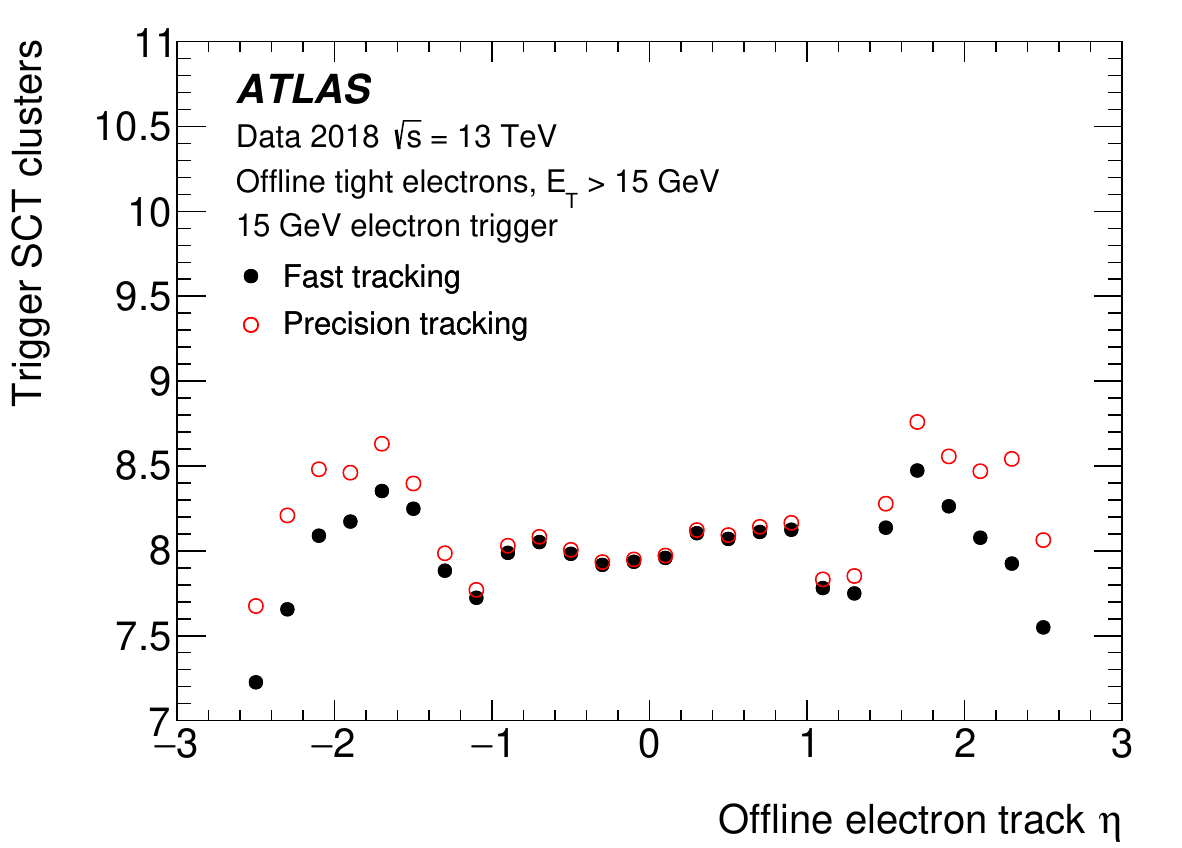}}
\subfloat[]{\includegraphics[width=0.5\textwidth]{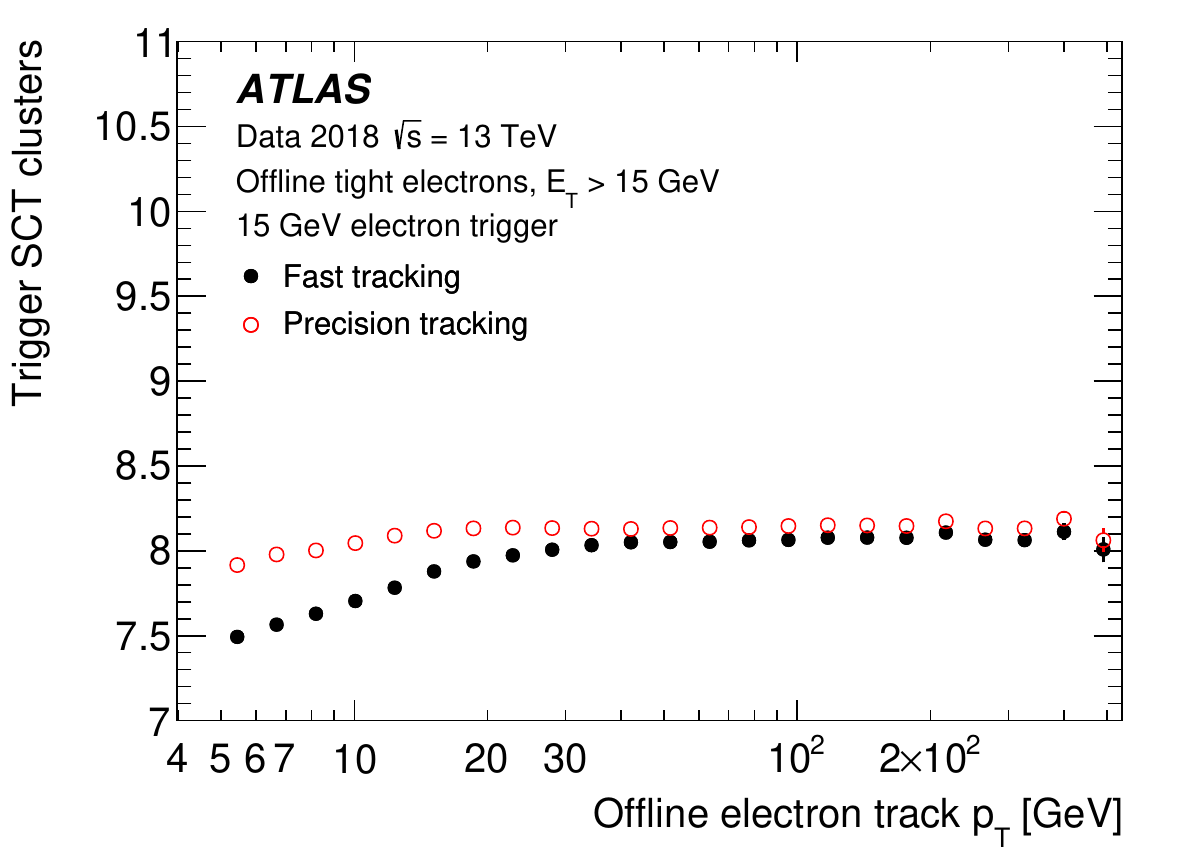}}
\caption{
The SCT cluster  multiplicity for the trigger tracks matched to offline electron candidate tracks as
a function of (a) the offline electron candidate pseudorapidity and (b) the offline electron candidate
transverse momentum from the dilepton performance trigger.}
\label{fig:egammasp}
\end{figure*}}
}

The resolutions in the track pseudorapidity and inverse transverse momentum are shown in Figure~\ref{fig:egtnpres}
as functions of the offline track pseudorapidity and transverse momentum themselves.
The performance of the resolution
at large transverse momentum is seen to be consistent with that observed in Figure~\ref{fig:egres}.
The resolutions at lower transverse momentum, and versus pseudorapidity are also consistent, lying
between the resolutions from the 10\gev and 26\gev triggers, but once again with a greatly reduced
statistical uncertainty.
 
\egammasp

The worse resolution for the inverse of the transverse momentum for the precision tracking
in comparison with the fast tracking is again seen at low transverse momenta and large pseudorapidities,
as in Figure~\ref{fig:egres}. This can
be understood as a consequence of the lower mean SCT hit multiplicity for the FTF tracks in the
endcaps.

Figure~\ref{fig:egammasp} shows the mean SCT cluster multiplicity for the precision and fast tracks
for the \et>15\gev electron trigger tracks as functions of both transverse momentum and pseudorapidity
of the matched offline electron candidate.
The general variation of the multiplicity  with pseudorapidity is the same as for the other signatures,
but at large pseudorapidities, and for transverse momenta lower than the calorimeter cluster threshold,
it is seen that the FTF tracks have a lower SCT hit multiplicity than the precision tracks.

Tracks from electron candidates
which have radiated an energetic photon may have significant kinks such that their trajectory will not
be helical. To account for this, the offline electron reconstruction runs a Gaussian sum
filter~\cite{ATLAS-CONF-2012-047} to allow for significant kinks in tracks due to bremsstrahlung,
whereas the electron track reconstruction used in the trigger during
\runii was running the standard track reconstruction.
As a consequence of the pixel requirements, when fitting tracks including the full complement
of correctly identified SCT hits, the  precision track candidates containing kinks will be less helical.
The transverse momentum reconstruction near the \beamline for the fitted helical track will therefore
be affected as the track fit is pulled away from the true trajectory to compensate for the
position of the outer SCT hits and so produce a fit with lower transverse momentum.
For the FTF tracks, if the outer SCT hits are rejected in the fit, as suggested by
Figure~\ref{fig:egammasp}, then the early sections  of
the track closest to the \beamline before any bremsstrahlung will be better reconstructed and so the \pt
reconstruction of the track nearest the \beamline will be somewhat better.

\newcommand{\ptbias}{
\begin{figure*}[th]
\centering
\subfloat[]{\includegraphics[width=0.5\textwidth]{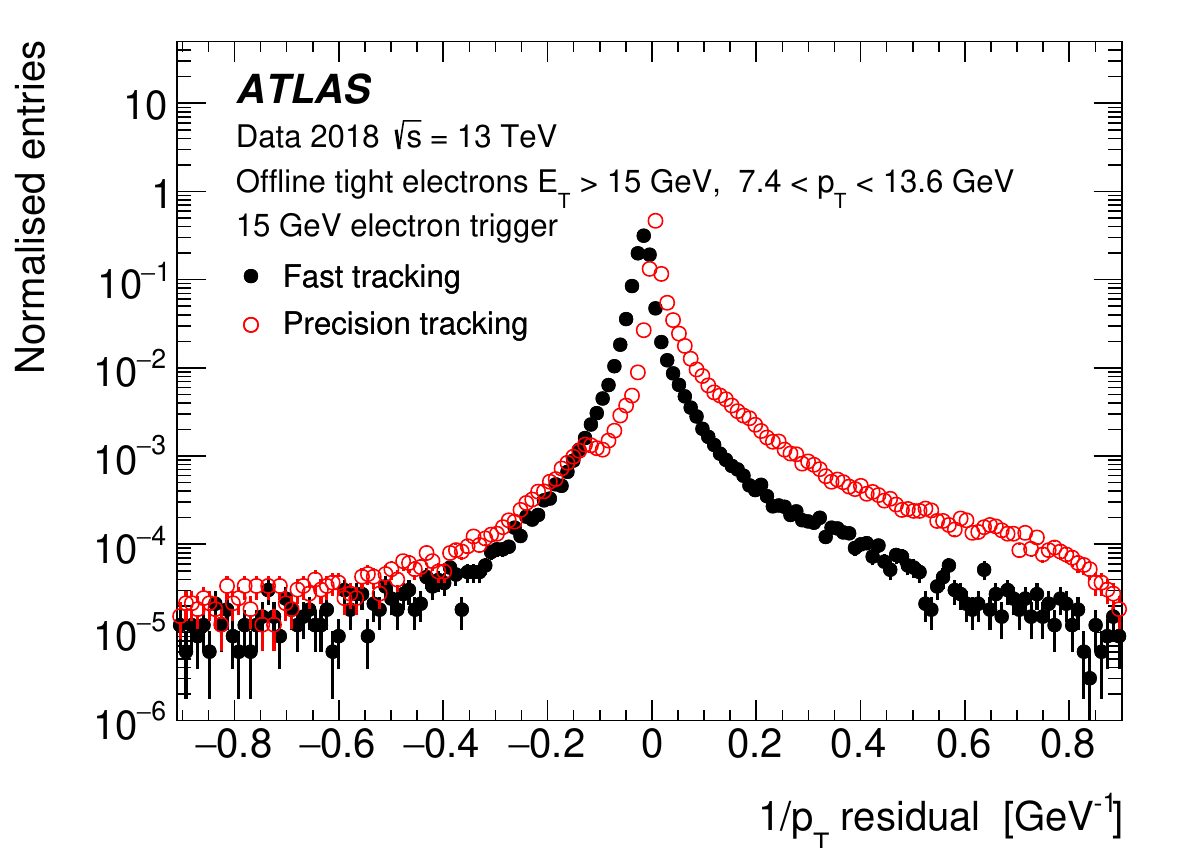}\label{fig:ptbiasa}}
\subfloat[]{\includegraphics[width=0.5\textwidth]{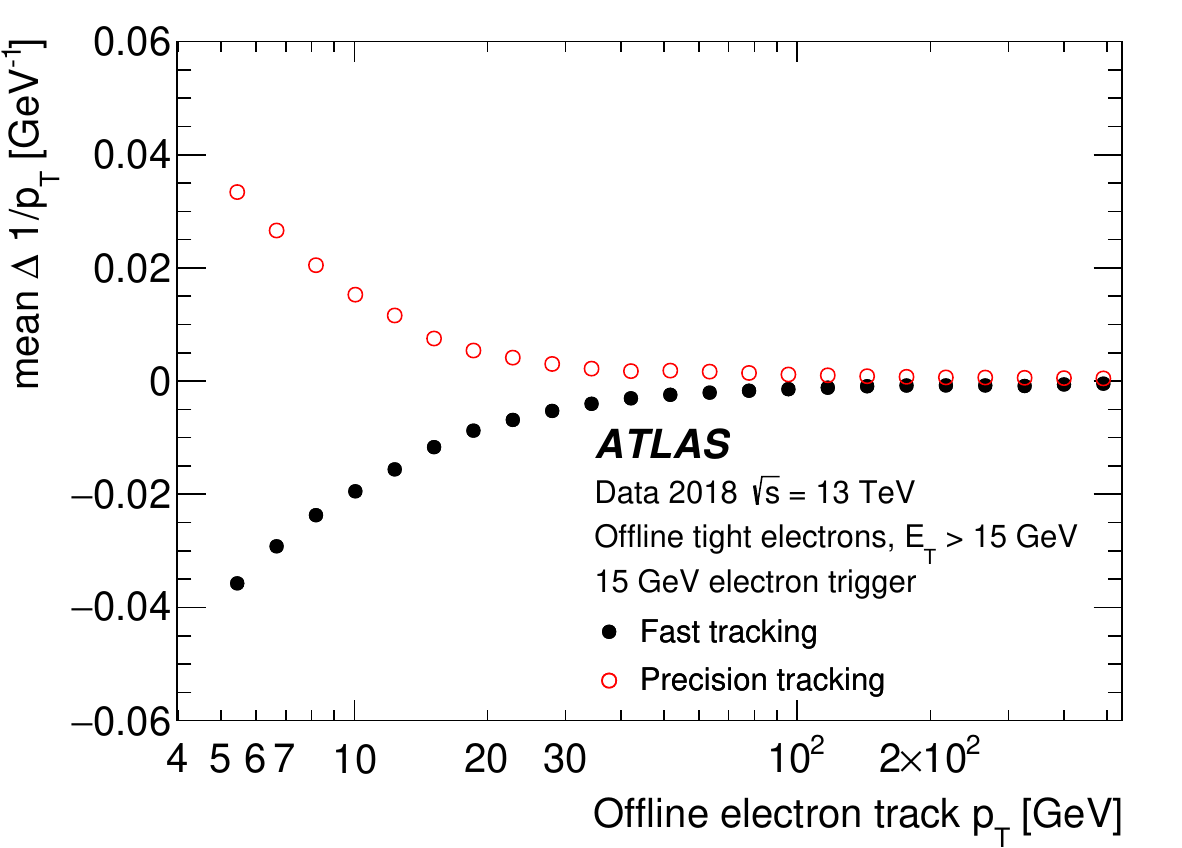}\label{fig:ptbiasb}}
\caption{The ID trigger residual in $1/\pt$ with respect to offline for the fast and precision tracking for electron
candidates with the offline track transverse momentum the range in $7.4<\pt<13.6$\gev selected by the \SI{15}{\GeV}
tag and probe trigger (a); and the mean of the residual as a function of the offline track \pt (b)
for the fast and precision tracking.}
\end{figure*}
}
 
\ptbias
 
This can be seen in Figure~\ref{fig:ptbiasa} which shows the residual in the  $1/\pt$ distribution for the
fast track finder and precision tracking with respect to the offline electron track candidate \pt for
candidates with the offline track \pt around 10\gev ($7.4<\pt\leq13.6$\gev). Here the fast track finder
residual distribution can be seen to be somewhat narrower than that from the precision tracking, which has a large tail
extending to larger $1/\pt$ values reconstructed in the trigger, {\em i.e.} to lower \pt. The central peak
for the fast track finder is shifted to somewhat lower values, whereas the peak in the precision tracking
is closer to zero, but with the large tail, pulling the mean to higher values.

This can be seen more concisely in the Figure~\ref{fig:ptbiasb} which shows the mean of the residual
distribution for the fast tracking finder and precision tracking as a function of the offline electron
candidate track \pt.
This clearly shows that for tracks with
very low \pt relative to the minimum electron \et, the offset of the mean is large and positive for
the precision tracking, meaning a bias towards lower track \pt than that measured offline.
For the fast track finder, the bias -- predominantly from the small shift in the peak position --
is towards higher \pt than
for the offline track, and so closer to the transverse energy of the electron candidate
as determined by the cluster measured in the calorimeter.
 
For future data taking it may be possible for the trigger to include aspects of the Gaussian sum
filter tracking, used offline, to improve the transverse momentum resolution of the precision
tracking.

\floatbarra


 
\subsection{Tracking in the tau signature}
\label{sec:tau}

{\begin{figure*}[p]
\centering
\subfloat[]{\includegraphics[width=0.5\textwidth]{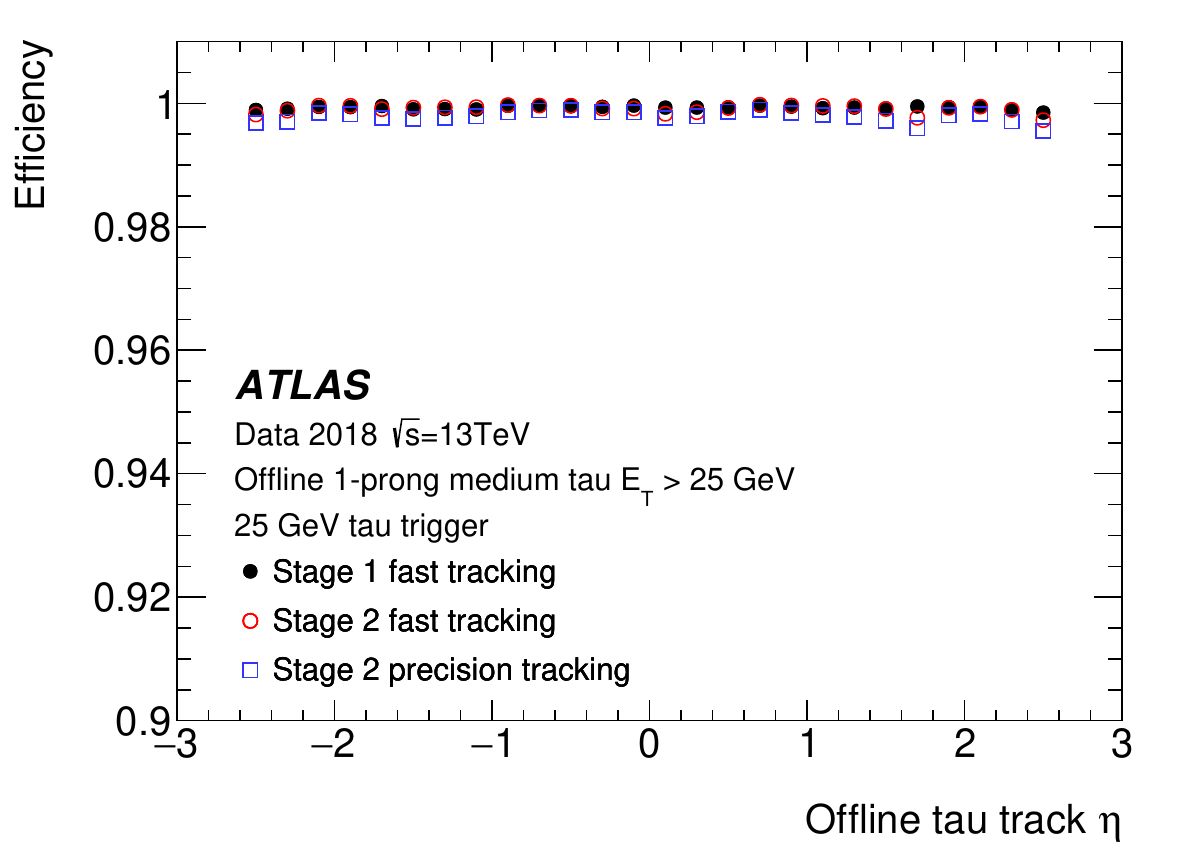}}
\subfloat[]{\includegraphics[width=0.5\textwidth]{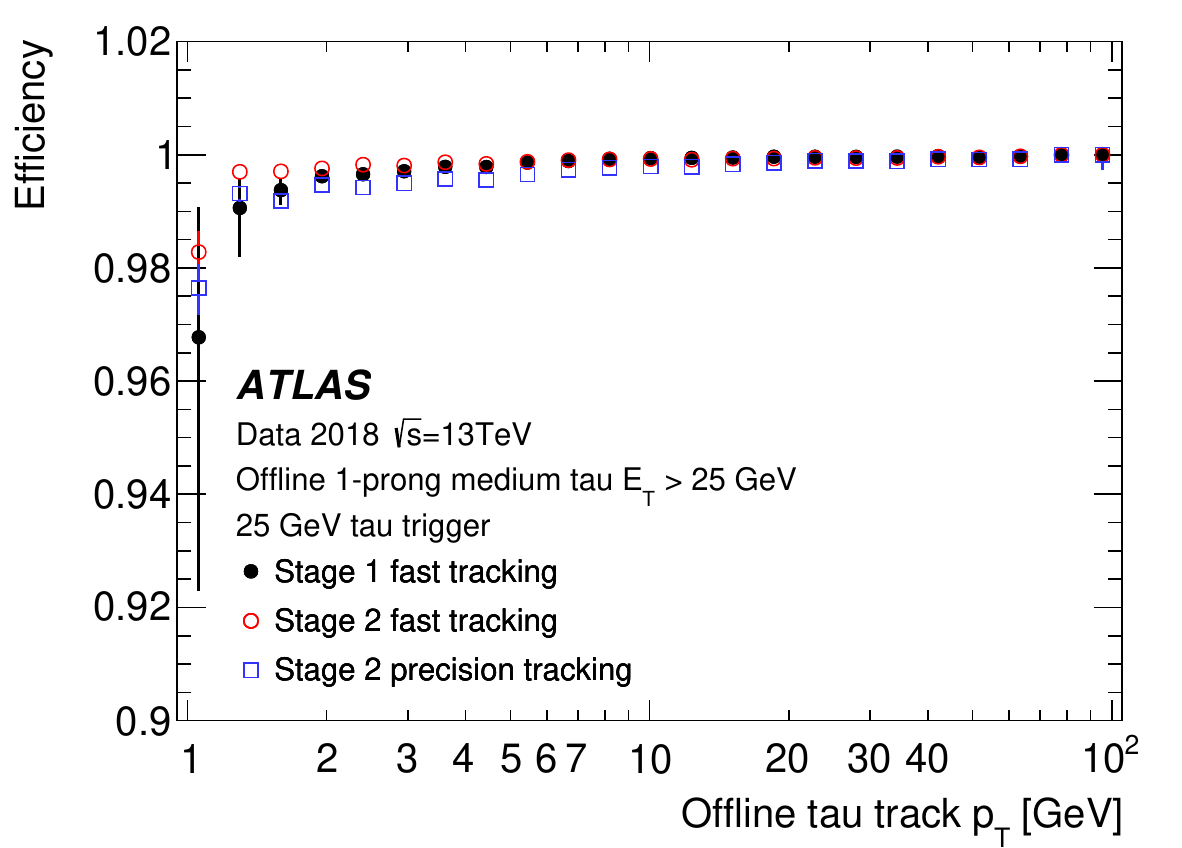}}\\
\subfloat[]{\includegraphics[width=0.5\textwidth]{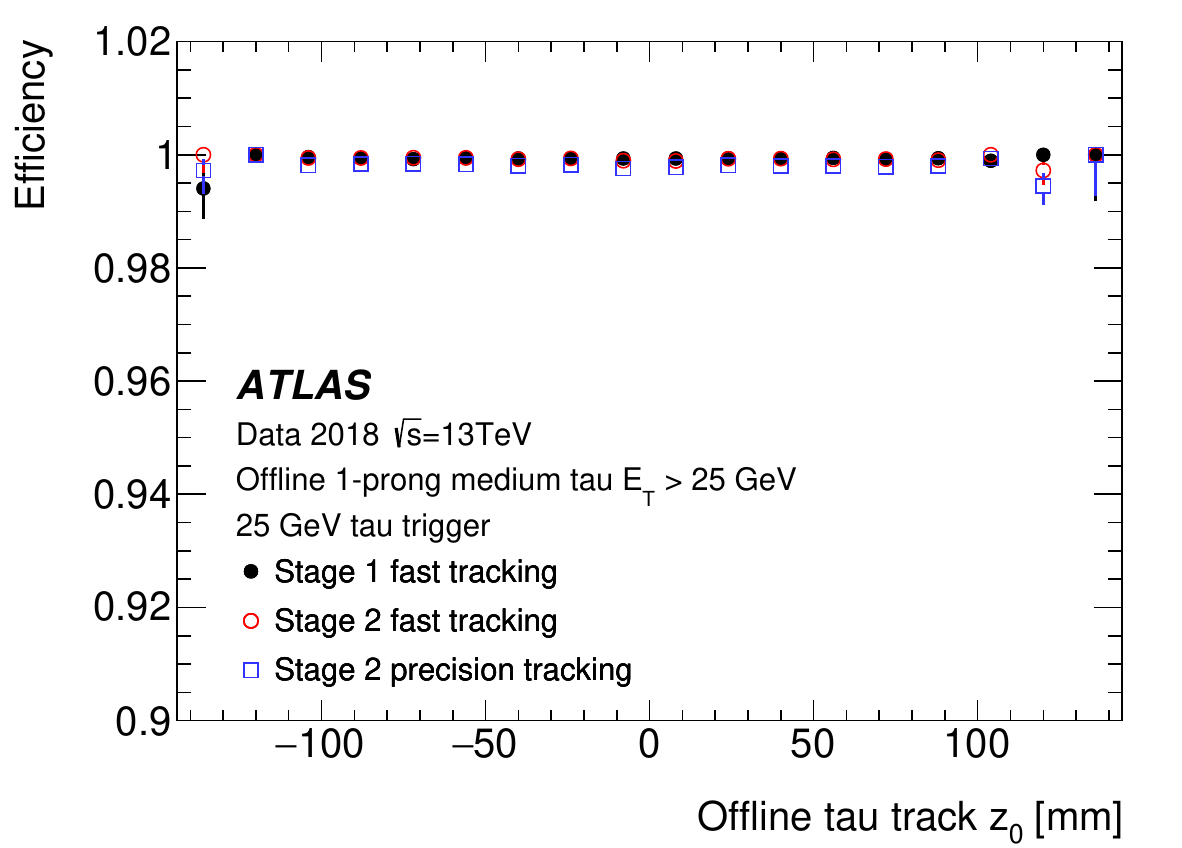}}
\subfloat[]{\includegraphics[width=0.5\textwidth]{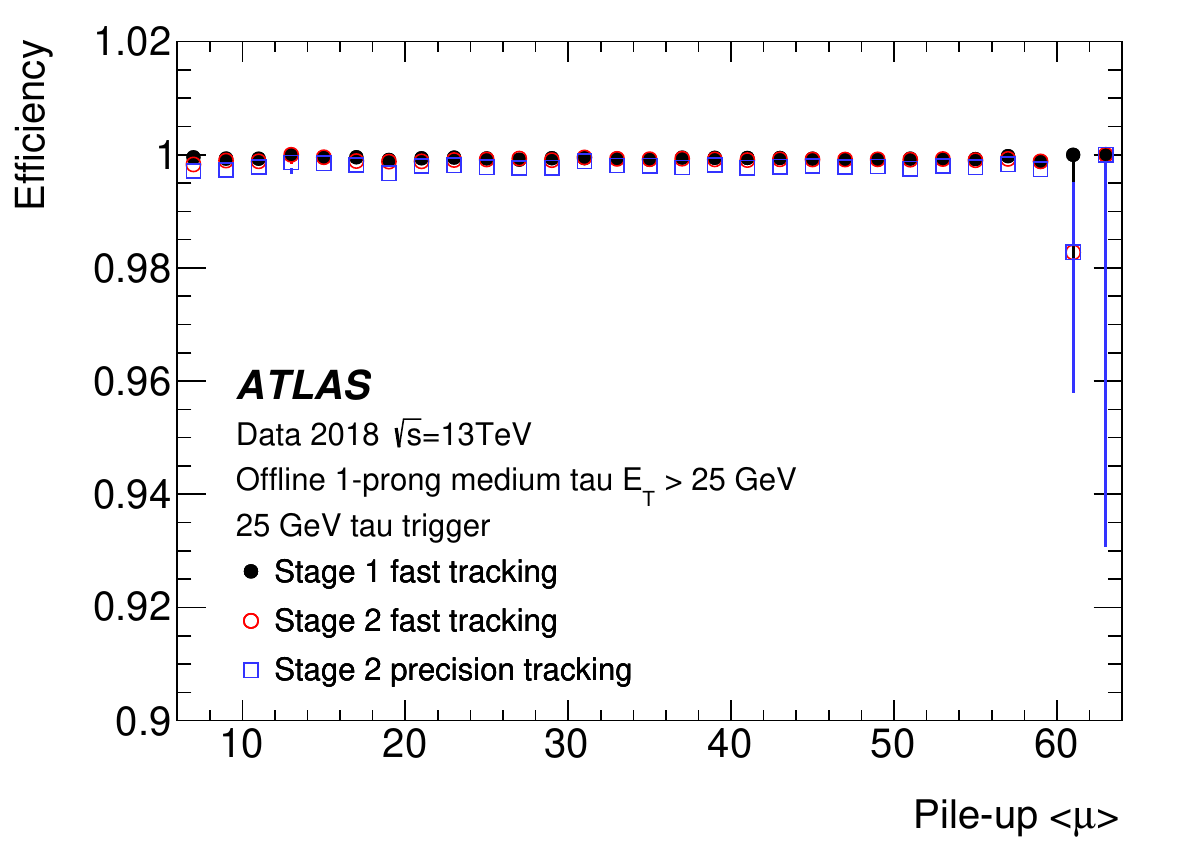}}
\caption{The track finding efficiency of the ID trigger for tracks from
tau decays with $\pt > 1\gev$.
The efficiency is evaluated for the \SI{25}{\GeV} tau performance trigger
which does not use ID tracking information for the selection.
The efficiency is shown as a function of
(a) offline track $\eta$,
(b) offline track \pt,
(c) offline track $z_{0}$ and
(d) mean number of interactions per bunch crossing.
Statistical Bayesian uncertainties are shown, using a 68.3\% confidence interval
and uniform beta function prior.}
\label{fig:taueff}
\end{figure*}}

Figure~\ref{fig:taueff} shows the efficiency of the tau two-stage tracking with respect to the
offline tracking for offline tracks with $\pt>\SI{1}{\GeV}$ originating from decays of offline tau lepton
candidates~\cite{PERF-2013-06,PERF-2014-06} with $\pt > \SI{25}{\GeV}$.
The multistage tracking used in the tau trigger signature~\cite{ATLAS-CONF-2017-061,TRIG-2016-01}
is described in detail
in Section~\ref{sec:multistage},
but it is worth recalling that the first stage runs in a narrow \roi in $\eta$--$\phi$. The narrow $\phi$ region means
that tracks at low \pt (up to around 5\gev) may bend significantly in the solenoid field, and if near the edge
of the \roi, may not be fully contained in the \roi and so could not be reconstructed.
This is partly taken into account when selecting reference tracks for the measurement of the efficiency for the different
stages, where offline tau candidate tracks which are not fully contained within the \roi are rejected.
The width of the \roi determines the lowest transverse momentum that can be reconstructed in an \roi.
However, since the \roi originates from  $(0,0)$ in the detector rather than at the \beamline, some lower momentum tracks
near the edges of an \roi in $\phi$ may still be subject to a reduction in efficiency if part of the track
falls outside the \roi.
Figure~\ref{fig:taueff} shows the efficiencies for the fast tracking from the first and second stages together with
that from the second stage precision tracking. Because of the \roi containment, the sample of tracks near threshold is
quite small, particularly for the first stage tracking, resulting in a lower efficiency and  a significantly larger
uncertainty.

The track resolutions for $\eta$, $d_{0}$ and $z_{0}$  with respect to offline track $\eta$ and \pt
are shown in Figure~\ref{fig:taures}. The precision tracking resolution is generally
better than the fast tracking resolution. The resolution from the fast tracking in both the first and second stages
has essentially the same performance for offline tau track candidates with $\pt>5$\gev but differs for $\pt<5$\gev.
This is an artifact of the \roi containment requirement for the different \roi discussed above.
It is worth noting that only the \dzero from the precision tracking reconstruction
is used in the final selection of tau candidates online, for which the resolution with respect to offline
is better than 20~$\mu$m for tracks with $\pt>5$\gev. When integrating over \pt, the mean resolution
lies in the range 15--20~$\mu$m for all pseudorapidities.

{\begin{figure*}[p]
\centering
\subfloat[]{\includegraphics[width=0.5\textwidth]{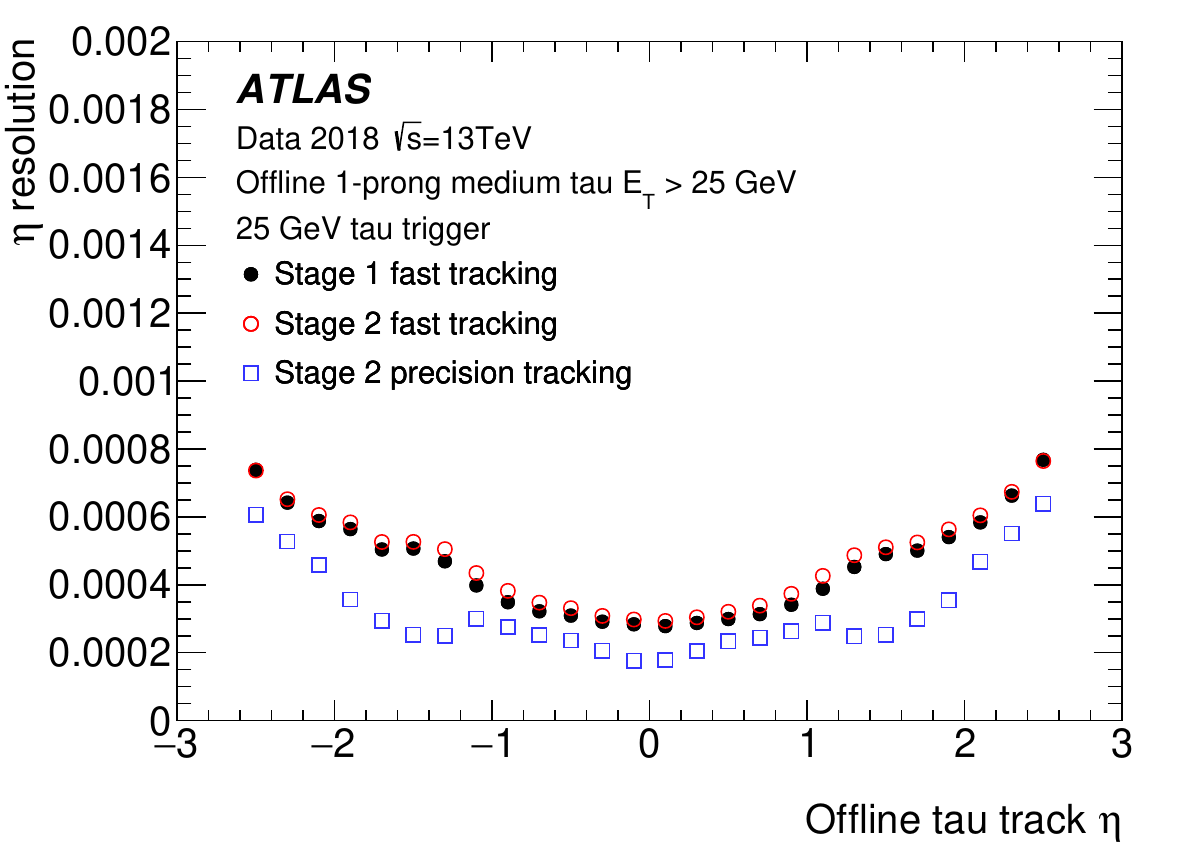}}
\subfloat[]{\includegraphics[width=0.5\textwidth]{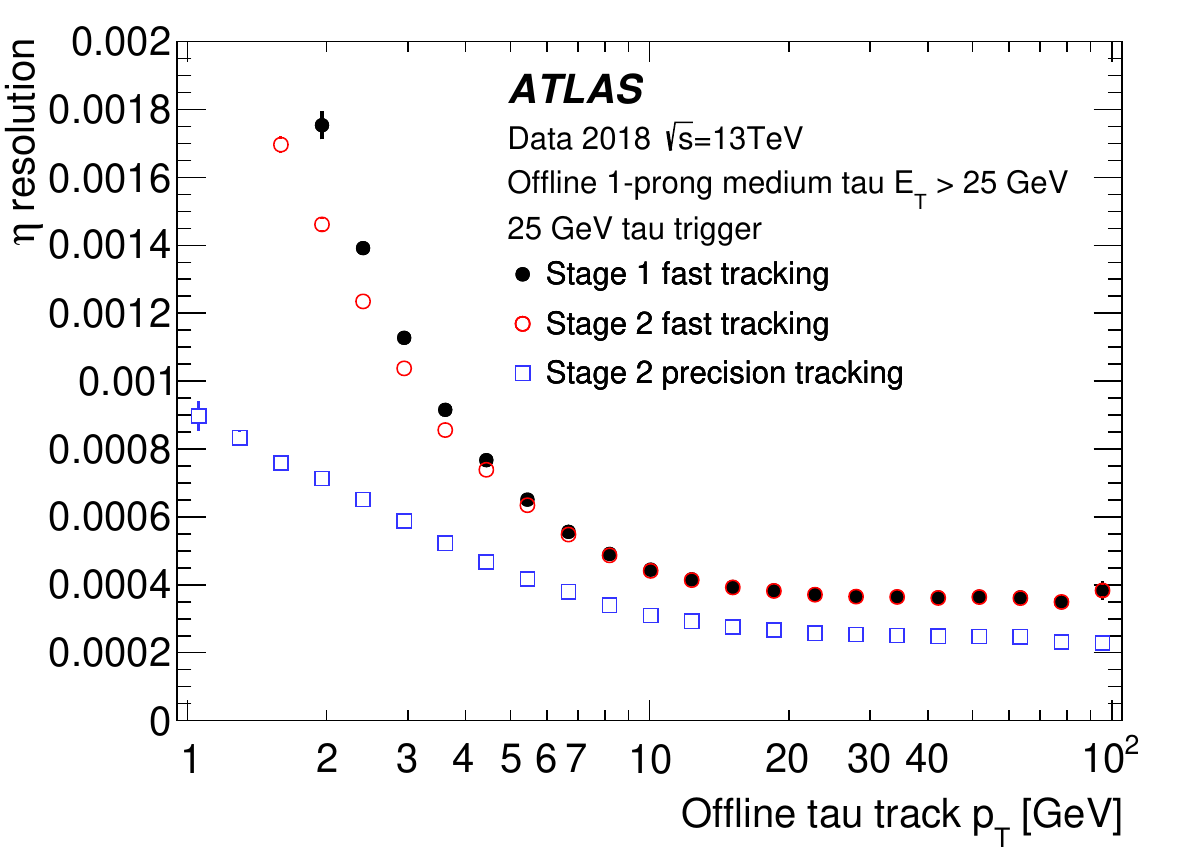}}\\
\subfloat[]{\includegraphics[width=0.5\textwidth]{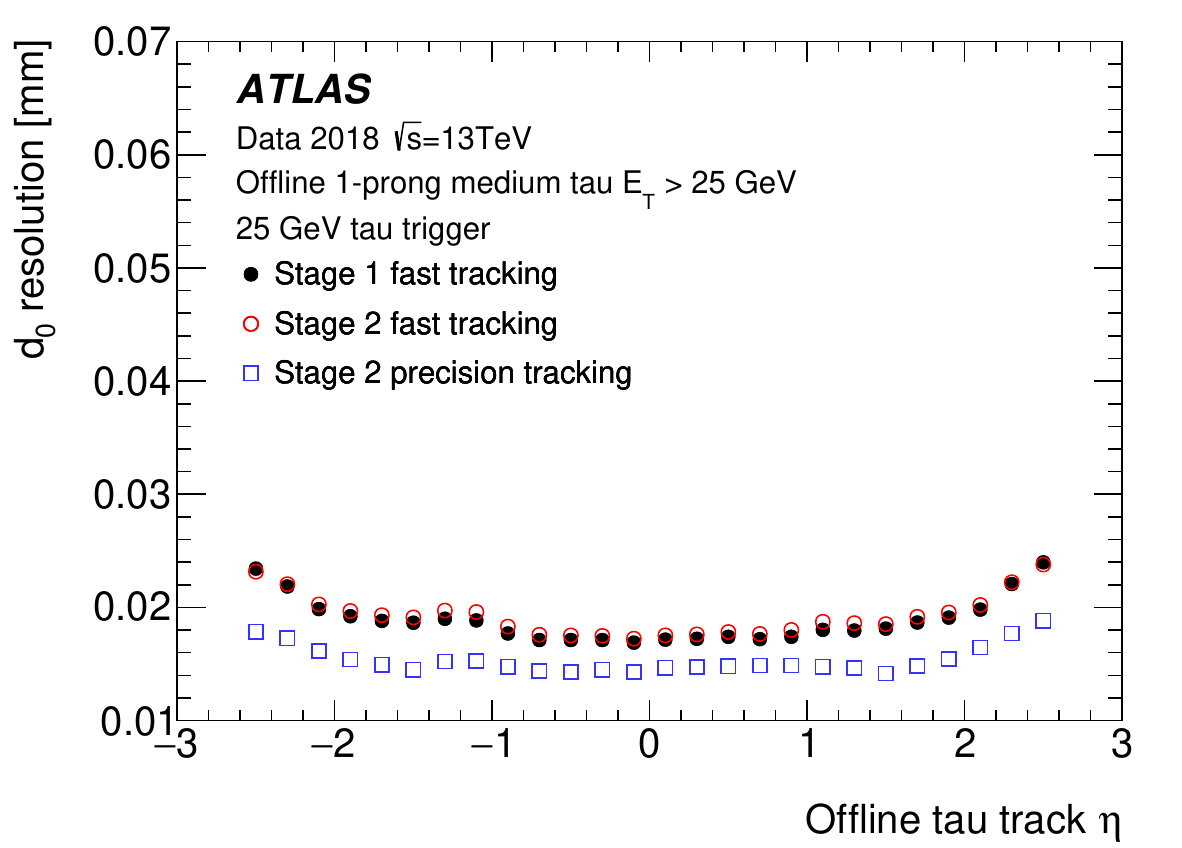}}
\subfloat[]{\includegraphics[width=0.5\textwidth]{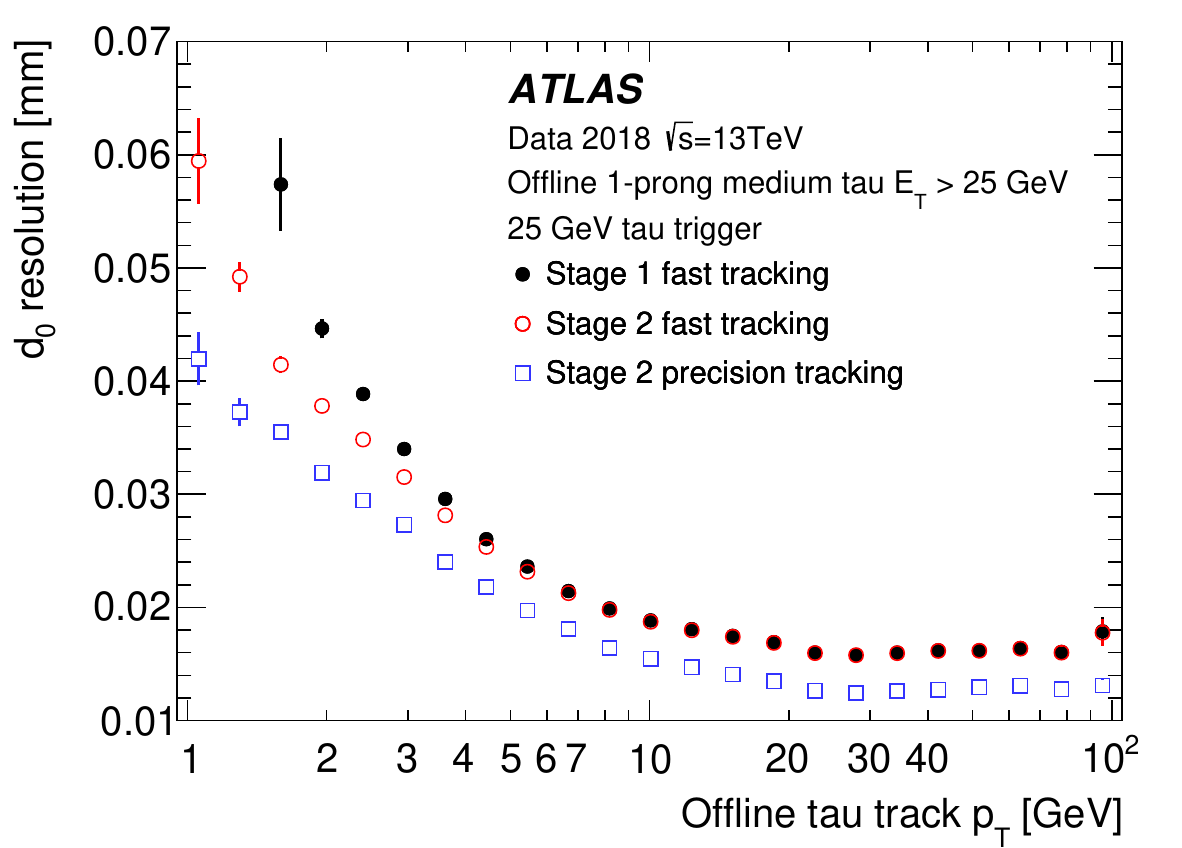}}\\
\subfloat[]{\includegraphics[width=0.5\textwidth]{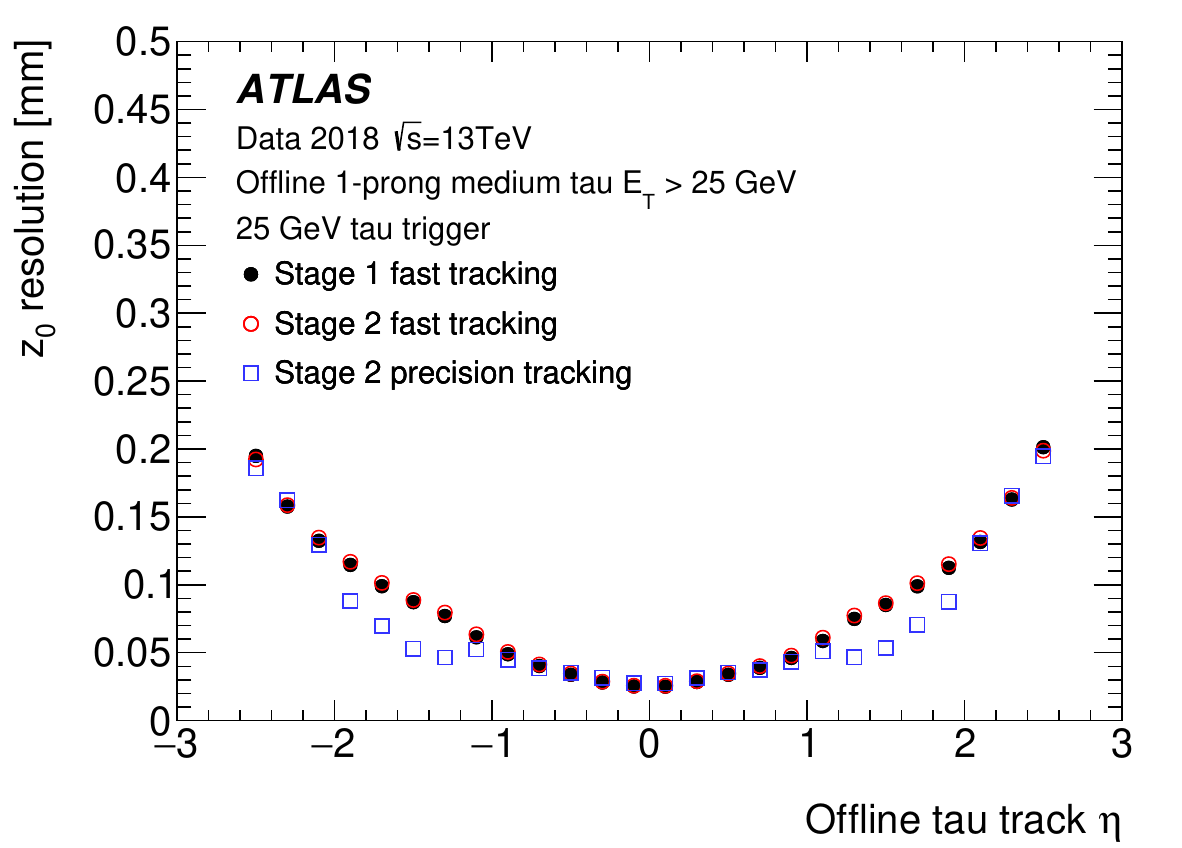}}
\subfloat[]{\includegraphics[width=0.5\textwidth]{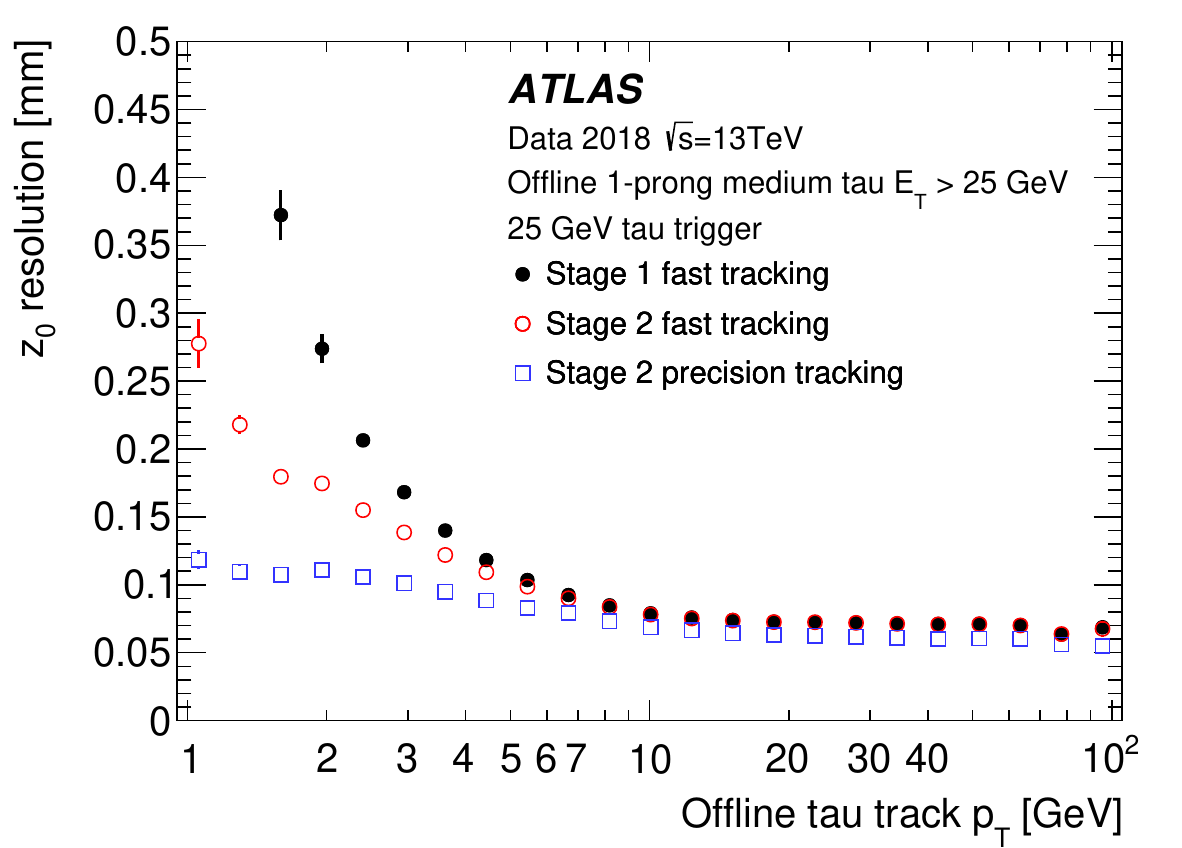}}
 
\caption{
The tau signature tracking resolution in (a, b) pseudorapidity,
(c, d) transverse impact parameter (c, d),
and (e, f) track $z$ position along the \beamline, all as functions of
the matching offline track pseudorapidity and transverse momentum.
}
\label{fig:taures}
\end{figure*}}

{\begin{figure*}[th]
\centering
\subfloat[]{\includegraphics[width=0.5\textwidth]{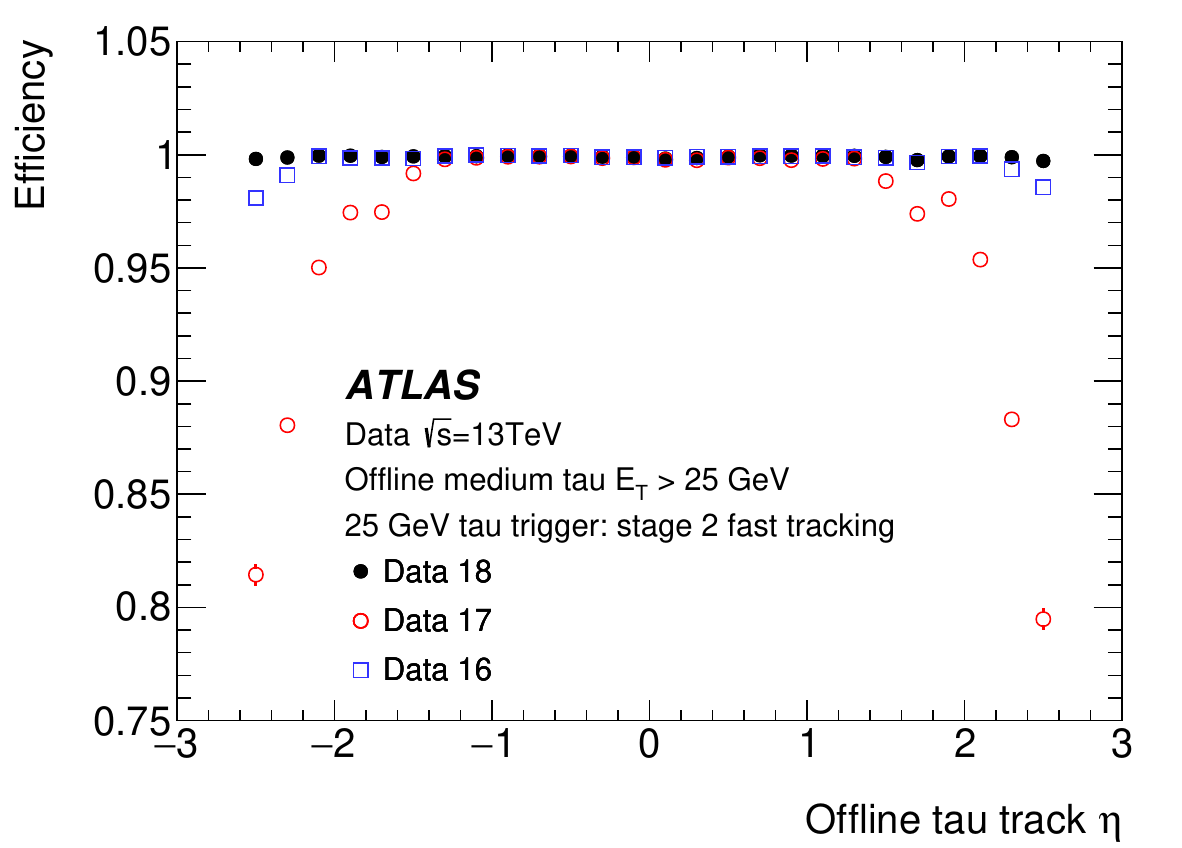}\label{fig:tauduffa}}
\subfloat[]{\includegraphics[width=0.5\textwidth]{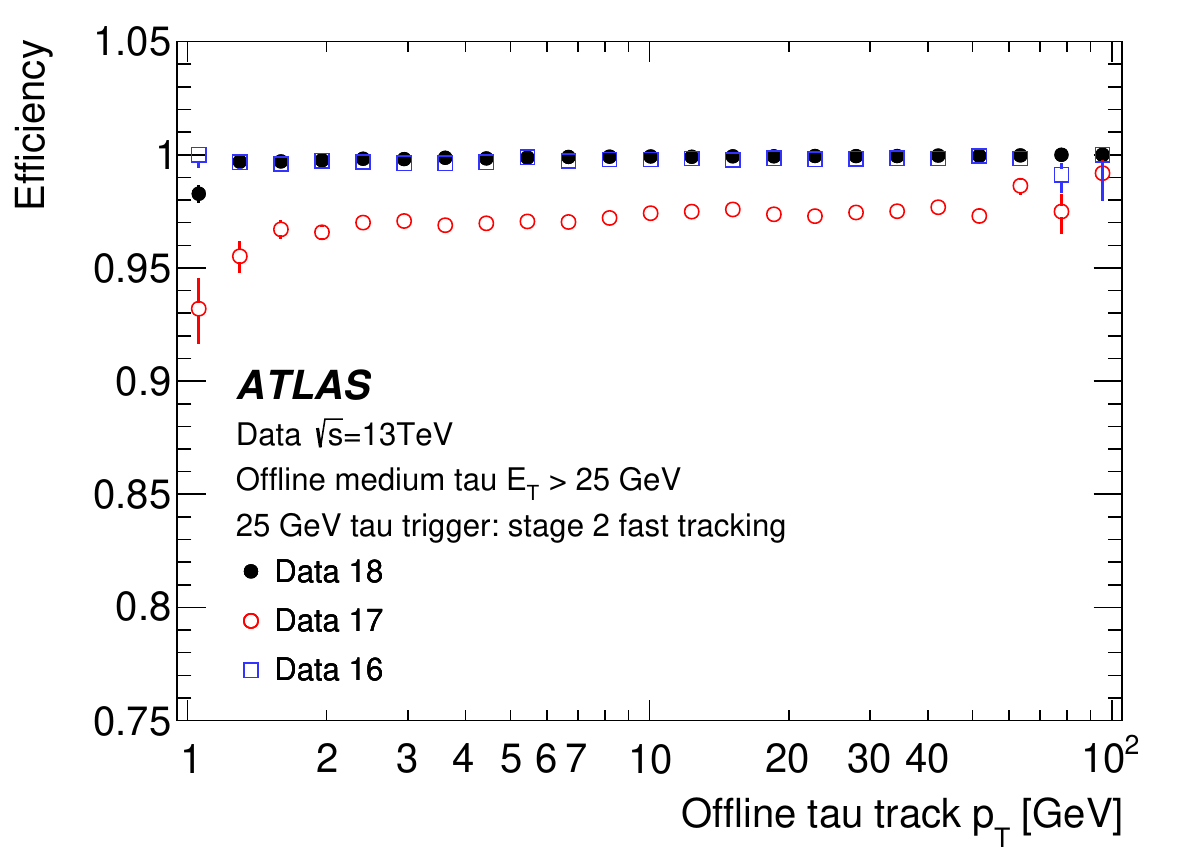}\label{fig:tauduffb}}
\caption{
The efficiency of the tau second stage fast tracking as a function of (a) pseudorapidity and (b)  transverse momentum,
comparing the performance in 2016, 2017, and 2018.}
\label{fig:tauduff}
\end{figure*}}

The effect of changes to the trigger between 2016, 2017, and 2018 can be seen for the second stage
fast tracking in Figure~\ref{fig:tauduff}.
The first stage tracking essentially underwent no changes between the three years, the differences being
predominantly in the second stage processing. As seen in Figure~\ref{fig:tauduffa}, during 2016 a small
inefficiency was observed at large pseudorapidities because of the tightening of the second stage \roi
about the $z$-position of the
leading track. The resolution of the predicted seed positions for seed pairs at large
pseudorapidities becomes important. Approximations used in the layer positions for the seed finding
were particularly affected by the worse resolution for seeds at large pseudorapidities, and as such
the predicted position of the seed pair at the \beamline was not precise enough to guarantee that the
seed pairs fell within the selection window of the \roi, and consequently a significant fraction of seeds
were being rejected. For 2017, the increased rate due to the higher \pileup occupancy required an
additional change to the seed finding used for the FTF tracking, which resulted in a further
deterioration of the efficiency at larger $|\eta|$, while the efficiency at central $\eta$ was unaffected.
 
Modifications to take account of the worse seed resolution at large pseudorapidities were under
development, but were not ready for the start of data taking in 2017.
For 2018 data taking, the seed
finding was reimplemented to avoid the approximations and restore the efficiency at large $|\eta|$.
The efficiency versus \pt, seen in Figure~\ref{fig:tauduffb}, illustrates the significant
loss of efficiency in 2017 at large $|\eta|$, which leads to an overall 3\% loss of efficiency,
improving to 2\% at higher \pt where the tracks become more central.
With the modifications to the seeding for 2018 running, the efficiency improves to better than 98\%
everywhere and better than 99.5\% for $\pt > 1.2$\gev.

\floatbarra


 
\subsection{Tracking in the \bjet signature}
 
As discussed in Section~\ref{sec:multistage}, for the \bjet trigger~\cite{TRIG-2016-01,ATL-COM-DAQ-2020-064}, the
tracking runs in two stages. The first stage runs the vertex tracking in narrow
$\eta$--$\phi$ \rois about the jet direction for those jets with $\pt>30$\gev,
and which are aggregated into a \superroi before running the track reconstruction.
The individual jet \rois about each jet are fully extended from
$-225 < z < 225$~mm along the \beamline.
These  tracks are then used to reconstruct the vertex for the second stage processing.
From 2017, to reduce the processing time, the pattern recognition for the first stage tracking
ran only for track candidates with $\pt>5$\gev.
 
The purpose of the vertex tracking is primarily to identify sufficient tracks with a high
enough transverse momentum to reconstruct the primary interaction vertex corresponding to the jets
to refine the \roi for the tracking used for the $b$-tagging.
Since the first stage vertex is used only to define the $z$ position of the
second stage \rois, the vertex resolution is not required to be better than approximately 1~mm.
The size of the jet \rois used in the first stage, and the higher threshold for track finding
were carefully tuned for 2017 data taking and chosen to keep the processing time to a
manageable level in events with high jet multiplicities whilst still maintaining a
high enough efficiency for the \bjet events of interest.
 
The second stage runs both the fast and the precision tracking in wider \rois centred on
the jet axes, with $|\Delta\eta|<0.4$ and $|\Delta\phi|<0.4$ relative to the \roi direction,
and with the centre of each \roi in $z$, at the position of the primary vertex identified
in the first stage. These second stage \rois have a smaller extent, $\Delta z < 10$~mm, about
this $z$ position.
 
Only the tracks from the precision tracking of the second stage are used for the \bjet selection.
The precision tracks are also used in the trigger for the global sequential
jet energy calibration~\cite{ATL-DAQ-PUB-2018-002}.

For the selection of the reference tracks used for the analysis of the efficiency and resolution of
the \bjet signatures,
a tight offline selection was used which takes into account the detector geometry and
performance as a function of pseudorapidity.  The selection criteria are:
\begin{itemize}
\item greater than eight silicon hits if $|\eta|\le 1.65$, or
eleven or more silicon hits if $|\eta|> 1.65$,
\item at least one hit in the IBL and next innermost pixel layer
if passing through active modules in those layers,
\item no missing pixel hits in any active modules in layers at a larger radius than the innermost pixel hit on a track,
\item at least one pixel hit from any of the remaining layers, should hits not be present in the IBL and next innermost pixel layer.
\end{itemize}
 
The efficiency is measured using dedicated triggers which run as standard
\bjet physics signatures,
but without selecting on the tracking information or running the $b$-tagging selection.

\newcommand{\bjetvtxeff}{
\begin{figure*}[tp]
\centering
\subfloat[]{\includegraphics[width=0.5\textwidth]{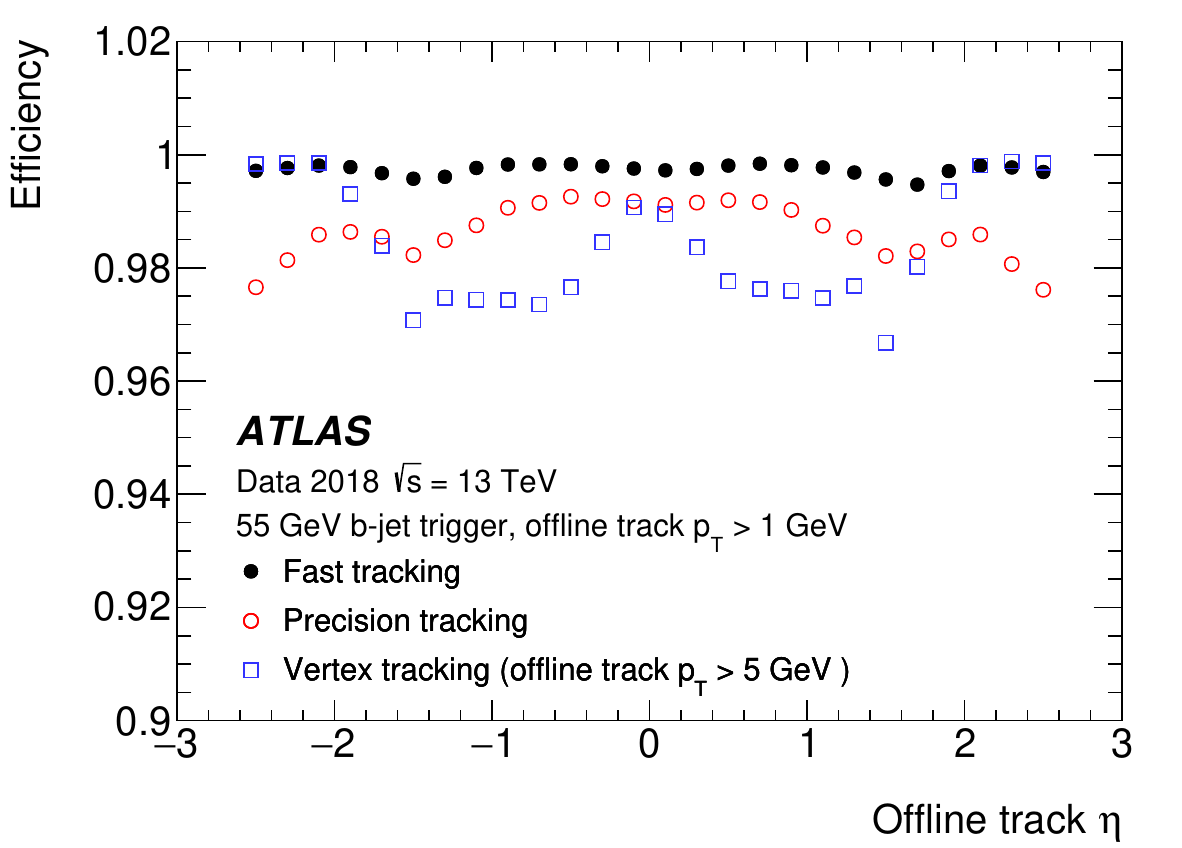}}
\subfloat[]{\includegraphics[width=0.5\textwidth]{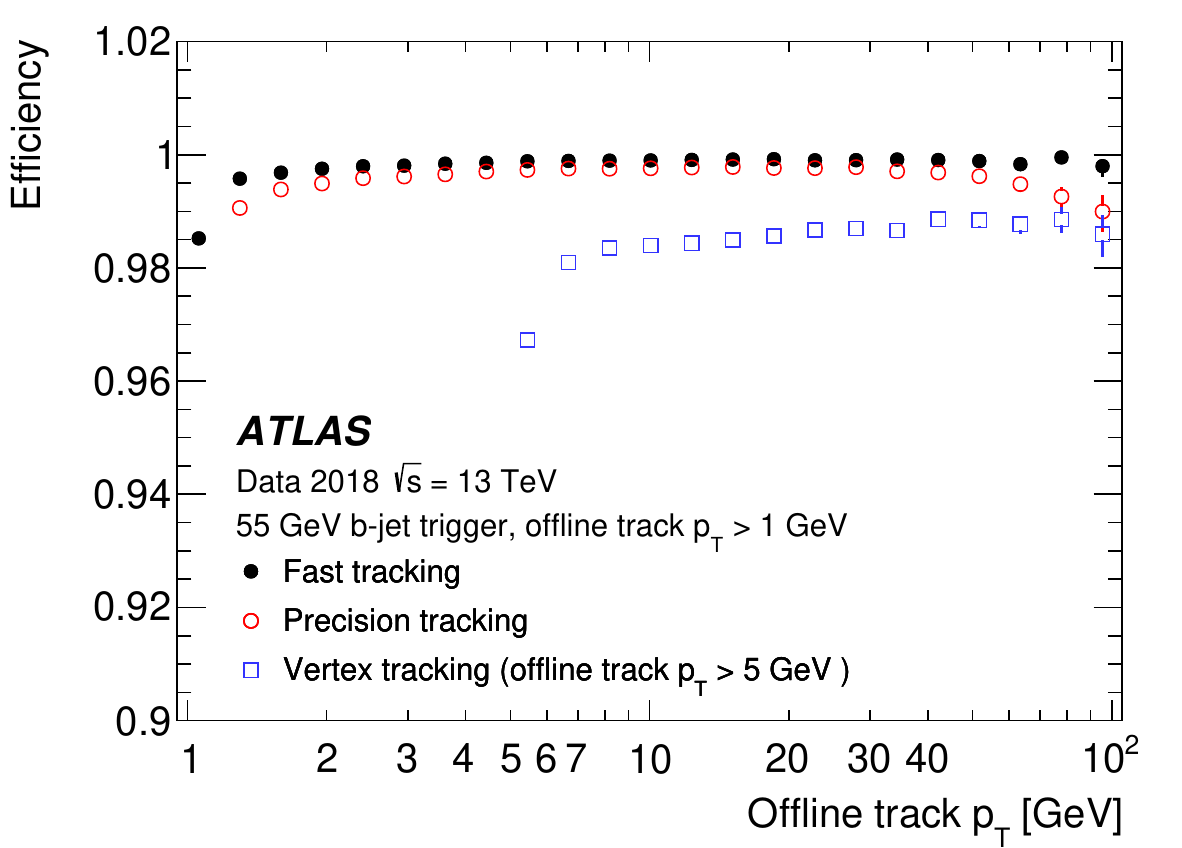}}\\
\subfloat[]{\includegraphics[width=0.5\textwidth]{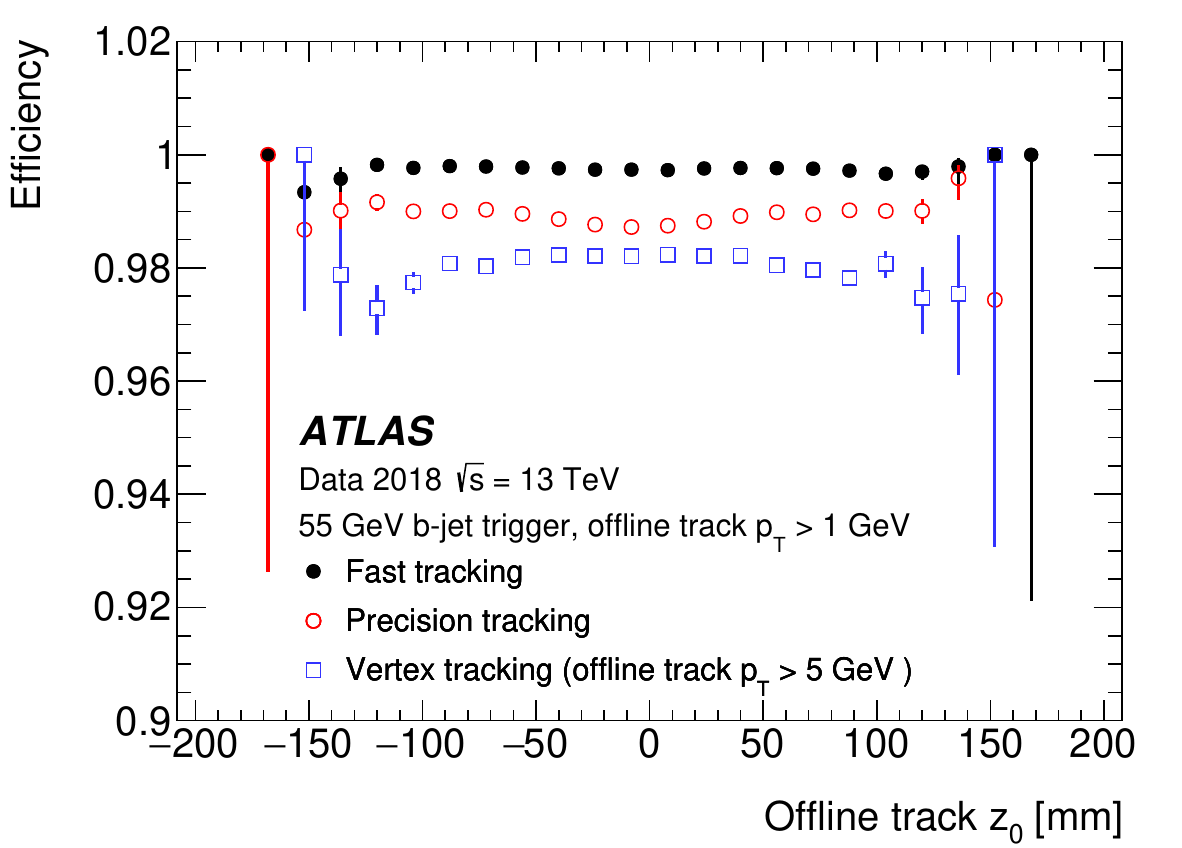}}
\subfloat[]{\includegraphics[width=0.5\textwidth]{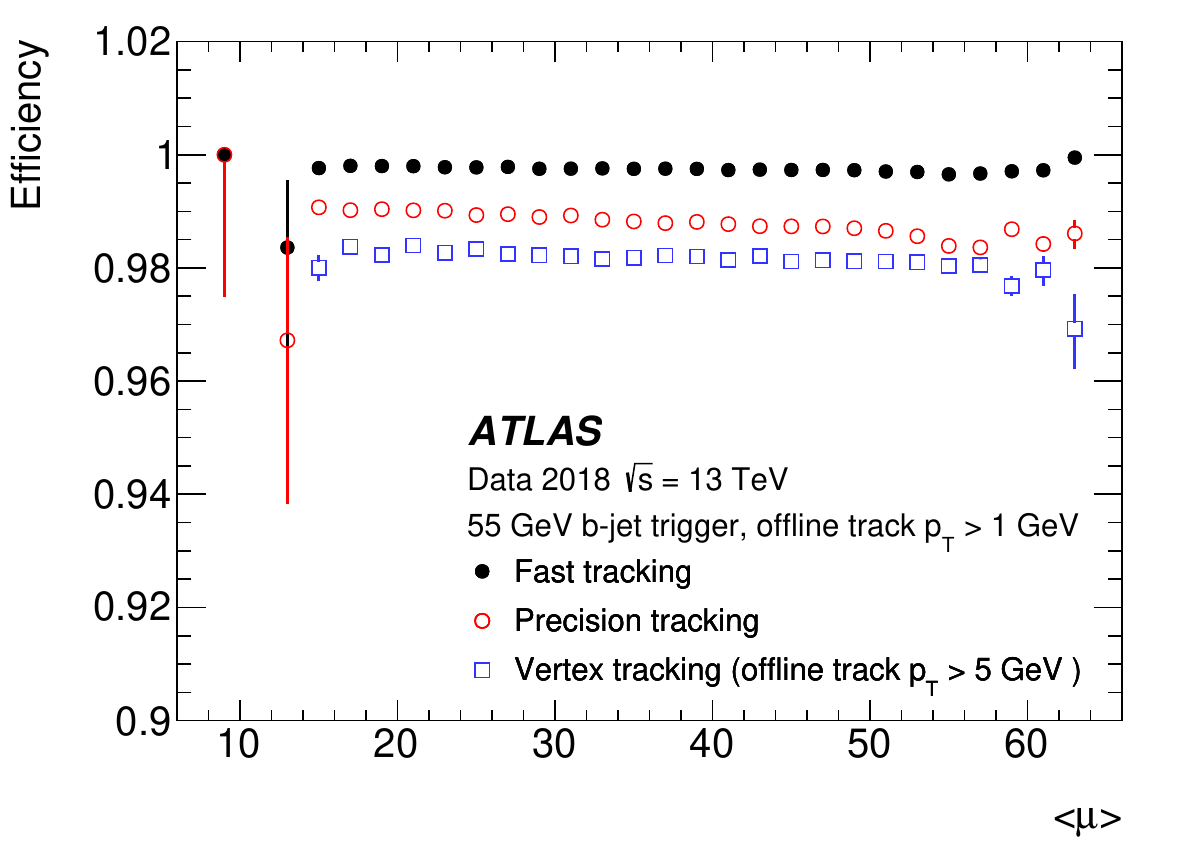}}\\
\caption{
The performance of the tracking in the \bjet signature, comparing the \roi based jet tracking with vertex tracking.
Shown are the efficiencies with respect to (a) the pseudorapidity, (b) transverse momentum, and (c) the $z_0$ position
at the point of closest approach to the beam for the offline reference tracks, and (d) with respect to the mean \pileup
interaction multiplicity.
Statistical Bayesian uncertainties are shown, using a 68.3\% confidence interval
and uniform beta function prior.}
\label{fig:bjetvtxtracking}
\end{figure*}
}

\newcommand{\bjetfigone}{
\begin{figure*}[thp]
\centering
\subfloat[]{\includegraphics[width=0.5\textwidth]{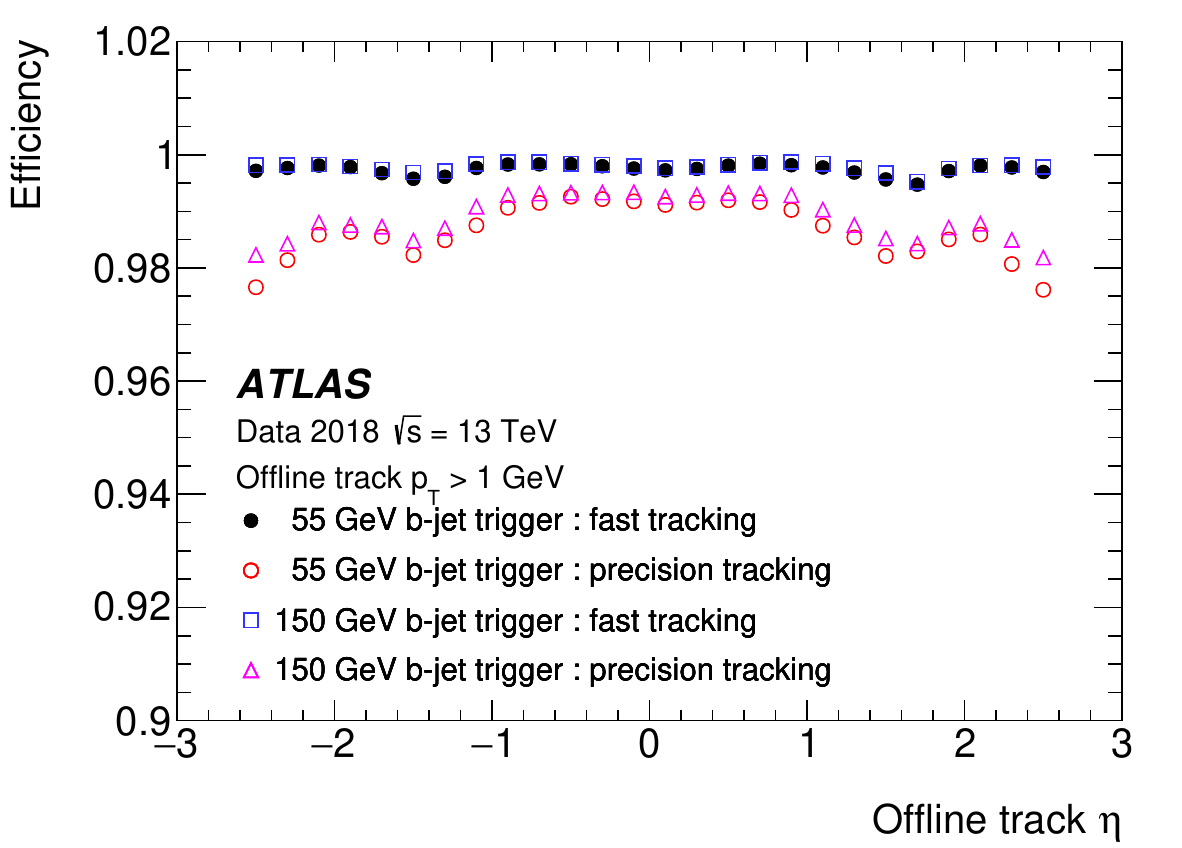}}
\subfloat[]{\includegraphics[width=0.5\textwidth]{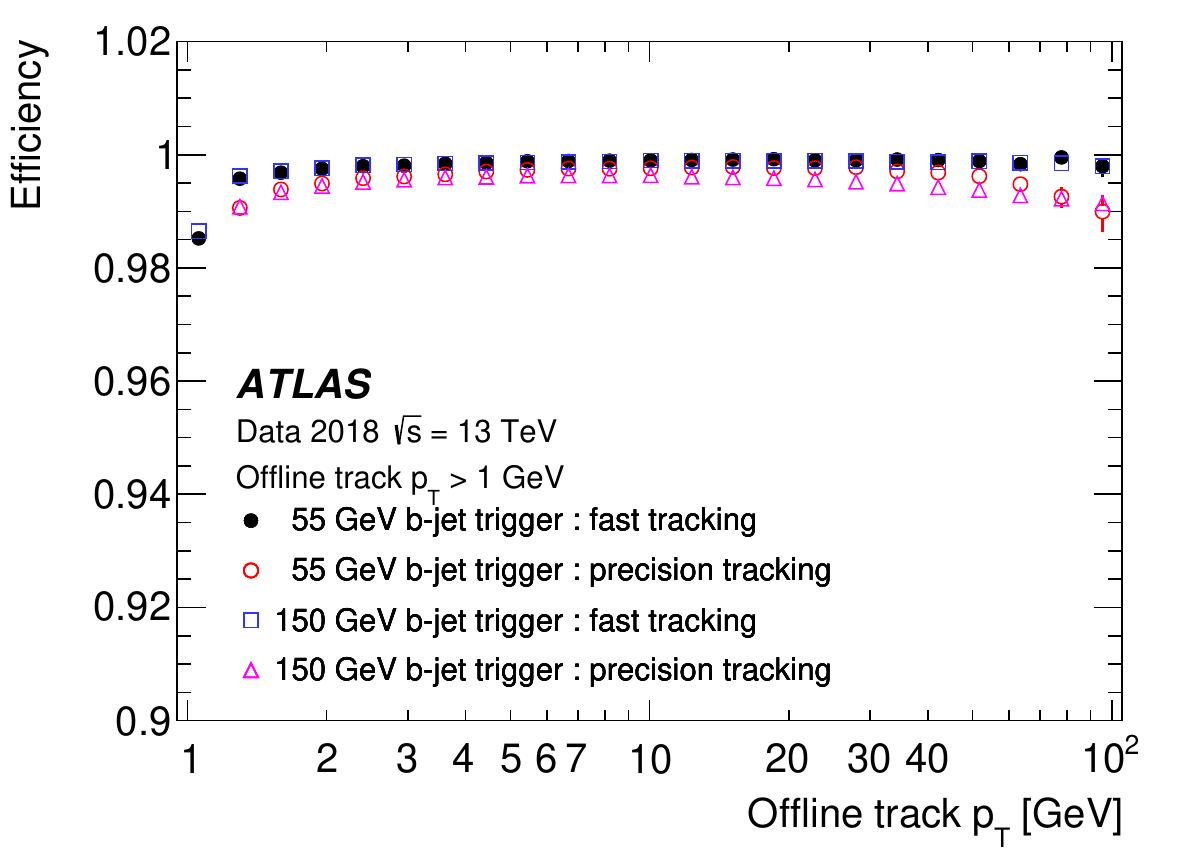}}\\
\subfloat[]{\includegraphics[width=0.5\textwidth]{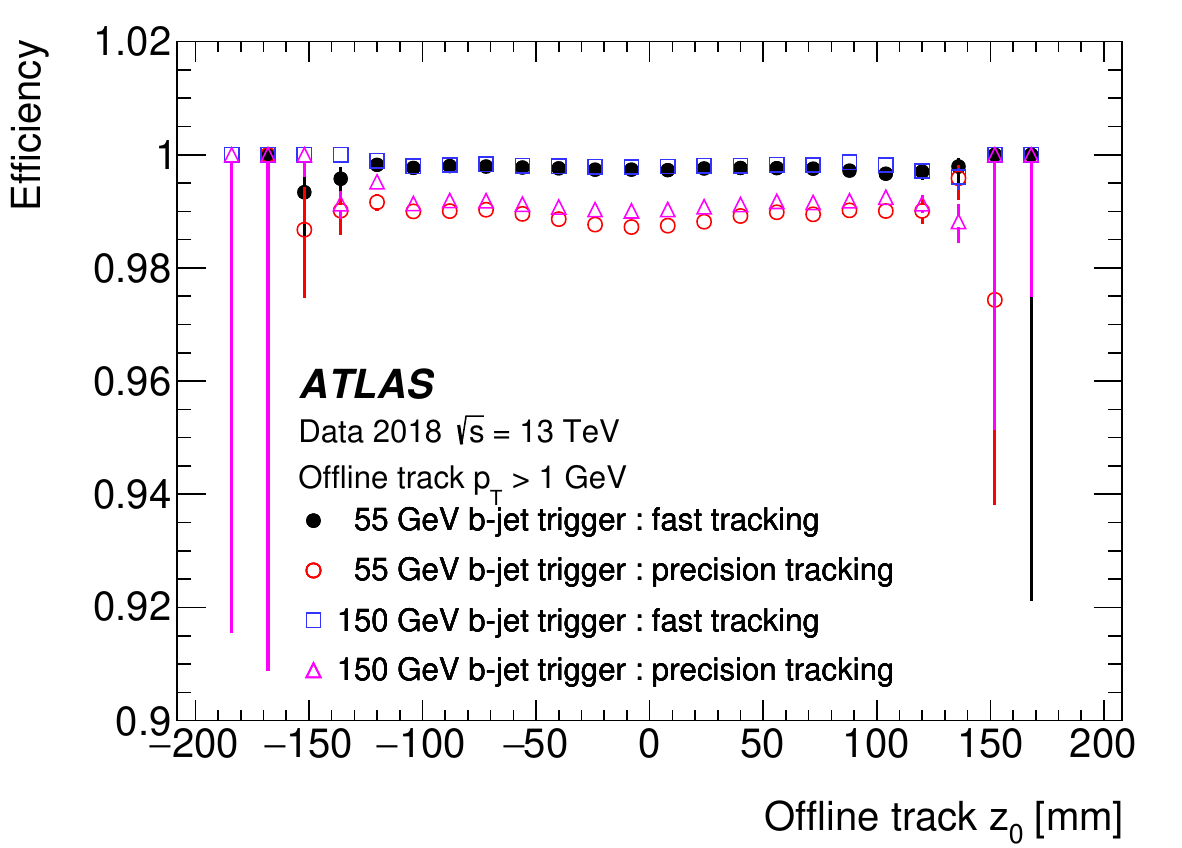}}
\subfloat[]{\includegraphics[width=0.5\textwidth]{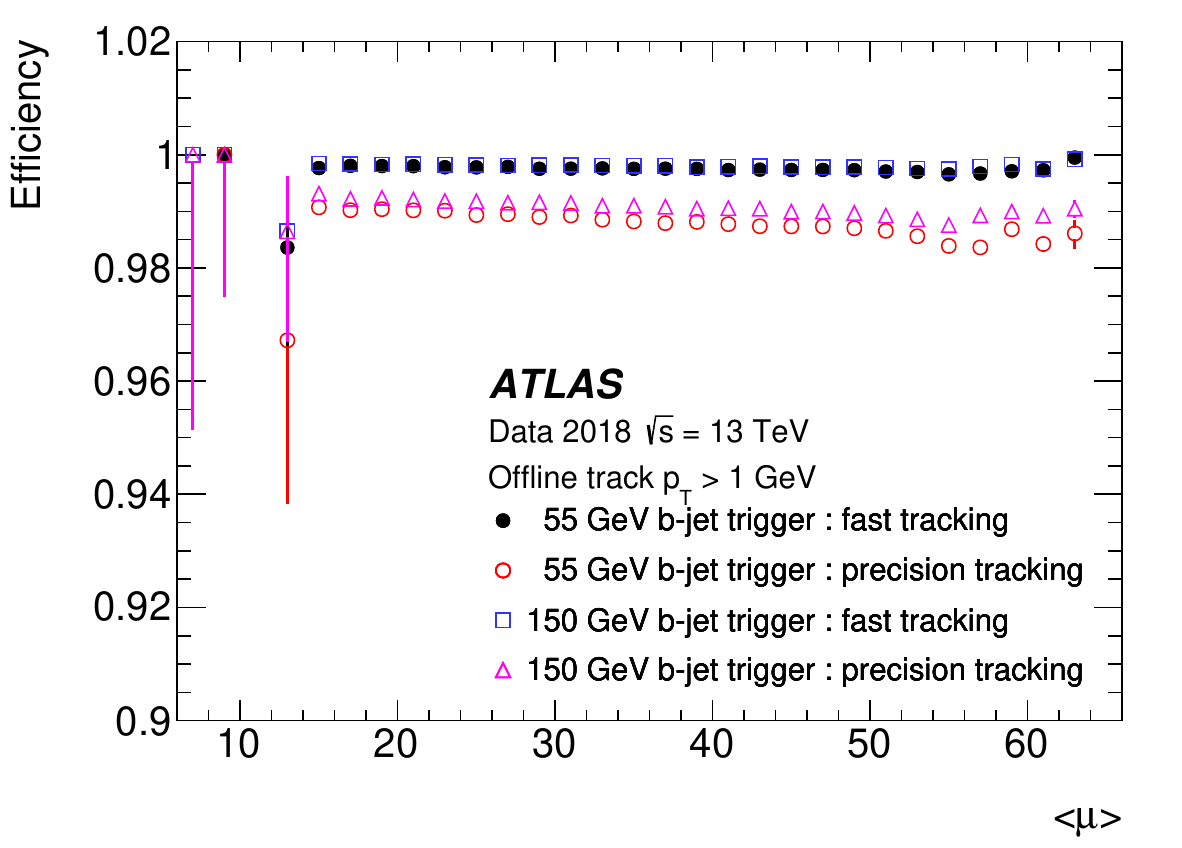}}
\caption{The \bjet trigger tracking efficiency for tracks with $\pt > \SI{1}{\gev}$ within jets shown as
a function of
(a) offline track $\eta$,
(b) offline track \pt,
(c) offline track $z_{0}$,
(f) the mean number of interactions per bunch crossing.
The efficiency is evaluated for the \SI{55}{\gev} \bjet trigger and is shown for the fast and precision tracking
in both the vertex finding, and jet tracking  stages of the \bjet ID trigger.
Statistical Bayesian uncertainties are shown, using a 68.3\% confidence interval
and uniform beta function prior.}
\label{fig:bjeteff}
\end{figure*}}

\newcommand{\bjetfigmult}{
\begin{figure*}[thp]
\centering
\subfloat[]{\includegraphics[width=0.5\textwidth]{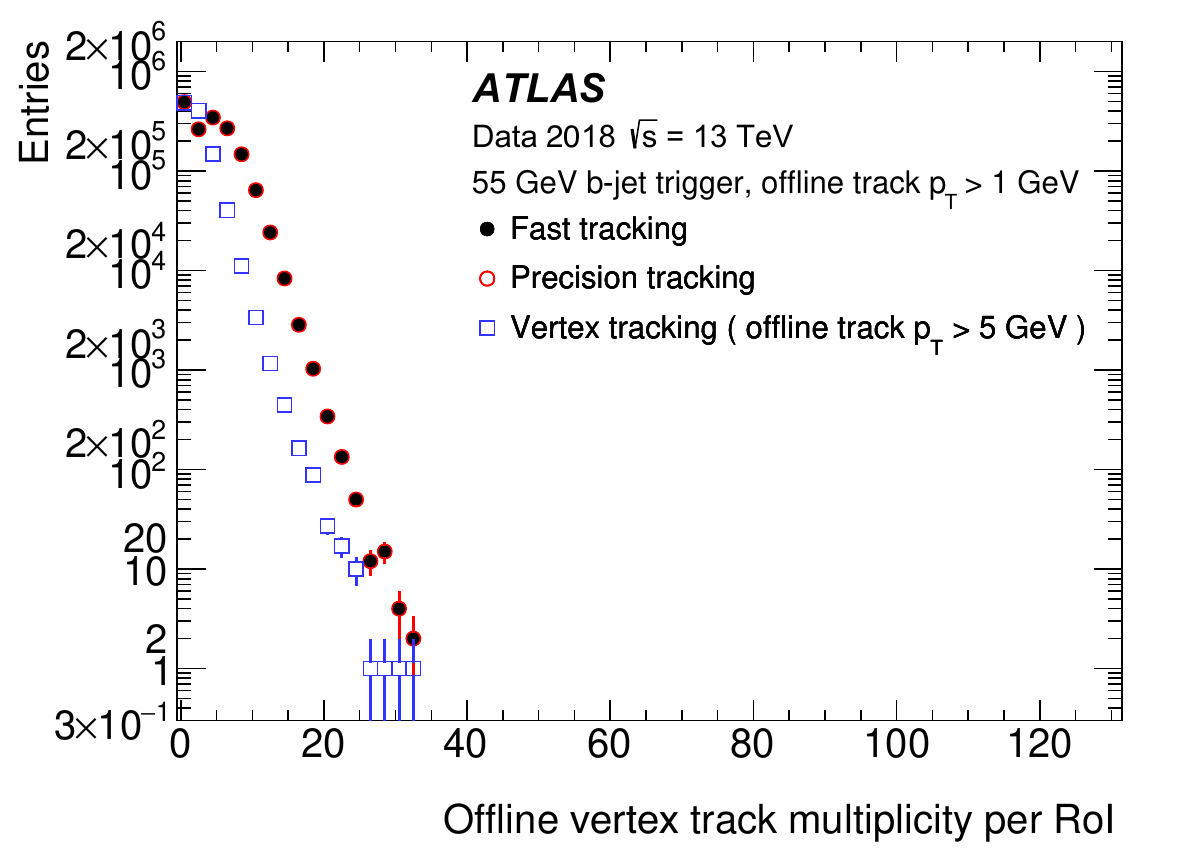}}
\subfloat[]{\includegraphics[width=0.5\textwidth]{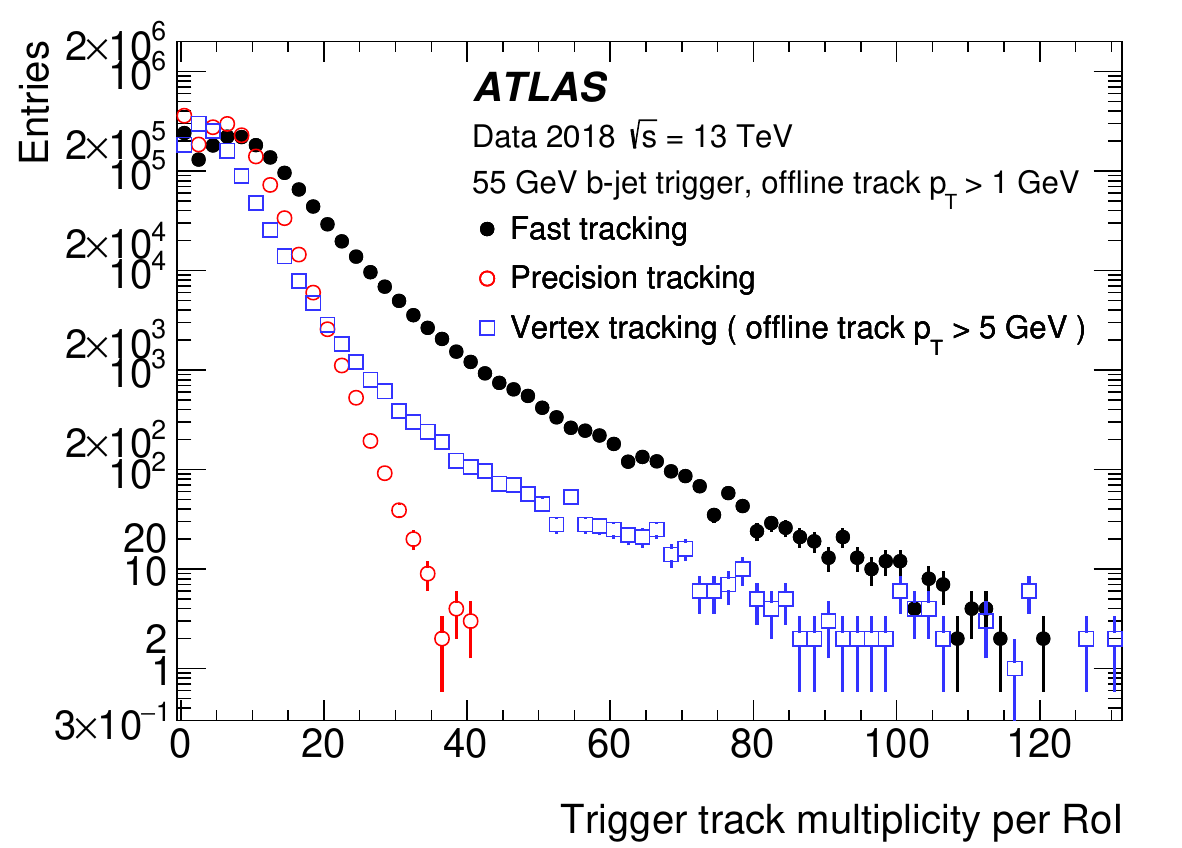}}
\caption{The multiplicity of tracks found by the offline reconstruction, associated with the primary vertex and
contained within the respective \bjet \roi (a),
and the multiplicity of trigger tracks per \roi (b).}
\label{fig:bjetmult}
\end{figure*}}

\newcommand{\bjetfigtwo}{
{\begin{figure*}[thp]
\centering
\subfloat[]{\includegraphics[width=0.5\textwidth]{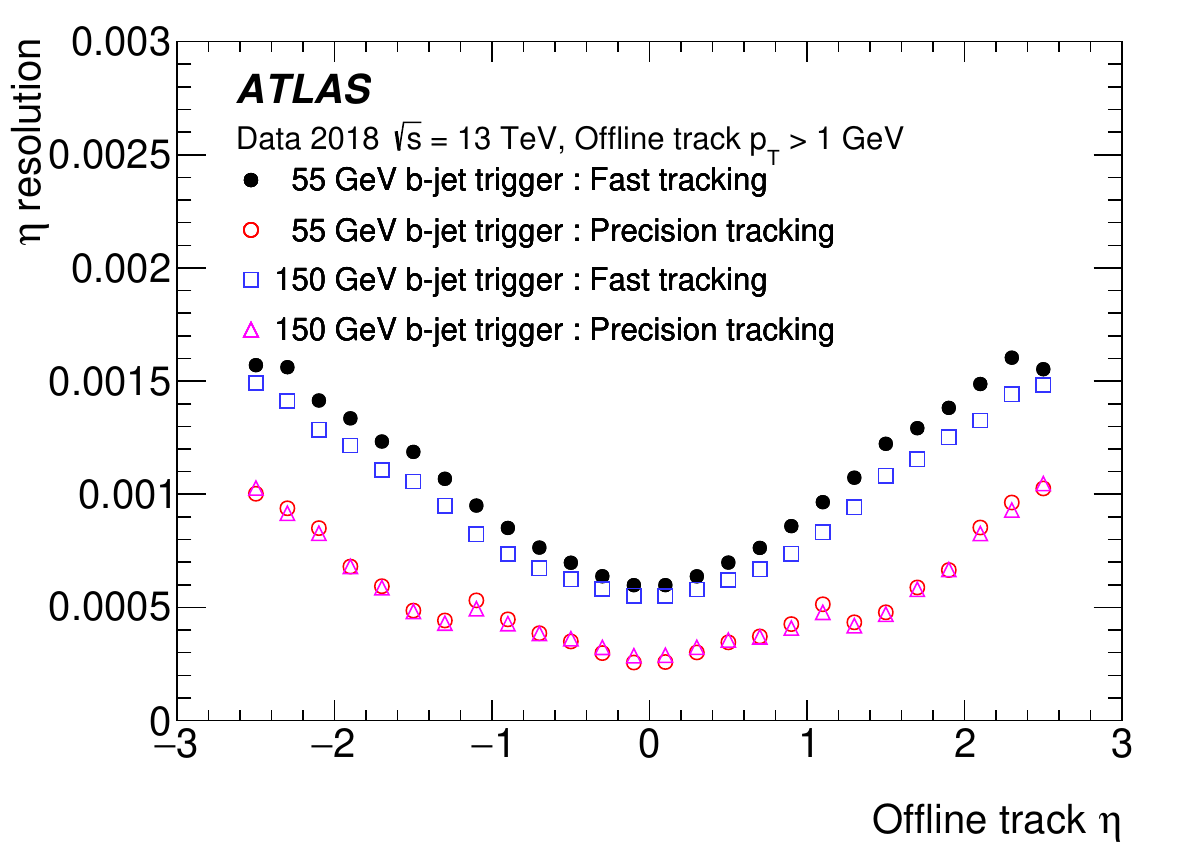}}
\subfloat[]{\includegraphics[width=0.5\textwidth]{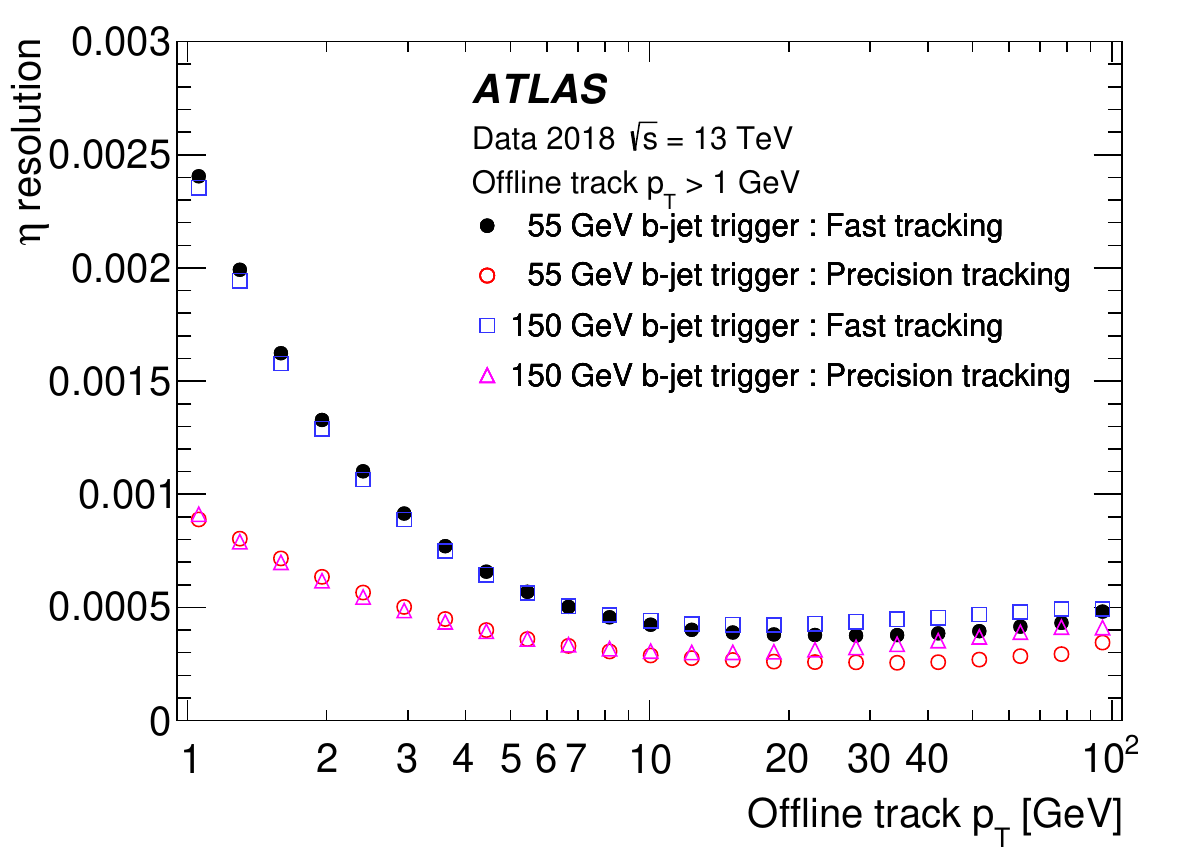}}\\
\subfloat[]{\includegraphics[width=0.5\textwidth]{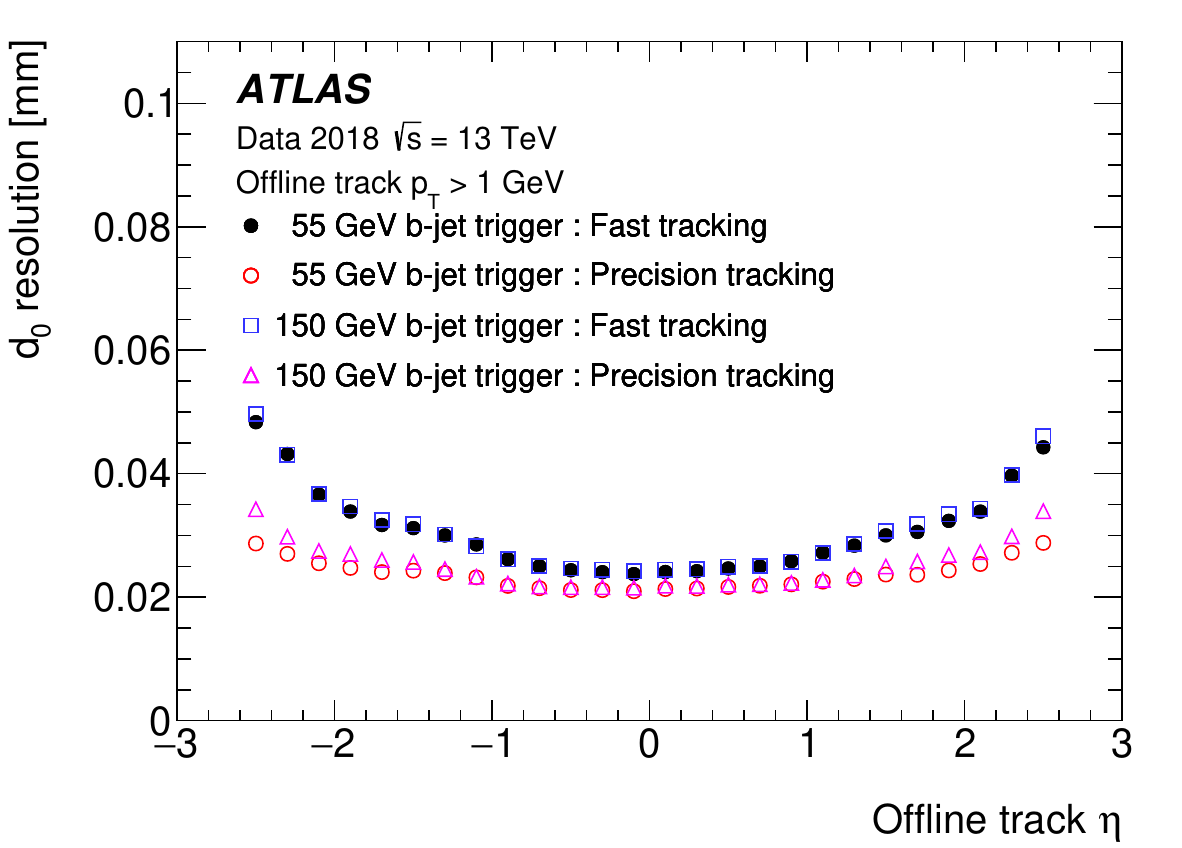}}
\subfloat[]{\includegraphics[width=0.5\textwidth]{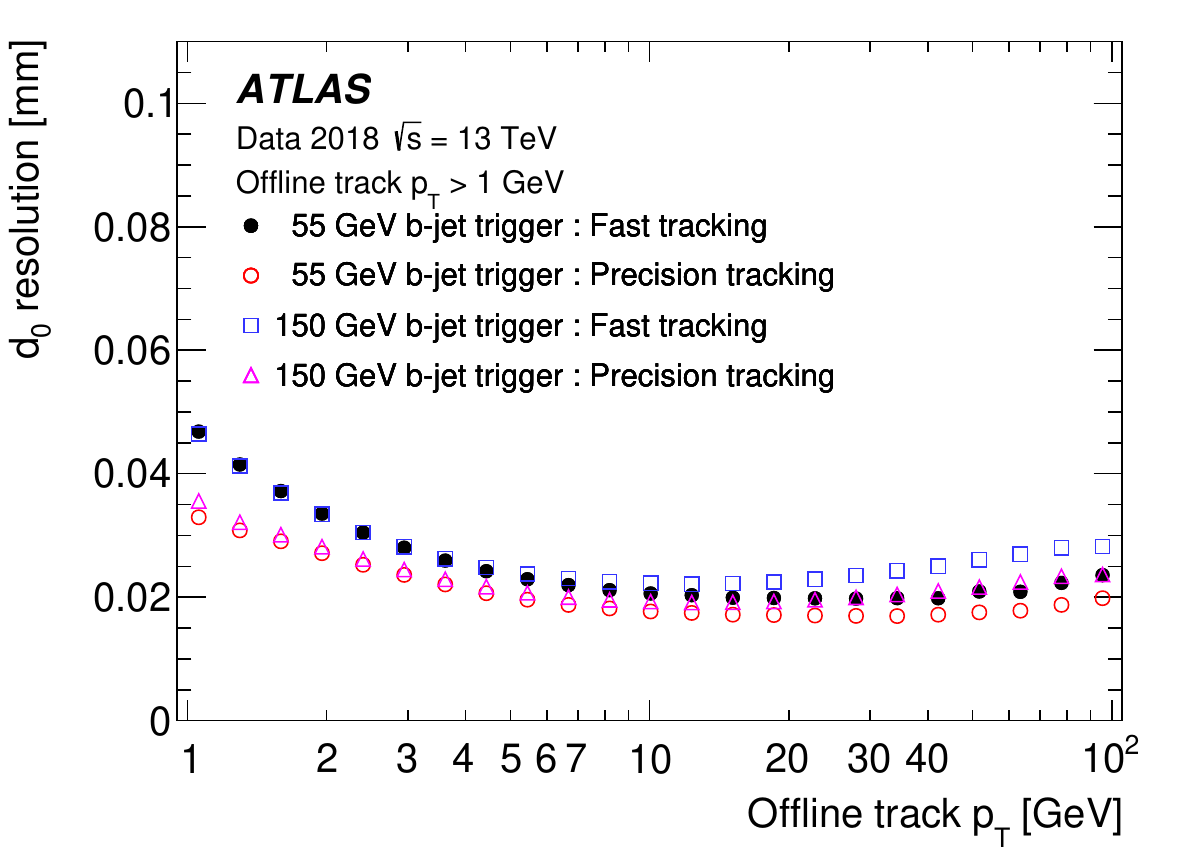}}\\
\subfloat[]{\includegraphics[width=0.5\textwidth]{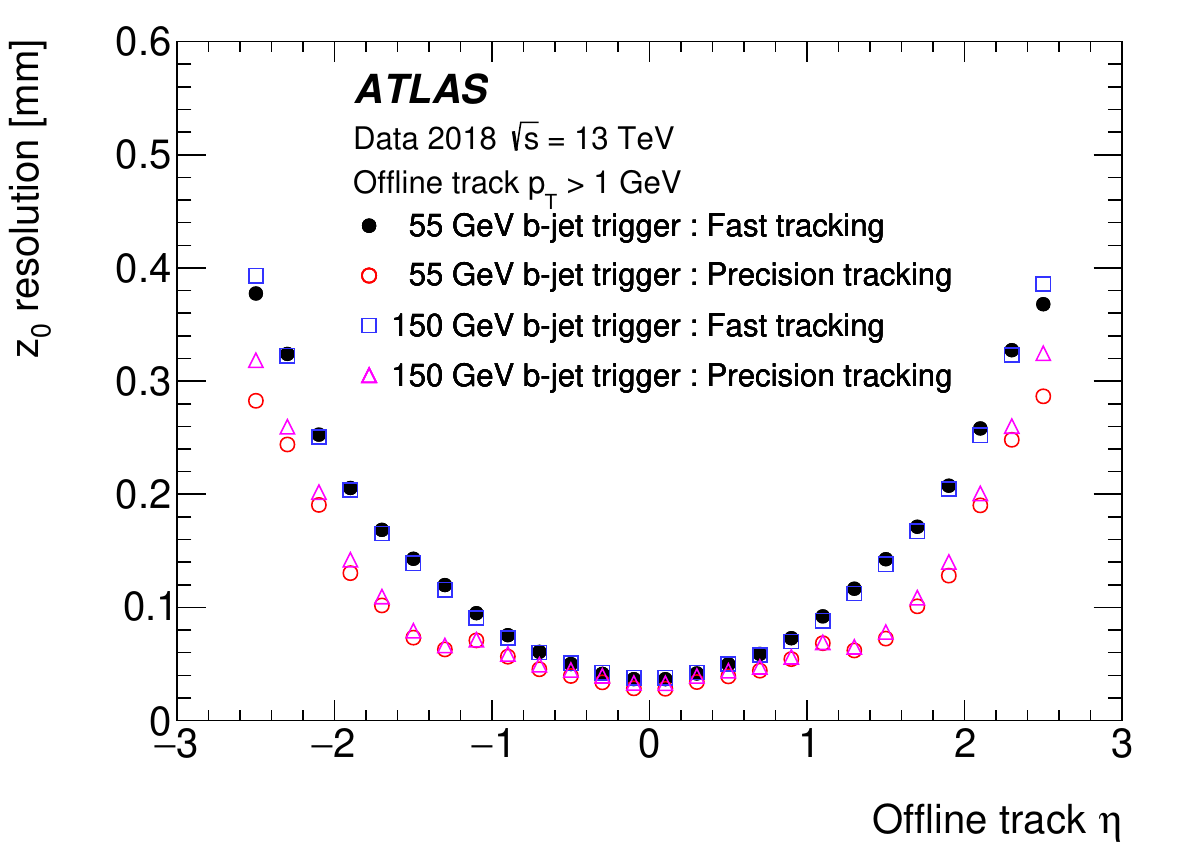}}
\subfloat[]{\includegraphics[width=0.5\textwidth]{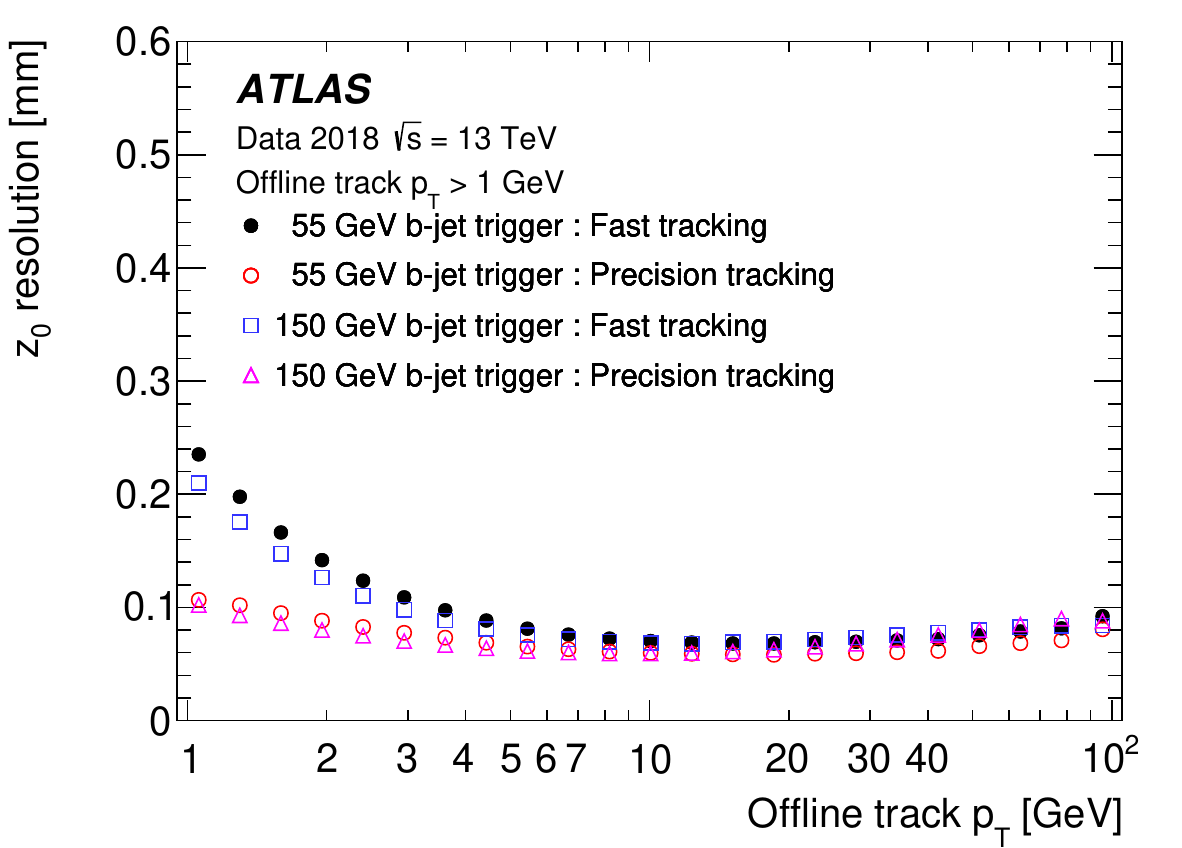}}
 
\caption{
The track resolutions for the fast
and precision tracking from the second stage of the  \bjet trigger for
the 55\gev and 150\gev signatures. Shown are the resolutions for the
trigger track pseudorapidity (a,b), transverse impact parameter (c,d)
and track $z$ position (e,f), as functions of the offline tracks
pseudorapidity and transverse momentum.
}
\label{fig:bjetres}
\end{figure*}}
}

\newcommand{\bjetfigfour}{
{\begin{figure*}[tp]
\centering
\subfloat[]{\includegraphics[width=0.5\textwidth]{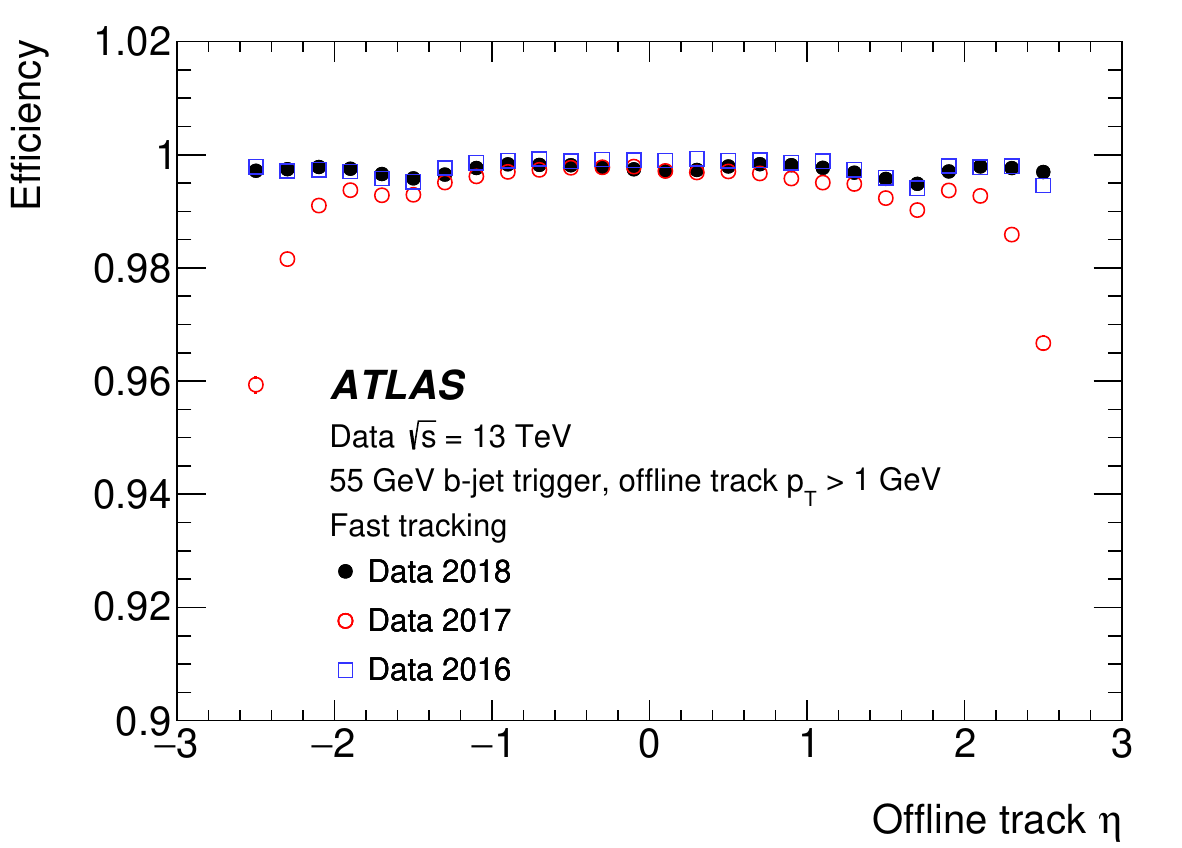}}
\subfloat[]{\includegraphics[width=0.5\textwidth]{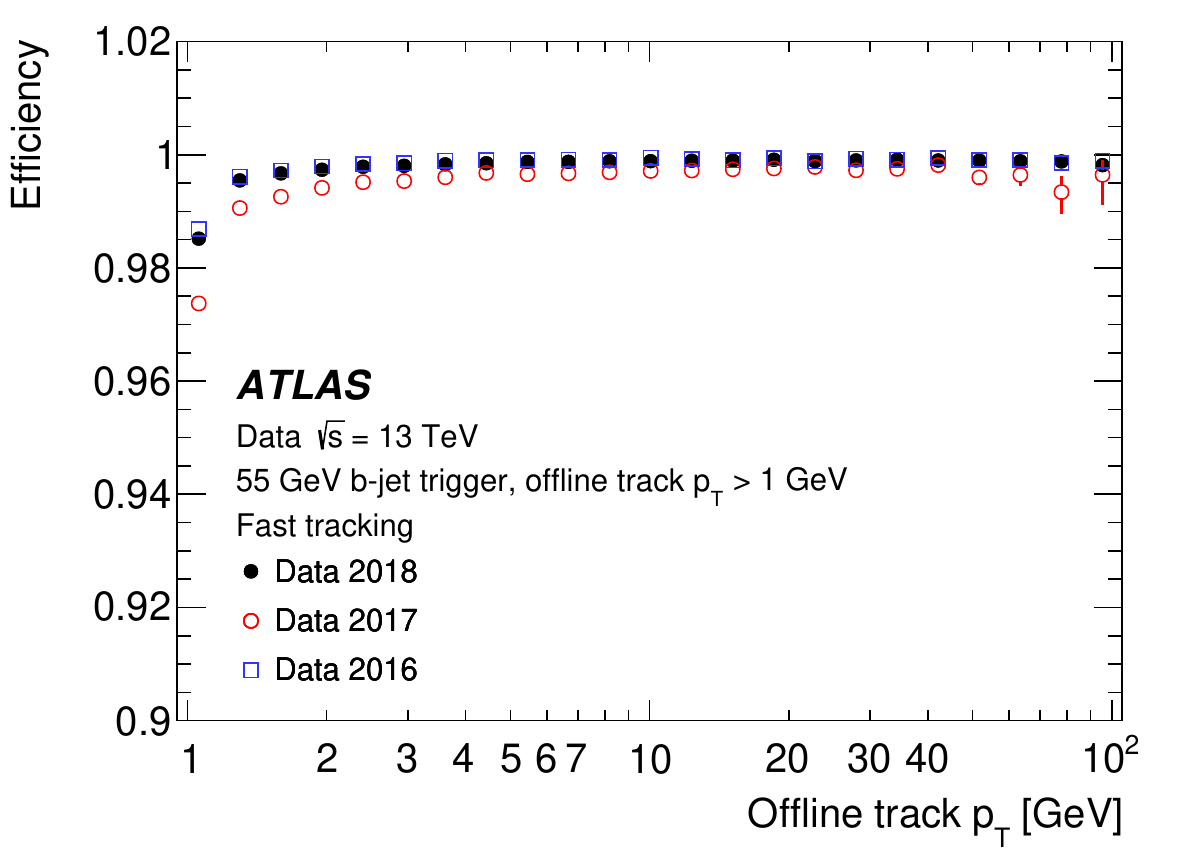}}\\
\subfloat[]{\includegraphics[width=0.5\textwidth]{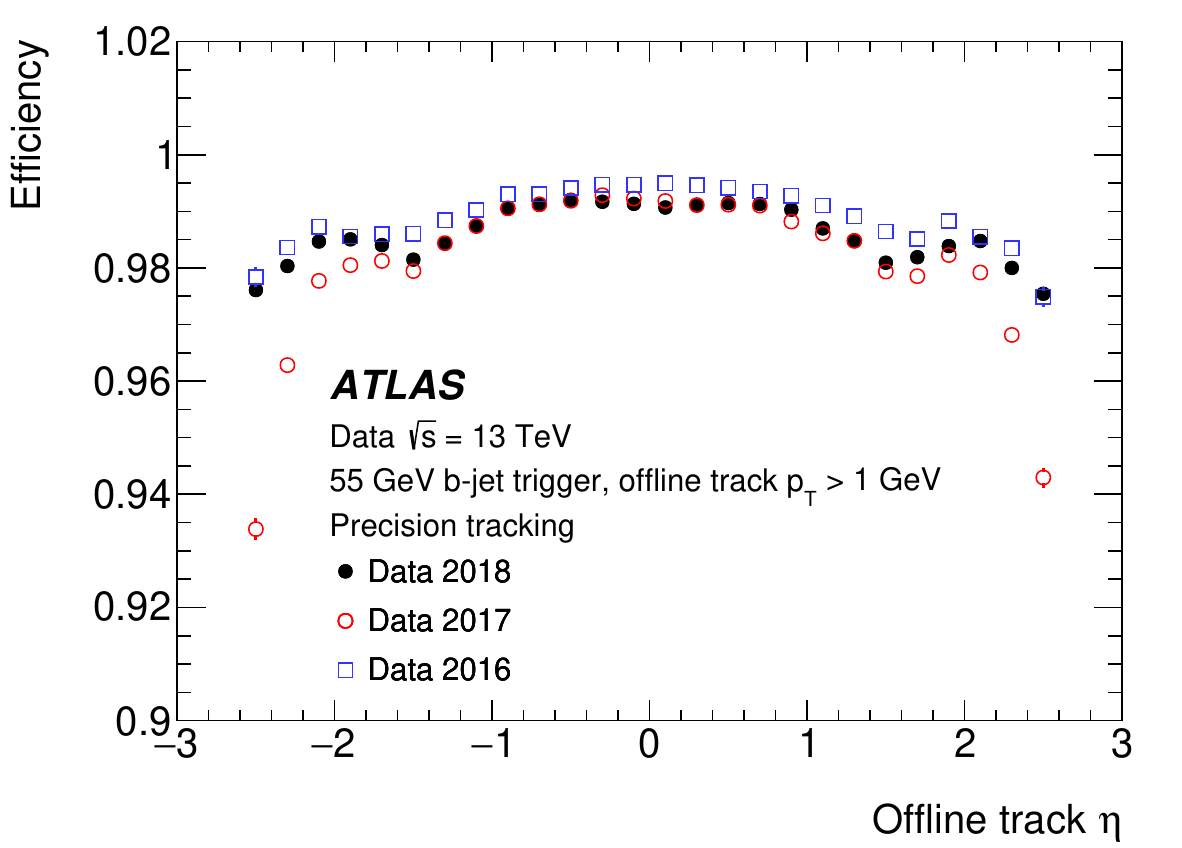}}
\subfloat[]{\includegraphics[width=0.5\textwidth]{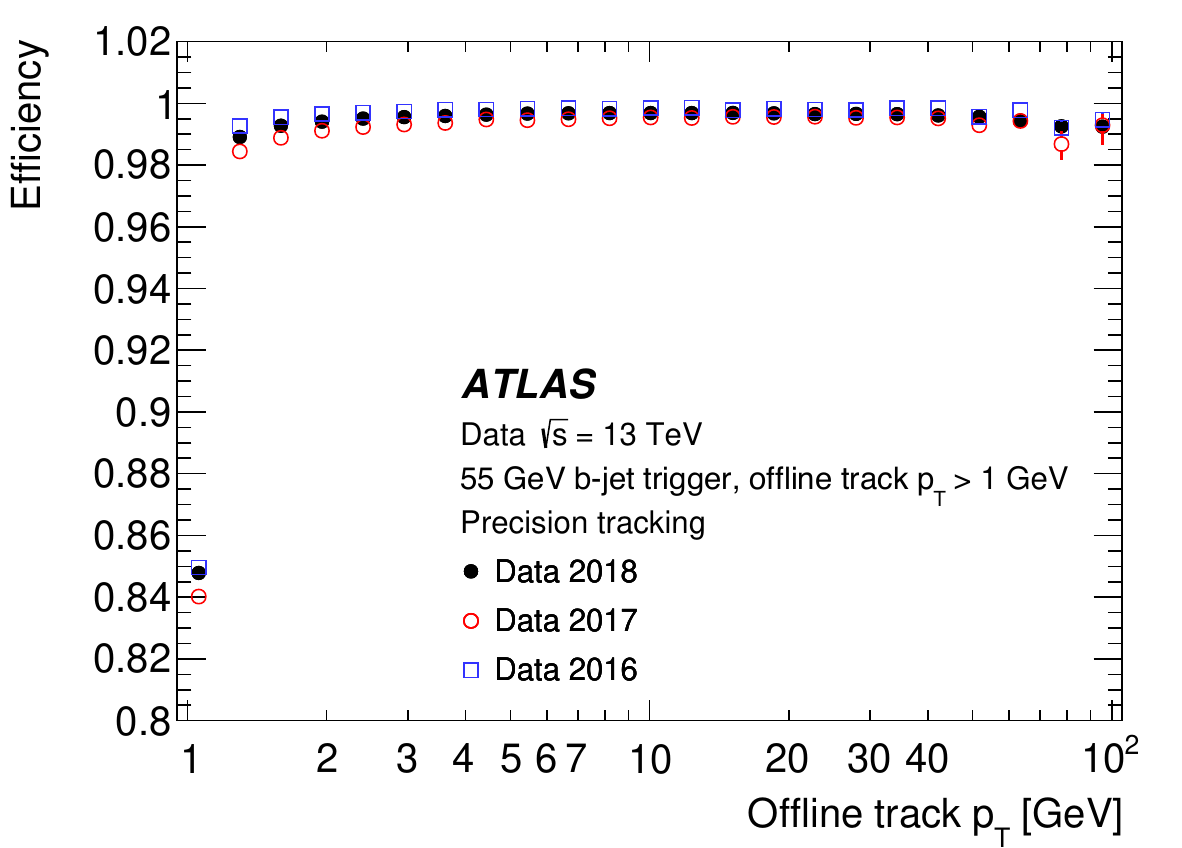}}
\caption{
The \bjet second stage tracking performance efficiency differences between 2016, 2017, and 2018. }
\label{fig:bjetalltracking}
\end{figure*}}
}

The efficiency for the tracking from both stages in the \bjet signature is shown in
Figure~\ref{fig:bjetvtxtracking} for the \SI{55}\gev threshold \bjet trigger, shown
as a function of the offline track parameters and \mubar.
Figure~\ref{fig:bjeteff} compares the precision and fast tracking
efficiencies for the jet tracking from  the
\SI{55}{\gev} and  \SI{150}\gev threshold triggers.
By design, the first-stage vertex tracking
only reconstructs tracks with $\pt > \SI{5}\gev$ and thus only offline tracks with $\pt>5$\gev were selected
for the vertex tracking reference sample used for  Figure~\ref{fig:bjetvtxtracking}.
For offline tracks above approximately 1.2\gev the second stage
fast tracking efficiency is better than 99.5\%. For tracks near 1\gev the fast tracking efficiency is better than 98\%
but the precision tracking efficiency is lower\emdash approximately 85\%.\footnote{This is not shown in this figure,
since the required axis range would render other more relevant features in the figure more difficult to discern, but can be
seen in Figure~\ref{fig:bjetalltracking}.}
This is expected and is a consequence of an explicit 1\gev cut placed on the tracks from the precision
tracking in the \bjet signature implemented for  processing-latency reasons. As such
the reconstruction for tracks near the threshold for which the trigger \pt would be reconstructed to be
lower than the 1\gev \pt threshold is terminated early.

\bjetvtxeff
 
\bjetfigone
 
\bjetfigmult
 
For offline tracks with transverse momenta
greater than approximately 10\gev the precision tracking efficiency starts to fall slightly,
and is more significant for the higher threshold \bjet triggers.
This is largely a consequence of the reduced
pixel hit multiplicity found on the precision tracks for the higher threshold \bjet triggers. Although the hits
are largely present on the tracks from the fast tracking, the tighter hit selection cuts on the precision tracks
used to improve the purity relative to the fast tracking leads to this reduction.
The effect of this can be seen in Figure~\ref{fig:bjetmult}, which compares
the multiplicity of trigger tracks with that of offline tracks contained within the corresponding \roi
and associated with the primary vertex.
The second stage precision and fast tracking must of course have an identical set of offline tracks
contained within their common \roi, whereas the multiplicity of the
second stage fast tracks themselves is significantly higher than the precision tracking multiplicity,
which is, in turn, much closer to that for the offline tracks. For this study, it should be noted that
the selection applied to the precision tracks is looser than for the offline tracks.
 
However, if using the same selection as used for the reference offline tracks for the determination of the
efficiency, but applying instead to the tracks from the precision tracking, the purity of the \bjet precision tracks
with respect to the offline tracking is seen to be better than 99\%.\footnote{Here, the purity is defined as the fraction
of the trigger tracks that have a matching offline track when applying the same matching criteria used
for determination of the efficiency.} In the trigger, additional track
selection can be applied by any subsequent algorithm within the trigger which is a client of the tracking,
to further improve the track purity. As such, any more detailed analysis of the
track purity would be very dependent on the specific requirements of that selection and is thus beyond
the scope of the current discussion.
 
The tighter hit selection in the precision track reconstruction and the significantly lower efficiency near 1\gev
are largely responsible for the lower overall efficiency of the
precision tracking relative to the fast tracking. There is a clear structure in pseudorapidity\emdash the efficiency
being slightly lower in the endcaps and for central $z$. The small fall in efficiency with increasing
\mubar is more significant for the precision tracking, again driven largely by the lower
efficiency at low \pt.

\bjetfigtwo
 
Figure~\ref{fig:bjetres} shows the track resolutions for the fast
and precision tracking from the second stage of the  \bjet trigger for
the 55\gev and 150\gev signatures. Shown are the resolutions for the
pseudorapidity, transverse impact parameter and track $z$-vertex position,
as functions of the offline track pseudorapidity and transverse momentum.
The resolutions are estimated using data from the full 2018 data set, and so
include small run-by-run variations in the residual distributions.
 
The precision tracking provides significantly
better resolution for all variables at lower track \pt.
However, there appears to be a slight
deterioration of the \dzero resolution with increasing \pt for the high \et
jet trigger. This is correlated with a slight loss of pixel hits and corresponding loss in
efficiency for the precision tracking at higher \pt in the high \et triggers
from Figure~\ref{fig:bjeteff}.

Finally, in discussing the performance of the \bjet trigger one should consider the changes to the
trigger during \runii. In 2017, to deal with the increasing \pileup interaction multiplicity,
the size of the first stage \roi in $\eta$--$\phi$ about each jet in the \superroi was reduced from
\mbox{$0.4 \times 0.4$} to the \mbox{$0.2 \times 0.2$} size discussed in Section~\ref{sec:multistage}.
The $z$ extent along the \beamline remained unchanged.
In addition
the minimum track \pt used in the pattern recognition for the reconstruction of the vertex tracks
was increased to 5\gev.
As in the case of the second stage tracking in the tau signature discussed in Section~\ref{sec:tau}
the track finding in the second stage
fast tracking was affected by the seeding. In the case of the \bjet trigger, the second stage \roi $z$-width
was larger than for the tau signature. The resulting reduction in the efficiency of the fast tracking in the second
stage jet \roi can be seen in Figure~\ref{fig:bjetalltracking} but this reduction is not as significant as for
the tau signature. In contrast to the tau case, this loss in
efficiency is seen to lie almost exclusively at lower \pt where the efficiency in 2017 is approximately 1\% lower
than in either 2016 or 2018. For 2018, the fixes to the seeding allowed the efficiency of the fast tracking in the
second stage \roi to increase, but not quite recover to the level of 2016, where  the LHC operated
with  significantly lower \pileup multiplicity.
 
\bjetfigfour

\floatbarra

\subsection{Vertex finding performance}
 
\newcommand{\vtxfigone}{
\begin{figure*}[t]
\centering
\subfloat[]{\includegraphics[width=0.5\textwidth]{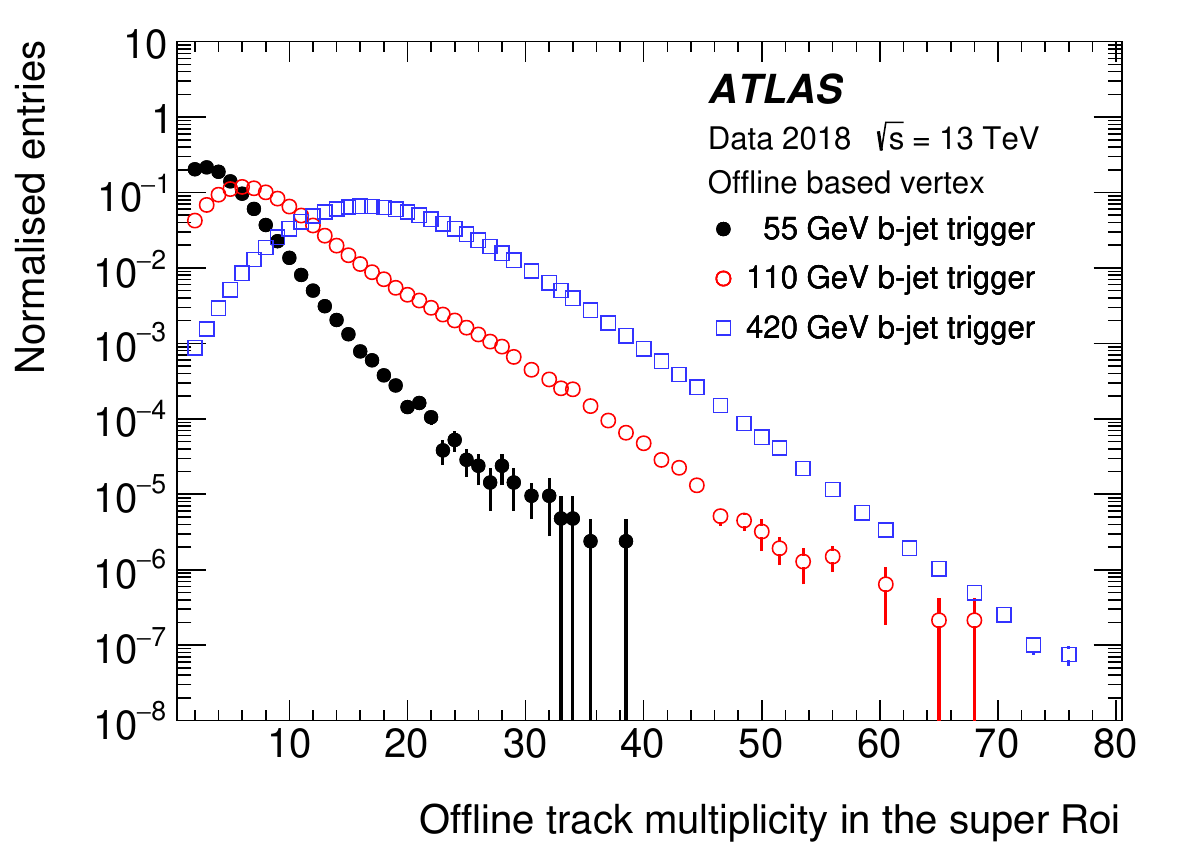}}
\subfloat[]{\includegraphics[width=0.5\textwidth]{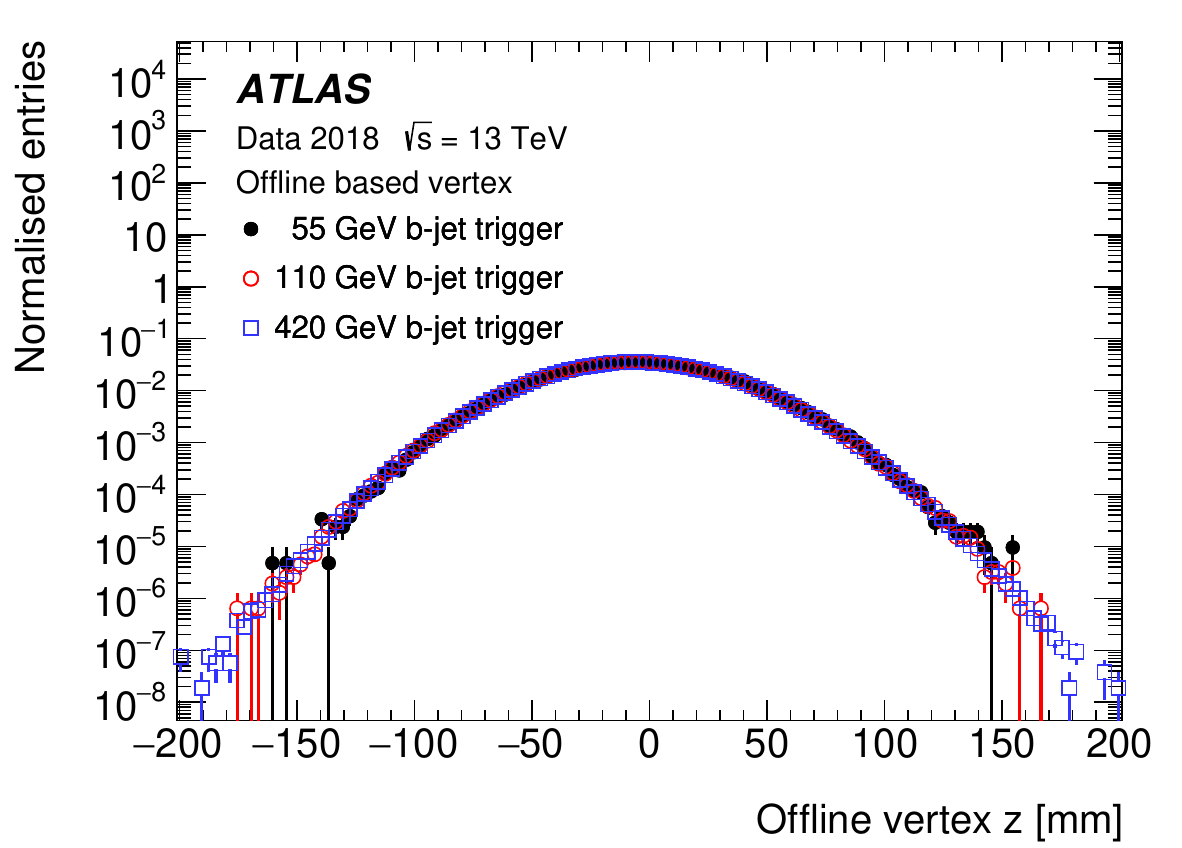}}
\caption{
The distribution of the offline track multiplicity for all selected offline tracks within the
\superroi used for the vertex tracking (a), and their distribution in track $z_0$ position at the \beamline (b).}
\label{fig:vtxdist}
\end{figure*}
}

\newcommand{\vtxfigtwo}{
\begin{figure}[thp]
\centering
\subfloat[]{\includegraphics[width=0.49\textwidth]{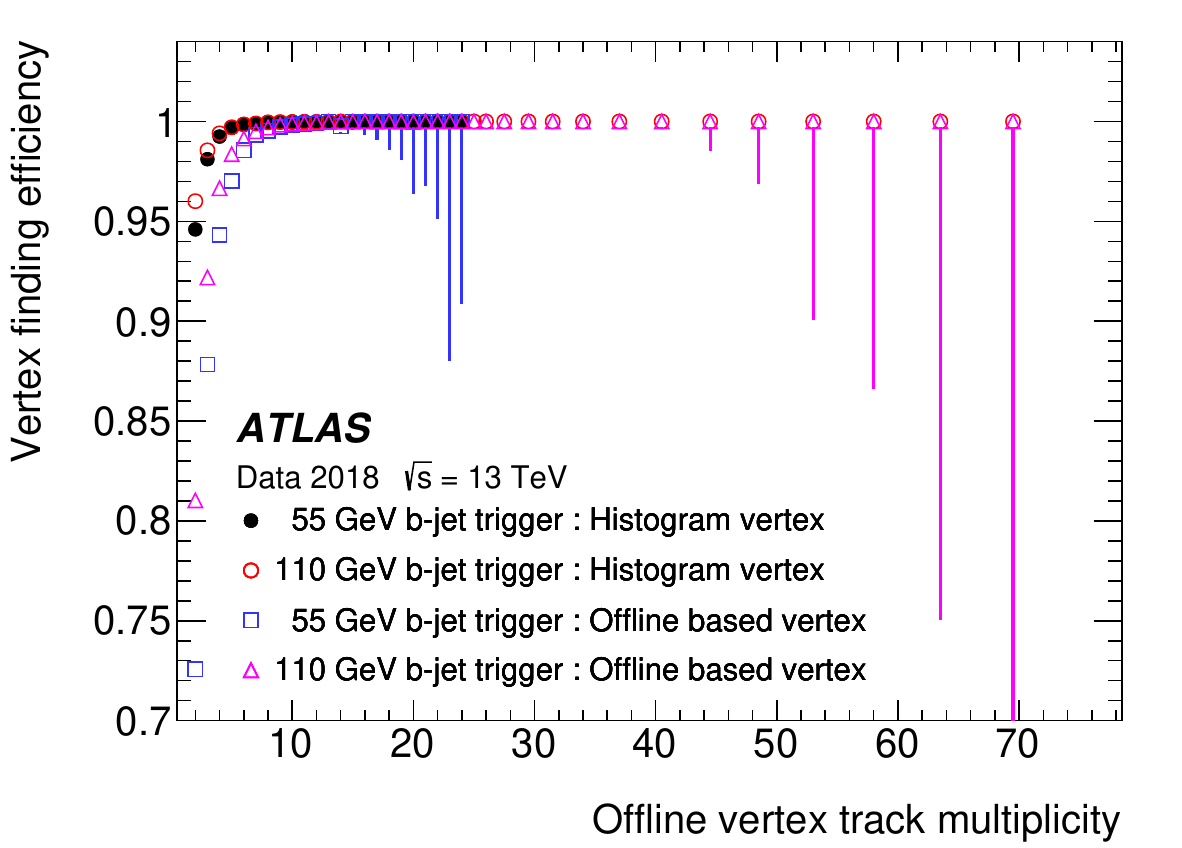}\label{fig:vtxeffa}}\figbreak
\subfloat[]{\includegraphics[width=0.49\textwidth]{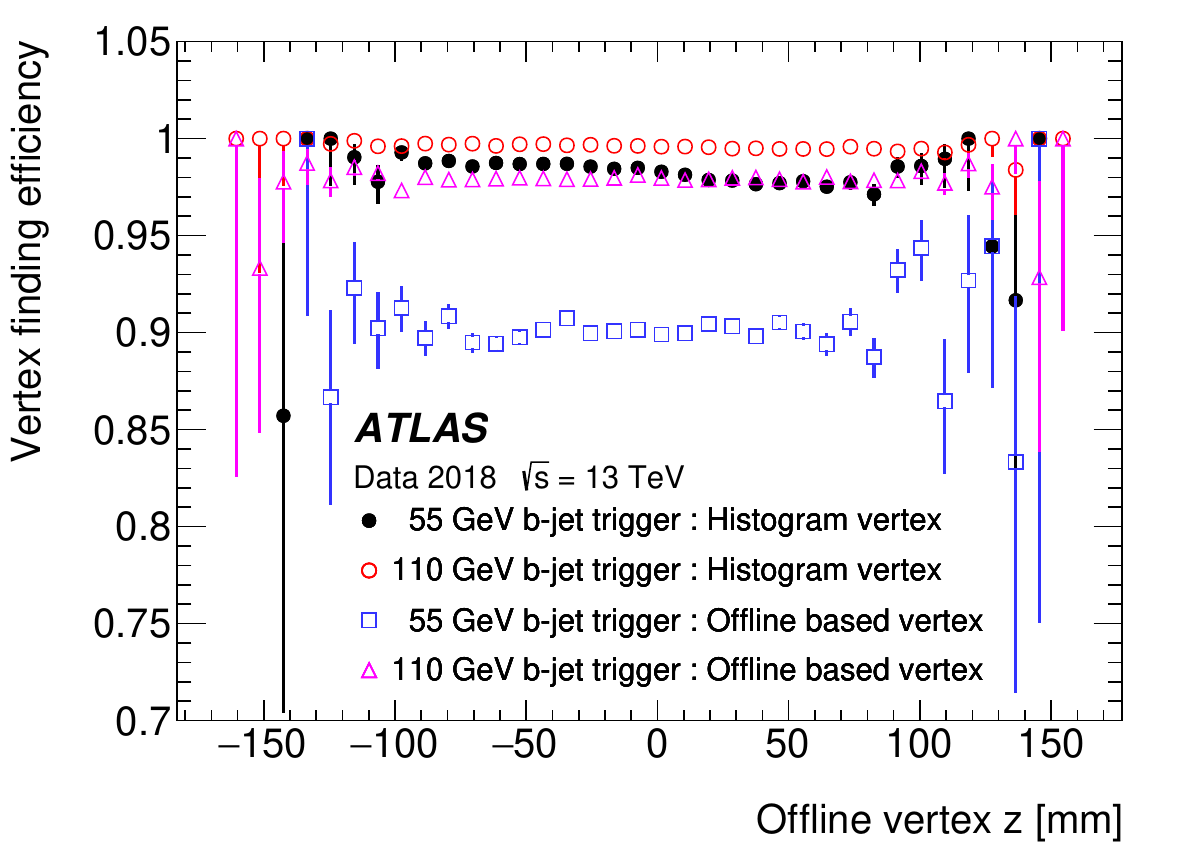}\label{fig:vtxeffb}}\\
\subfloat[]{\includegraphics[width=0.49\textwidth]{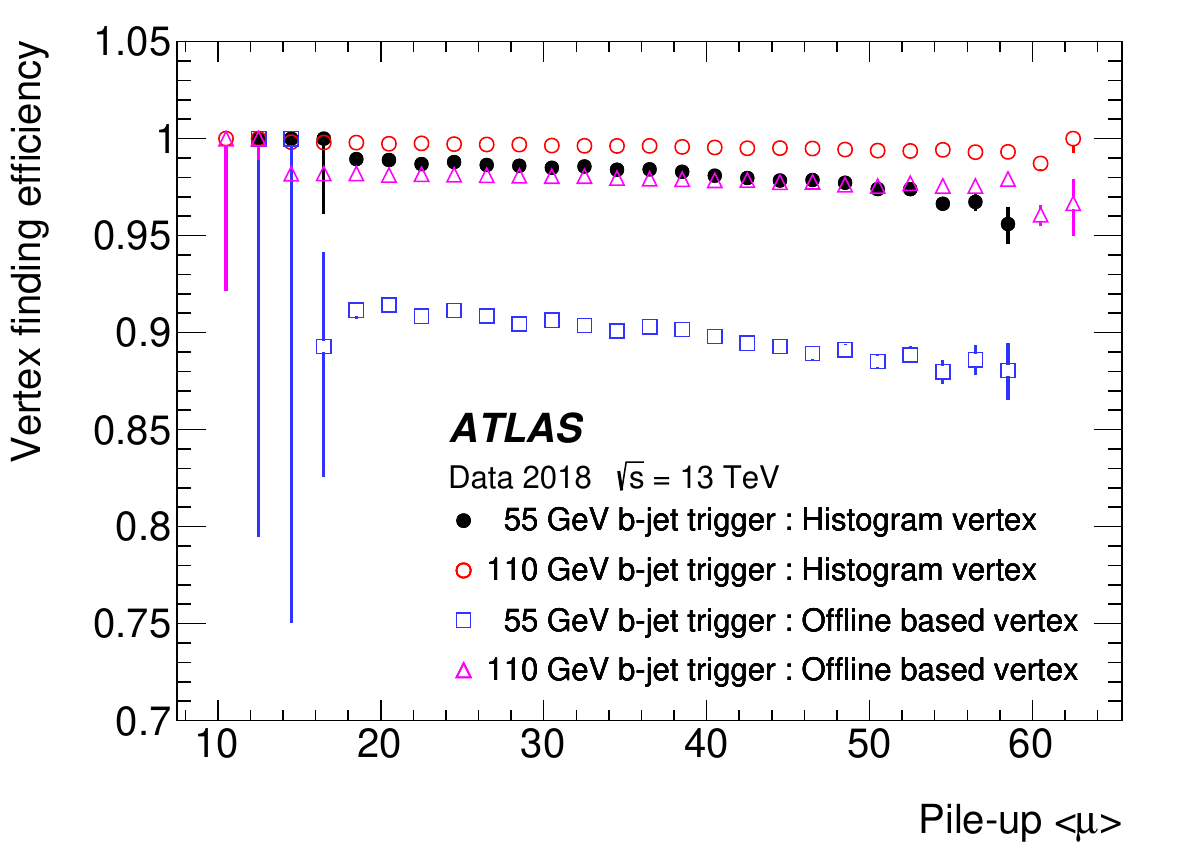}\label{fig:vtxeffc}}
\caption{The vertex algorithm efficiencies for low and intermediate \et \bjet triggers.
Shown are the efficiencies
(a) versus the offline track multiplicity for tracks in the \superroi;
(b) versus the $z_0$ position of the offline vertex; and
(c) versus the mean \pileup interaction multiplicity.
Statistical Bayesian uncertainties are shown, using a 68.3\% confidence interval
and uniform beta function prior.}
\label{fig:vtxeff}
\end{figure}
}

\newcommand{\vtxfigthree}{
\begin{figure*}[th]
\centering
\subfloat[]{\includegraphics[width=0.5\textwidth]{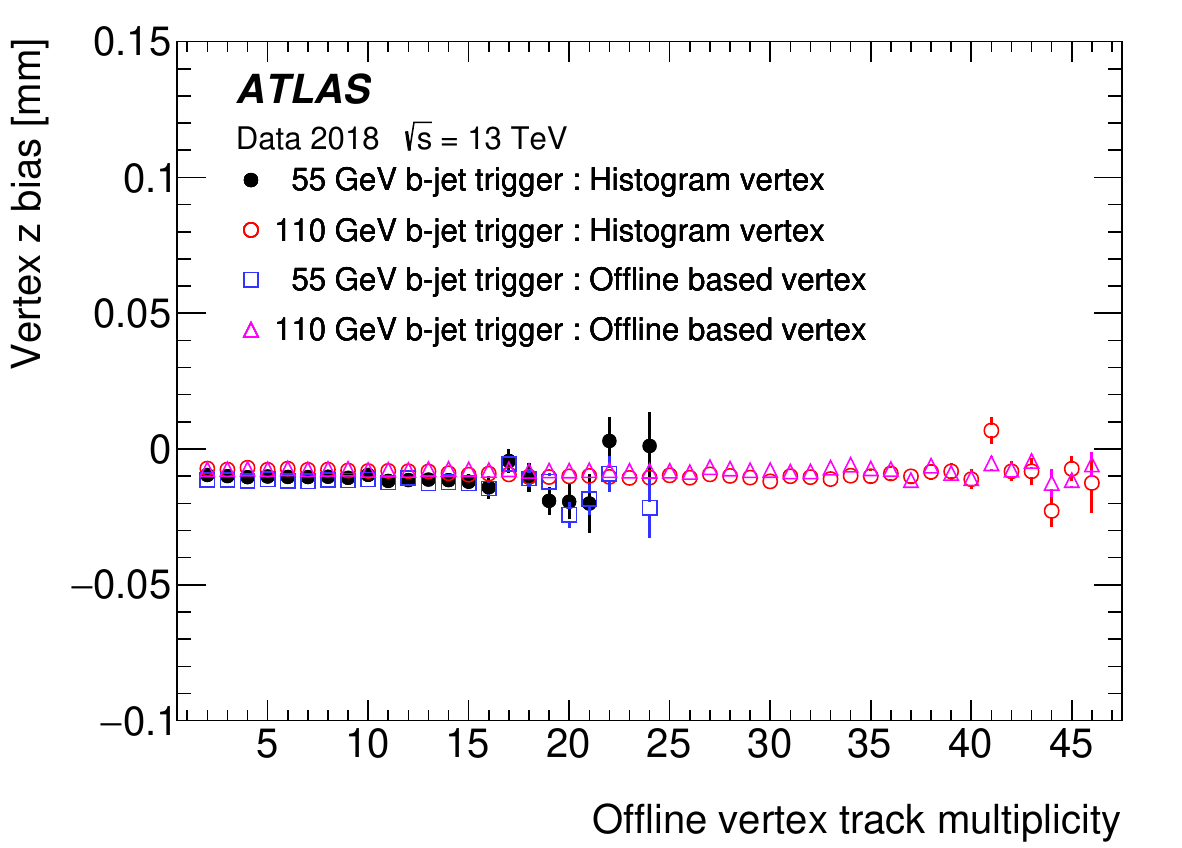}}
\subfloat[]{\includegraphics[width=0.5\textwidth]{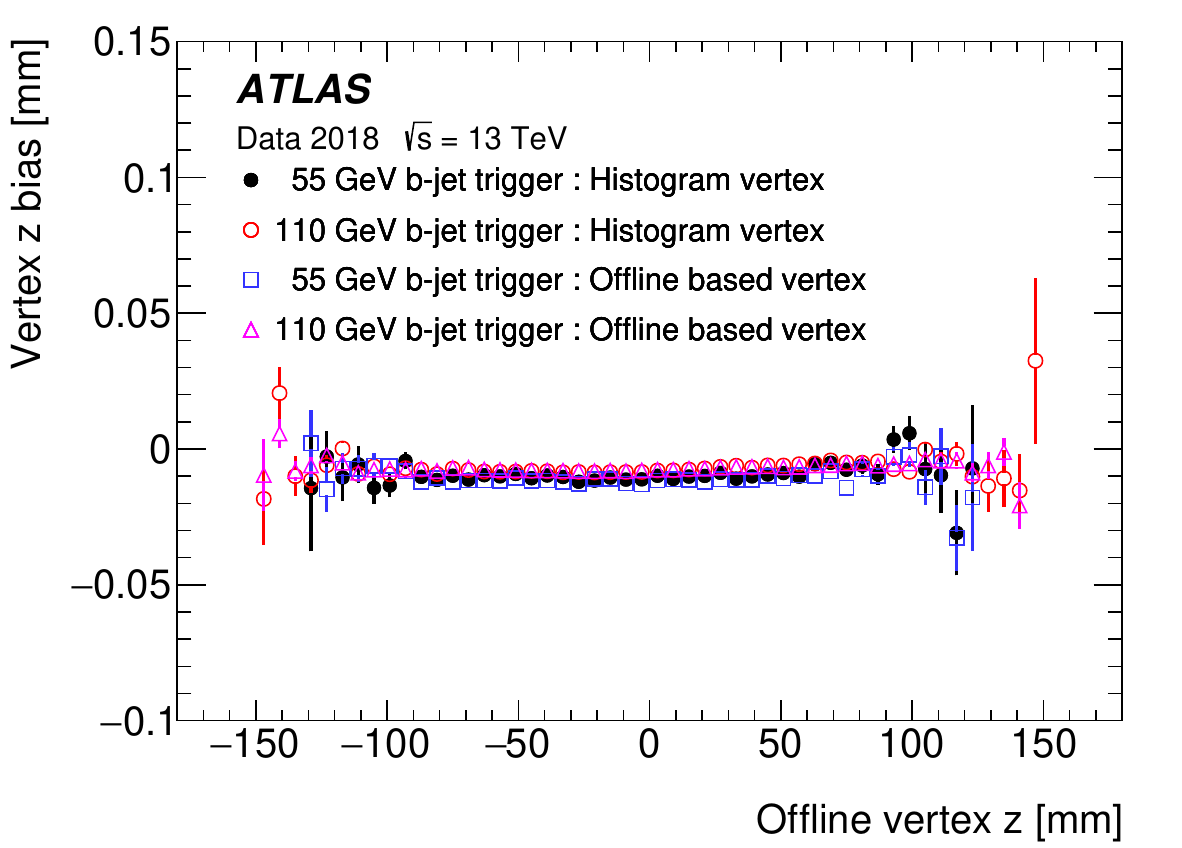}}\\
\subfloat[]{\includegraphics[width=0.5\textwidth]{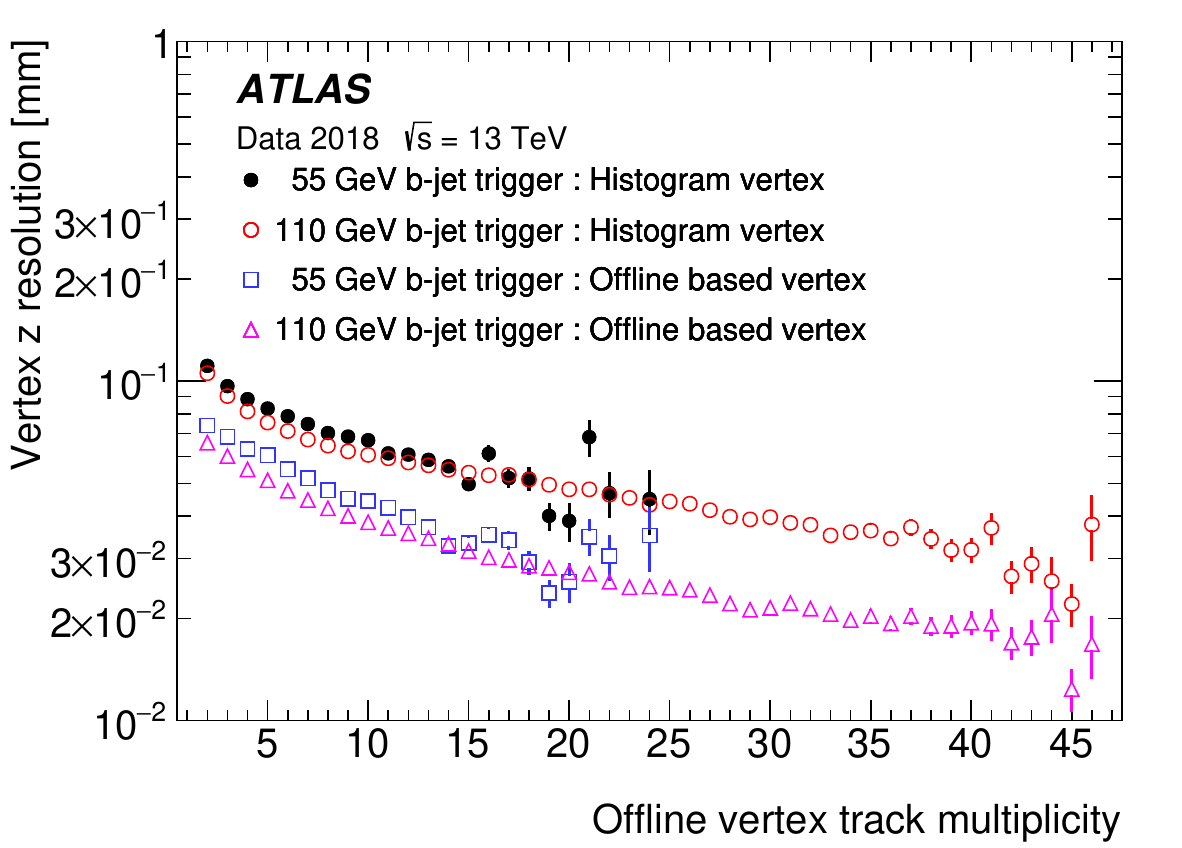}}
\subfloat[]{\includegraphics[width=0.5\textwidth]{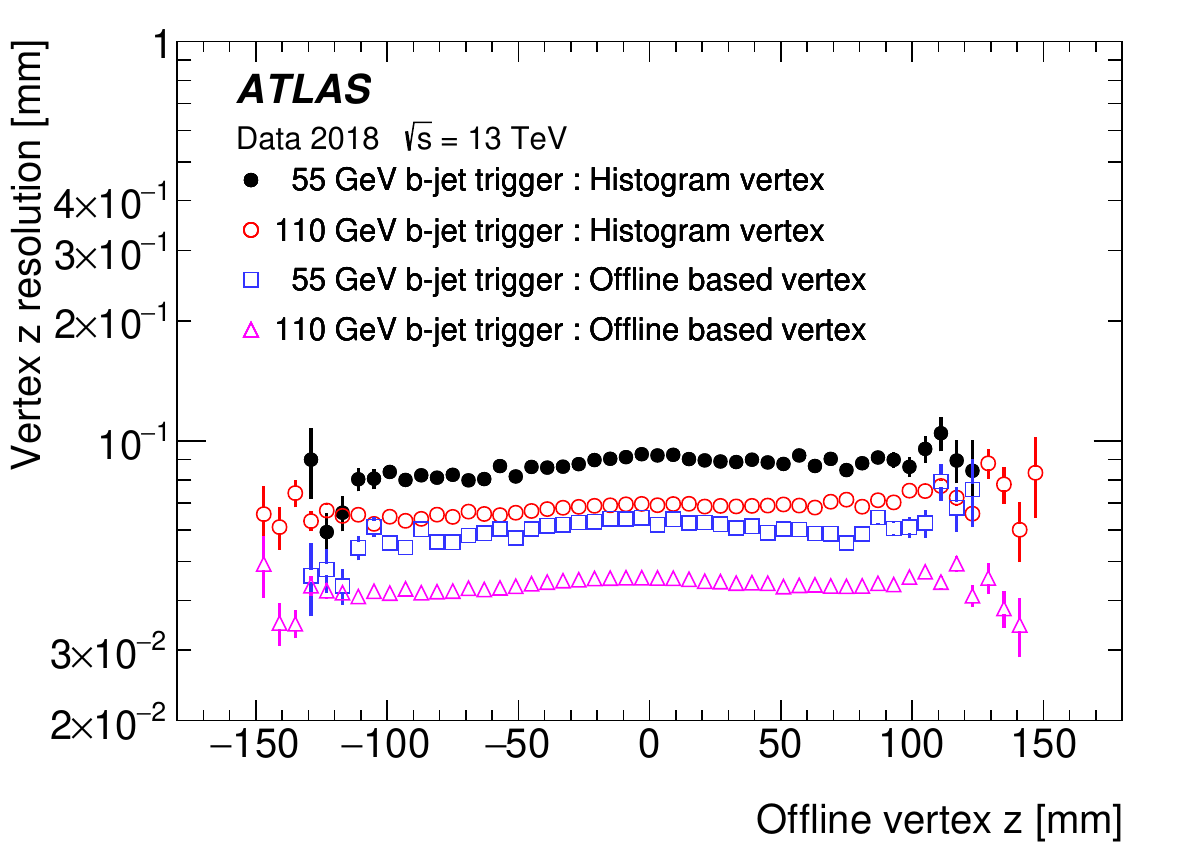}}
\caption{The trigger vertex algorithm bias and resolution in the $z$ position along the beamline,
for low and intermediate \et \bjet triggers.
Shown are the (a) bias, and (c) resolution
with respect to the multiplicity of offline
tracks within the \superroi which are matched to the vertex,
and the (b) bias, and (d) resolution,
with respect to the offline vertex $z$ position.  }
\label{fig:vtxres}
\end{figure*}
}

\newcommand{\vtxfigfour}{
\begin{figure*}[thp]
\centering
\subfloat[]{\includegraphics[width=0.49\textwidth]{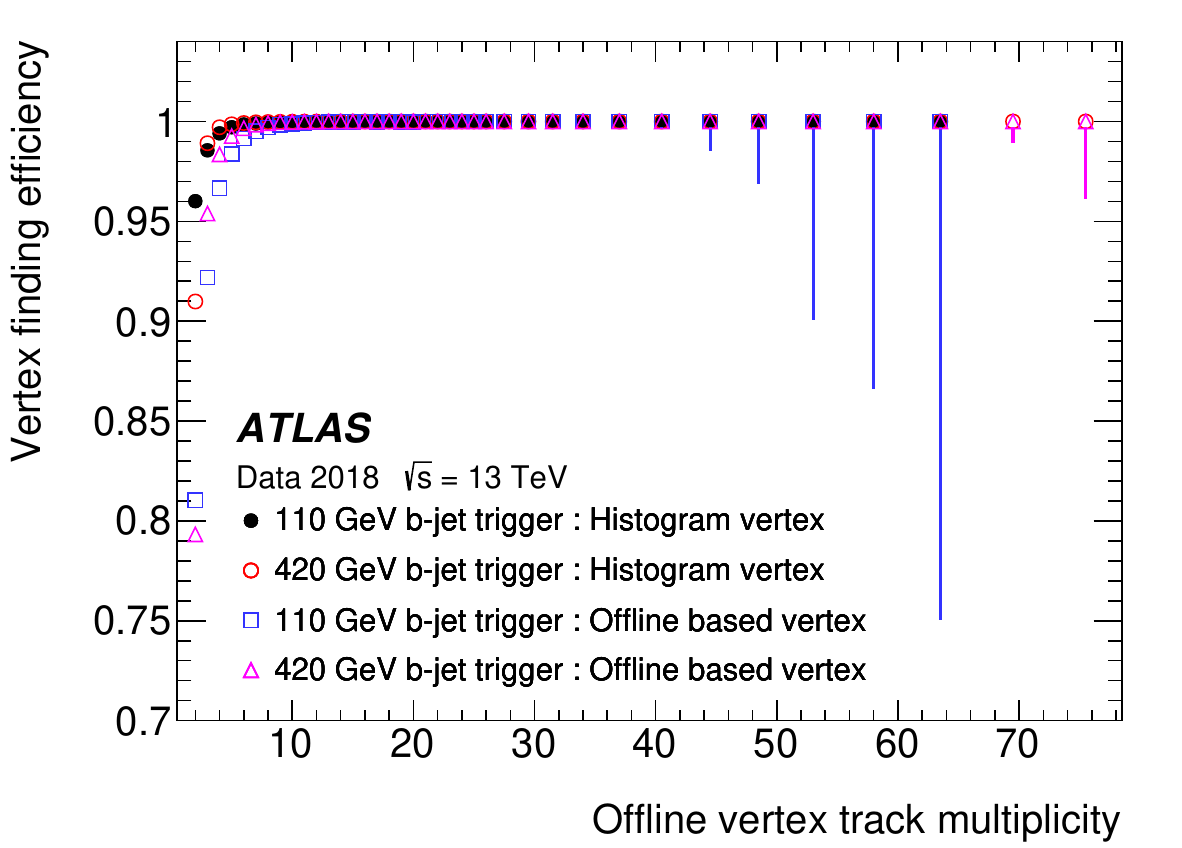}}
\subfloat[]{\includegraphics[width=0.49\textwidth]{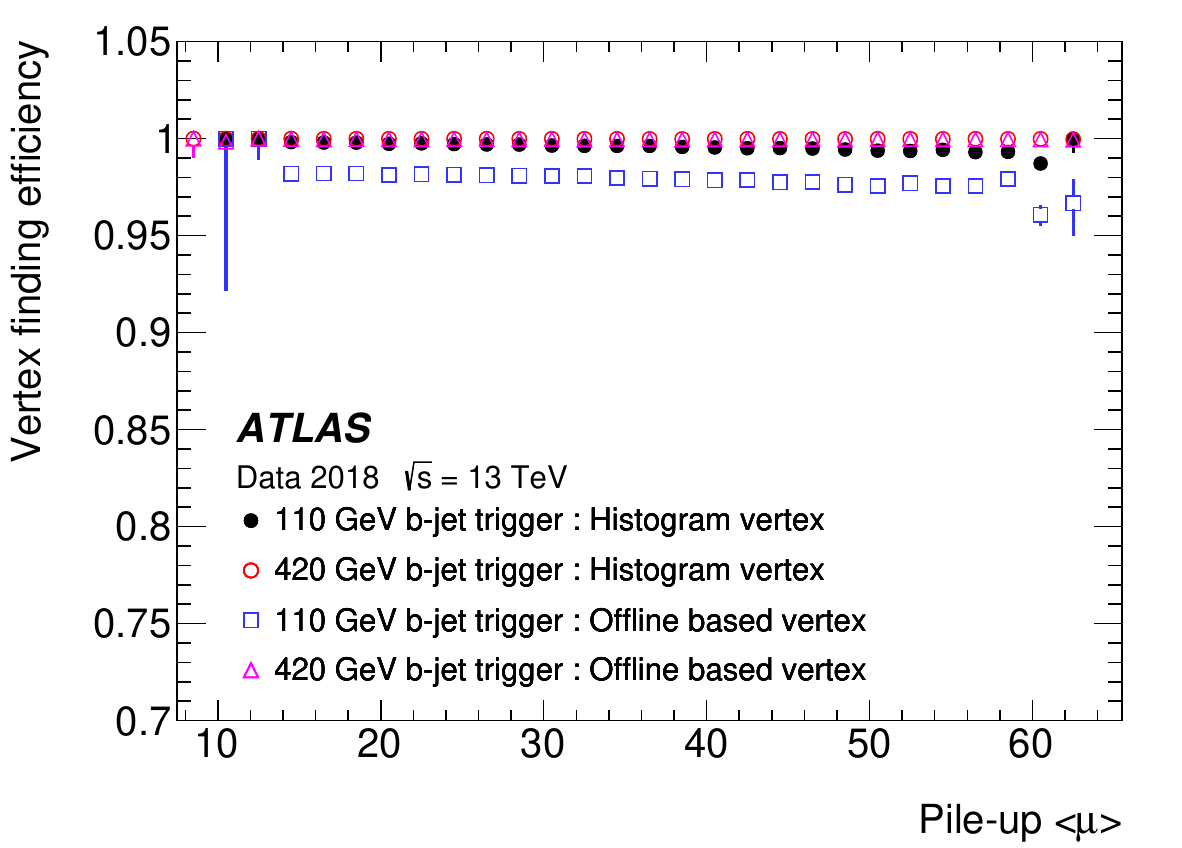}}
\caption{The vertex algorithm efficiencies for intermediate and high \et \bjet triggers.
Shown are the efficiencies
(a) versus the offline track multiplicity for tracks in the \superroi;
and (b) versus the mean \pileup interaction multiplicity.
Statistical Bayesian uncertainties are shown, using a 68.3\% confidence interval
and uniform beta function prior.}
\label{fig:vtxeffhi}
\end{figure*}
}

\newcommand{\vtxfigfive}{
\begin{figure*}[th]
\centering
\subfloat[]{\includegraphics[width=0.5\textwidth]{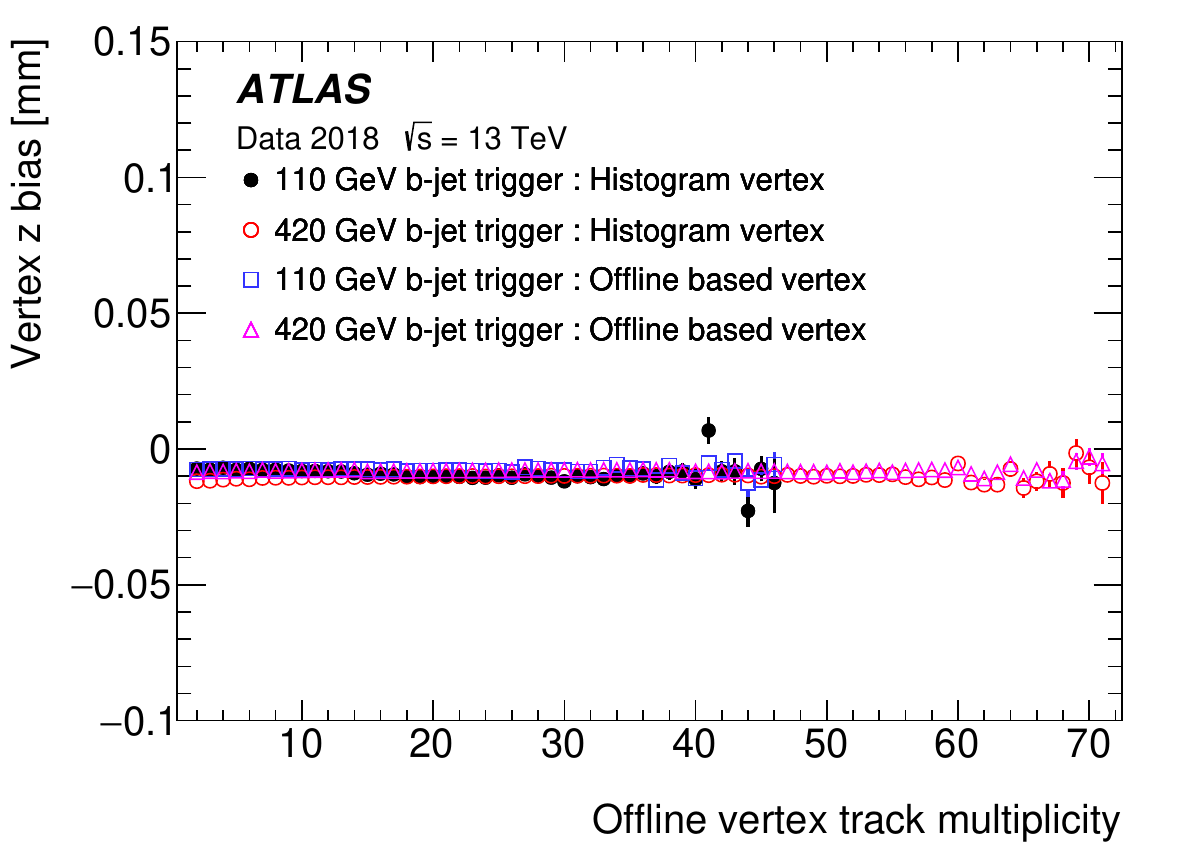}}
\subfloat[]{\includegraphics[width=0.5\textwidth]{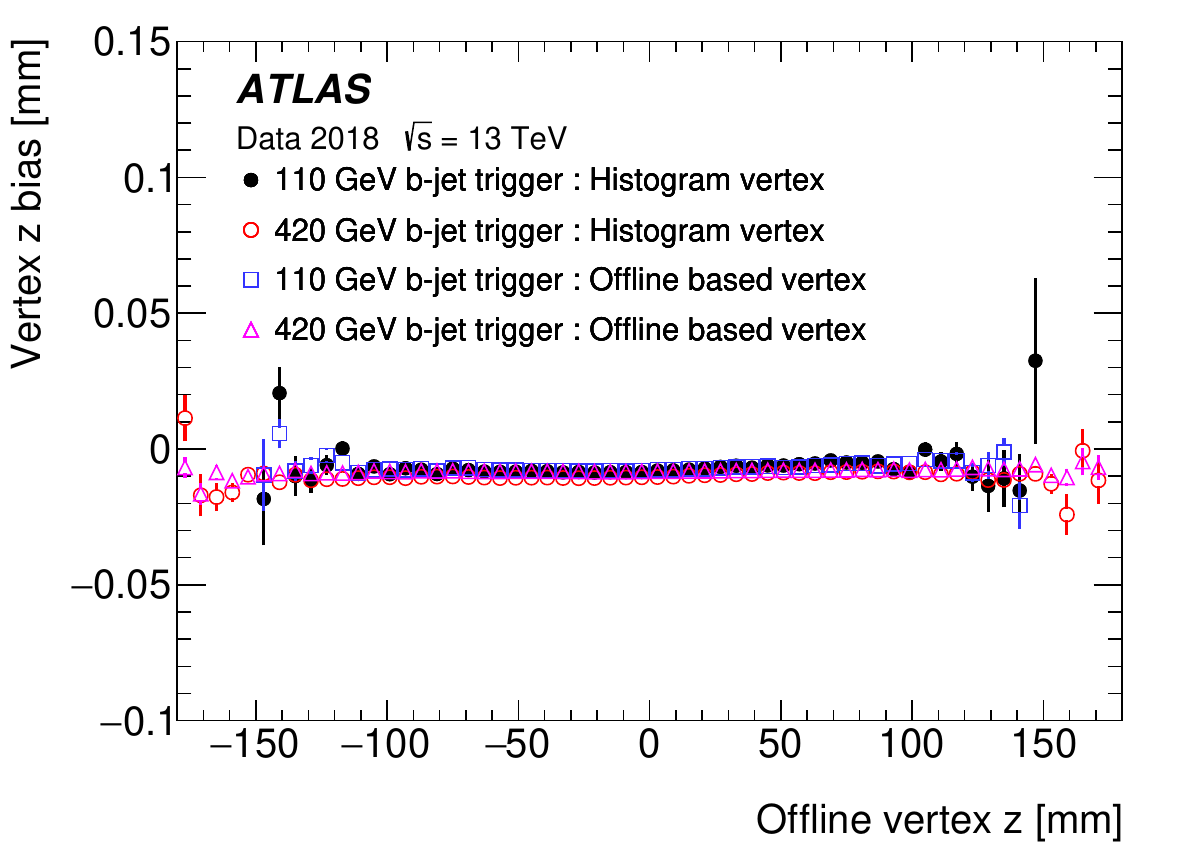}}\\
\subfloat[]{\includegraphics[width=0.5\textwidth]{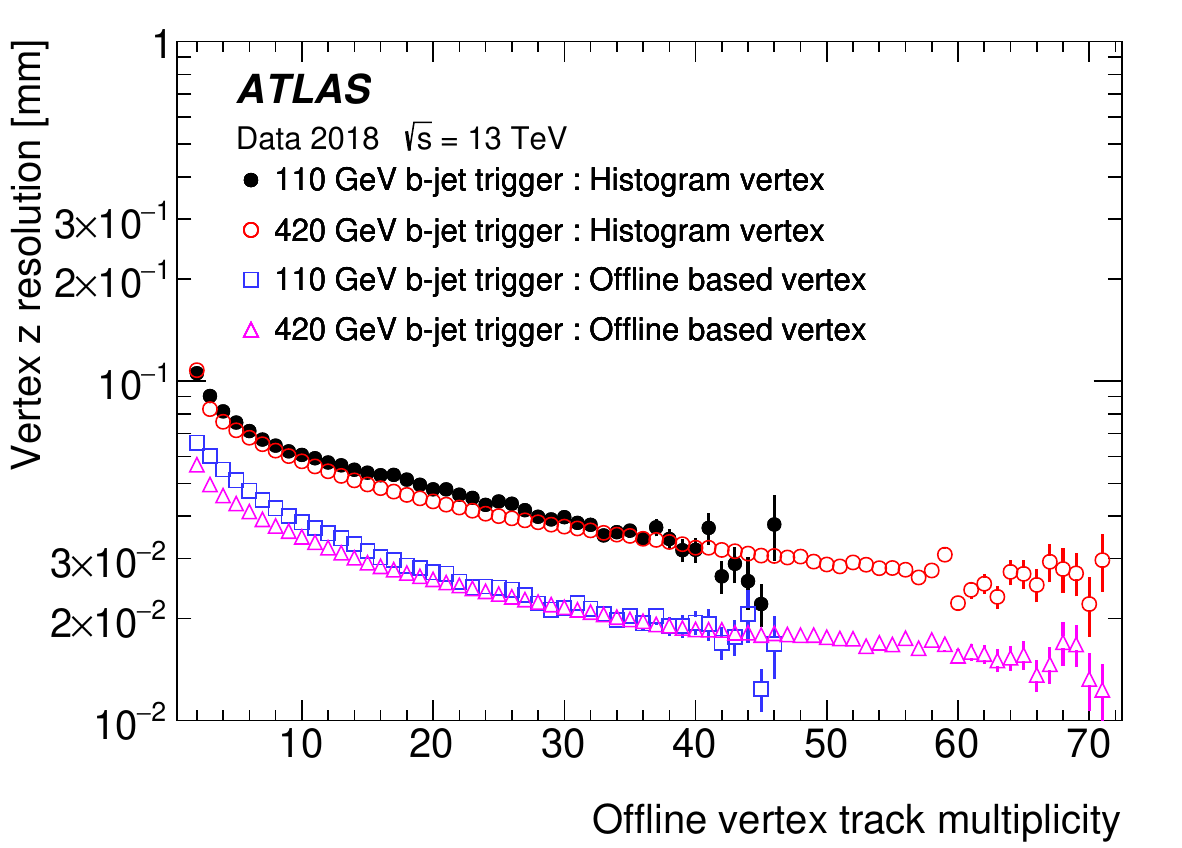}}
\subfloat[]{\includegraphics[width=0.5\textwidth]{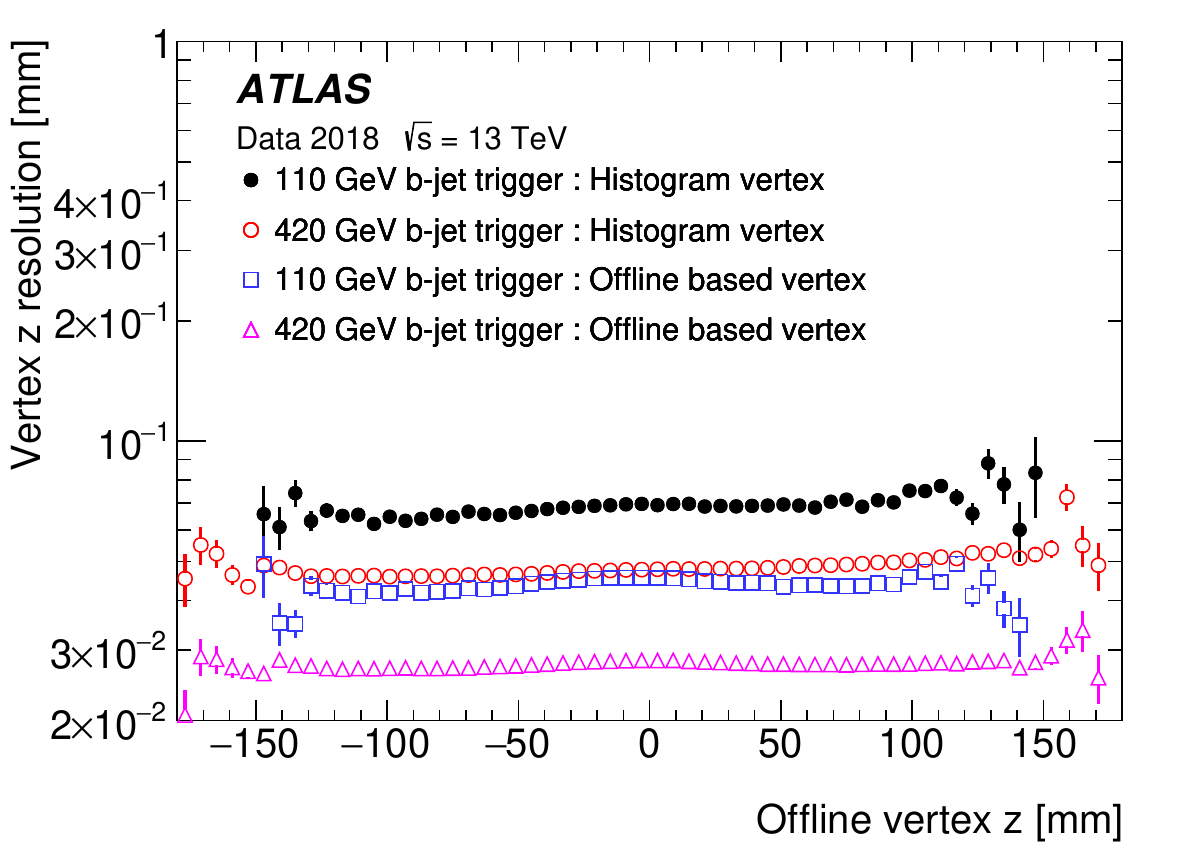}}
\caption{The bias and resolution in the $z$ position along the \beamline for the trigger vertex reconstruction
for intermediate and high \et \bjet triggers.
Shown are the bias (a), and resolution (c) with respect to the multiplicity of offline
tracks on the vertex matched to within the \superroi, and the bias (b), and resolution (d),
with respect to the offline vertex $z$ position.  }
\label{fig:vtxreshi}
\end{figure*}
}
 
\vtxfigone
 
During \runii, the \bjet vertex tracking stage ran with both vertex algorithms discussed in
Section~\ref{sec:vertexalg}.  For the online vertex selection, the more robust, but
less precise histogramming vertex algorithm was used to identify  vertex candidates
for those events where the offline based vertex  algorithm failed to find a vertex.
For the measurement of the performance, the online vertex efficiency is calculated for the single
offline vertex candidate
with the highest sum of the squared
transverse momenta from the bunch crossing.

Since the offline
vertex algorithm~\cite{Boutle_2017} runs over all offline tracks in the event, the multiplicity of
offline tracks associated with the vertex can be quite high. In contrast, as the trigger runs only
on those tracks in the jet \superroi, the multiplicity of tracks to be considered is much lower.
As the offline vertex was intended to be used in a high occupancy environment, the performance
in lower multiplicity conditions, may not be expected to be optimal.
 
In order to characterise the vertex in terms of track multiplicity, only those offline tracks associated
with the vertex and which lie within the \superroi are counted. It is possible that the primary vertex
identified by the offline reconstruction does not in fact have any tracks on the vertex that lie
within the \superroi pointing to the jets. It is observed that this possibility increases with the \pileup
multiplicity since the chance of a \pileup vertex with a high enough track multiplicity to be chosen as
the primary vertex in preference to the vertex from which the jets originate increases with the number
of separate interactions in the event.
Since, in this case, the offline algorithm used to select the offline primary vertex would in fact
have selected a \pileup vertex candidate rather than the hard interaction corresponding to the jets,
such events are excluded from the study.
 
As an additional selection, only those events in which there are at least two trigger jets in the \superroi
within the  tracking pseudorapidity acceptance ($|\eta|<2.5$) and with $\et>30$\gev are considered.
The \et requirement is a consequence of the \superroi, which is created only for trigger jets with \et above
30\gev.

Figure~\ref{fig:vtxdist} shows the distribution of the $z$ position of the offline primary vertex
candidates together with the multiplicity of offline tracks within the \superroi
with $\pt>5$\gev  for three \bjet triggers with \ET thresholds of 55\gev, 110\gev and 420\gev. The multiplicity is for
all selected offline tracks in the \superroi, not simply those associated with the offline primary vertex.
For the higher \et triggers,
the track multiplicities are significantly higher as expected.
Although not shown here, the mean \pt of the tracks within the jets is also harder for the higher
\et triggers. The distribution of the track $z$-positions is approximately Gaussian, and is largely
independent of the jet threshold, again as expected.

\vtxfigtwo
 
The efficiency for identifying the vertex candidates in the trigger is shown in Figure~\ref{fig:vtxeff}
for the 55\gev and 110\gev triggers,
as a function of the offline track multiplicity in the \superroi, the $z$ position of the offline vertex,
and the mean offline \pileup interaction multiplicity of the event. Figure~\ref{fig:vtxeffhi} compares
the efficiencies for the 110\gev and 420\gev triggers.
 
Shown in Figure~\ref{fig:vtxeffa}, the vertex finding efficiency rises steeply with track multiplicity
and reaches
full efficiency for the histogram based algorithm for events with more than four vertex tracks
within the \superroi. The offline based algorithm reaches full efficiency only for
events with more than eight vertex tracks within the \superroi. This is due to the offline based
algorithm having a tighter selection on track quality. 
Consequently, some vertices with only a few tracks in the \superroi may not have any tracks
remaining with which to form a vertex once the tracks with lower quality are removed.
 
\vtxfigfour
 
The efficiency as a function of the $z$ position, seen in Figure~\ref{fig:vtxeffb}, is approximately
constant for the offline based  vertex algorithm, whereas the simple histogramming algorithm shows a
clear bias towards negative $z$ for the lower \et jet trigger. This results from the very simple
implementation of the histogramming algorithm, which selects the bin with the highest
occupancy starting from the most negative $z$ bin. If it subsequently finds a bin with the
same occupancy, it keeps the original bin. In low \et events, the chance of some
additional \pileup interaction having the same track multiplicity in the \superroi can
be reasonably high and consequently, if such a vertex is at a more negative $z$ than the interaction
of interest then the incorrect vertex can be reconstructed online, leading to a lower efficiency
for vertices with a more positive-$z$ position.

For the higher \et jet triggers shown in Figures~\ref{fig:vtxeff} and ~\ref{fig:vtxeffhi}, the
larger jet \et  results in significantly higher track multiplicities and larger average track
\pt such that the probability of the track multiplicity
for a \pileup vertex matching or exceeding that for the interaction of interest is significantly lower.
 
It should be noted that the overall lower efficiency for the online based vertex algorithm,
and particularly for the lower \et jet trigger is due to the lower overall track multiplicity in
the events passing the trigger,  such that the trigger is still on the rising edge of the efficiency
distribution for many events.
For the high \et triggers where the multiplicity is higher, the algorithm is further along the
rising edge for most events. For very high \et jet triggers, where the \superroi track multiplicity
is much higher, the efficiency approaches 100\%. For the histogramming algorithm, the
efficiency rises earlier.

Both algorithms show a decrease
in efficiency as \mubar\ increases, seen in Figure~\ref{fig:vtxeffc}, with
the offline based algorithm having a lower overall
efficiency due to the track multiplicity of the vertex discussed above.
The decrease with increasing \mubar\ follows from the increased possibility for the offline
selection to misidentify the jet vertex candidate within the \superroi.
In this case, as the \pileup interaction multiplicity
increases within an event, so do the chances of there being additional tracks from additional
vertices within the \superroi that compete with the tracks from the desired primary vertex.
For higher jet \et thresholds, the track multiplicities of the desired
vertex candidates is also higher, and the fall in efficiency with increasing \pileup is
less pronounced. For triggers with higher still thresholds, the efficiency approaches 100\% for both algorithms,
and exhibits no residual loss as a function of \pileup. This can be seen in Figure~\ref{fig:vtxeffhi},
which compares the efficiency versus \pileup \mubar\ for the 420\gev jet trigger in addition.
 
\vtxfigthree
 
\vtxfigfive

Figures~\ref{fig:vtxres} and~\ref{fig:vtxreshi} illustrate the vertex $z$ bias
and resolution for both online algorithms, where the bias is calculated as the mean of the residuals between the
trigger and offline $z$ positions.
Both the bias and resolution are shown as a function of either the offline track multiplicity or
the offline $z$ vertex position, in
Figure~\ref{fig:vtxres} for the low and intermediate \et jet triggers, and in Figure~\ref{fig:vtxreshi}
for the intermediate and high \et jet triggers.
In all cases the bias is reasonably constant and is seen to be approximately
10~$\mu$m in the negative direction. This arises from a small systematic
bias in the track $z$ position reconstruction. This bias is itself significantly
smaller than the vertex $z$ resolution, even at high track multiplicities, and has
no discernible impact on the subsequent trigger reconstruction or selection.
For vertices with more positive $z$, this bias appears to decrease by a few $\mu$m
for the histogramming algorithm, and less for offline based algorithm.
The resolution of the vertex $z$ position for both online algorithms is
seen to clearly improve
with increasing track multiplicity, with the offline based algorithm showing significantly
better resolution. For low multiplicity vertices the histogramming algorithm resolution
is approximately 100~$\mu$m whereas for the offline based algorithm it is 70~$\mu$m.
For both algorithms the resolution improves logarithmically with track multiplicity,
such that for 50 tracks the resolutions for the histogramming and offline based
algorithms are approximately 30~$\mu$m and 20~$\mu$m respectively.
 
For  Figure~\ref{fig:vtxreshi}
it should be noted that for the 420\gev high \et jet trigger, even for low track multiplicities,
the resolution is somewhat better than for the low and intermediate \et triggers,  being
approximately 10~$\mu$m better than for the 110\gev trigger, and around 20~$\mu$m better
than for the 55\gev trigger. This is a consequence of the larger average \pt of tracks from the
higher \et jet triggers, such that the track $z$ positions themselves have an intrinsically
better $z$ resolution.
 
The resolution is largely constant with the offline vertex $z$ position with perhaps a
small deterioration of the resolution around $z=0$~mm, and a slight trend towards better
resolutions for more negative $z$ for the lower \et triggers, notably for the
histogramming algorithm.

\floatbarra



\FloatBarrier

 
\section{Conclusion}
 
The ATLAS inner detector trigger was modified significantly
in order to operate with the extremely demanding conditions  in \runii.
The many improvements in the inner detector trigger are described, and the performance
is reported here for the 2016, 2017, and 2018  data recorded by ATLAS
from 13~\tev proton--proton collisions with  25~ns bunch spacing at the LHC.
The performance has been
excellent and exceeded expectations, even at the very high interaction multiplicities
observed at the end of data taking in 2018.
The detailed efficiencies and resolutions of the trigger in a wide range of physics
signatures are presented for the full \runii high-luminosity data set, showing the continued
high efficiency, and in particular the robustness in the face of increasing
\pileup interaction multiplicities.
The excellent performance
of the inner detector trigger algorithms
demonstrates how the inner detector trigger continues to lie at the
heart of the trigger performance, playing an essential  r\^{o}le
in the ATLAS physics programme, \mbox{and  will continue to do so in the future.}


\section*{Acknowledgements}


The authors thank CERN for the very successful operation of the LHC, as well as the
support staff from the institutions without whom ATLAS could not be
operated efficiently.
 
The authors also gratefully acknowledge the support of
ANPCyT, Argentina;
YerPhI, Armenia;
ARC, Australia;
BMWFW and FWF, Austria;
ANAS, Azerbaijan;
SSTC, Belarus;
CNPq and FAPESP, Brazil;
NSERC, NRC and CFI, Canada;
CERN;
ANID, Chile;
CAS, MOST and NSFC, China;
Minciencias, Colombia;
MSMT CR, MPO CR and VSC CR, Czech Republic;
DNRF and DNSRC, Denmark;
IN2P3-CNRS and CEA-DRF/IRFU, France;
SRNSFG, Georgia;
BMBF, HGF and MPG, Germany;
GSRI, Greece;
RGC and Hong Kong SAR, China;
ISF and Benoziyo Center, Israel;
INFN, Italy;
MEXT and JSPS, Japan;
CNRST, Morocco;
NWO, Netherlands;
RCN, Norway;
MEiN, Poland;
FCT, Portugal;
MNE/IFA, Romania;
JINR;
MES of Russia and NRC KI, Russian Federation;
MESTD, Serbia;
MSSR, Slovakia;
ARRS and MIZ\v{S}, Slovenia;
DSI/NRF, South Africa;
MICINN, Spain;
SRC and Wallenberg Foundation, Sweden;
SERI, SNSF and Cantons of Bern and Geneva, Switzerland;
MOST, Taiwan;
TAEK, Turkey;
STFC, United Kingdom;
DOE and NSF, United States of America.
In addition, individual groups and members have received support from
BCKDF, CANARIE, Compute Canada and CRC, Canada;
COST, ERC, ERDF, Horizon 2020 and Marie Sk{\l}odowska-Curie Actions, European Union;
Investissements d'Avenir Labex, Investissements d'Avenir Idex and ANR, France;
DFG and AvH Foundation, Germany;
Herakleitos, Thales and Aristeia programmes co-financed by EU-ESF and the Greek NSRF, Greece;
BSF-NSF and GIF, Israel;
Norwegian Financial Mechanism 2014-2021, Norway;
NCN and NAWA, Poland;
La Caixa Banking Foundation, CERCA Programme Generalitat de Catalunya and PROMETEO and GenT Programmes Generalitat Valenciana, Spain;
G\"{o}ran Gustafssons Stiftelse, Sweden;
The Royal Society and Leverhulme Trust, United Kingdom.
 
The crucial computing support from all WLCG partners is acknowledged gratefully, in particular from CERN, the ATLAS Tier-1 facilities at TRIUMF (Canada), NDGF (Denmark, Norway, Sweden), CC-IN2P3 (France), KIT/GridKA (Germany), INFN-CNAF (Italy), NL-T1 (Netherlands), PIC (Spain), ASGC (Taiwan), RAL (UK) and BNL (USA), the Tier-2 facilities worldwide and large non-WLCG resource providers. Major contributors of computing resources are listed elsewhere~\cite{ATL-SOFT-PUB-2021-003}.


\printbibliography

\clearpage 
 
\begin{flushleft}
\hypersetup{urlcolor=black}
{\Large The ATLAS Collaboration}

\bigskip

\AtlasOrcid[0000-0002-6665-4934]{G.~Aad}$^\textrm{\scriptsize 101}$,    
\AtlasOrcid[0000-0002-5888-2734]{B.~Abbott}$^\textrm{\scriptsize 128}$,    
\AtlasOrcid[0000-0002-7248-3203]{D.C.~Abbott}$^\textrm{\scriptsize 102}$,    
\AtlasOrcid[0000-0002-2788-3822]{A.~Abed~Abud}$^\textrm{\scriptsize 36}$,    
\AtlasOrcid[0000-0002-1002-1652]{K.~Abeling}$^\textrm{\scriptsize 53}$,    
\AtlasOrcid[0000-0002-2987-4006]{D.K.~Abhayasinghe}$^\textrm{\scriptsize 93}$,    
\AtlasOrcid[0000-0002-8496-9294]{S.H.~Abidi}$^\textrm{\scriptsize 29}$,    
\AtlasOrcid[0000-0002-8279-9324]{O.S.~AbouZeid}$^\textrm{\scriptsize 40}$,    
\AtlasOrcid{N.L.~Abraham}$^\textrm{\scriptsize 156}$,    
\AtlasOrcid[0000-0001-5329-6640]{H.~Abramowicz}$^\textrm{\scriptsize 161}$
% ...
\bigskip
\\

% ...
$^{29}$Physics Department, Brookhaven National Laboratory, Upton NY; United States of America.\\
% ...
$^{36}$CERN, Geneva; Switzerland.\\
% ...
$^{40}$Niels Bohr Institute, University of Copenhagen, Copenhagen; Denmark.\\
% ...
$^{53}$II. Physikalisches Institut, Georg-August-Universit\"{a}t G\"ottingen, G\"ottingen; Germany.\\
% ...
$^{93}$Department of Physics, Royal Holloway University of London, Egham; United Kingdom.\\
% ...
$^{101}$CPPM, Aix-Marseille Universit\'e, CNRS/IN2P3, Marseille; France.\\
$^{102}$Department of Physics, University of Massachusetts, Amherst MA; United States of America.\\
% ...
$^{128}$Homer L. Dodge Department of Physics and Astronomy, University of Oklahoma, Norman OK; United States of America.\\
% ...
$^{156}$Department of Physics and Astronomy, University of Sussex, Brighton; United Kingdom.\\
% ...
$^{161}$Raymond and Beverly Sackler School of Physics and Astronomy, Tel Aviv University, Tel Aviv; Israel.\\
% ...

\end{flushleft}


\end{document}